\documentclass[fleqn]{aa}

\usepackage[normalem]{ulem}
\usepackage{lastpage}
\usepackage{widetext}
\usepackage{graphicx}
\usepackage{dcolumn}
\usepackage{bm}
\usepackage{multirow}
\usepackage{url}
\usepackage{color}
\usepackage{xcolor}
\usepackage{float}
\usepackage{cancel}
\usepackage{amssymb}
\usepackage{amsmath}
\usepackage{comment}
\usepackage{txfonts}
\usepackage[T1]{fontenc}
\usepackage{ae,aecompl}
\usepackage{orcidlink}
\usepackage{academicons}
\usepackage{hyperref}
\usepackage{xcolor}
\usepackage{graphicx}
\usepackage{epstopdf}
\usepackage{amssymb}
\usepackage{latexsym}
\usepackage{tikz-cd}
\usepackage{twoopt}

\newcommand{\be}{\begin{equation}}
\newcommand{\dd}{{\rm d}}
\newcommand{\lp}{\left(}
\newcommand{\rp}{\right)}

\newcommand{\vr}{\textbf{r}}

\newcommand{\ee}{\end{equation}}

\newcommand{\vk}{\textbf{k}}

\begin{document}

%%%%%%%%%%%%%%%%%%%%%%%%%%%%%%%%%%%%%%%%%%%%%%%%%%
\title{Secondary halo bias through cosmic time I: Scaling relations and the connection with the cosmic web}
\titlerunning{Balaguera-Antolínez, Montero-Dorta $\&$ Favole}
\author{
Andrés Balaguera-Antolínez $^{1,2}$\thanks{balaguera@iac.es}\orcidlink{0000-0001-5028-3035},
Antonio D. Montero-Dorta$^{3}$ and Ginevra Favole$^{1,2}$}
\institute{
 Instituto de Astrof\'{\i}sica de Canarias, s/n, E-38205, La Laguna, Tenerife, Spain \and
 Departamento de Astrof\'{\i}sica, Universidad de La Laguna, E-38206, La Laguna, Tenerife, Spain \and
 Departamento de Física, Universidad Técnica Federico Santa María, Casilla 110-V, Avda. España 1680, Valparaíso, Chile
  }
\authorrunning{Author}
% 0000000000000000000000000000000000000000000000000000000000000000000000000000000000
% 0000000000000000000000000000000000000000000000000000000000000000000000000000000000
% 0000000000000000000000000000000000000000000000000000000000000000000000000000000000

\date{Received /Accepted}

\abstract 
{The spatial distribution of dark matter halos carries cosmological and astrophysical information. Cosmological information can be considered to be contained in the connection between halo main properties and the large-scale halo bias, while the astrophysical information would be encoded in the scaling relations between halo properties. The combination of these two contributions leads to the effect of secondary halo bias.
}
{Our goal is to measure the signal of secondary halo bias as a function of a variety of intrinsic and environmental halo properties and to characterize its statistical significance as a function of cosmological redshift.}
{Using fixed and paired $N$-body simulations of dark-matter halos -- the \texttt{UNIT} simulation -- with masses above $\sim 10^{11}M_{\odot}h^{-1}$ identified over a wide range of cosmological redshifts ($0<z<5$), we explored the behavior of the scaling relations among different halo properties. We included novel environmental properties based on the halo distribution as well as the underlying dark-matter field. We implemented an object-by-object estimator of large-scale effective bias and tested its validity against standard approaches. With a bias assigned to each tracer, we performed a statistical analysis aimed at characterizing the distribution of the bias and the signal of the secondary halo bias.}
{We show how the halo scaling relations linking direct probes of the halo potential well do not depend on the environment. On the contrary, links between the halo mass and the so-called set of secondary halo properties are sensitive to the cosmological environment, mainly to under-dense regions. We show that the signal of secondary bias is derived statistically from secondary correlations beyond the standard link to the halo mass.}{We show that the secondary bias arises through nonlocal and/or environmental properties related either to the halo distribution or to the properties of the underlying dark-matter field. In particular, properties such as the tidal field (a measure of the anisotropy of the density field) and the local Mach number (a measure of the local kinetic temperature of the halo distribution) generate the signals of the secondary bias with the highest significance. We propose applications of the assignment of individual bias for the generation of mock catalogs containing the signal of secondary bias, as well as a series of cosmological analyses aimed at mining large galaxy datasets.}


\maketitle
%=============================================================================================================
%=============================================================================================================
%=============================================================================================================
\section{Introduction}

 It is well known that the clustering of dark matter halos strongly depends on their internal properties. In this context, the halo mass is usually considered to be responsible for one of the primary dependencies, as a direct manifestation of the more fundamental dependence of the halo bias on the peak height of density fluctuations, $\nu$ \citep[e.g.,][]{Press1974,1984ApJ...284L...9K,1986ApJ...304...15B, ShethTormen1999,Sheth2001,ShethTormen2002}. More massive halos are more strongly clustered than less massive halos, as is indeed described by the $\Lambda$-cold dark matter ($\Lambda$-CDM) structure formation formalism. However, a number of additional secondary dependencies at a fixed halo mass have been discovered mostly using cosmological simulations \citep[see e.g.,][]{Sheth2004,gao2005,Wechsler2006,Gao2007,2007MNRAS.377L..15P,2008ApJ...687...12D,Angulo2008,2008Li,Faltenbacher2010, 2017JCAP...03..059L,2018Salcedo,2019MNRAS.482.1900H,2018MNRAS.474.5143M,SatoPolito2019,Johnson2019, 2019MNRAS.489.2977R,MonteroDorta2020B,Tucci2021,MonteroDorta2021_mah}. As an example, lower mass halos that assemble a significant portion of their mass early on are more strongly clustered than those that form at later times, an effect named {{halo assembly bias}}. These types of secondary dependencies have been measured for a number of internal properties, including halo concentration, spin, shape, anisotropic velocity dispersion, and a variety of definitions of the halo age and proxies of the assembly history \citep[see e.g.,][just to name a few]{Gao2007,2008ApJ...687...12D,2008ApJ...687...12D,Faltenbacher2010,SatoPolito2019,MonteroDorta2021_mah}. 

Throughout this work, we refer to the complete set of secondary dependencies of the halo bias at a fixed halo mass as a {{secondary halo bias}}. These also include environmental properties and parameters characterizing the tidal field around halos, which are known to display such signals\footnote{Note that it is common to refer to all secondary dependencies as halo assembly bias, instead of just the dependence on the assembly history. We avoid this nomenclature since it is not established that all secondary dependencies of halo bias originate from the same set of physical mechanisms.} \citep[see e.g.,][]{2018MNRAS.476.3631P}. Despite the abundance of measurements, the physical origins of secondary halo bias are still not completely characterized, although several results point precisely toward tidal fields as the main drivers. \citet[][]{Dalal2008} introduced the notion that low-mass halo assembly bias, as opposed to its high-mass counterpart, could be due to a subpopulation of halos whose accretion was halted early on. \citet[][]{2009MNRAS.398.1742H} subsequently claimed that tidal effects produced by a neighboring massive halo could be the dominant driver for this suppressed growth. These results are align well with the works of \citet[][]{Borzyszkowski2017} and \citet[][]{Musso2018}. Based on zoom-in simulations of a small number of halos, \cite{Borzyszkowski2017} concluded that low-mass assembly bias is due with the existence of same-mass ``stalled" and ``accreting" halos, typically associated with filaments and nodes, respectively. \citet[][]{Musso2018} extended this analysis to take the whole geometry of tides into account. It has also been shown that the secondary bias signals (i.e., not only assembly bias) correlate with the anisotropy of the tidal tensor at a fixed halo mass \citep[see e.g.,][]{2018MNRAS.476.3631P, 2019MNRAS.489.2977R}. Importantly, the exact relations between these mechanisms and the different halo properties are still unclear. 

The measurement of halo bias (primary or secondary) is usually performed on the basis of the two-point correlation function, either in configuration or in Fourier space \citep[see e.g.,][]{1999ApJ...520..437K,2012MNRAS.420.3469P,MonteroDorta2021_mah}. For this paper, however, we implemented an object-by-object estimator of the halo bias \citep[][]{2018MNRAS.476.3631P}, which allows for a robust statistical characterization of the halo bias and its dependencies on a number halo properties. This bias estimation was employed to explore, for the first time, the signal of the secondary bias in the \texttt{UNIT} simulation
\citep[][]{2019MNRAS.487...48C}, which followed the evolution of a cosmological volume of $1($Gpc$/h)^{3}$ with outputs from redshift $z\sim 5$. We dissected the secondary bias by analyzing a wide range of both intrinsic halo and environmental properties.
The assessment of the secondary bias performed in this work includes novel halo properties such as the {relative Mach number} and the {neighbor statistics}, as well as the evaluation of properties that are inherent to the dark matter field, extending recent analysis on the same subject \citep[][]{2021A&A...654A..67W}. Our analysis is not only relevant in the context of secondary bias. We also performed several numerical tests to assess the strength of the dependencies of the halo scaling relations on secondary properties and their impact on the signal of the halo primary and secondary bias.

This work is the first of a series of papers where we intend to shed light on several key aspects of secondary bias, by taking advantage of the singular cosmological characteristics of the {{paired-fixed amplitude}} \texttt{UNIT} simulation, and the novel angle that the object-by-object bias estimate brings. The outline of this paper is the following. In \S~\ref{sec:sim} we introduce the \texttt{UNIT} simulation and describe the different halo properties employed in the analysis. In particular, we characterize the correlation among the different halo properties and the impact of the environment on their scaling relations. In \S\ref{sec:bias} we show the measurements of the secondary bias based on an object-by-object estimator and show the scaling relations between the large-scale bias and the halo properties. Section~\ref{sec:ori} presents a methodology for the reconstruction of a secondary bias and potential applications for the generation of halo mock catalogs. Finally, in \S\ref{sec:conclusions} we discuss our results, present the main conclusions of this research, and comment on forthcoming research in line with the present work.

%=============================================================================================================
%=============================================================================================================
%=============================================================================================================
\section{The UNIT simulation}\label{sec:sim}
The \texttt{UNIT} suite of simulations (UNITSim hereafter) \footnote{\url{http://www.UNITSims.org/}} \citep[][]{2019MNRAS.487...48C} is a set of $N$-body simulations of dark matter particles in cosmological volumes evolved from  $z=99$ until $z=0$ using the code \texttt{Gadget} \citep[][]{SPRINGEL200179}. Initial conditions are generated using the \texttt{FastPM} algorithm \citep[][]{2016MNRAS.463.2273F} imposing a linear matter power spectrum in a $\Lambda$CDM Universe with cosmological parameter taken from \cite[][]{2016A&A...594A..13P}. The UNITSim  provides three different volumes. We use a cosmological volume of $1(\mathrm{Gpc}\,h^{-1})^{3}$, run with $4096^{3}$ dark matter particles with mass resolution of $1.2\times 10^{9}M_{\odot}h^{-1}$. The dark matter field is represented by a mesh of $2048^{3}$ cells, which we have reduced to $512^{3}$ by averaging in configuration space \footnote{Applying a low pass filter can give rise to the so-called Gibbs ringing, embodied in the presence of negative densities in the low-resolution field; we avoid this path as we explicitly use the density field in configuration space: for applications involving two point statistics, filtering the high frequencies in Fourier space would be a valid approach.}, yielding a spatial resolution of $\sim 1.95$Mpc$h^{-1}$. We shall refer to this as our fiducial resolution.

%------------------------------------------------------------------------
\begin{figure*}
%dens_fields_UNITSim.py
\includegraphics[trim = .3cm 0.5cm 0cm 0cm ,clip=true, width=0.5\textwidth]{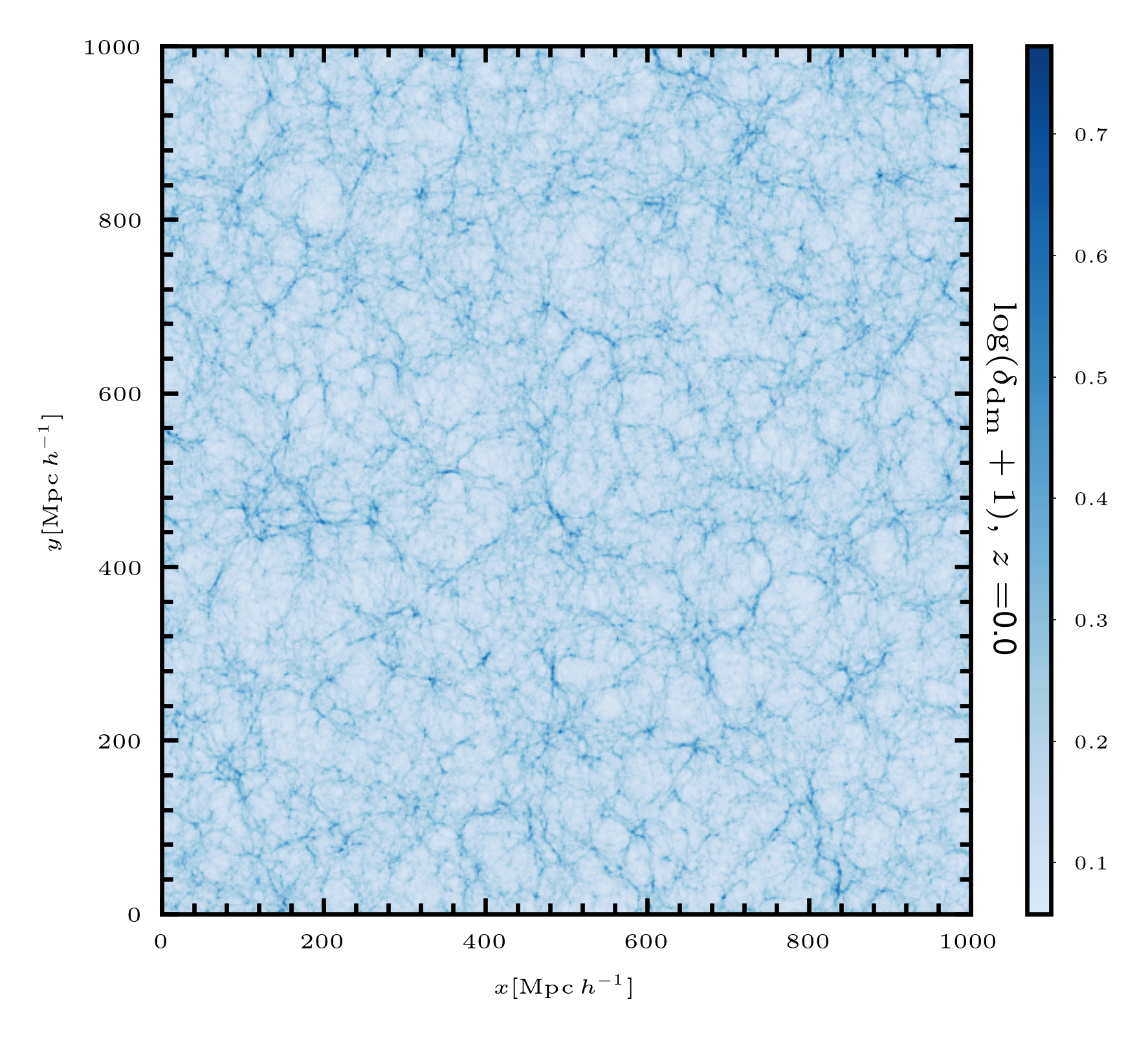}
\includegraphics[trim = .45cm 0.5cm 0cm 0cm ,clip=true, width=0.5\textwidth]{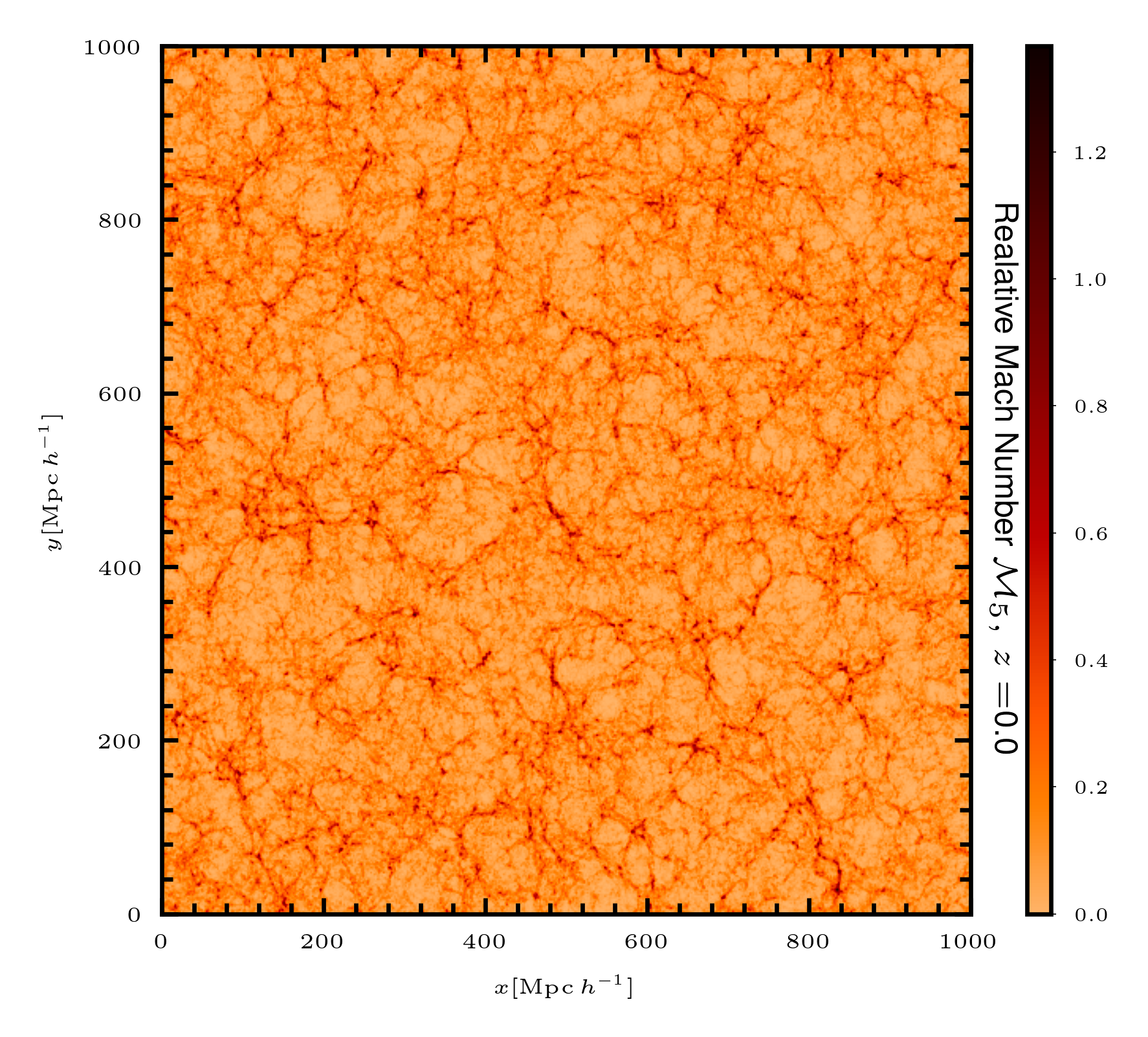}
\includegraphics[trim = .3cm .1cm 0cm 0cm ,clip=true, width=0.5\textwidth]{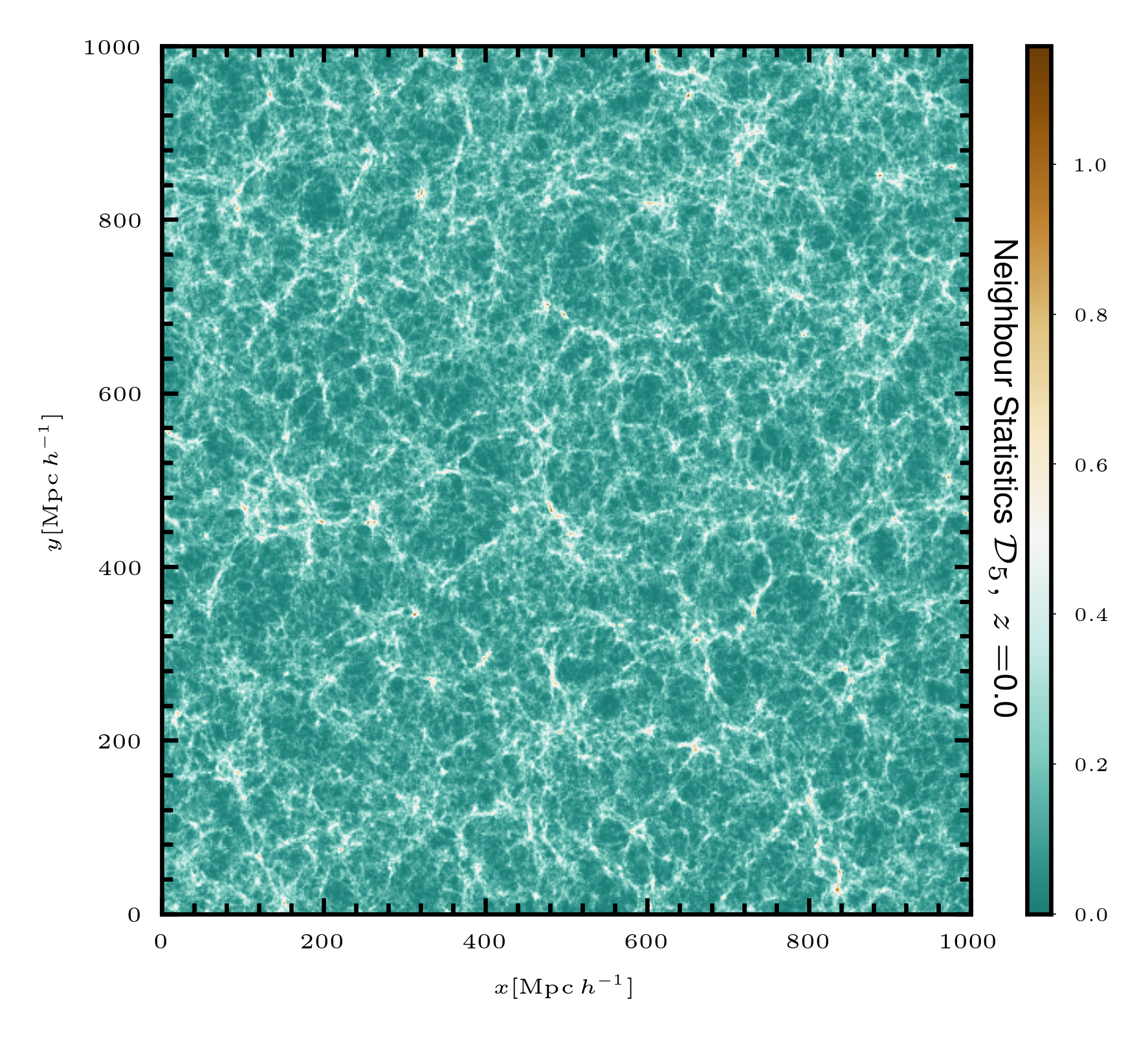}
\includegraphics[trim = .45cm .1cm 0cm 0cm ,clip=true, width=0.5\textwidth]{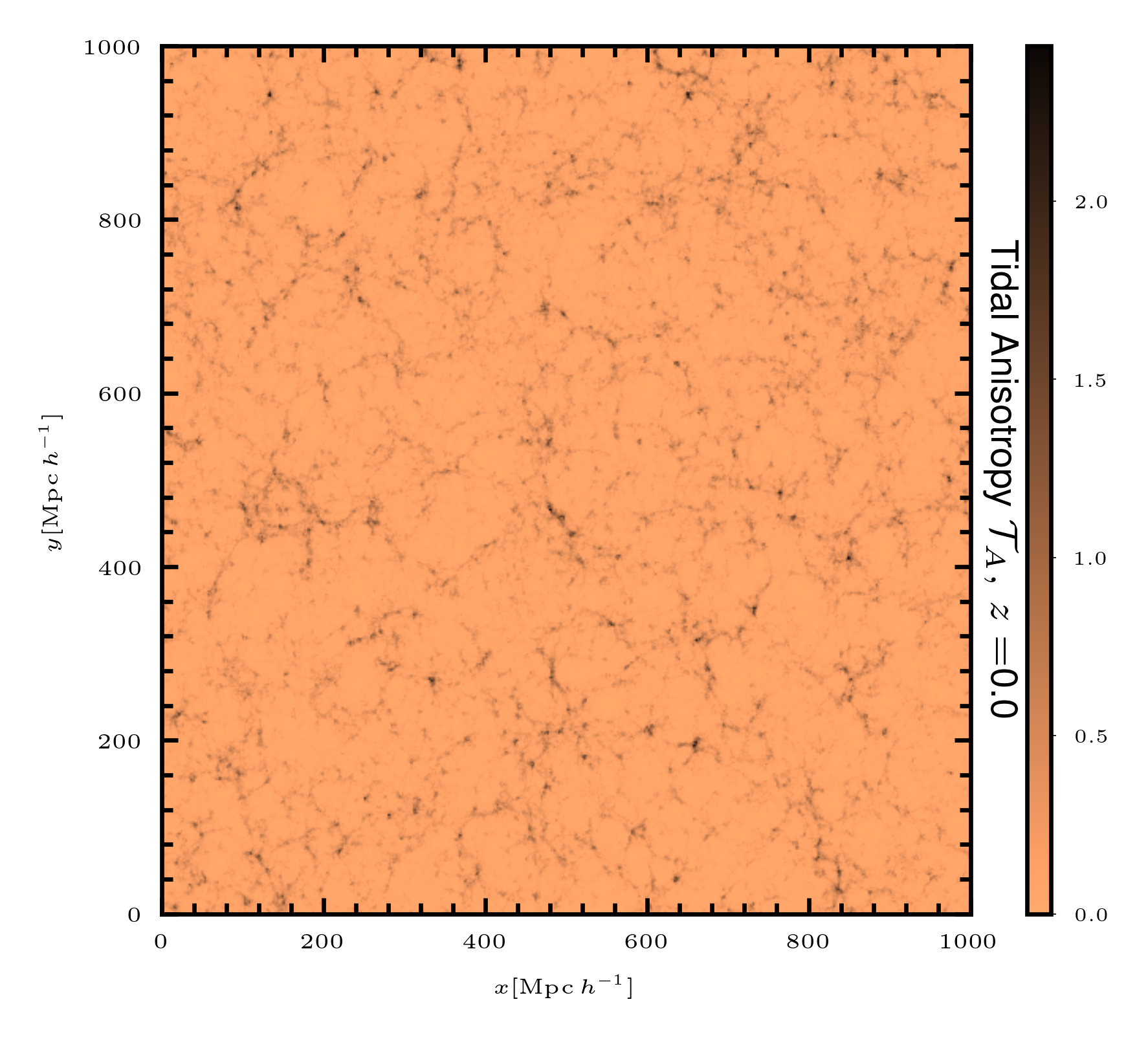}
\caption{\small{Slices of the $\sim 80$ Mpc$h^{-1}$ thickness in the UNITSim (at $z=0$) showing the dark matter density field (upper left), the local halo Mach $\mathcal{M}_{5}$ number (upper right, see \S\ref{sec:mach}), the neighbor statistics $\mathcal{D}_{5}$ (lower left, see \S\ref{sec:nei}), and the tidal anisotropy $\mathcal{T}_{A}$  (see \S\ref{sec:tidal}. )}}\label{fig:slices}
\end{figure*}
%------------------------------------------------------------------------

%------------------------------------------------------------------------
\begin{figure}
%dm_dist_cwt.py
\includegraphics[trim = .0cm .2cm 0cm 0cm ,clip=true, width=0.45\textwidth]{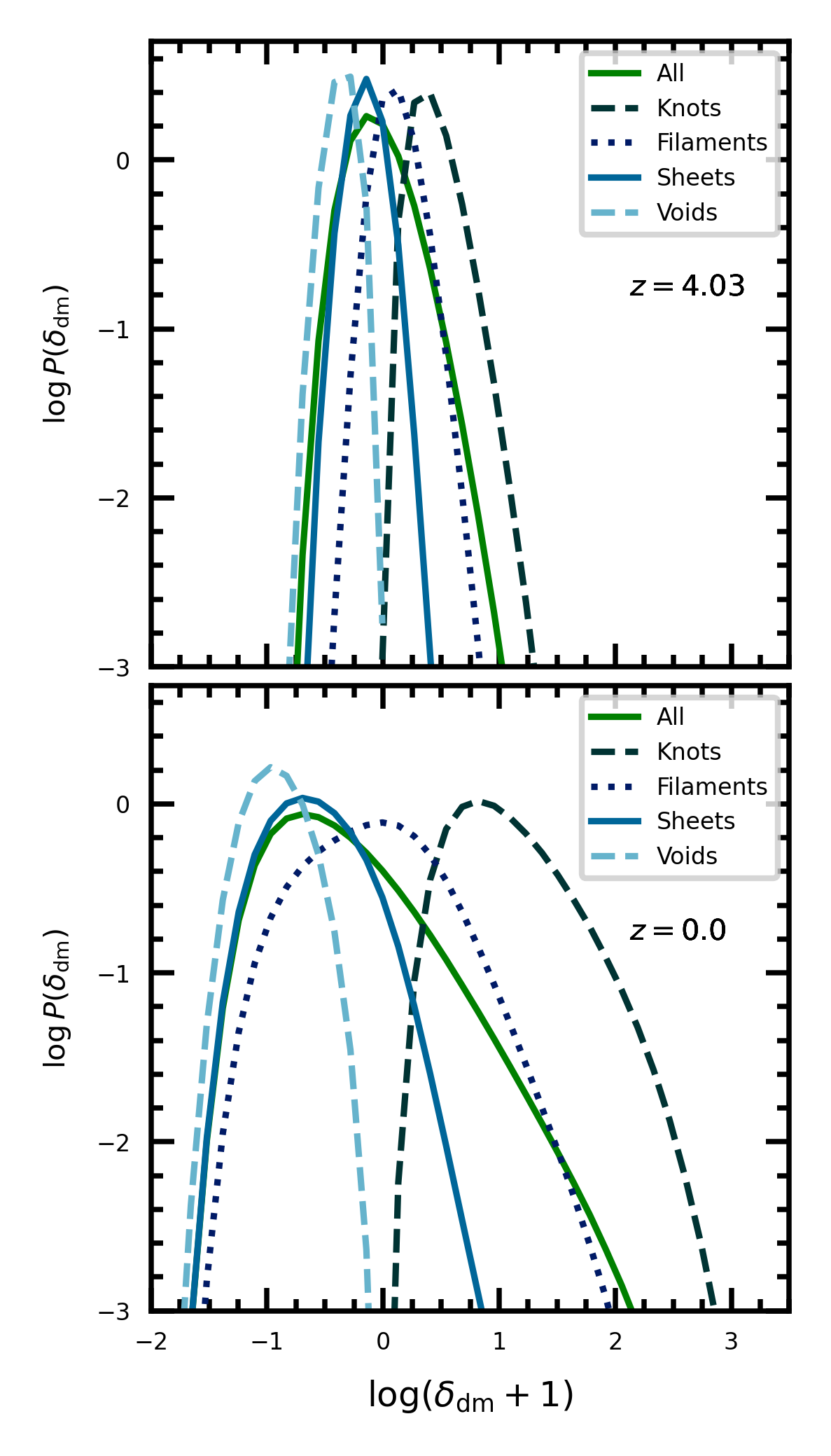}
\caption{\small{Probability distribution of the dark matter density in the UNITSim for two redshifts, in the four cosmic-web types defined in \S\ref{sec:cwc}. As we have used a threshold $\lambda_{th}=0$ in the cosmic-web classification, we identified under-densities with voids and overdensities with knots.}}\label{fig:dist_cwt_dm}
\end{figure}
%------------------------------------------------------------------------

Importantly, the UNITSim is a paired-fixed amplitude simulation \citep[][]{2016MNRAS.462L...1A}. On the one hand, ``fixed-amplitude'' means that, while the phases $\phi_{k}$ of the complex overdensity field in Fourier space $\delta(\vec{k})=A_{k}e^{i\phi_{k}}$ are randomly distributed in the range $[0, 2\pi)$, their amplitudes $A_{k}$ follow a distribution of the form $\delta^{D}(A_{k}-\sqrt{VP_{0}(k)/(2\pi)^{3}})$, instead of a Rayleigh distribution with mode $P_{0}(k)$, as it is the case of  standard simulations based on Gaussian random field (GRF hereafter). ``Pairing'', on the other hand, means that one realization is in reality a set of two simulations, in which the phases of the initial density fields are shifted by an amount $\pi/2$ (or the initial overdensity differing by a minus sign). This means that two paired simulations display the same initial power spectrum by construction, with full lack of correlation toward large scales. Gravitational evolution induces mode coupling, which translates into a variance in the power spectrum that is shown to be much smaller than that obtained from standard simulations with the same initial conditions, when analyzed from a variety of tracers \citep[][]{2018ApJ...867..137V,2019MNRAS.487...48C}. This feature makes this type of simulations suitable for the modeling of cosmological probes and the assessment of systematic errors in large-scale structure analyses \citep[see e.g.,][]{2018ApJS..236...43G,2019MNRAS.490.3667Z,2021MNRAS.508.4017M}. It has been also demonstrated that these types of simulations do not introduce a bias (as compared to standard simulations) in the one- or two-point statistics \citep[][]{2016MNRAS.462L...1A,2018ApJ...867..137V} nor in galaxy/halo properties and scaling relations (means and scatter). The scatter in the distribution of  probes such as density distributions, void abundances, and halo bias is still compatible with that observed in standard simulations \citep[see e.g.,][]{2018ApJ...867..137V,2020MNRAS.496.3862K}. This point is of particular interest for this paper, as we will explore the behavior of halo scaling relations and derive many of our conclusions based on the scatter observed from one simulation out of the pair. All the calculations shown in this paper are performed using the same simulation, and we have verified that almost indistinguishable results are obtained using the paired one. 
Fixed-paired simulations have been also shown to be suitable for the generation of halo mock catalogs based on learning algorithms \citep[][]{2023A&A...673A.130B}.

%=============================================================================================================
%=============================================================================================================
%=============================================================================================================
\section{Halo and environmental properties: Correlation and scaling relations}

In this section we present the  halo properties employed in the analysis, and we discuss the intricate relations between them as a function of redshift. The set of halo properties used in this work is divided in three sectors, namely i) intrinsic properties (i.e., those directly provided by the halo-finder), ii) nonlocal properties (i.e., those computed from the phase-space properties of neighbor halos) and iii) environmental properties (i.e., computed from the underlying dark matter density field.

%=============================================================================================================
\subsection{Intrinsic halo properties}\label{sec:hprops}

In the UNITSim, halo catalogs and their properties have been obtained using the \texttt{ROCKSTAR} (Robust Overdensity
Calculation using K-Space Topologically Adaptive Refine) halo finder algorithm \citep[][]{2013ApJ...762..109B}. The robustness of this halo finder has been studied in a number of comparison projects \citep[see e.g.,][]{2011MNRAS.415.2293K,2014MNRAS.441.3562E, 2016MNRAS.462..893R,2020MNRAS.493.4763M,2021MNRAS.500.3309M}. In particular, based on scale-free cosmologies,
\citet[][]{2021MNRAS.501.5064L,2024MNRAS.527.5603M} assessed the level of convergence to the physical limit of the statistical properties of dark matter halos defined with the \texttt{ROCKSTAR} algorithm, showing its better performance with respect to other halo finders, a claim that is arguably extrapolated to non scale-free simulations as the UNITSim. In general, we expect that our results can be extrapolated to those obtained with other halo-finders, although a thorough comparison in this direction is beyond the scope of this work.

Throughout this paper, we shall refer to  ``scaling relations'' (or link) between halo properties. Such term must be understood as the probability that a halo has a property $a$ conditional to a property $b$ in the range $b,b+\dd b$, i.e, $\mathcal{P}(a|b)\dd b$.

%=============================================================================================================
\subsubsection{Halo mass}\label{sec:mass}
The halo finder provides a number of definitions of halo mass,  based on a spherical overdensity $\Delta$ \citep[][]{2013ApJ...762..109B}. In this work we use the virial mass $M_{\rm vir}=(4/3)\pi  \bar{\rho}(z)\Delta(z)R_{\rm vir}^{3}$, obtained by the halo finder using spherical overdensities of $\Delta(z) = (18\pi^{2}+82x -39x^{2})/\Omega_{\rm mat}(z)$ where $x\equiv \Omega_{\rm mat}(z)-1$ \citep[][]{Bryan_1998}. Here $R_{\rm vir}$ denotes the radius at which the enclosed mass equals the virial mass \citep[see e.g.,][]{2016MNRAS.457.4340K}, a quantity similarly provided by the halo finder. We select parent halos with $M_{\rm vir}>2\times 10^{11}M_{\odot} h^{-1}$ (which corresponds to $\sim 300$ dark matter particles). The selection of $M_{\rm vir}$ as a mass-proxy is consistent with other halo properties such as spin, concentration and shape parameters, which are computed based on $R_{\rm vir}$ and hence $M_{\rm vir}$. 

The halo mass distribution at $z=0$ is shown in the panel (a) of Fig.~\ref{fig:prop_distribution} as a function of the cosmic-web type (to be defined in \S\ref{sec:cwc}). We have verified that the halo abundance as a function of virial mass is in good agreement with fitting functions found in literature \citep[see e.g.,][]{2008ApJ...688..709T}.
%------------------------------------------------------------------------
\subsubsection{Maximum circular velocity}
\begin{figure*}[htbp]
\centering
%prop_dist_cwt.py:
\includegraphics[trim = 0.25cm 0.16cm 0cm 0cm ,clip=true, width=0.32\textwidth]{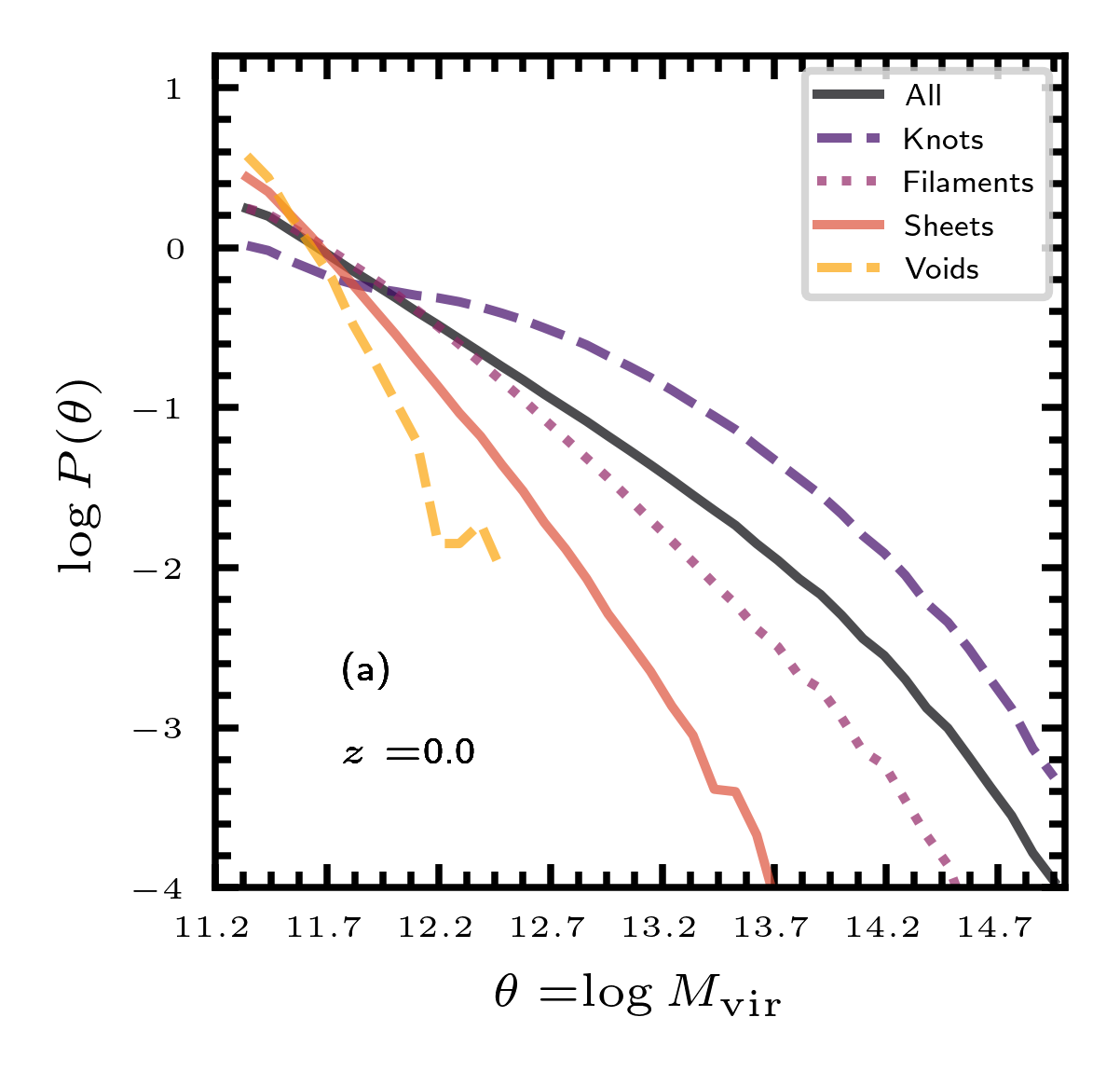}    
\includegraphics[trim = 0.25cm 0.16cm 0cm 0cm ,clip=true, width=0.32\textwidth]{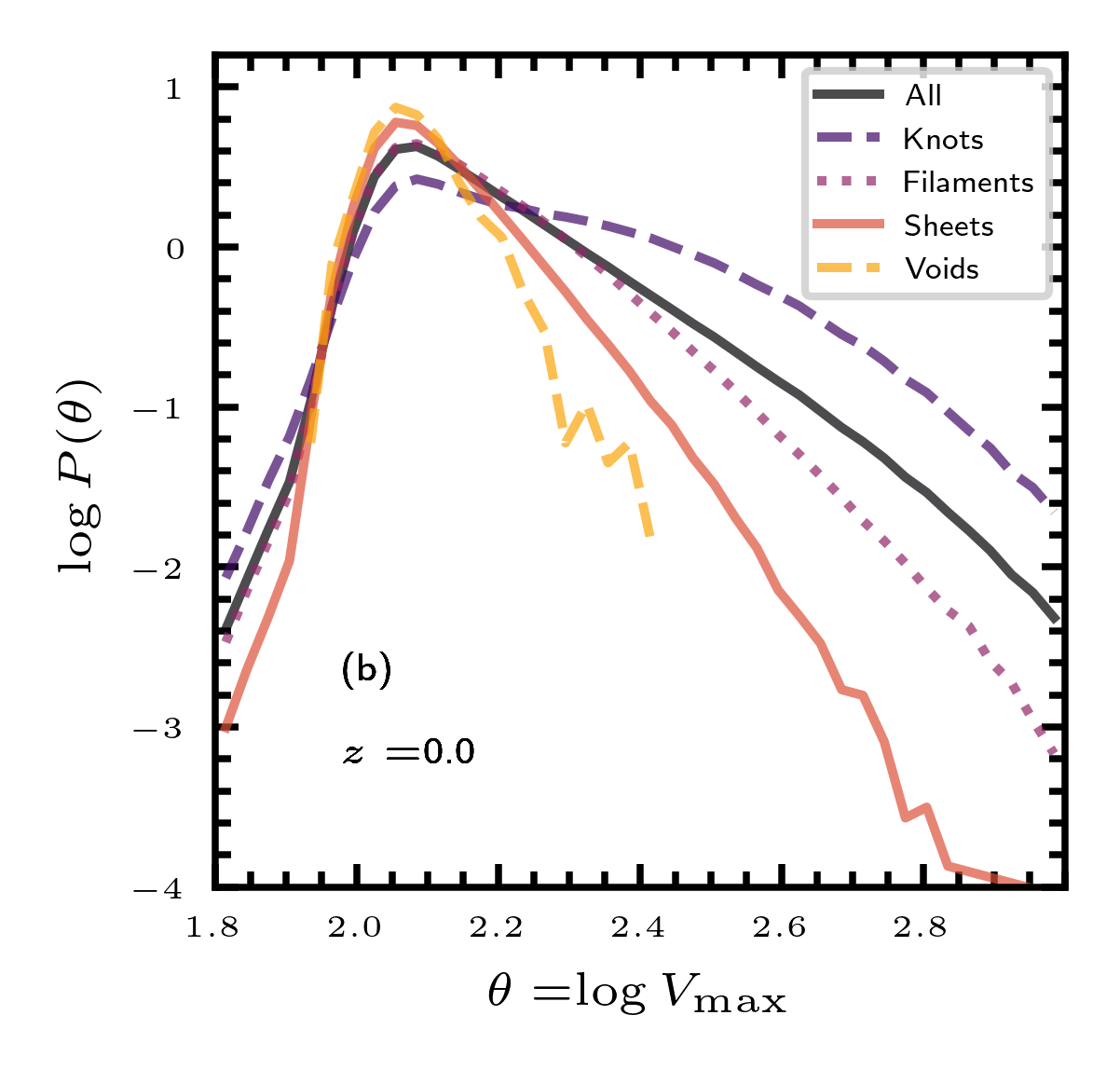}    
\includegraphics[trim = 0.25cm 0.16cm 0cm 0cm ,clip=true, width=0.32\textwidth]{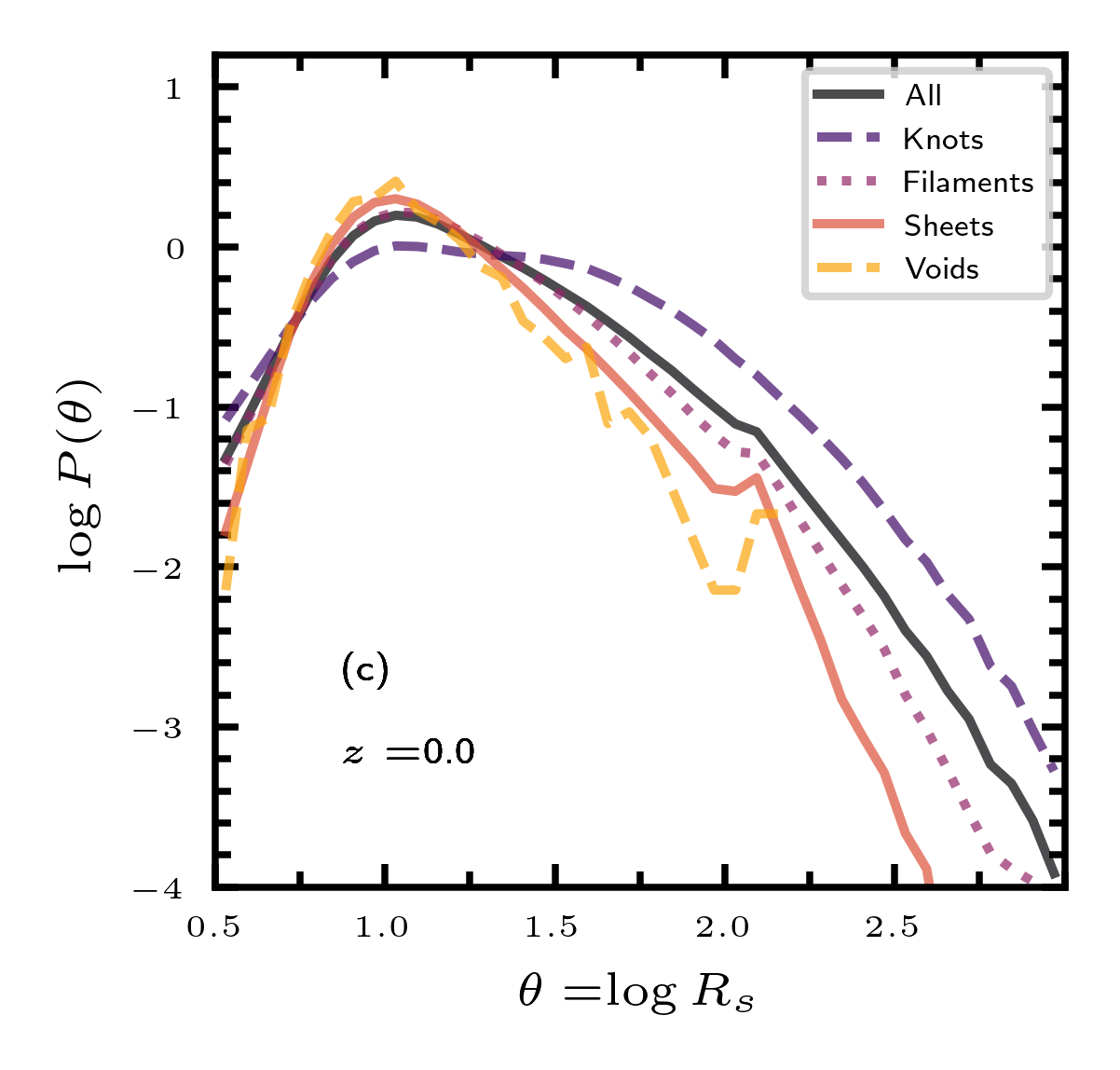}    
\includegraphics[trim = 0.25cm 0.16cm 0cm 0cm ,clip=true, width=0.32\textwidth]{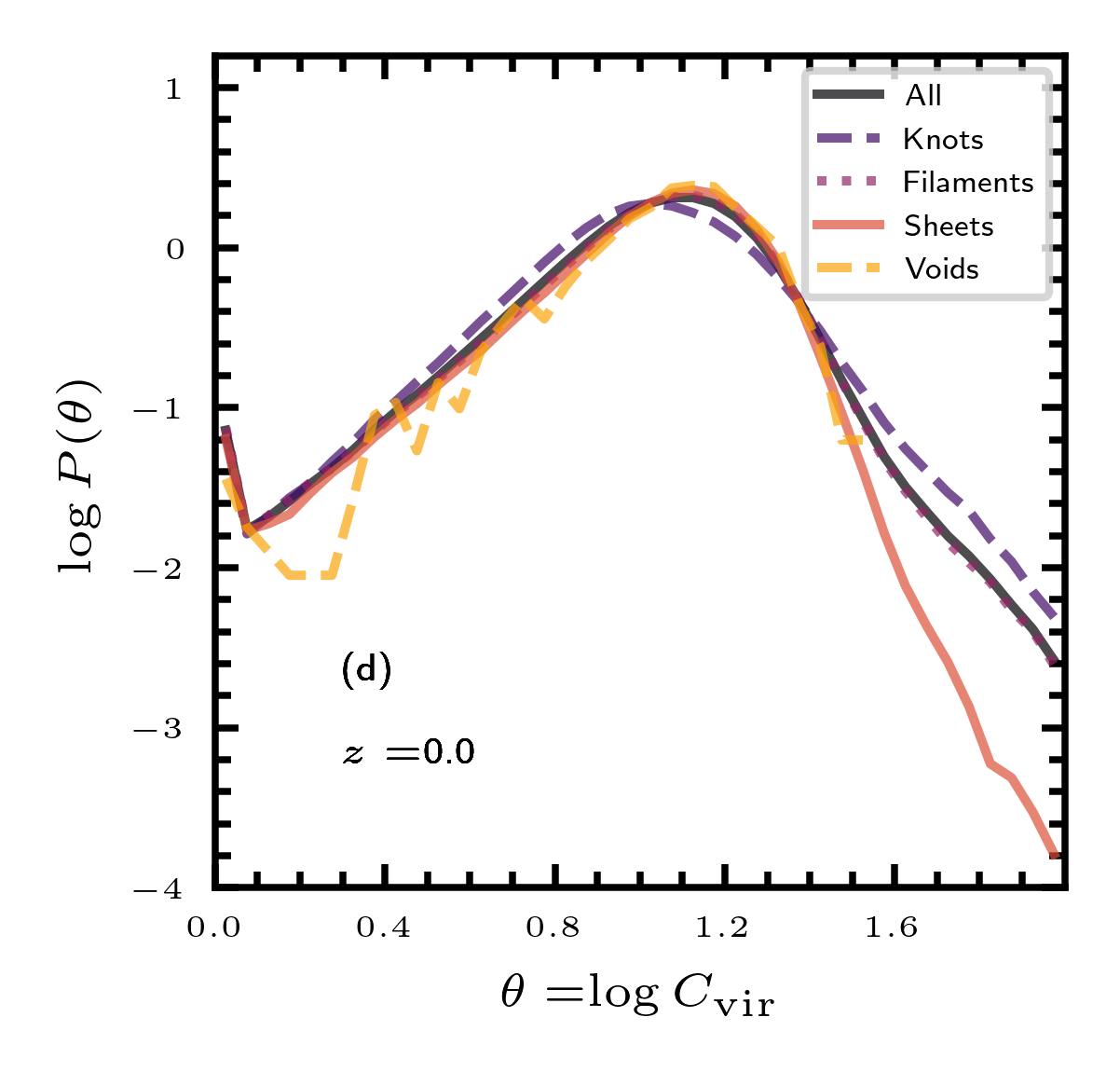}    
\includegraphics[trim = 0.25cm 0.16cm 0cm 0cm ,clip=true, width=0.32\textwidth]{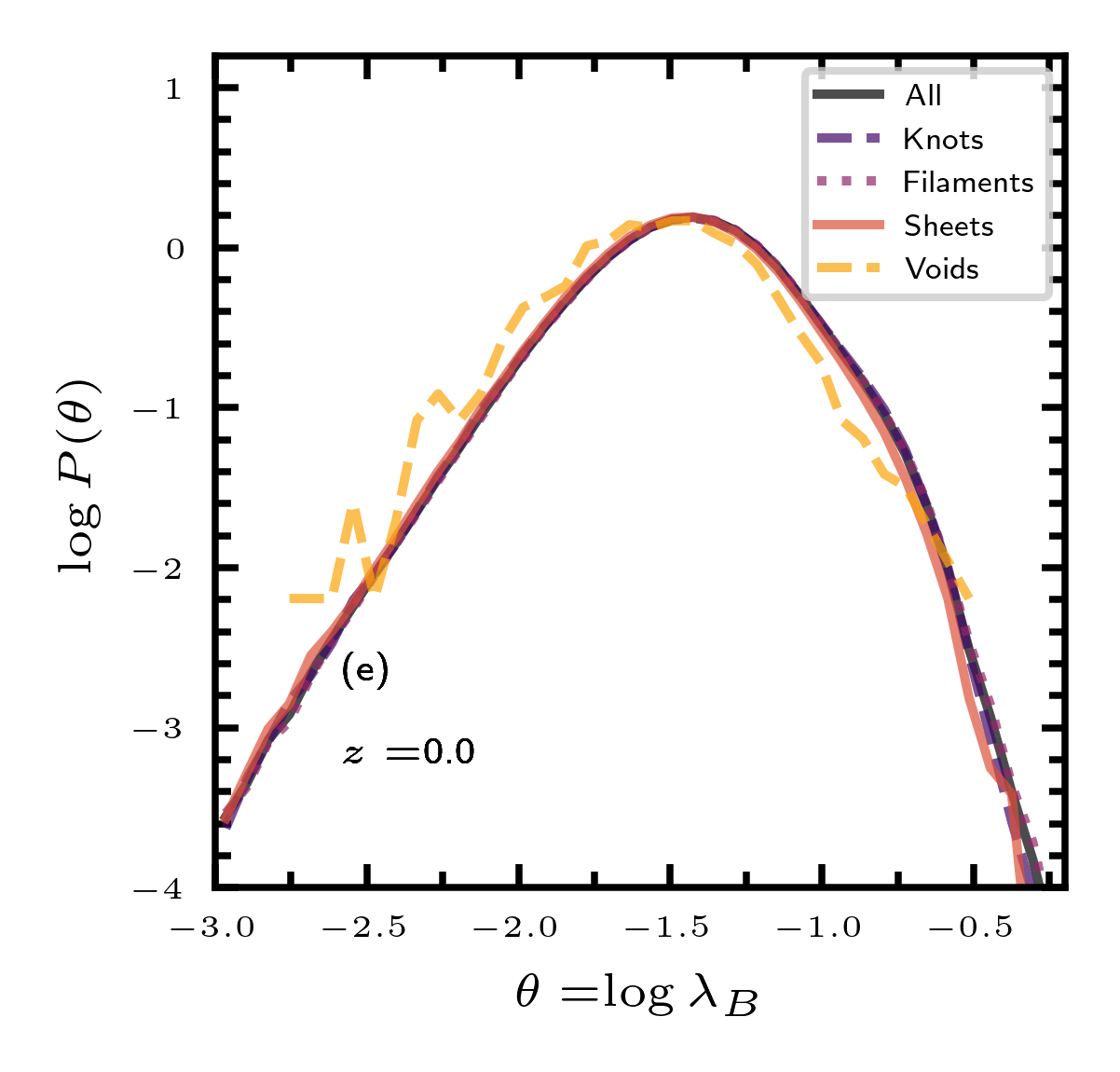}    
\includegraphics[trim = 0.25cm 0.16cm 0cm 0cm ,clip=true, width=0.32\textwidth]{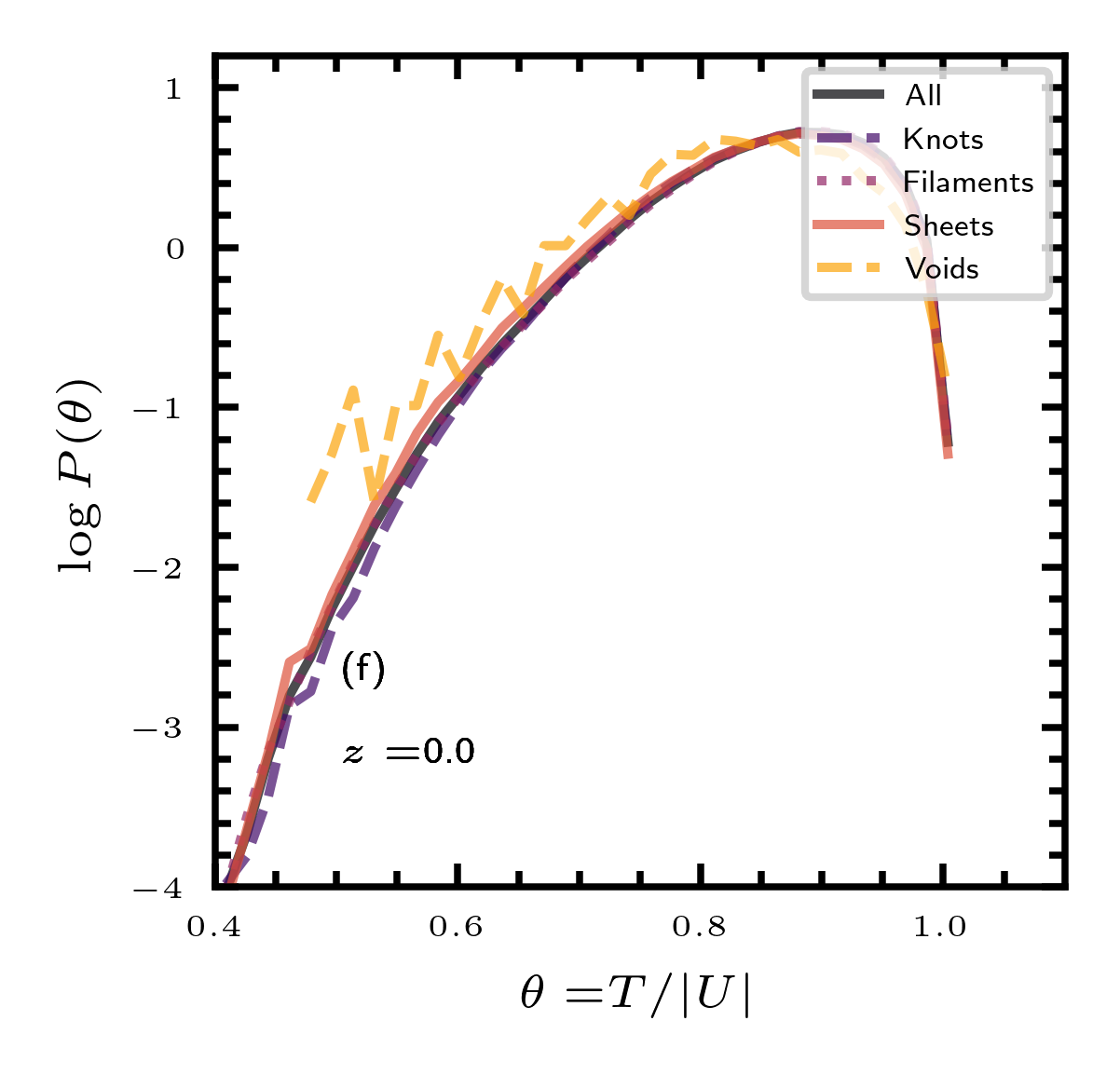} 
\includegraphics[trim = 0.25cm 0.16cm 0cm 0cm ,clip=true, width=0.32\textwidth]{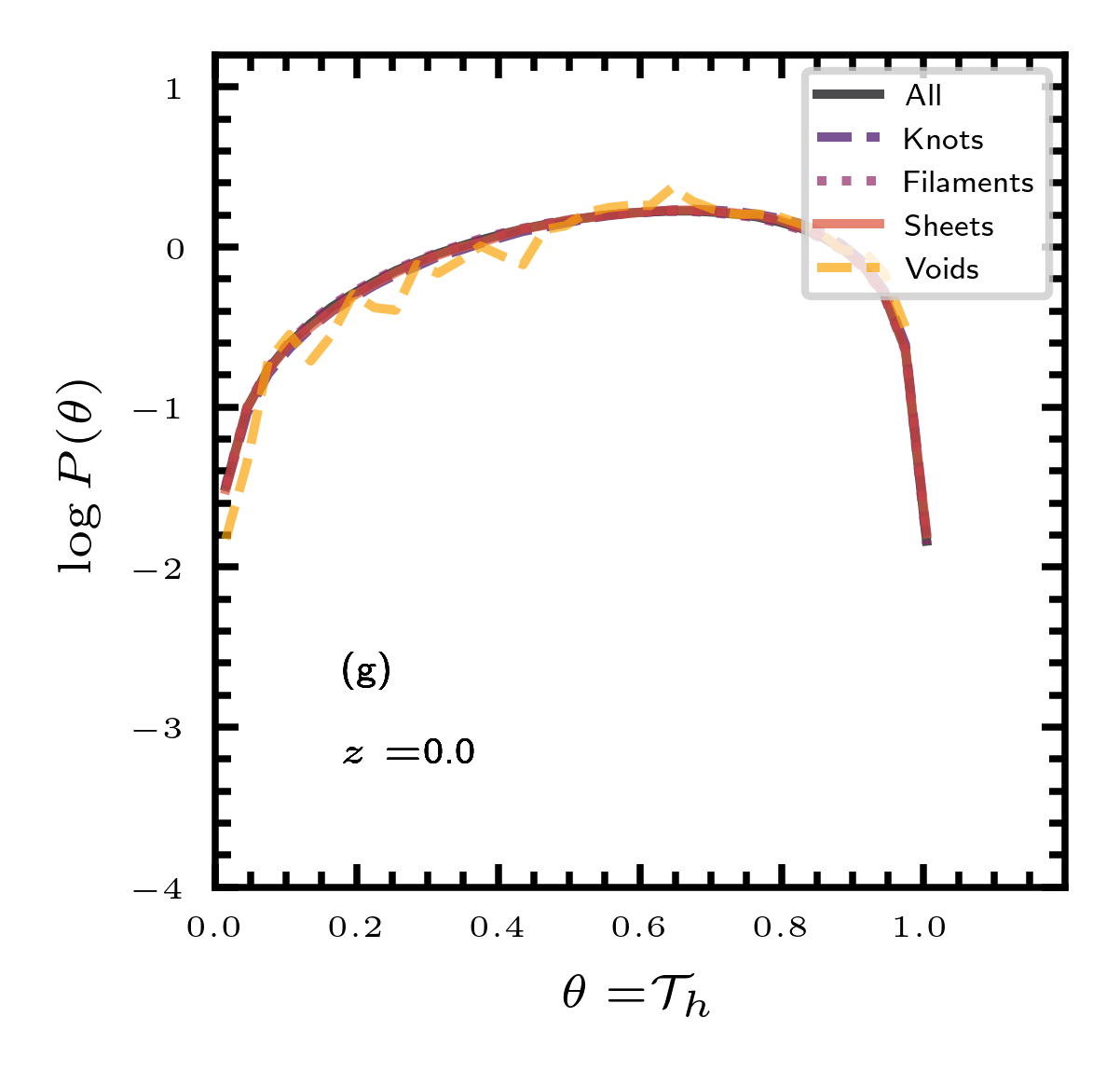}    
\includegraphics[trim = 0.25cm 0.16cm 0cm 0cm ,clip=true, width=0.32\textwidth]{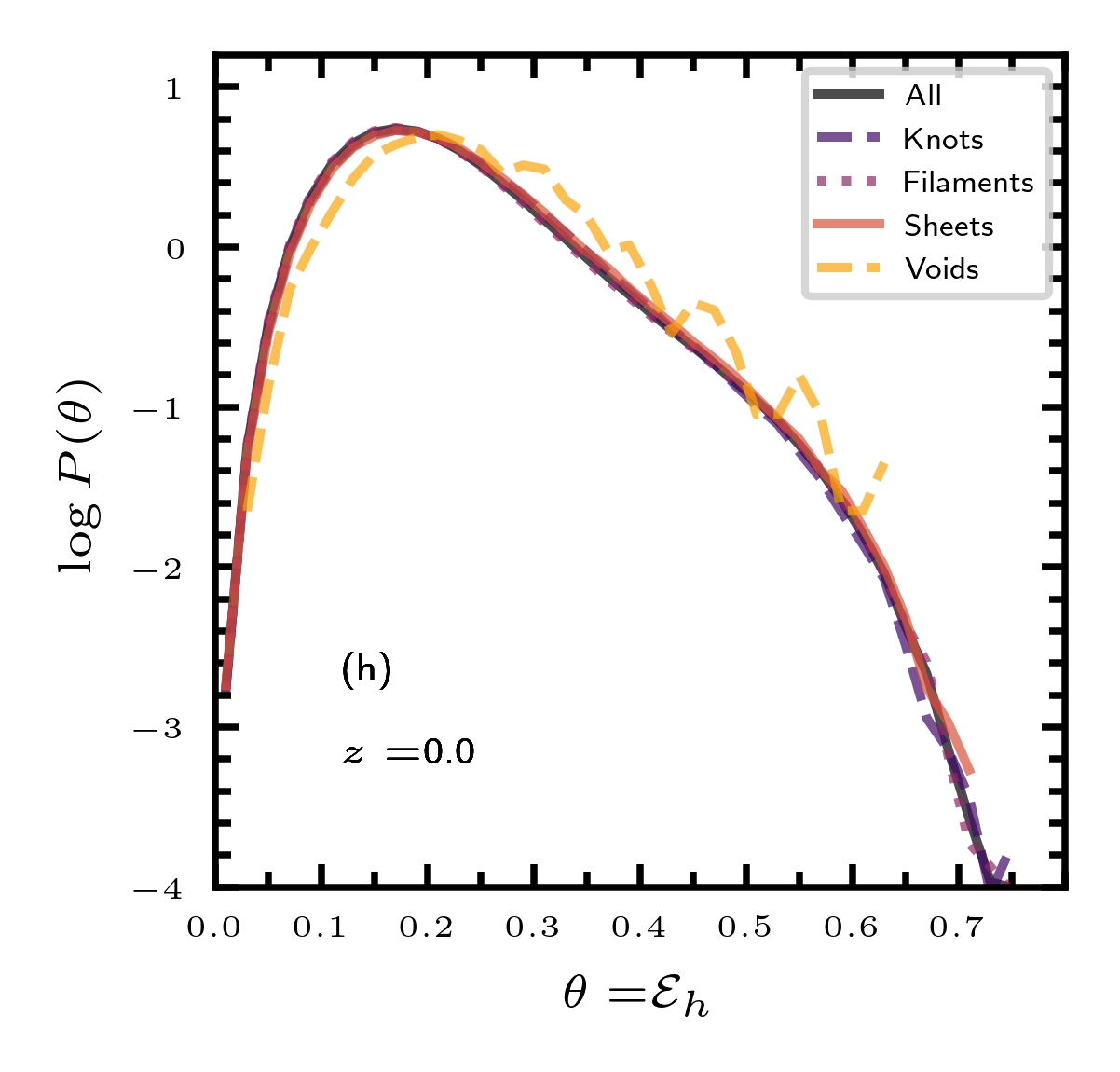}    
\includegraphics[trim = 0.25cm 0.16cm 0cm 0cm ,clip=true, width=0.32\textwidth]{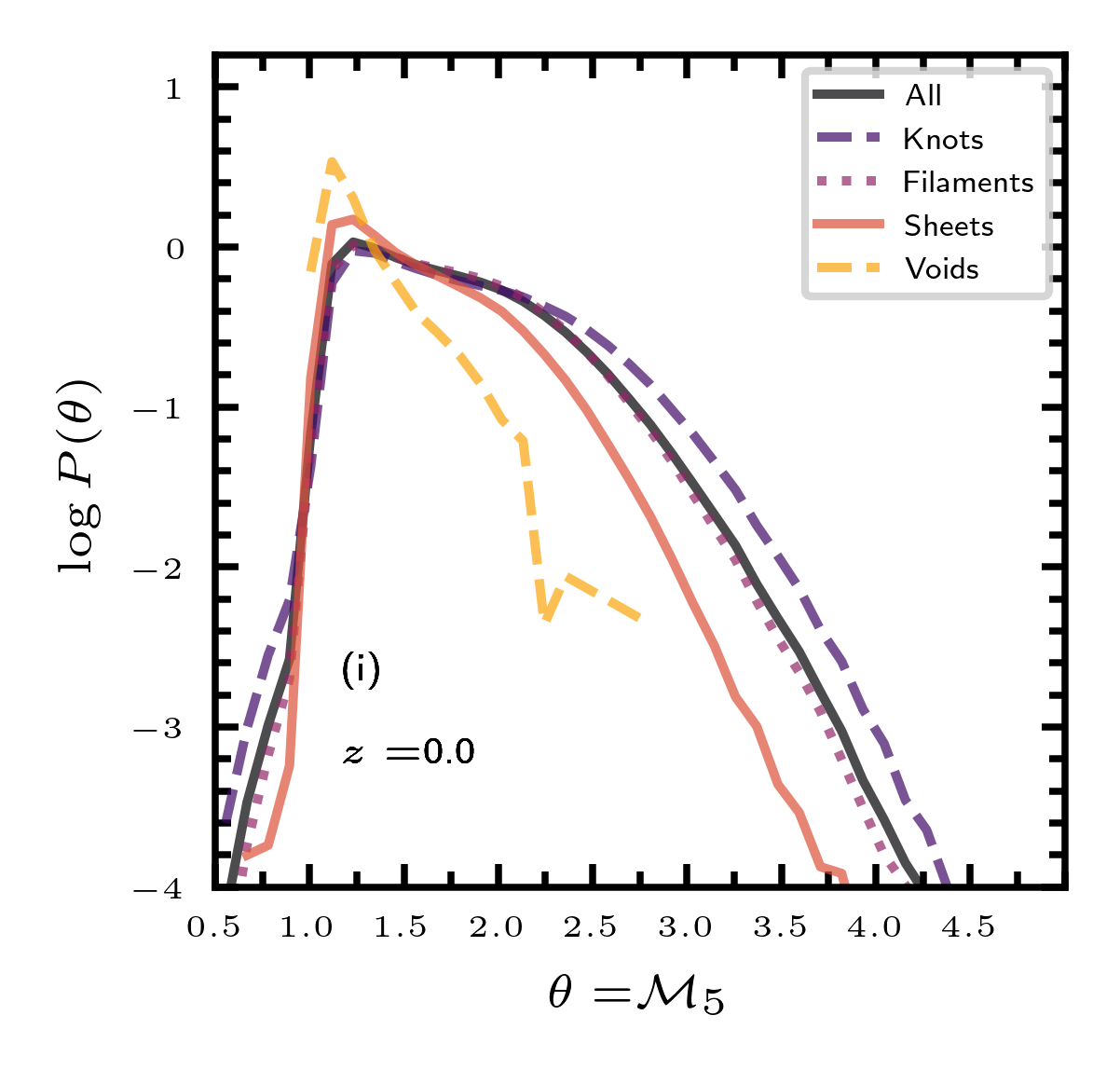}    
\includegraphics[trim = 0.25cm 0.16cm 0cm 0cm ,clip=true, width=0.32\textwidth]{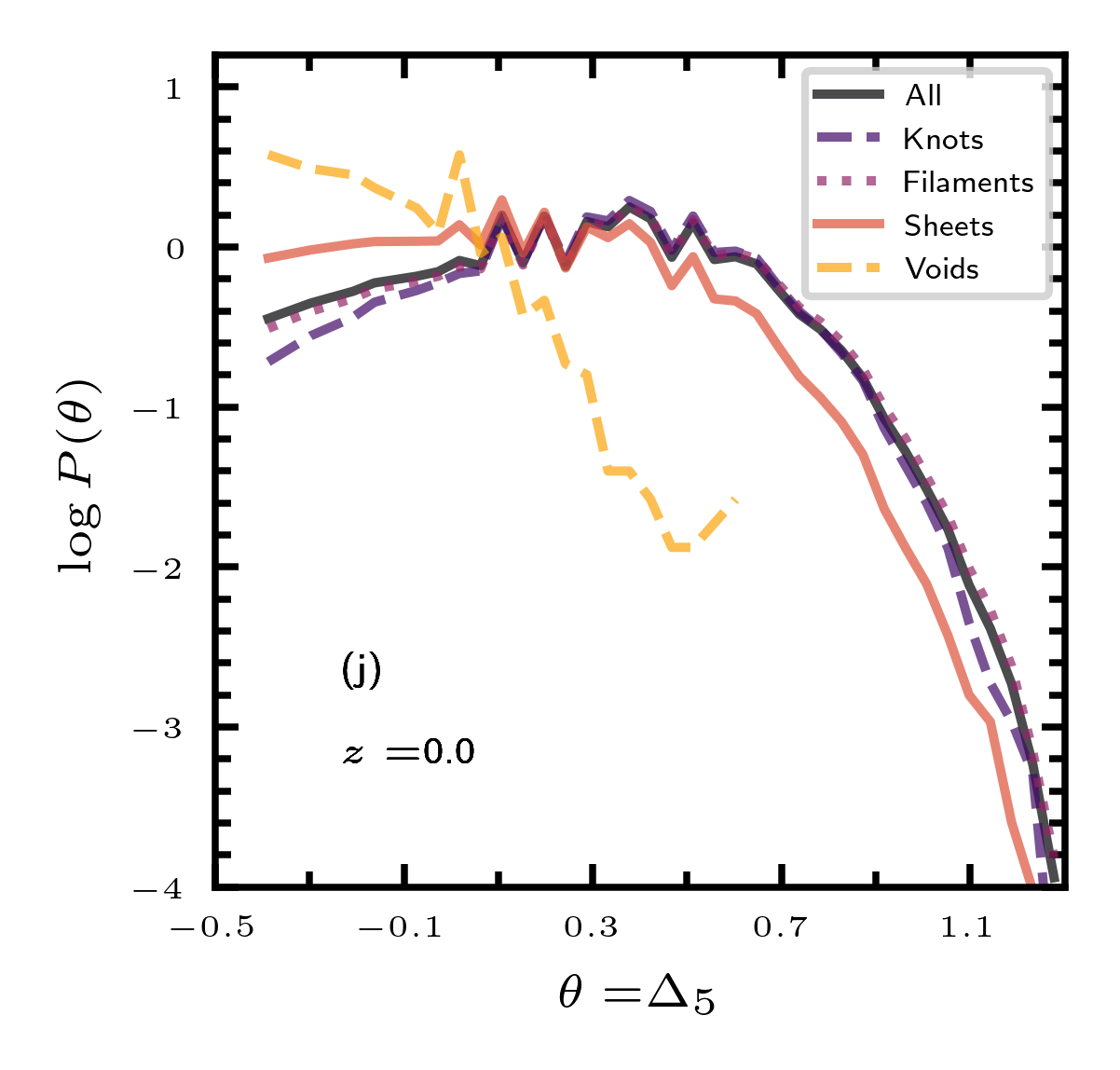} 
\includegraphics[trim = 0.25cm 0.16cm 0cm 0cm ,clip=true, width=0.32\textwidth]{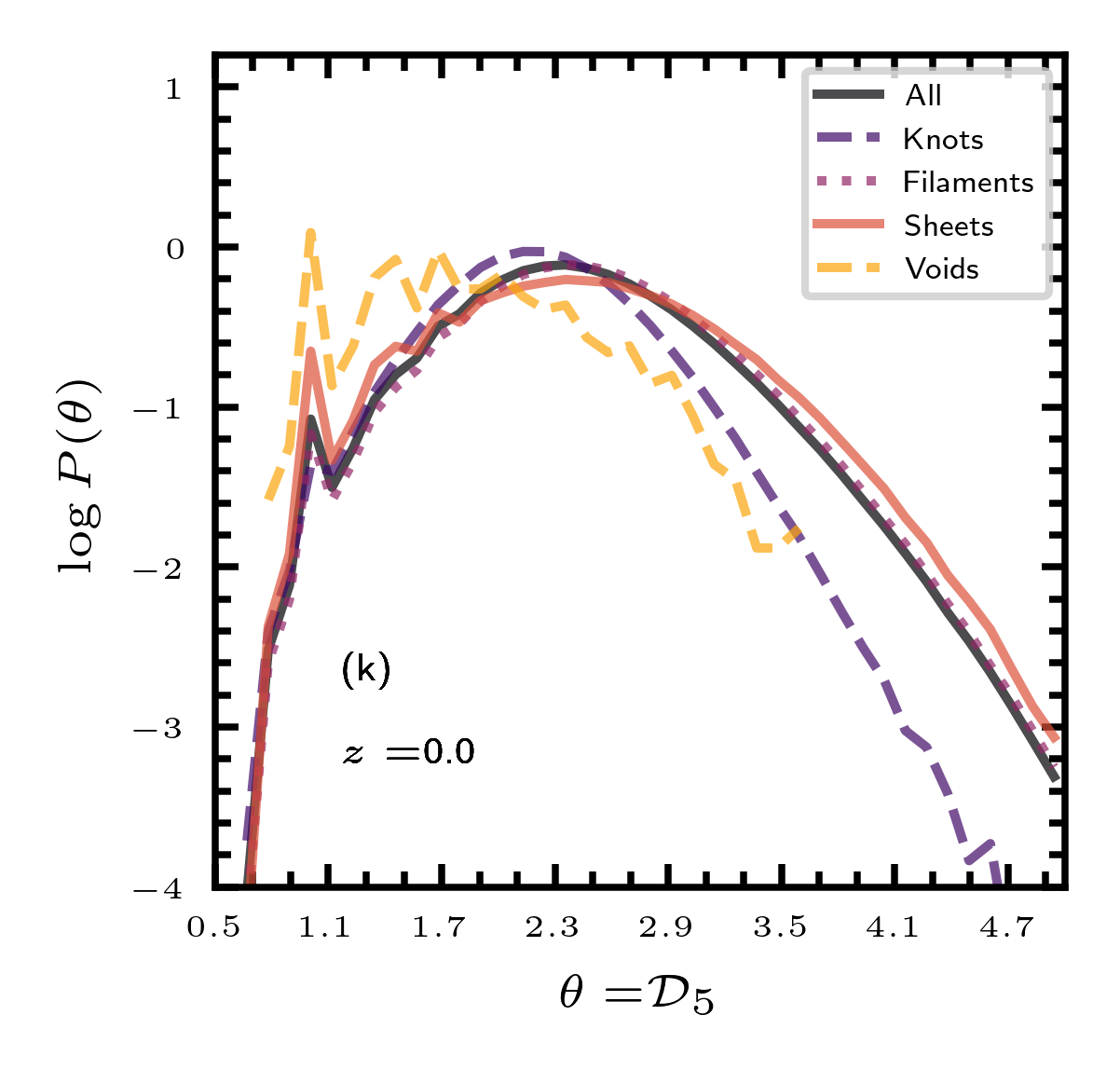}    
\includegraphics[trim = 0.25cm 0.16cm 0cm 0cm ,clip=true, width=0.32\textwidth]{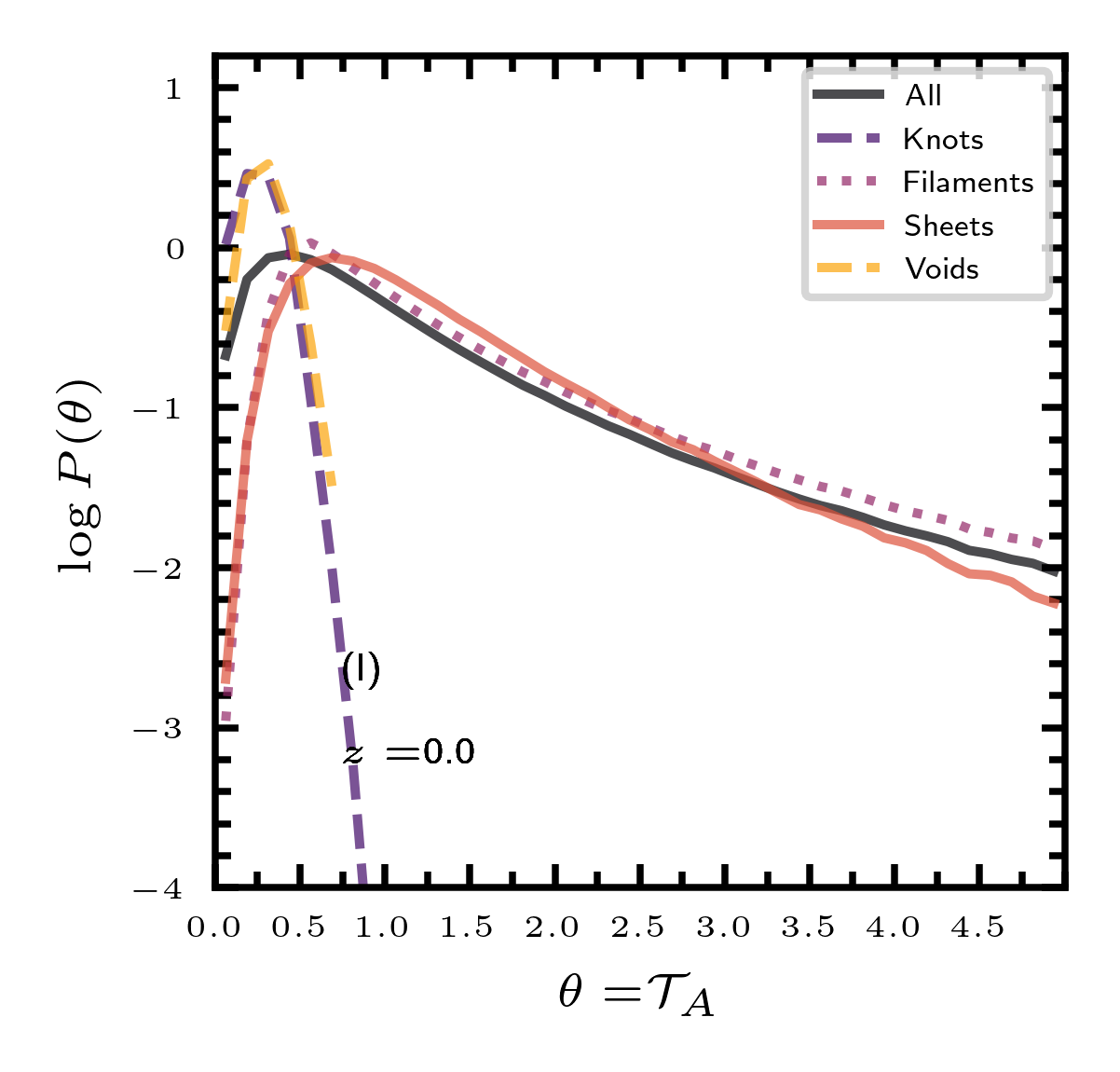}    
\caption{\small{Distribution of intrinsic $\{M_{\rm vir},V_{\rm max}, R_{s},C_{\rm vir}, \lambda_{B}, T/|U|, \mathcal{T}_{h},\mathcal{E}_{h}\}$, nonlocal $\{\mathcal{M}_{5}, \mathcal{D}_{5}, \Delta_{5}\},$ and environmental ($\mathcal{T}_{A}$) halo properties in the UNITSim at $z=0$. In all panels, the black solid line represents the distribution from the full population. We also present the property distribution in the cosmic-web types defined in \S\ref{sec:cwc}.}.  }\label{fig:prop_distribution}
\end{figure*}

This quantity is computed as $V_{\rm max}\equiv {\rm Max}(V_{c})$, where $V_{c}(r)=(G_{N} M(r)/r)^{1/2}$ is the halo circular velocity at a radius $r$ and $M(r)$ is the mass enclosed within that radius (and $G_{N}$ is the gravitational constant). A simple version of the virial theorem \citep[see e.g.,][]{1978vtsa.book.....C} predicts $V_{\rm max}\propto M_{\rm vir}^{1/3}$, a dependency that we have confirmed in the UNITSim at different redshifts \citep[see also][]{2006ApJ...646..815S,2016MNRAS.462..893R,2023MNRAS.523.1919C}. 
Panel (b) of Fig.~\ref{fig:prop_distribution} contains the distribution of $V_{\rm max}$ at $z=0$, while its scaling relation with the virial mass is shown in panel (a) Fig.~\ref{fig:prop_mass}. The maximum circular velocity can be regarded, along with the halo mass, as direct probes of the potential well of dark matter haloes (see appendix \S\ref{sec:pca}).  As such, this quantity is sometimes preferred over the mass  in order to model the statistics of galaxies inside halos \citep[see e.g.,][]{2019ApJ...887...17Z,2017ApJ...834...37L,2022RNAAS...6...80M}.

The halo finder also provides the velocity dispersion of dark matter halos, computed from velocity statistics of dark matter in halos. As for $V_{\rm max}$, this quantity scales approximately as $\propto M_{\rm vir}^{1/3}$ \citep[see e.g.,][]{2010MNRAS.407..581R}. Even though we do not show results in terms of this quantity, it is important to indicate that the behavior of the halo velocity dispersion as a function of halo mass is key for the assignment of phase-space coordinates for galaxies in dark matter halos in the framework of the Halo model \citep[see e.g][]{2006MNRAS.365..842C,2023A&A...673A.130B}.

%==================================================================================================
%==================================================================================================
%------------------------------------------------------------------------
\subsubsection{Scale radius and halo concentration}\label{sec:mconc}
High-resolution $N$-body simulations have demonstrated that the density profile of dark matter halos can be described by the Navarro-Frenk-White profile \citep[][]{1996ApJ...462..563N}. As such, \texttt{ROCKSTAR} uses this profile \citep[see e.g.,][]{2011MNRAS.415.2293K,2020MNRAS.493.4763M} to provide values of scale radius $R_{s}$ and the halo concentration $C_{\rm vir}$. For the scale radius, we have used the estimates based on fits to the NFW profile, although the halo finder also provides estimates based on the method by \citep[][]{Klypin_2011} (which also relies on the NFW profile). Besides of shaping the density profile (it marks the scale at which the logarithmic slope of the NFW profile equals $2$), the scale radius $R_{s}$ has been shown to correlate with halo anisotropy parameters \citep[see e.g.,][]{2008gady.book.....B,2005ApJ...634..775B} and hence can encode information of the dynamical state of dark matter halos. Panel (c) of Fig.\ref{fig:prop_distribution} shows the distribution of this property at $z=0$ in the UNITSim. Panel (b) of Fig.~\ref{fig:prop_mass} shows the scaling relation between the scale radius (in kpc$/h$) and the halo mass for different redshifts. 

The halo concentration, computed as $C_{\rm vir}\equiv R_{\rm vir}/R_{s}$, is a key property, not only for characterizing the halo density profile, but also because it has been shown to be a proxy for halo formation time at fixed halo mass \citep[][]{2002ApJ...568...52W,2016MNRAS.460.1214L,2020MNRAS.498.4450W}. 
Panel (d) of Fig.~\ref{fig:prop_distribution} shows the distribution (at $z=0$) of halo concentration in the UNITSim, while panel (c) of Fig.~\ref{fig:prop_mass} displays the mean scaling relation in terms of halo mass  \citep[see e.g.,][for a thorough analyusis of this scaling relation]{2007MNRAS.378...55M,2016MNRAS.460.1214L,2018ApJ...859...55C}. This mass-concentration scaling relation shows interesting features, as it displays a positive slope at high redshift (i.e, more massive halos are more concentrated), decreasing toward low redshift, to become fully inverted for redshifts below $z\sim 1$. This behavior can be best understood when the halo concentration is expressed as a function of the peak height $\nu$ (to be defined in \S\ref{sec:peak}) in conjunction with cosmological parameters describing the statistical properties of primordial density fluctuations; we have verified the $C_{\rm vir}-\nu$ relation displays an approximately universal shape as a function of redshift \citep[see e.g.,][]{2012MNRAS.423.3018P,2019ApJ...871..168D}. 

%------------------------------------------------------------------------
\subsubsection{Dimensionless spin parameter}
\texttt{ROCKSTAR} computes dimensionless spin parameters using both the \citet[][]{1971A&A....11..377P} and the \citep[][]{2001ApJ...555..240B} prescription. The latter defined as $\lambda_{B}=J/(\sqrt{2}M_{\rm vir}V_{c}(R_{\rm vir})R_{\rm vir})$ where $J$ is the total halo angular momentum. We use this version of halo spin, and we have verified that consistent results are obtained for the test carried within this paper when using the Peebles' definition. Panel (e) in Fig.~\ref{fig:prop_distribution} presents the distribution of halo spin (at $z=0$) for different cosmic web types, from which we can observe  deviations (in the form of kurtosis and skewness) from a log-normal distribution \citep[][]{2001ApJ...555..240B}. Panel (d) of Fig.~\ref{fig:prop_mass} displays its behavior with respect to halo mass at different redshifts. The behavior observed for the spin at different redshifts is in principle compatible with previous analysis \citep[see e.g.,][]{2007MNRAS.378...55M, 2008ApJ...678..621K,2015MNRAS.450.1486A}, displaying a weak correlation with the halo mass across redshifts, passing from a correlation coefficient of $\sim-0.02$ at $z\sim 5$ to $\sim 0.01$ at $z=0$. Nevertheless, we see that the behavior of the scaling relation, despite the small correlation, has a remarkable deviation from a constant behavior, contrary to what has been reported in the literature. This is a matter of future research.

%------------------------------------------------------------------------
\subsubsection{The virial parameter}
We refer as the ``virial parameter'' to the ratio between kinetic to gravitational potential energy, $T/|U|$ \footnote{In forthcoming sections we also use the symbol $\mathcal{V}$ to identify this parameter.}. This quantity can be used as a proxy for the degree of relaxation of a halo \citep[][]{2008ApJ...678..621K}, as the virial theorem \citep[see e.g.,][]{1978vtsa.book.....C} predicts $T/|U|\sim 1/2$ for relaxed isolated halos. Nevertheless, since the internal dynamics of dark matter halos involve pressure terms (from velocity anisotropies) and surface terms (inflow-outflow of material), deviations from the value $T/|U|\sim 1/2$ do not strictly imply a lack of relaxation \citep[][]{1999MNRAS.302..321H,2006ApJ...646..815S,2007MNRAS.376..215B,2016MNRAS.457.4340K}. Halos in the UNITSim have $T/|U|>1/2$, as can be checked from panel (f) of Fig.~\ref{fig:prop_distribution}. Panel (e) of Fig.~\ref{fig:prop_mass} presents its dependency with respect to the virial mass for different redshifts.

\subsubsection{Halo triaxiality and ellipticity}
\begin{figure*}[htbp]
%scaling_relations_detailed.py v1
%scaling_relations_detailed_SHARED.py
\includegraphics[trim = 0.2cm 0.2cm 0cm 0cm ,clip=true, width=\textwidth]{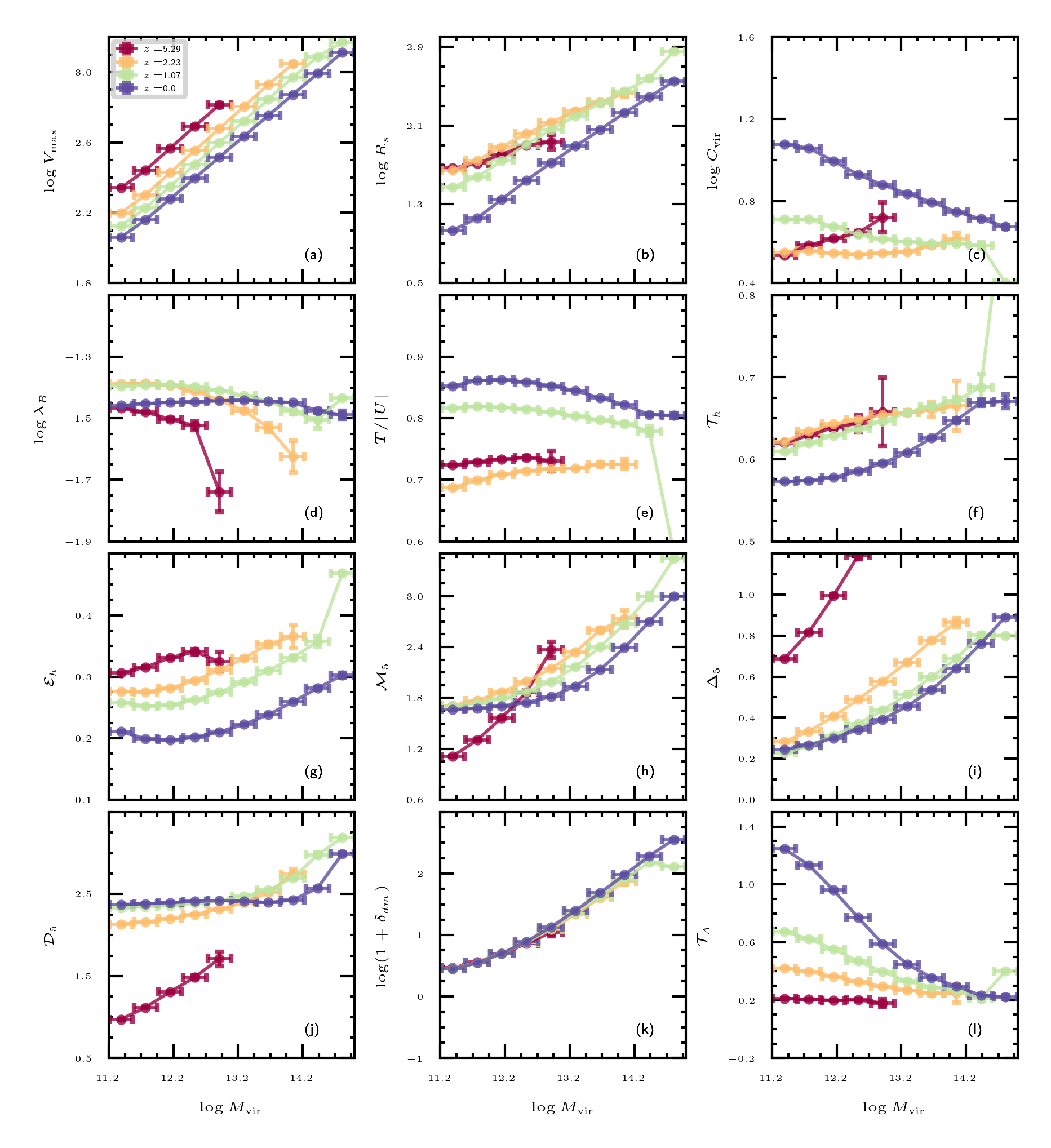}    
\caption{\small{Mean scaling relations between halo virial mass and a number of halo properties described in \S\ref{sec:hprops}, at different redshifts. Shown intrinsic properties are $V_{\rm max}$ (panel a), $R_{s}$ (panel b) $C_{\rm vir}$ (panel c), $\lambda_{B}$ (panel d), $T/|U|$ (panel e), $\mathcal{T}_{h}$ (panel f) , $\mathcal{E}_{h}$ (panel g). Nonlocal properties: local Mach number $\mathcal{M}_{5}$ (panel h), $\Delta_{5}$ (panel i), $\mathcal{D}_{5}$ (panel j). Environmental properties: local dark matter density (panel k) and tidal anisotropy $\mathcal{T}_{A}$ (panel l). The error bars denote the error in the mean in each mass bin.}}\label{fig:prop_mass}
\end{figure*}

Ellipsoidal shape parameters are defined in terms of the ratio between the halo semi-axes $(a,b,c)$ as the triaxiality parameter $\mathcal{T}_{h}$
\be
\mathcal{T}_{h}\equiv\frac{1-q^{2}}{1-s^{2}},
\ee
and the halo ellipticity:
\be
\mathcal{E}_{h}\equiv \frac{1-s^{2}}{1+s^{2}+q^{2}},
\ee
where $q=b/a$ and $s=c/a$ ($a>b>c$). The set $(a,b,c)$ is computed by the halo finder as the sorted eigenvalues of the mass distribution tensor using particles inside the virial radius \cite[see e.g.,][]{Zemp_2011}. The triaxiality parameter $\mathcal{T}_{h}$ defines ellipsoidal geometries such as the oblate ($\mathcal{T}_{h}\to 0$) and prolate ($\mathcal{T}_{h}\to 1$) spheroids. On the other hand, halo ellipticity $\mathcal{E}_{h}$ has as special limits the spherical ($\mathcal{E}_{h}\to 0$) and nonspherical (or aspherical) configurations \citep[see e.g.,][]{1996MNRAS.281..716C, 2008gady.book.....B}. An example of the distribution of this geometry-proxies at $z=0$ is displayed in panels (g,h) of Fig.~\ref{fig:prop_distribution}. The dependence with the virial mass at different redshifts is shown in panels (f,g) of Fig.~\ref{fig:prop_mass}, from where we see how, at a fixed mass, low redshift halos are inclined to be more rounded with oblate-shapes, in agreement with previous results \citep[see e.g.,][]{2006ApJ...646..815S}. The behavior of these scaling relations are linked to the merging history of halos \citep[see e.g.,][]{2006MNRAS.367.1781A}. The exact details on the connection between mergers and its impact on halo properties are to be addressed in forthcoming research.
%------------------------------------------------------------------------
\subsubsection{Peak height}\label{sec:peak}
We assign to each halo a value of peak height $\nu(M,z)=\delta_{sc}/\sigma(M,z)$, where $\delta_{sc}=(3/20)(12\pi)^{2/3}$ is the critical collapse threshold  linearly extrapolated to redshift $z=0$ and 
\be \label{eq:sigma}
\sigma^{2}(M,z)=\frac{1}{2\pi^{2}}\int_{0}^{\infty}\dd k k^{2}P(k,z)|W(k;R)|^{2}
\ee
is the variance in the matter distribution on scales $R\propto (M_{\rm vir}/\bar{\rho})^{1/3}$ \citep[see e.g.,][]{1980lssu.book.....P}. $P(k,z)$ represents the linear matter power spectrum at the corresponding redshift (obtained from the power spectrum defining the initial conditions of the UNITSim, scaled by the corresponding growth factor \citep[][]{1977MNRAS.179..351H}). $W(k;R)$ is the Fourier transform of a top-hat filter with scale $R$ and $\bar{\rho}(z)$ is the mean matter density at redshift $z$. In this work, we use $\log \nu$ as a primary halo property, that can be exchanged with the halo mass according its definition.  The mass scale set by $\nu(M_{\star},z)=1$ defines the mass of collapsing halos at redshift $z$. At $z=0$, $M_{\star}\sim 3.1\times 10^{12}M_{\odot} h^{-1}$.
The peak height $\nu$ is the natural variable in which the peak-formalism can predict halo abundances and bias \cite[][]{1984ApJ...284L...9K,1996MNRAS.282..347M,1999MNRAS.308..119S}. These quantities display approximately a universal shape (independent of mass and redshift) when expressed in terms of it \citep[][]{2008ApJ...688..709T,2010ApJ...724..878T,2011ApJS..195....4M}. Also, trends in scaling relations (such as the mass-concentration relation,\S\ref{sec:mconc}) can be well described with the same functional form in terms of this parameter \citep[see e.g.,][]{2019ApJ...871..168D}. Importantly, it is common to assume that the redshift evolution of secondary bias scales with $\nu(M,z)$ (so it vanishes at fixed $\nu(M,z)$ \citep[see e.g.,][]{Wechsler2006, Gao2007}). Peak height $\nu(M,z)$ is, therefore, a cornerstone for theoretical models of halo bias, including its secondary dependencies.
%=================================================================================================
%==================================================================================================
%==========================================================================
%==========================================================================

%==================================================================================
\subsection{Nonlocal halo properties}
The following set of properties are computed based on the halo distribution on small scales. Throughout this paper we refer to these as ``nonlocal'', as they are computed based on the phase-space properties of tracers enclosed on spheres of radius $R=5 $Mpc $h^{-1}$ around each halo.  As such, this set can provide information  of the small-scale clustering of dark matter haloes.

%==================================================================================
\subsubsection{Relative Mach number}\label{sec:mach}
The Mach number $\mathcal{M}_{R}$, introduced in the cosmological context by \citet[][]{1990ApJ...348..378O}, is a measure of the ``temperature'' of the cosmic fluid. In its original definition \citep[see also][]{1990ApJ...348..378O,1992ApJ...395....1S,2012JCAP...06..026M,2013MNRAS.432..307A,2022MNRAS.512...27M}, this quantity measures the amount of bulk motion of patches of the Universe (characterized by a scale $R$) with respect to the velocity dispersion of the random motions inside those regions. This parameter contains cosmological information, as it can used to probe the shape of the matter power spectrum \citep[][]{1998ApJ...494...20S}. It has also been shown to display correlations with galaxy and halo properties \citep[][]{2001ApJ...553..513N, 2022MNRAS.512...27M}.

In this work, we define ``relative Mach number'' by computing the relative velocity $\vec{v}_{ij}=\vec{v}_{i}-\vec{v}_{j}$ of a dark matter halo (labeled $i$) with peculiar velocity vector $\vec{v}_{i}$, and its neighboring halos with velocities $\vec{v}_{j}$ inside spheres of radius $R$, namely:
\be\label{eq:mach}
\mathcal{M}^{(i)}_{R}\equiv \frac{\langle |\vec{v}_{ij}|\rangle_{j,R} }{\sigma_{i}},
\ee
where $\langle \rangle_{j,R}$ denotes averages inside spheres of radius $R$ and $\sigma_{i}$ is the dispersion in the variable $|\vec{v}_{ij}|$ inside those spheres. As in the original definition of the Mach number, high (low) values of $\mathcal{M}^{(i)}_{R}$ can be associated with warm $\mathcal{M}^{(i)}_{R}\geq1$ (cool, $\mathcal{M}^{(i)}_{R}<1$) halo environments. Under the assumption of no velocity bias (i.e, dark matter halos and dark matter particles share the same velocity field), this measures the ''local temperature'' of the dark matter density field. In panel (i) of Fig.~\ref{fig:prop_distribution} we present the distribution of local halo Mach number at $z=0$ for different cosmic-web types. In addition, panel (h) of Fig.~\ref{fig:prop_mass} displays the mean scaling relation between halo mass and Mach number at different redshifts. The trends at various redshift imply that tracers live in ``warm'' environments, and the scaling relations can in principle be described by two different power-laws with a break at mass scales varying with redshift, as well at the overall amplitude (the high-redshift population of halos are warmer than low-redshift one). In particular, at $z=0$, the relative Mach number for masses below $\sim 10^{13}M_{\odot}/h$ varies little. The information encoded in Fig.~\ref{fig:dist_cwt_dm} reveals that the halo population, up to the mentioned mass scale, is mainly living in filaments; above that mass, the fraction of halos living in knots dominate the halo sample. Hence, the relative Mach number varies faster with mass in high density regions. This is also seen in panel (h) of Fig.~\ref{fig:prop_mass_cwt}, where the scaling relation is split in different cosmic-web environments (to be defined in \S\ref{sec:cwc}).

%==================================================================================
\subsubsection{Neighbor statistics}\label{sec:nei}
Along with the relative Mach number previously defined, we introduce the ``neighbor statistics'' $\mathcal{D}_{R}$, as a probe for the statistics of pair separations around each tracer. This estimator is computed as
\be\label{eq:dach}
\mathcal{D}^{(i)}_{R} \equiv  \frac{ \langle \lambda_{ij} \rangle_{j,R}}{\sigma_{i,\lambda}},
\ee
where $\langle \lambda_{ij} \rangle_{j,R}$ is the mean separation of halos at a position $\vec{r}_{j}$ and within spheres of radius $R$ from the main tracer, while $\sigma_{\lambda_{R}}$ is its population variance. The neighbor statistics aims at compressing the two first moments of the distribution of pair separations, and as such, is expected to contain information on the small-scale clustering of halos.
{Panel (k) of Fig. \ref{fig:prop_distribution} shows the distribution of $\mathcal{D}_{R}$ across the halo population at $z=0$ in different cosmic-web environments, while panel (j) in Fig.~\ref{fig:prop_mass} displays the mean scaling relation between halo mass and $\mathcal{D}$ at different redshifts. Similarly, panel (j) in Fig.~\ref{fig:prop_mass_cwt} shows (at $z=0$) the scaling relation as a function of cosmic-web environments (to be defined in \S\ref{sec:cwc}), evidencing a quantitative difference between this statistics in low (voids) and high (density regions), specially at low halo masses. 

Similar quantities have been explored in the context of secondary bias, such as a scale dependent correlation
\citep[][]{2018A&A...615A.109K} or ``neighbor distance'' \citep[][]{Salcedo2022}. The impact of massive neighbors in the signal of large-scale bias will be addressed in a forthcoming publication.

%------------------------------------------------------------------------
\subsubsection{Halo overdensity}
We compute a local halo overdensity $\Delta_{R}$ by collecting the number of tracers $N_{R}$ used to measure $\mathcal{M}_{R}$ (or $\mathcal{D}_{R}$) and defining $\Delta_{R}=\log (N_{R}/\bar{N})$, where $\bar{N}$ is the expected number of randomly distributed tracers in spheres of volume $\propto R^{3}$. The distribution of this variable within the halo population at $z=0$ is shown in panel (j) of Fig.~\ref{fig:prop_distribution}, while its dependency with the halo mas is shown in panel(i) of Fig.~\ref{fig:prop_mass}, reflecting the expectation that more massive tracers tend to be surrounded by more neighbors compared to low-mass halos.

%==========================================================================
%==========================================================================
\begin{figure}
\includegraphics[trim = .0cm 0.2cm 0cm 0cm ,clip=true, width=0.49\textwidth]{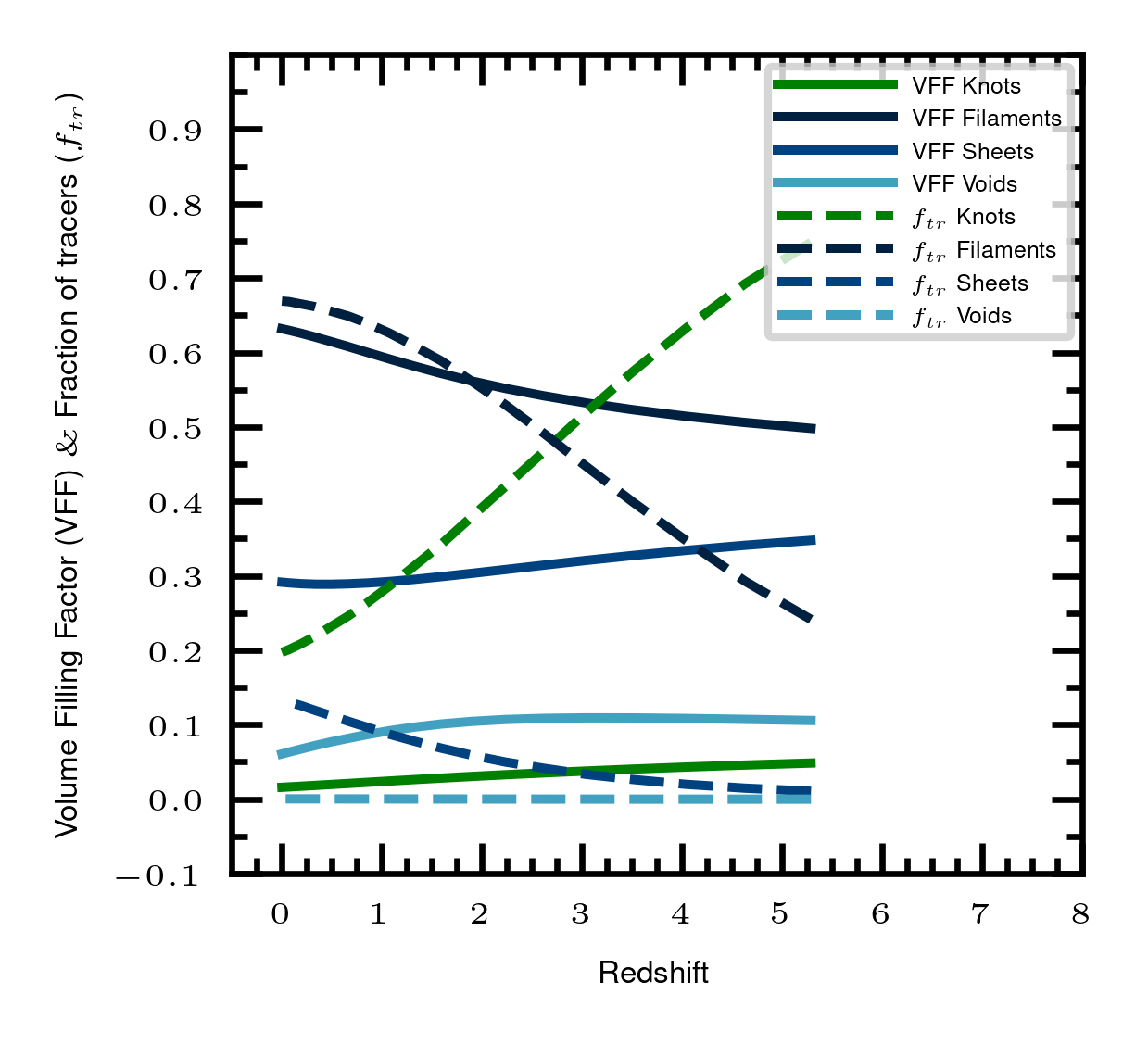}
\caption{\small{Volume filling fraction (VFF) and the fraction of tracers $f_{\rm tr}$ in each cosmic-web type defined in \S\ref{sec:cwc} as a function of redshift.}}
\label{fig:vff_cwt}
\end{figure}
%==================================================================================================
%==================================================================================================

%==================================================================================
%==========================================================================
%==========================================================================
\begin{figure}
\includegraphics[trim = .0cm 0.2cm 0cm 0cm ,clip=true, width=0.5\textwidth]{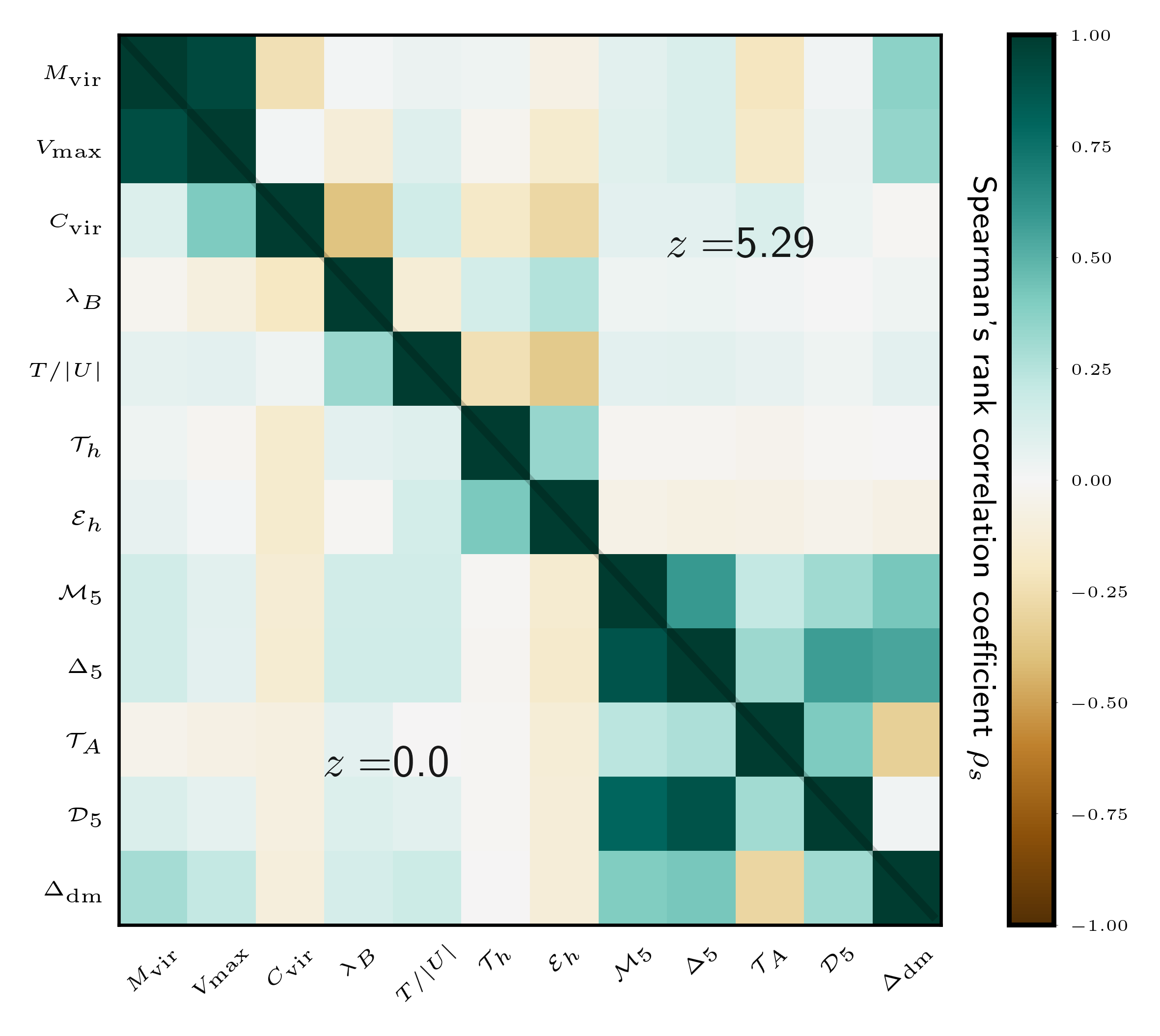}
\caption{\small{Spearman's rank correlation coefficient $\rho_{s}$ between the halo effective bias and a number of halo properties for two different redshifts.}}
\label{fig:spear1}
\end{figure}
%==========================================================================
%==========================================================================

%==========================================================================
%==========================================================================
\begin{figure}
\includegraphics[trim = .0cm 0.2cm 0cm 0cm ,clip=true, width=0.5\textwidth]{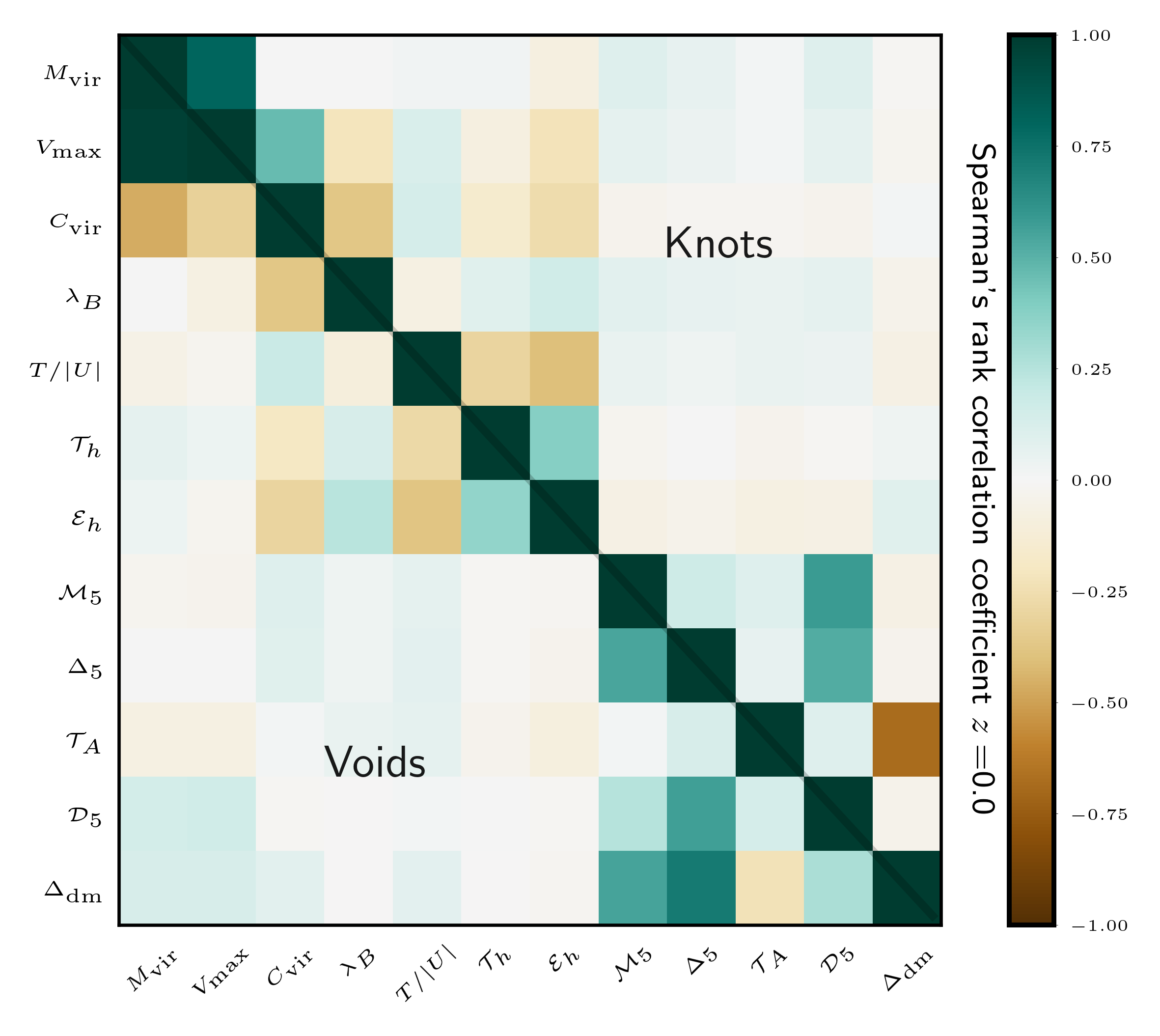}
\caption{\small{Spearman's rank correlation coefficient $\rho_{s}$ between the halo effective bias and a number of halo properties at $z=0$. The upper diagonal shows the correlation computed in knots, lower diagonal in voids.}}
\label{fig:spear_cwt}
\end{figure}
%==========================================================================
%==========================================================================

\subsection{Environmental properties}\label{sec:envprops}

By ``environmental properties'', we refer to quantities computed from the underlying dark matter field from which the halo catalog has been built. These properties can be classified as local, such as the dark matter density in a cell $\delta_{\rm dm}$\footnote{For clarity, in the presentation of some results, we use the label $\Delta_{\rm dm}=\log (1+\delta_{\rm dm})$.}, and nonlocal, like the tidal field \citep[see e.g.,][]{2020MNRAS.491.2565B}. The panel (k) of Fig.\ref{fig:prop_mass} shows the link between halo mass and the local overdensity $\Delta_{\rm dm}$ at different redshifts.

%==================================================================================
%==================================================================================
\begin{figure*}[htbp]
\centering
%scaling_relations_cwt_zfixed.py v1
%scaling_relations_cwt_zfixed_SHARED.py
\includegraphics[trim = 0.2cm 0.2cm 0cm 0cm ,clip=true, width=\textwidth]{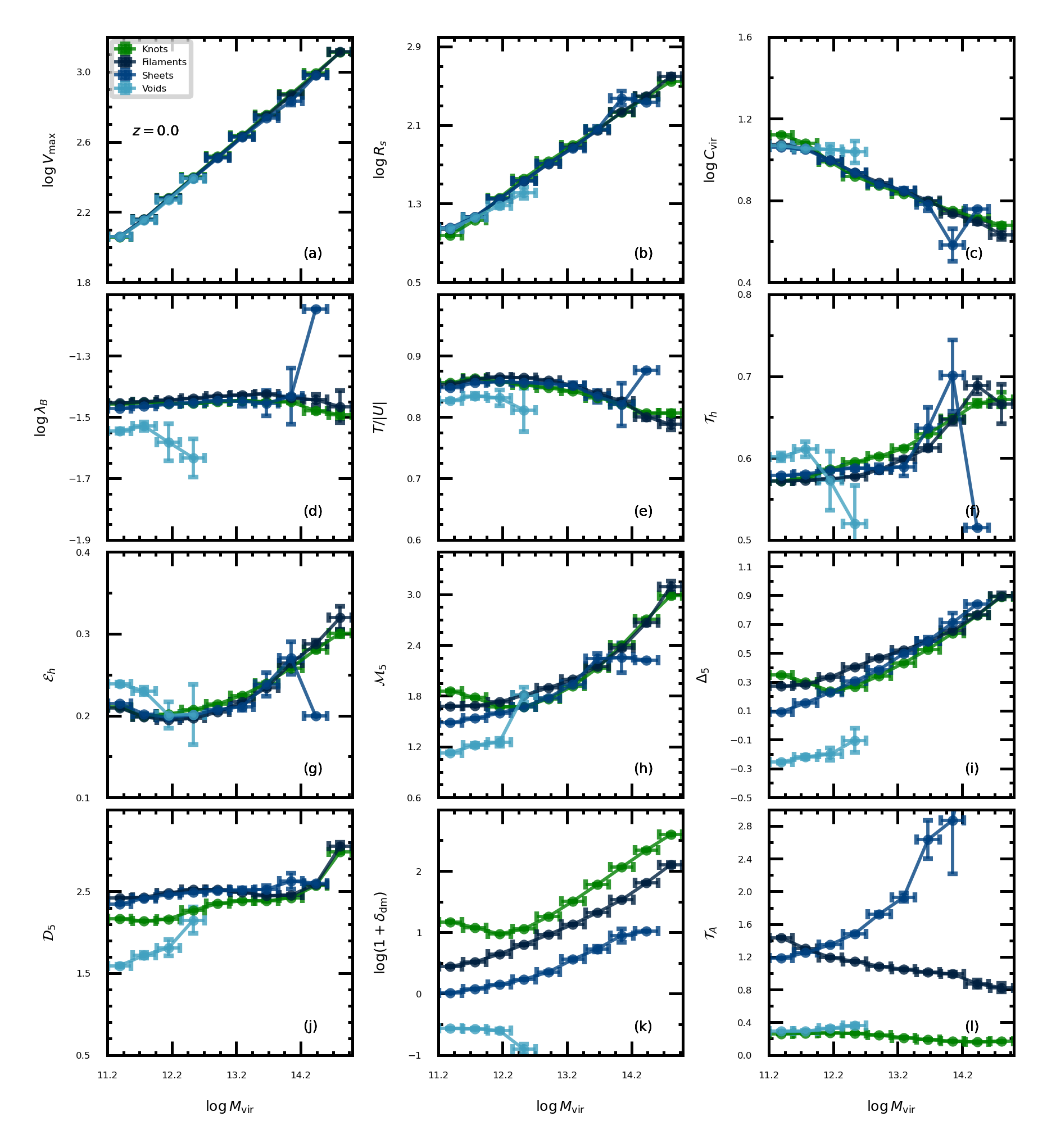} 
\caption{\small{Mean scaling relations between the halo virial mass and a number of halo properties described in \S\ref{sec:hprops}, at $z=0$ evaluated in different cosmic-web types. Shown intrinsic properties are $V_{\rm max}$ (panel a), $R_{s}$ (panel b) $C_{\rm vir}$ (panel c), $\lambda_{B}$ (panel d), $T/|U|$ (panel e), $\mathcal{T}_{h}$ (panel f), $\mathcal{E}_{h}$ (panel g). Nonlocal properties: local Mach number $\mathcal{M}_{5}$ (panel h), $\Delta_{5}$ (panel i), $\mathcal{D}_{5}$ (panel j). Environmental properties:  local dark matter density (panel k) and tidal anisotropy $\mathcal{T}_{A}$ (panel l). The error bars denote the error in the mean in each mass bin.}}\label{fig:prop_mass_cwt}
\end{figure*}
% ==================================================================================
% ==================================================================================

%==================================================================================
\subsubsection{Cosmic-web classification} \label{sec:cwc}

The cosmic web is a well recognized pattern \citep[see e.g.,][]{GottIII_2005} arising from the dynamical evolution of density fluctuations. Several methods have been designed to define 
different environments within the cosmic structure \citep[see e.g.,][and references therein]{2000A&A...363L..29S,2010MNRAS.407.1449G, 2010MNRAS.408.2163A,2011MNRAS.414..384S,2018MNRAS.473.1195L}. Here we use the approach by \citet[][]{2007MNRAS.375..489H} and compute the Hessian of the gravitational potential (or tidal field) $\mathcal{T}_{ij}=\partial_{ij}\Phi$ (where $\Phi$ is the comoving gravitational potential obtained from Poisson's equation $\nabla ^{2}\Phi=\delta_{\rm dm}$) \footnote{We compute the tidal field directly from the overdensity field $\delta_{\rm dm}(\vr)$, as the inverse Fourier transform $\mathcal{T}_{ij}=F^{-1}[(k_{i}k_{j})\delta_{\rm dm}(\vk)/k^{2}]$ using the \texttt{FFTW} algorithm \citep[][]{FFTW98}. The eigenvalues have been computed using the \texttt{gsl} public library \citep[][]{contributors-gsl-gnu-2010}.} and its eigenvalues $\lambda_{i},(i=1,2,3)$ \citep[see e.g.,][]{2007MNRAS.375..489H, 2009arXiv0912.3448V, 2009MNRAS.396.1815F, 2016MNRAS.455..438A,2017ApJ...848...60Y,2018MNRAS.476.3631P} satisfying $\sum_{i}\lambda_{i}=\delta_{\rm dm}$. Using these eigenvalues, each cell is classified as part of a {\it{knot}} ($\lambda_{1}>\lambda_{\rm th}$, $\lambda_{2}>\lambda_{\rm th}$, $\lambda_{3}>\lambda_{\rm th}$), {\it{filament}} ($\lambda_{1}<\lambda_{\rm th}$, $\lambda_{2}>\lambda_{\rm th}$, $\lambda_{3}>\lambda_{\rm th}$), {\it{sheet}} ($\lambda_{1}<\lambda_{\rm th}$, $\lambda_{2}<\lambda_{\rm th}$, $\lambda_{3}>\lambda_{\rm th}$), or {\it{void}} ($\lambda_{1}<\lambda_{\rm th}$, $\lambda_{2}<\lambda_{\rm th}$ and $\lambda_{3}<\lambda_{\rm th}$), where $\lambda_{\rm th}$ is an arbitrary threshold. We use $\lambda_{\rm th}=0$, which has the practical advantage of allowing us to associate knots (voids) with over-(under-) dense regions. 
The relevance of the cosmic-web classification resides in the dynamical properties of each environment and the density these are characterized with \citep[][]{1970A&A.....5...84Z,1978SvA....22..653D,1996Natur.380..603B}, as well as the connection with perturbation theory \citep[][]{10.1093/mnras/stac671}. Halo bias has been observed to change within cosmic-web environment \citep[see e.g.,][]{2017ApJ...848...60Y,2019MNRAS.483L..58B,10.1093/mnras/stac671,2022A&A...661A.146B,2023arXiv230915306W} , a claim we support with this paper. In future work we shall also explore the dependencies of halo properties with respect to the geometrical properties of peaks in the density field \citep[][]{Peacock1985}, which has been shown to correlate with halo bias as well \citep[see e.g.,][]{2021ApJ...921...66S}.

In Fig.~\ref{fig:vff_cwt} we show the fraction of cells (with volume defined by the fiducial resolution, see \S\ref{sec:sim}) in the UNITSim volume (i.e., the volume filling fraction, VFF) classified in the different cosmic-web environments, together with the fraction of tracers, $f_{tr}$, which, at each redshift, are located in cells classified accordingly. The effects of gravitational evolution and the formation of the cosmic-web is condensed in this plot: as the Universe evolves, less cells are classified as knots, while the fraction occupied by filaments increases (filaments are, as expected, the dominant component across redshift). The trends in the number of tracers follow that of the VVF, except for sheets: while the fraction of cells classified as such decreases toward low redshift, the fraction of tracers in such environments rises. This as a consequence of the density regimes probed by sheets (see Fig.~\ref{fig:dist_cwt_dm}) which reveals the interplay between the growth of voids ($\delta_{\rm dm}<0$) and the formation of halos in regions with $\delta_{\rm dm}>0$.

%==================================================================================
\subsubsection{Tidal anisotropy}\label{sec:tidal}

Based on the set of eigenvalues $\{\lambda_{i}\}$ defined in \S\ref{sec:cwc}, the degree of anisotropy of the dark matter density field can be quantified by computing geometrical quantities such as the ellipticity $e=(\lambda_{1}-\lambda_{3})/2\delta_{\rm dm}$ or the prolateness $p=(\lambda_{1}+\lambda_{3}-2\lambda_{2})/2\delta_{\rm dm}$. We can also use the invariants of the tidal field, which have been shown to correlate with halo statistics at the level of number counts \citep[see e.g.,][]{2019MNRAS.483L..58B,10.1093/mnras/stac671}. In this work, we compute the so-called tidal anisotropy parameter
\be
\mathcal{T}_{A}\equiv \frac{\sqrt{(\lambda_{1}-\lambda_{2})^{2}+(\lambda_{1}-\lambda_{3})^{2}+(\lambda_{3}-\lambda_{2})^{2}}}{2(2+\lambda_{1}+\lambda_{2}+\lambda_{3})}.
\ee
Notice that this definition differs from others often found in the literature e.g., \citet[][]{2009MNRAS.398.1742H,2018MNRAS.476.3631P, 10.1093/mnras/stt141,2019MNRAS.489.2977R,2022MNRAS.509.1614F}, where a tidal anisotropy is computed based on the halo distribution and thus measures the level of anisotropy of such field (smoothed on a certain scale). Also, notice that in the denominator of the equation above we use a factor $2+\delta_{\rm dm}$ (instead of $1+\delta_{\rm dm}$), in order to avoid divergences. According to the threshold chosen for the eigenvalues $\lambda_{i}$, filaments and sheets are expected to display higher values of tidal anisotropy compared to knots and voids. Hence, very high-mass halos, mainly found in high density regions (knots), are expected to reside in regions with low anisotropy, as can be confirmed by panel (l) of Fig.\ref{fig:prop_distribution} and panel (l) of Fig.~\ref{fig:prop_mass}.

%==================================================================================

\subsection{Correlations among halo properties}\label{sec:corr}

We assess the level of correlation among the different halo properties using the Spearman's rank correlation coefficient. As an example, in Fig.~\ref{fig:spear1} we present the correlation between all properties at high and low redshift ($z=5$ and 0, respectively). The highest correlations are found, as expected, among  properties probing the depth of the halo potential well (e.g virial mass, maximum circular velocity), as well as the correlation among environmental properties (e.g., Mach number, neighbor statistics, tidal anisotropy). The transition from high to low redshift depicted in this figure shows how the correlation between intrinsic and environmental properties increases with decreasing redshift, a change that is evident for properties such as halo concentration and spin, as can be also seen from panels (c) and (d) Fig.~\ref{fig:prop_mass}.

Halo properties display, in general, a different degree of variability across the halo population. In appendix \ref{sec:pca}, we perform a principal component analysis across redshift to demonstrate that even though intrinsic properties linked to the depth of the halo potential well (mass, maximum circular velocity) contain most of the variability (understood here as information), nonlocal and environmental properties can contain similar levels of variability. In that line, halo properties such as spin, concentration and geometry can be regarded as ``secondary'' due to their lower degree of variability. Statistically, such a low variability can be associated with a large scatter in their link with, for example, halo mass, as can be inferred from Fig.~\ref{fig:spear1}.

%==================================================================================
\subsection{Scaling relations' conditional to cosmic-web environments}

Based on the cosmic-web classification described in \S\ref{sec:cwc}, we assess now the level of correlation and the statistical behavior of the halo scaling relations in different cosmological environments. Recall, first, that the panels in Fig.~\ref{fig:prop_distribution} display the distribution of halo properties (intrinsic and nonlocal) evaluated in different cosmic-web environments, with the noticeable feature that these distributions are much more sensitive to under-dense regions (voids). To complement these results, Fig.~\ref{fig:spear_cwt} shows the Spearman's rank correlation coefficient $\rho_{s}$ at $z=0$ in extreme environments, i.e, voids and knots, probing under- and over-dense regions, respectively. This figure depicts again the two main regions in the parameter space (intrinsic and environmental properties) displaying the largest correlations. The variations in the correlation coefficient in these two cosmic-web types are a signature that the links among halo properties can be sensitive to the cosmological environment they live in. In Fig.~\ref{fig:prop_mass_cwt} we present the mean scaling relations between halo mass and other properties at $z=0$, evaluated in different cosmic-web types. Some conclusions based on this figure are (we have verified that these conclusions are representative of the behavior at different cosmological redshifts):
\begin{itemize}
\item Properties probing the depth of the halo potential well (e.g., halo mass, maximum circular velocity) are linked through a rather stable scaling relation against changes in the cosmic web type, as shown in panel (a). We have checked that this is also the case for the velocity dispersion.

\item Statistically significant differences among scaling relations of halo secondary properties are mainly found in voids, as can be seen for the scale-radius $R_{s}$ (panel b), the halo concentration $C_{\rm vir}$ (panel c) and the spin $\lambda_{B}$ (panel d). In particular, halo spin scaling relation not only displays a lower amplitude in voids (i.e, low-mass halos in voids have lower spin than halos of the same mass in higher density environments) but also a different trend (higher mass halos in voids tend to have lower spin compared with halos of the same mass in knots) compared to other cosmic-web classifications. 

\item The ratio $T/|U|$ computed for tracers in voids is lower than those values seen in high density region, as shown in panel (e) of Fig.~\ref{fig:prop_mass_cwt}. If we assume that relaxation is represented as $T\to 2|U|$, this can be interpreted as a signal that halos in voids are more relaxed as these tracers are less prone to growing by merging.

\item  The dependency of ellipticity and triaxiality with halo mass are not (statistically speaking) sensitive to the cosmic-web type, except, again, in voids: halos with masses up to $\sim10^{12}M_{\odot}h^{-1}$ in such low-density environments tend to display higher ellipticity and triaxiality (i.e, halos tend to be more prolate) than halos of the same mass in other environments, as shown in panels (f, g) of Fig.~\ref{fig:prop_mass_cwt}.

\item The mean scaling relations between nonlocal properties and halo mass also display (statistically) large differences among cosmic-web types. In particular, halos in voids found themselves in a rather cool environment compared with other cosmic-web types, as shown in (panel h). Also, voids are sparsely populated, as shown by the local halo overdensity (panel i) and  the neighbor statistics (panel j) and in regions with a mean halo density below the mean (local overdensity).

\item The comparison between the $z=0$ curve in panel (k) of Fig.~\ref{fig:prop_distribution} and panel (k) of Fig.~\ref{fig:prop_mass_cwt} shows how the population of halos, expressed through their halo mass, is sensitive to the cosmic-web environment. This is a key aspect exploited in bias mapping techniques aimed at generating mock catalogs \citep[see e.g.,][]{2023A&A...673A.130B}.

\item Environmental properties such as local dark matter density (panel k) and tidal anisotropy (panel l) show large differences, which is a trivial consequence of their dependence on the cosmic-web classification. In particular, sheets and filaments are characterized by a large anisotropy. 
\end{itemize}

These conclusions are in agreement with previous analysis \citep[see e.g.,][]{2006ApJ...646..815S,2021PhRvD.103f3517H}, and highlight the importance of the environment in shaping the link between different halo properties. The behavior of the halo scaling relations might not only change due to the environment, but also through the underlying cosmological model \citep[see e.g.,][]{2018A&A...619A.122H}. Also, halo properties and scaling relations have been shown not only to correlate with the cosmic environments, but also with properties of its components such as the thickness of filaments \citep[][]{10.1093/mnras/stz441,2020A&A...641A.173G,2021MNRAS.503.2280G}, or the mass of collapsing regions defined as knots \citep[][]{2015MNRAS.451.4266Z, 2023A&A...673A.130B,2021MNRAS.500.2417D}

As a general remark, we can argue that the differences seen in the scaling relations, specially those linked to spin and geometry, can be understood as a signature of the coupling between halo dynamical properties and the tidal field, as suggested by the tidal-torque theory \citep[see e.g.,][]{1970Afz.....6..581D,1984ApJ...286...38W,2002MNRAS.332..325P}\citep[see also e.g.,][for an analysis of spin in the cosmic-web]{2012MNRAS.427.3320C,2013ApJ...762...72T,2014MNRAS.440L..46A}. Nevertheless, there could be other evolutionary effects (aside the growth in the strength of such coupling) that influence the shape of the scaling relations. We remind the reader that our results are based on a very particular cosmic-web classification, and in general, the properties of halos in particular environments such as  voids can vary when using different techniques to define them \citep[see e.g.,][]{2008MNRAS.387..933C,2018MNRAS.476.3195C}.  

A deep understanding the halo scaling relations demands a complex analysis and is beyond the goal of this paper, as it deserves a project on its own. We shall pursue this in future work. In the coming sections we focus on the main subject of this paper, namely,  the dependencies of large-scale halo bias on the halo properties and the signal of secondary bias.

%==========================================================================
%==========================================================================
\begin{figure*}
%density_fields_UNITSim.py
\includegraphics[trim = .0cm 0.2cm 0cm 0cm ,clip=true, width=1.03\textwidth]{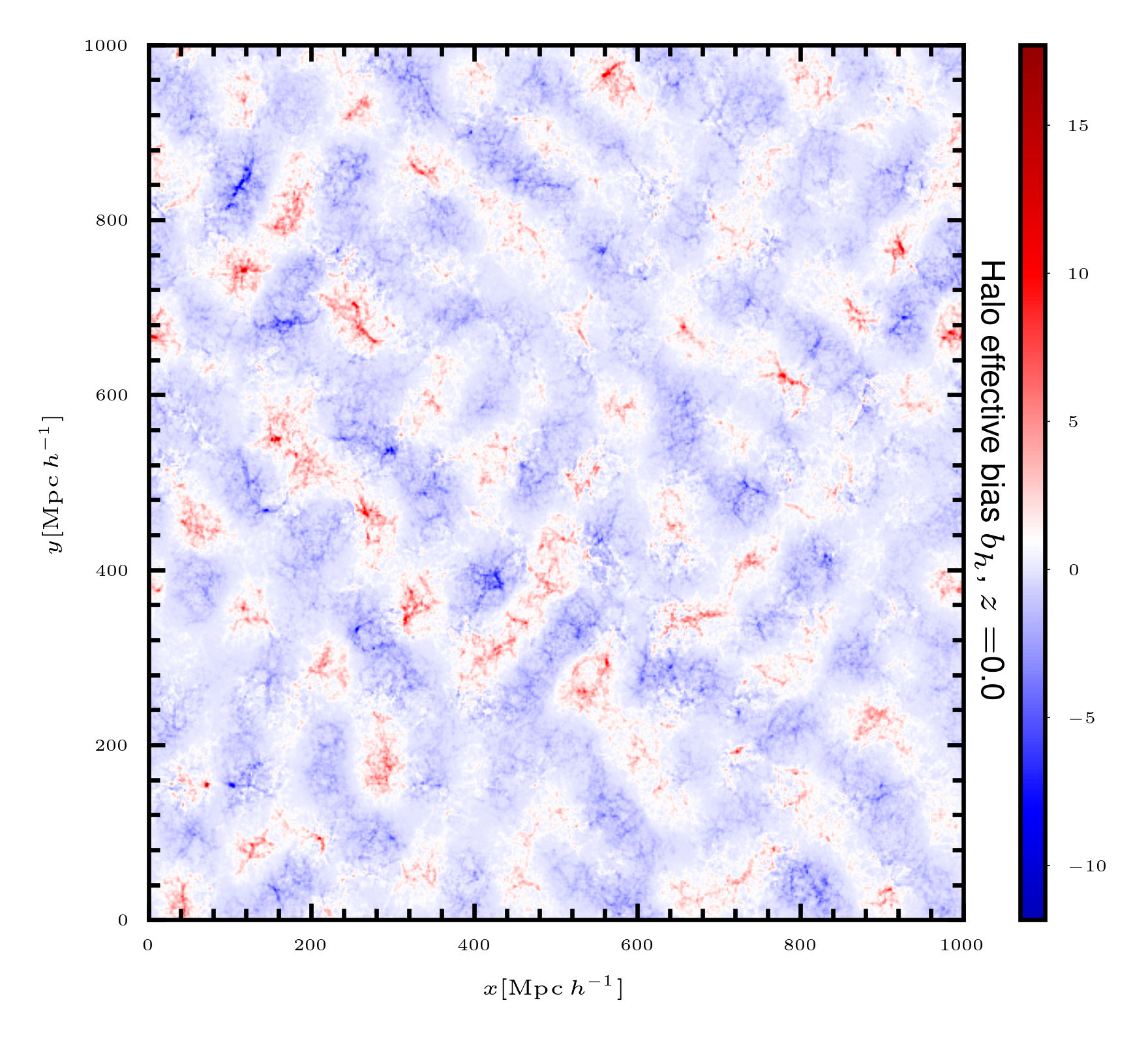}
\caption{\small{Slice of thickness $\sim 80$ Mpc $h^{-1}$ of the UNITSim (at $z=0$) showing the spatial distribution of the effective halo bias, as obtained form the interpolation of the bias assigned to each halo according to Eq.(\ref{eq:bias_object}) on $512^3$ mesh. }}\label{fig:bias_field}
\end{figure*}
%==========================================================================
%==========================================================================

\begin{figure*}
%mass_bias_scaling_relation.py
\includegraphics[trim = .1cm 0cm 0cm 0cm ,clip=true, width=0.49\textwidth]{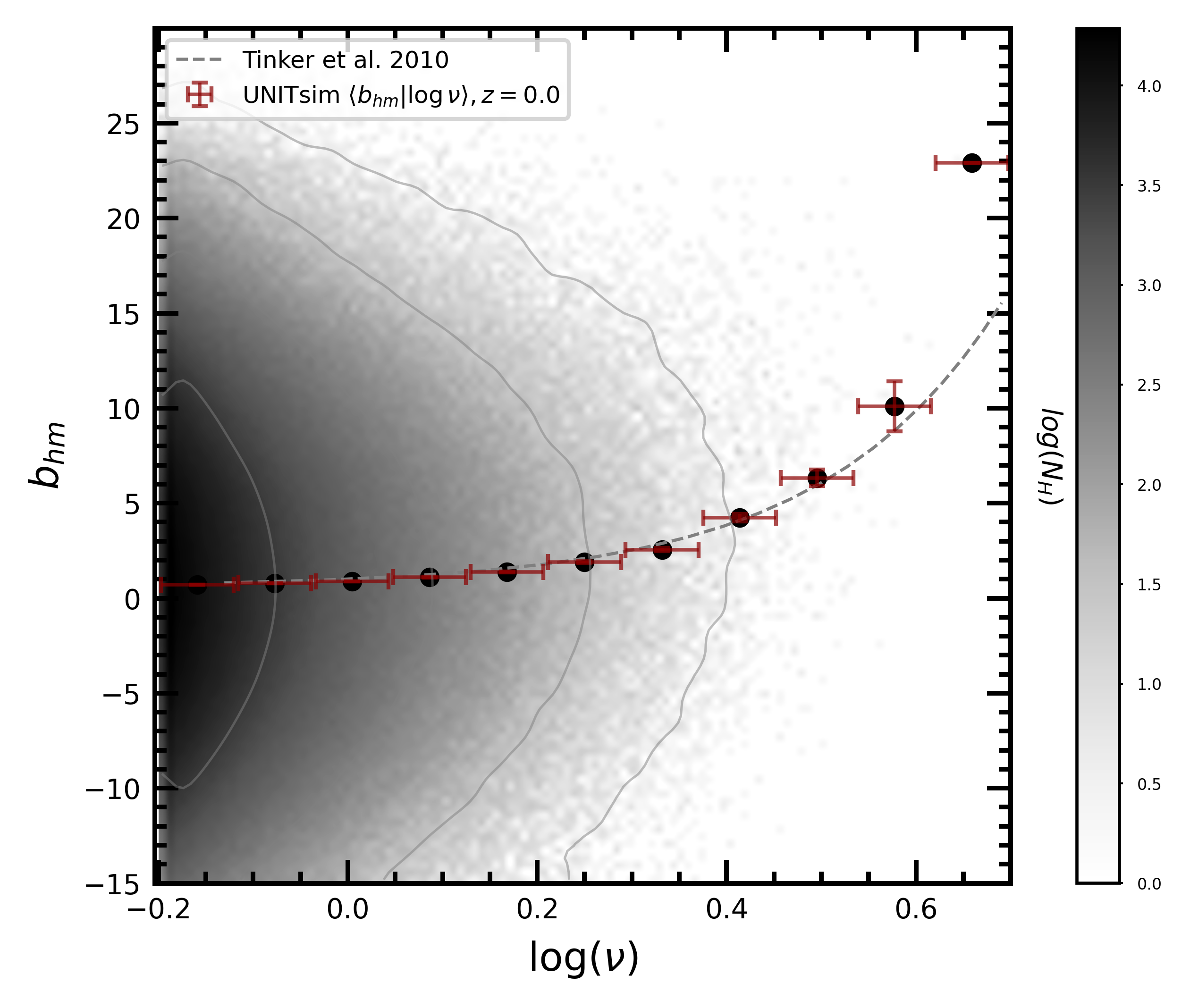}
\includegraphics[trim = .1cm 0cm 0cm 0cm ,clip=true, width=0.49\textwidth]{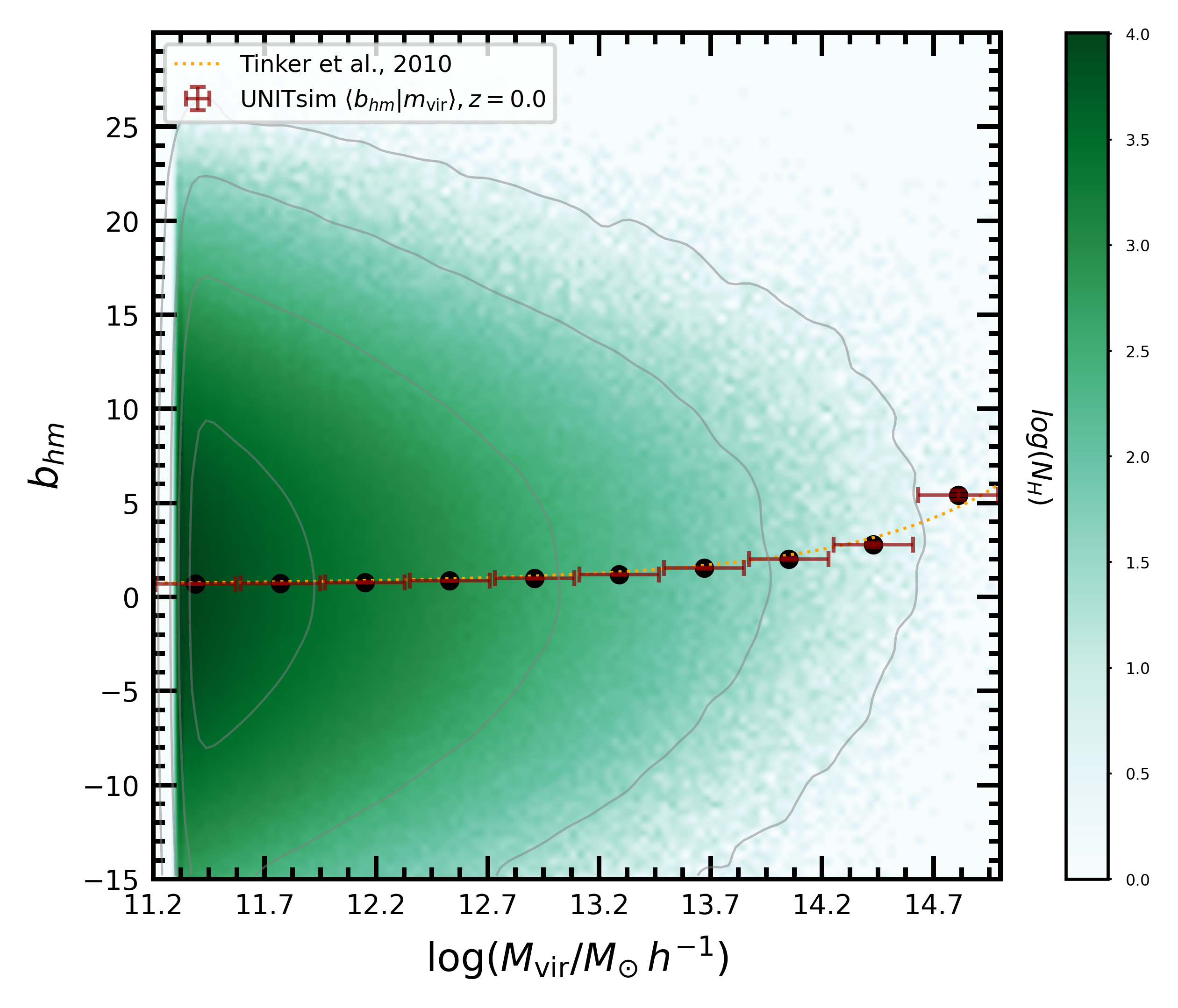}
\caption{Halo effective bias $b_{hm}$ computed with Eq.(\ref{eq:bias_object}) at $z=0$. Left panel: Bias as a function of the halo peak height; the points denote the mean bias in different $\nu$ bins, and the error bars denote the standard error of the mean in each bin. The dotted line shows the prediction of \citet[][]{2010ApJ...724..878T}. Right panel: Halo bias $b_{hm}$ as a function of the halo virial mass. In both cases, the contours indicate a region of an equal number of tracers $N_{H}$.}
\label{fig:bias_nu}
\end{figure*}
%==========================================================================

\section{Halo bias across cosmic time}\label{sec:bias}

\subsection{The large-scale halo bias}

It is very well established that the bias of dark matter tracers needs to be modeled beyond the linear scale-independent scheme \citep[see e.g.,][and references therein]{2018PhR...733....1D}, in which the halo bias $b$ simply relates dark matter $\delta_{\rm dm}$ with halo $\delta_{h}$ overdensities (determined on a mesh characterized by some scale $R$) through an expression of the form $\delta_{h}=b\delta_{\rm dm}$. Scale dependencies induced by the process of halo formation, halo merging history, nonlinear evolution of the dark matter density field \cite[see e.g.,][and references therein]{1999ApJ...525..543M,2000ApJ...540...62S,2001MNRAS.320..289S,2007PhRvD..75f3512S,2007IJMPD..16..763Z,2010ApJ...724..878T,2011A&A...525A..98V,2012MNRAS.420.3469P,2013PhRvD..87h3002S,2013PhRvD..88h3507B,2015MNRAS.450.1486A,2019MNRAS.482.1900H,2019arXiv191012452N}, and the discrete presentation of dark matter halo and matter density fields generalizes the concept of halo bias to a nonlocal and stochastic quantity, characterized by a conditional probability distribution $\mathcal{P}(\delta_{\rm h}|\delta_{\rm dm})_{R}$ \cite[see e.g.,][]{1993ApJ...413..447F,1998ApJ...500L..79T,1999ApJ...520...24D,2000ApJ...544...63B,2005A&A...430..827S}.

The effective large-scale halo bias measured from two-point statistics, even if assessed on scales where linear perturbation theory applies, is not simply the factor $b$ defining a link of the form $\delta_{h}=b\delta_{\rm dm}$. Instead, it is the combination (or renormalization) of linear-order coefficients involved in the perturbative expansion of the overdensity field, aimed to describe the tracer density field, \citep[see e.g.,][]{1998MNRAS.301..797H,2006PhRvD..74j3512M,2009JCAP...08..020M,2018PhR...733....1D,2020MNRAS.492.1614W}. Measurements of higher order effective bias (e.g., quadratic bias) need to be accomplished with other probes such as cross-correlations \citep[see e.g.,][]{2009PhRvD..80f3528S}, higher order statistics \citep[see e.g.,][]{2014MNRAS.440..555P}, direct approaches such as the so-called ``separate universe technique'' \citep[see e.g.,][]{2015MNRAS.448L..11W,2016JCAP...02..018L,2017MNRAS.468.2984P} or effective filed theories of large scale structure \citep[see e.g.,][]{Senatore_2015,2017JCAP...03..059L}.

%========================================================
%========================================================
\begin{figure}
\includegraphics[trim = .0cm 0.1cm 0cm 0cm ,clip=true, width=0.5\textwidth]{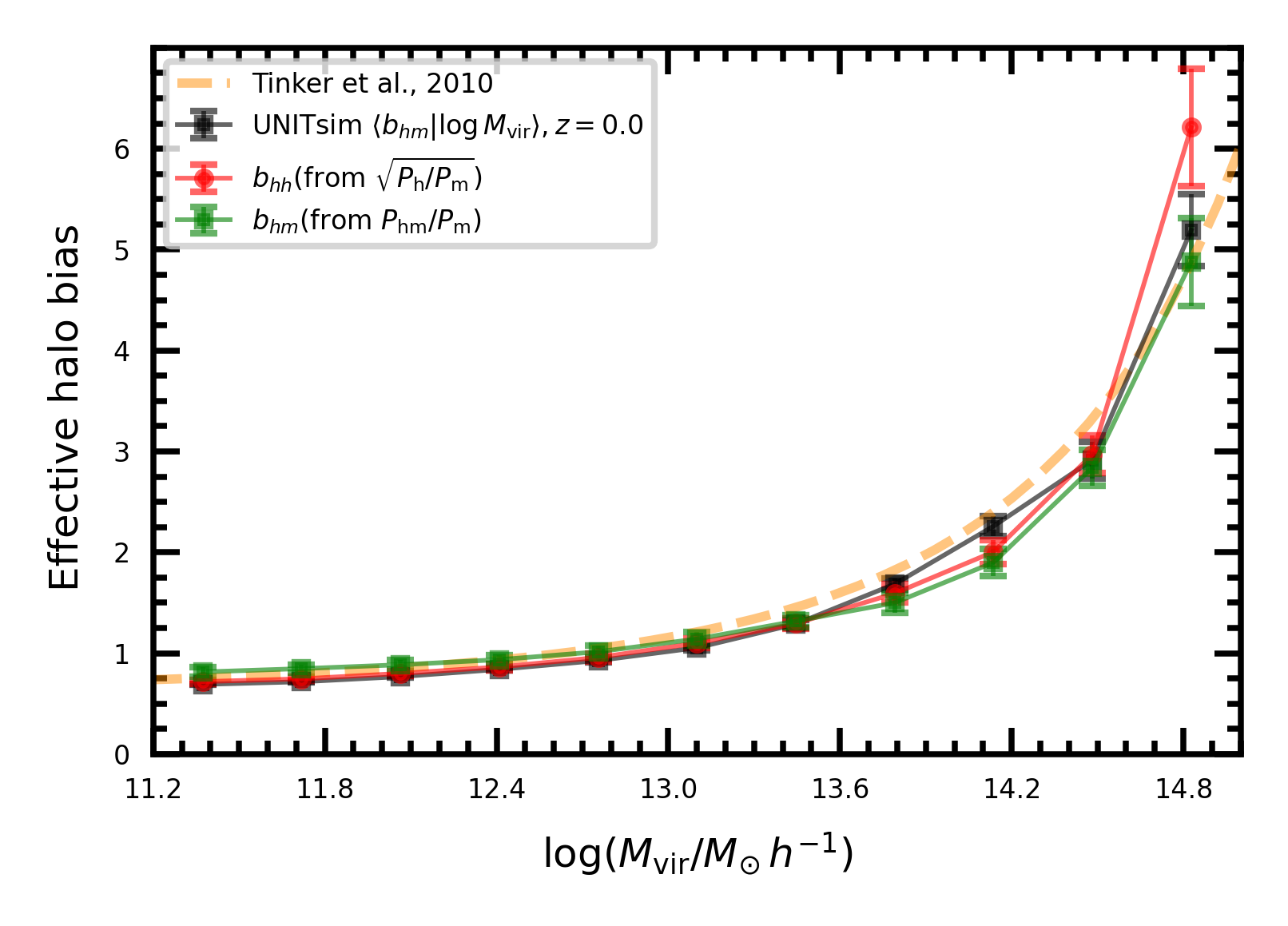}
\caption{Effective halo bias in the UNITSim at $z=0$, measured in bins of the halo viral mass. Symbols denote the mean halo-mass bias $\langle b_{hm}|\log M_{\rm vir}\rangle$ measured with Eq.(\ref{eq:bias_object}), the halo-to-matter ratio $b_{hh}$ and the cross power spectrum $b_{hm}$. The dashed lines corresponds to the prediction of \citep[][]{2010ApJ...724..878T}, computed with an spherical overdensity at virialization of $\Delta_{\rm vir}=330$.}
\label{fig:bias_mass}
\end{figure}
%========================================================
%========================================================

%========================================================
%========================================================
\begin{figure}
\includegraphics[trim = .0cm 0.1cm 0cm 0cm ,clip=true, width=0.5\textwidth]{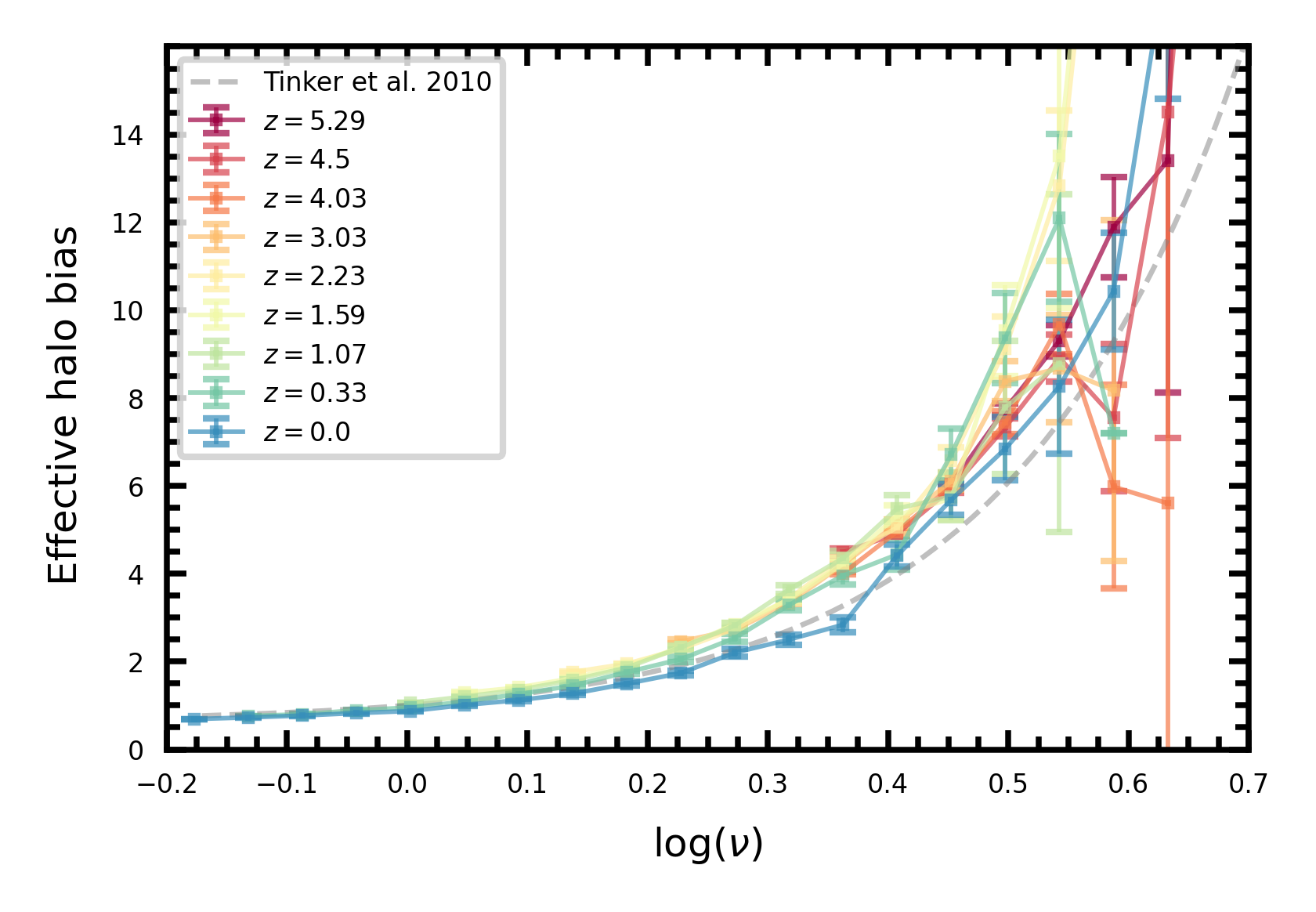}
\caption{Mean halo bias as a function of the peak height  at different redshifts in the UNITSim. The dashed line shows the fitting formula of \cite[][]{2010ApJ...724..878T}. Error bars denote the error of the mean in each mass bin.}
\label{fig:bias_nu_allz}
\end{figure}
%========================================================
%========================================================

%========================================================
%========================================================
\begin{figure}
\includegraphics[trim = .0cm 0.2cm 0cm 0cm ,clip=true, width=0.5\textwidth]{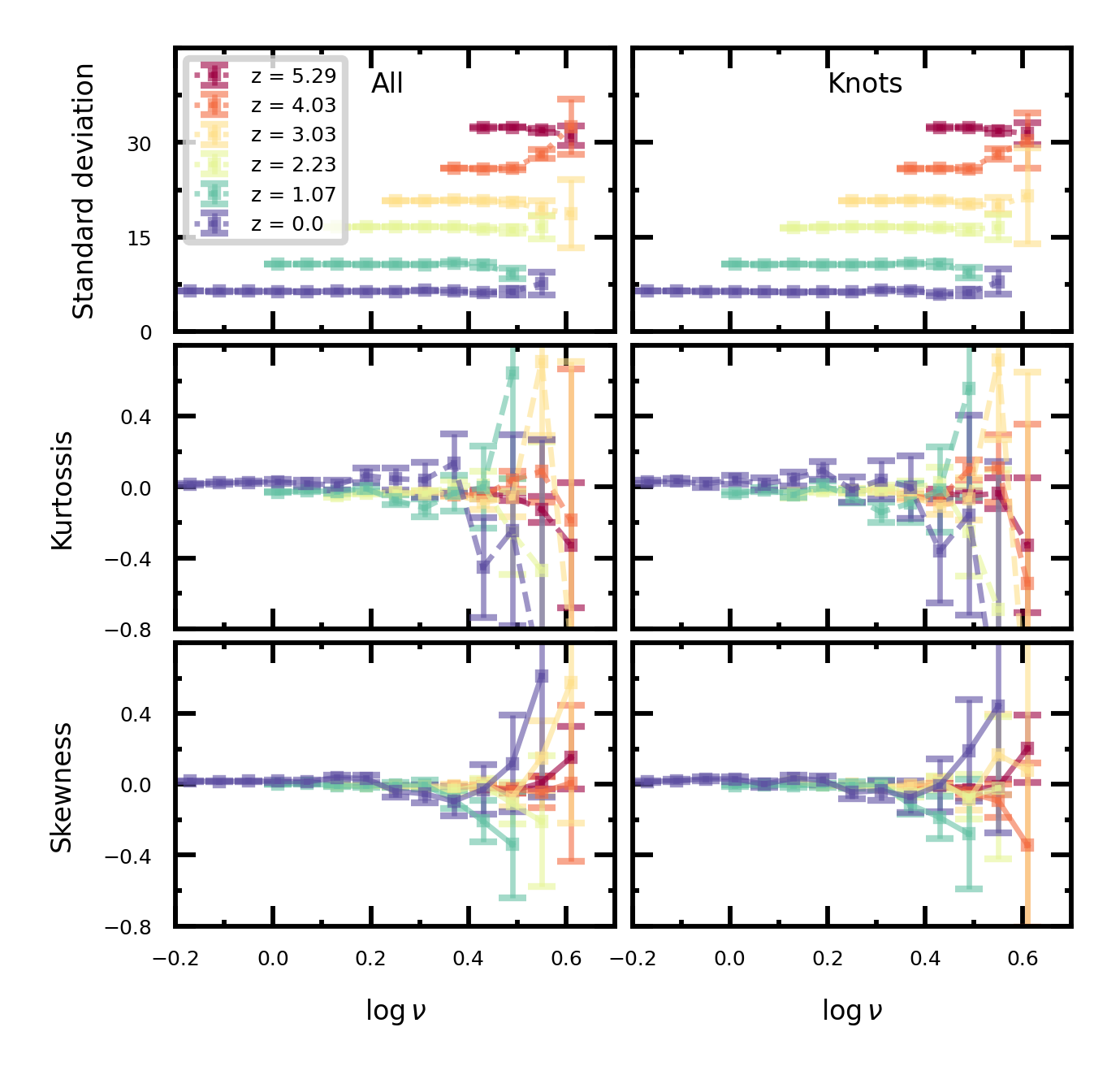}
\caption{Second and third order moments (variance, skewness, and kurtosis) of the halo scaling relation between the effective bias and peak height $P(b_{h}|\nu)$, computed for different redshifts. The right column shows the behavior in high density regions (knots).}
\label{fig:nu_stats}
\end{figure}
%========================================================
%========================================================

\subsection{Estimators of effective bias}\label{sec:individual_bias}

The large-scale effective halo bias can be measured via a number of methods, such as number-count-in-cells \citep[e.g.,][]{1980lssu.book.....P,1994ApJ...433....1B,10.1111/j.1365-2966.2011.18682.x}, or two- and three-point statistics in Fourier \citep[see e.g.,][]{1999ApJ...520..437K, 2007PhRvD..75f3512S,2010ApJ...724..878T, 2012MNRAS.420.3469P} and configuration space \citep[see e.g.,][]{1999ApJ...523...32C}. These methods characterize the clustering strength of a halo population either with respect to the underlying dark matter (absolute bias) or to a reference population (relative bias) \citep[see e.g.,][]{2014A&A...563A.141B}. 
In Fourier space, estimators of the large-scale bias can be based on auto-power spectrum, namely, ratio between halo $P_{hh}(k)$ and dark matter power spectrum $P_{dm}(k)$:
\be\label{eq:bhh_k}
b_{hh}(k_{j})=   \sqrt{\frac{P_{hh}(k_{j})}{P_{\rm dm}(k_{j})}}, 
\ee
or based on the halo-dark matter cross-power spectra $P_{hm}(k)$:
\be\label{eq:bhm_k}
b_{hm}(k_{j})= \frac{P_{hm}(k_{j})}{P_{\rm dm}(k_{j})}, 
\ee
the latter being free of shot-noise corrections and hence in principle more suitable for sparse samples. In both cases, the estimate of effective large-scale bias can be obtained by averaging the values of $b_{hm}(k_{j})$ (or $b_{hh}(k_{j})$) over a range of wavenumbers $k_{j}<k_{max}$ in which the ratio between the halo and the dark matter power spectra is constant, i.e., 
\be
b=\frac{\sum_{j}^{k_{j}<k_{max}} N_{k}^{j} b(k_{j})} {\sum_{j}^{k_{j}<k_{max}} N_{k}^{j}},
\ee
where $N^{j}_{k}$ is the number of Fourier modes in the $j-$th spherical shell. It is important to notice that for a fixed maximum wavenumber, these estimators are not expected to provide the same large-scale bias, as nonlinear evolution affects differently the halo and dark matter field \citep[see e.g.,][]{2007PhRvD..75f3512S,2014MNRAS.440..555P}.

\citet[][]{2018MNRAS.476.3631P} used the bias estimator of Eq.~(\ref{eq:bhm_k}) to write the effective large-scale bias as a sample mean, obtained from a tracer catalog in which each element contributes with an ``object-by-object'' bias of the form
\be\label{eq:bias_object}
b^{(i)}_{hm}=\frac{\sum_{j,k_{j}<k_{max}}N^{j}_{k}\langle {\rm e}^{-i\vk \cdot \vr_{i}} \delta_{\mathrm{dm}}^{*}(\vk) \rangle_{k_{j}}}{\sum_{j,k_{j}<k_{max}} N^{j}_{k}P_{\rm dm}(k_{j})},
\ee
where $\delta_{\mathrm{dm}}(\vk)$ is the Fourier transform of the dark matter density field \citep[see also][]{2019MNRAS.489.2977R,2019MNRAS.482.1900H,2020MNRAS.495.3233P}. The sum is carried over the range of wavenumbers in which the ratio between the halo and the dark matter power spectra is constant. We have taken such limit as $k\leq 0.08\,h$Mpc$^{-1}$. This limit is rather conservative for high redshift, where nonlinear evolution of the halo bias takes place at much smaller scales. At low-redshift ($z\sim 0$), the ratio between halo and dark matter power spectrum is already scale-dependent at $k\sim 0.1\,h$ Mpc$^{-1}$.

The large-scale effective bias of a subsample of $N_{h}(\Delta \theta)$ tracers in a bin of property $\Delta \theta$ can be obtained as a sample mean:
\be \label{eq:allbias}
\langle b_{hm}  | \Delta \theta \rangle =\frac{1}{N_{\rm h}(\Delta \theta)}\sum_{i \in \Delta \theta}^{N_{h}(\Delta \theta)}b^{(i)}_{hm }.
\ee
where $N_{h}(\Delta \theta)$ denotes the number of halos in the  $\Delta \theta$ bin. The effective bias estimator of Eq.~(\ref{eq:bias_object}) has a number of advantages, some of which will be mined in this paper. The first of them is that, by assigning a bias to each object, we can statistically analyze the effective bias in terms of different (primary or secondary) halo properties without the explicit need to divide the sample in bins and compute two-point statistics (correlation function or power spectra). It also opens the possibility to perform tests on models of redshift space distortion and applications in the reconstruction of halo properties for mock catalogs (Balaguera-Antolínez et al, in prep). Importantly, Eq.~(\ref{eq:bias_object}) has been already implemented in the context of both halo and galaxy assembly bias \citep[see e.g.,][]{2018MNRAS.476.3631P,2019MNRAS.489.2977R,2021MNRAS.504.5205C}.

Figure \ref{fig:bias_field} shows a slice of thickness $\sim 80$ Mpc $h^{-1}$ through the UNITSim at $z=0$, with the color code indicating the average (within the slice shown) of the bias-weighted halo density field. It is interesting to notice how a threshold in bias can define relatively regularly distributed structures with typical sizes of $\lesssim 60$ Mpc. This can be considered as  a consequence of setting a limit $k_{\rm max}$ to compute the effective bias: in practice, such limit acts as a low-pass filter that reduces the spatial resolution of the dark matter and halo fields from $\sim 1.95$ Mpc $h^{-1}$ (see \S\ref{sec:sim}) to $\pi / k_{\rm max}\sim 40$ Mpc $h^{-1}$. In appendix \S\ref{sec:scales_bias} we show how the bias-weighted distribution of separation of tracers reflects this fact in the form of a ringing effect.

Figure \ref{fig:bias_nu} shows the scaling relation between the effective halo bias computed with Eq.(\ref{eq:bias_object}) and both peak height, $\nu$ (left panel), and halo mass (right panel). In both panels, we show the fitting formula of \citet[][T10 hereafter]{2010ApJ...724..878T}, which depends on the spherical overdensity used to define dark matter halos, $\Delta_{\rm vir}$ \citep[][]{Bryan_1998} (see \S\ref{sec:mass}). We have employed $\Delta_{\rm vir} \sim 331$ (at $z=0$), which corresponds to the mean spherical overdensity of the halo population in the \texttt{UNIT}sim \citep[][]{2013ApJ...762..109B}. The differences observed with respect to the prediction from T10 are expected, as the latter are generated using a different definition of halo mass \citep[see e.g.,][]{2001A&A...367...27W,2011ApJS..195....4M, 2012MNRAS.425.2244B}.

We have compared the estimates of effective halo bias from the estimators of Eq.~(\ref{eq:bhh_k}) (\ref{eq:bhm_k}) and (\ref{eq:bias_object}). This is presented in Fig.~\ref{fig:bias_mass}  \footnote{All power spectra have been computed using a $512^{3}$ mesh with cloud-in-cell mass assignment \citep[][]{1988csup.bookH} and its corresponding correction in Fourier space and Poisson shot-noise subtraction. When averaging in Fourier bins, use weighted averages with number of modes in each bin as the weights.}, showing measurements obtained using Fourier modes to a maximum $k_{\rm max}={\rm min}(0.08, k_{sn})\,h$Mpc$^{-1}$, where $k_{sn}$ is the wavenumber at which the signal of power spectrum equals the Poisson shot-noise. The results based on Eq.~(\ref{eq:bhh_k}) are in agreement with those based on Eq.~(\ref{eq:bias_object}), while the measurement of $b_{hm}$ based on Eq.~(\ref{eq:bhm_k}) differs by a factor $\sim 1 \sigma$. \footnote{Both the estimates and error bars assigned to the bias obtained from the power-spectrum-based estimators are computed as weighted averages using the number of Fourier modes in each shell as weights.}
The overall agreement among these estimates confirms the robustness of Eq.~(\ref{eq:bias_object}) in terms of measuring effective halo bias. From now on, we shall refer to the effective halo bias with the symbol $b_{h}$.

\subsection{Characterization of the effective halo bias distribution}

In Fig.~\ref{fig:bias_nu_allz}, we present the measurement of mean halo bias as a function of the peak height for different redshifts. There is an approximate universality as a function of redshift, spoiled only at high redshift (high $\nu$) with deviations on the order of $1\sigma$ compared to T10. The universality of the halo bias can be extended to the full bias distribution.
To verify that, in Fig.~\ref{fig:nu_stats} we present the estimates of the higher moments (variance, skewness and kurtosis) of the $\mathcal{P}(b_{h}|\nu)$ distribution. These moments are (within the error bars) compatible with zero, which added to the fact that the correlation between peak height and effective bias is small ($\rho_{s}\leq 0.03$ at all redshifts explored, see Fig.~\ref{fig:spear1}) leads to the conclusion that effective halo bias follows a normal distribution, i.e, 
$\mathcal{P}(b_{h}|\nu,z)\sim \mathcal{N}(\langle b_{h}| \nu \rangle, \sigma_{\nu}(z))$, 
where the standard deviation $\sigma_{\nu}(z)$ is independent of $\nu$ \footnote{The variance scales with redshift as $\sigma_{\nu}(z)\approx (1+bz)^{c}\sigma_{\nu}(z=0)$, where $b=1.4\pm0.1$ and $c=0.41\pm 0.06$, obtained using $\sim 20$ UNITSim snapshots}. Figure \ref{fig:nu_stats} also shows the higher moments computed in high-density regions (knots), with no evident modifications with respect to the behavior in the full sample. We have verified that this is also the case in other cosmic-web environments (i.e, sheets, filaments).
This behavior can complement the idea of an approximate universality of halo bias when expressed in terms of the peak height, as the formalism of structure-formation  predicts \citep[][]{1986ApJ...304...15B,ShethTormen1999,ShethTormen2002}.

%========================================================
%========================================================
\begin{figure}
%UNITSim_bias_hprop_pearson_spearman_coeff_redshift.py
\includegraphics[trim = .0cm 0.2cm 0cm 0cm ,clip=true, width=0.5\textwidth]{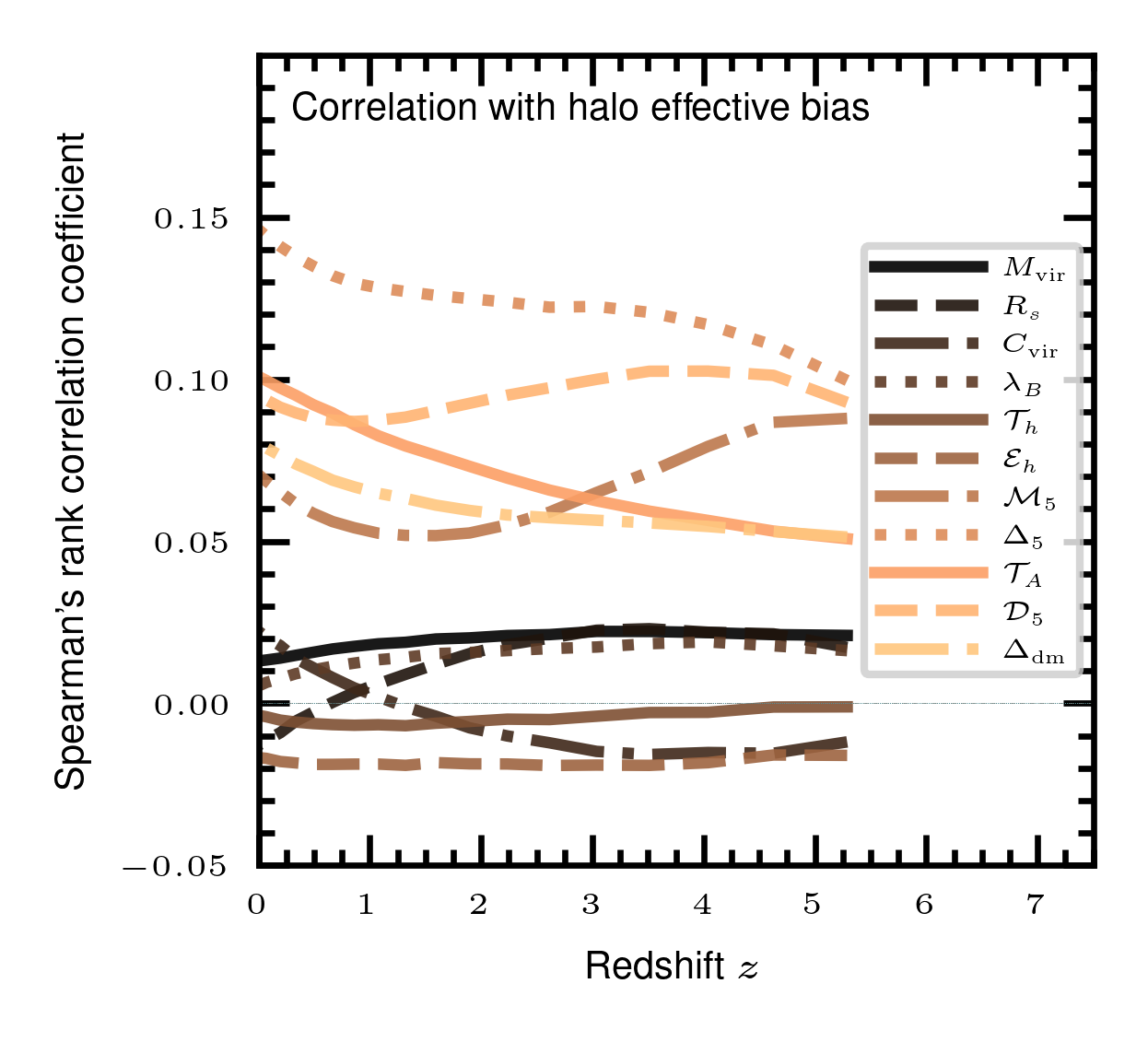}
\caption{\small{Spearman's rank correlation coefficient between the halo effective bias and a number of halo properties as a function of redshift.}}
\label{fig:bias_spear}
\end{figure}
%========================================================
%========================================================
\begin{figure}
%UNITSim_bias_hprop_pearson_spearman_coeff_redshift.py
\includegraphics[trim = .0cm 0.5cm 0cm 0cm ,clip=true, width=0.5\textwidth]{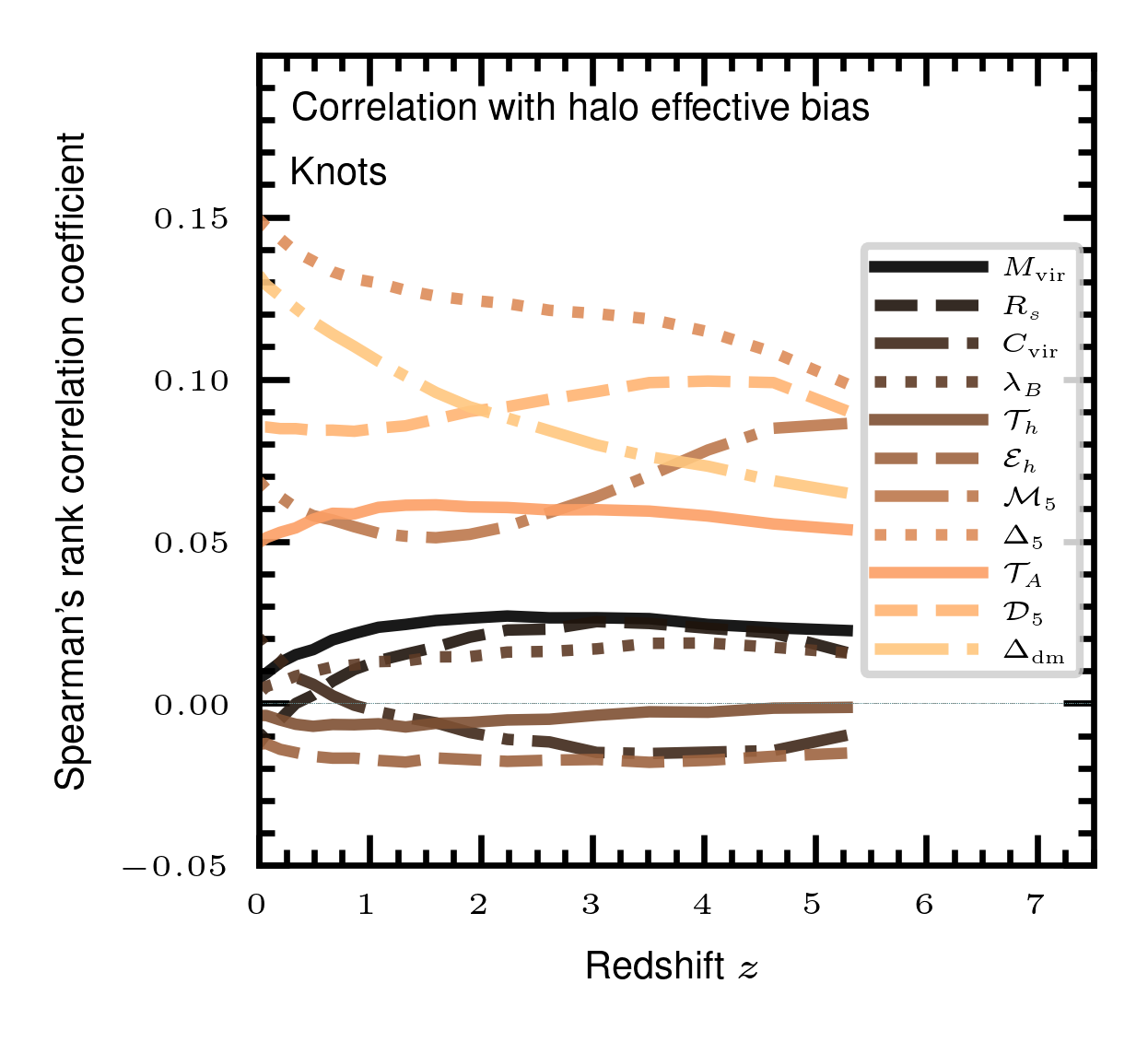}
\includegraphics[trim = .0cm 0.2cm 0cm 0cm ,clip=true, width=0.5\textwidth]{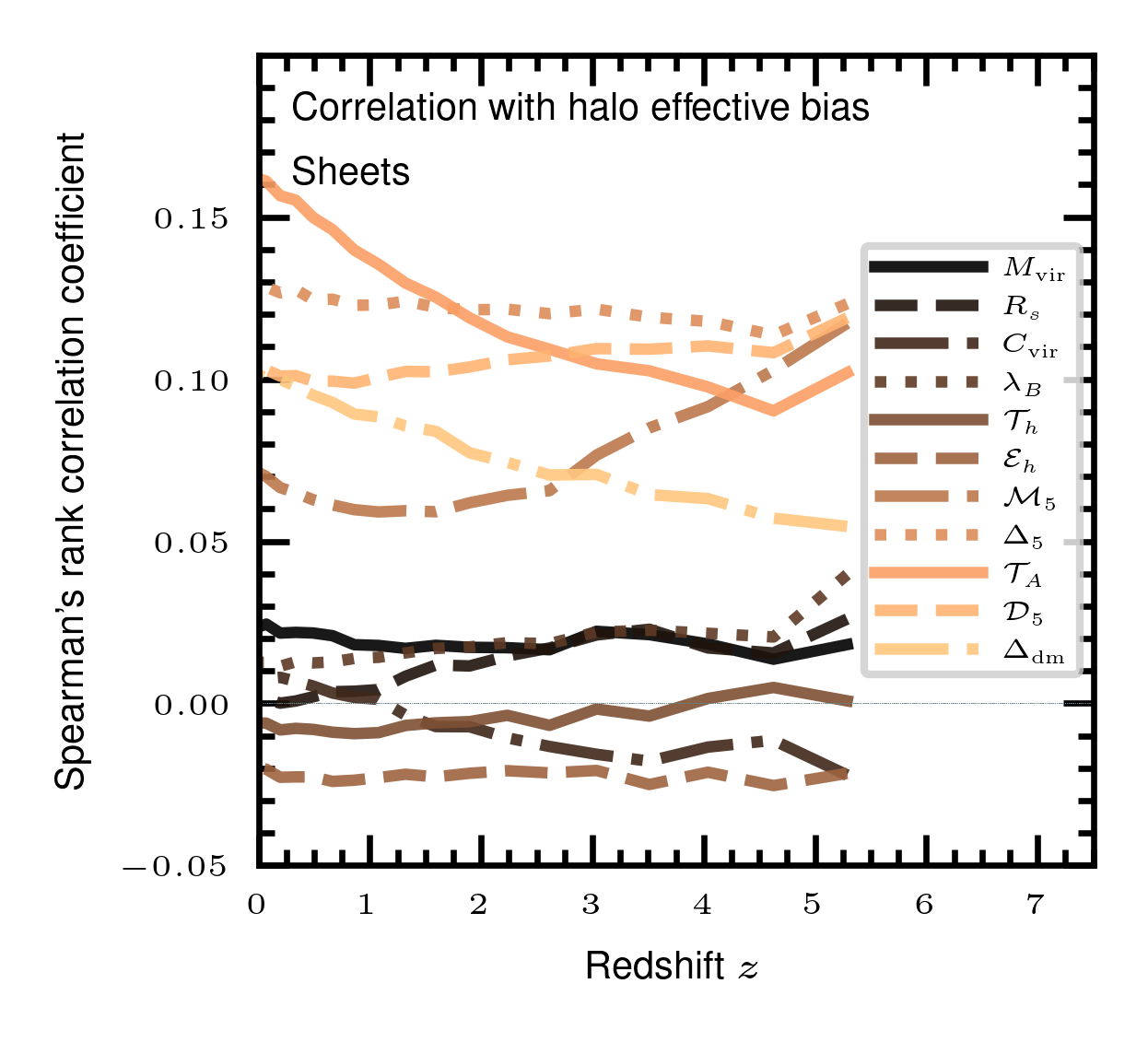}
\caption{\small{Spearman's rank correlation coefficient between the halo effective bias and a number of halo properties as a function of redshift.}}
\label{fig:bias_spear_cwt}
\end{figure}

The scaling relations between halo bias and any other halo property $\theta$ (internal or external) can be expressed based on their link to peak height (or halo mass) \footnote{As the scaling relation between peak height and halo mass is $\mathcal{P}(\nu|M_{\rm vir})\dd M_{\rm vir}=\delta^{D}(\nu-\nu_{0}(M_{\rm vir}))\dd M_{\rm vir}$ where $\nu_{0}(M_{\rm vir})=\delta_{sc}/\sigma(M_{\rm vir})$ (see Eq.~(\ref{eq:sigma})), the statistical properties of the halo bias as a function of halo mass are directly inherited from those of peak height.} using
\be\label{eq:scaling}
\mathcal{P}(b_{h}|\theta) = \int_{-\infty}^{\infty}  \mathcal{P}(b_{h}|\nu) \mathcal{P}
(\nu|\theta)\dd \nu,
\ee
such that the mean effective bias as a function of a property $\theta$ can be written as
\be \label{eq:mscaling}
\langle b_{h}|\theta\rangle = 
\frac{\int_{-\infty}^{\infty} b_{h} \mathcal{P}(b_{h}|\theta)\dd b_{h}}{\int_{-\infty}^{\infty}\mathcal{P}(b_{h}|\theta)\dd b_{h}}
=
\frac{\int_{-\infty}^{\infty} \,\mathcal{P}(\nu)\mathcal{P}(\nu|\theta)\langle b_{h}| \nu\rangle \dd\nu}{\int \dd \nu \mathcal{P}(\nu)\mathcal{P}(\nu|\theta)},
\ee
where $\mathcal{P}(\nu)=\int \dd b_{h} \mathcal{P}(b_{h}|\nu)$ is the distribution function of dark-matter halos as a function of their peak height. Equations (\ref{eq:scaling}) and (\ref{eq:mscaling}) (the latter is the continuous representation of Eq.(\ref{eq:bias_object})) represent the link between large scale structure and astrophysics, which is present at the level of dark matter halos and galaxy clusters: it implies that in order to generate predictions on large-scale bias of a halo population (containing cosmological information) as a function of a halo property, we need to characterize the halo scaling relations, mainly shaped by astrophysical processes \citep[][]{2006MNRAS.365..842C,2011ARA&A..49..409A,2012ARA&A..50..353K, 2014MNRAS.440.2077M,2014A&A...563A.141B, 2014MNRAS.441.3562E,2020MNRAS.494.3728S}. 
Notice that Eq.(\ref{eq:scaling}) is a symbolic separation, as the scaling relations of halos contain cosmological information \citep[][]{2002ApJ...568...52W,2016MNRAS.460.1214L,2019ApJ...871..168D}

It is also key to notice that previous attempts to assign an individual bias were based on Eq.(\ref{eq:mscaling}) \citep[see e.g.,][]{2001A&A...368...86S,2011MNRAS.413..386B}; such approach misses the information contained in the higher moments of the bias distribution.
%========================================================
%========================================================
\begin{figure*}[htbp]
\centering
%bias_vs_prop_scaling_relations_cwt.py 
\includegraphics[trim = 0.25cm 0.2cm 0cm 0cm ,clip=true, width=0.32\textwidth]{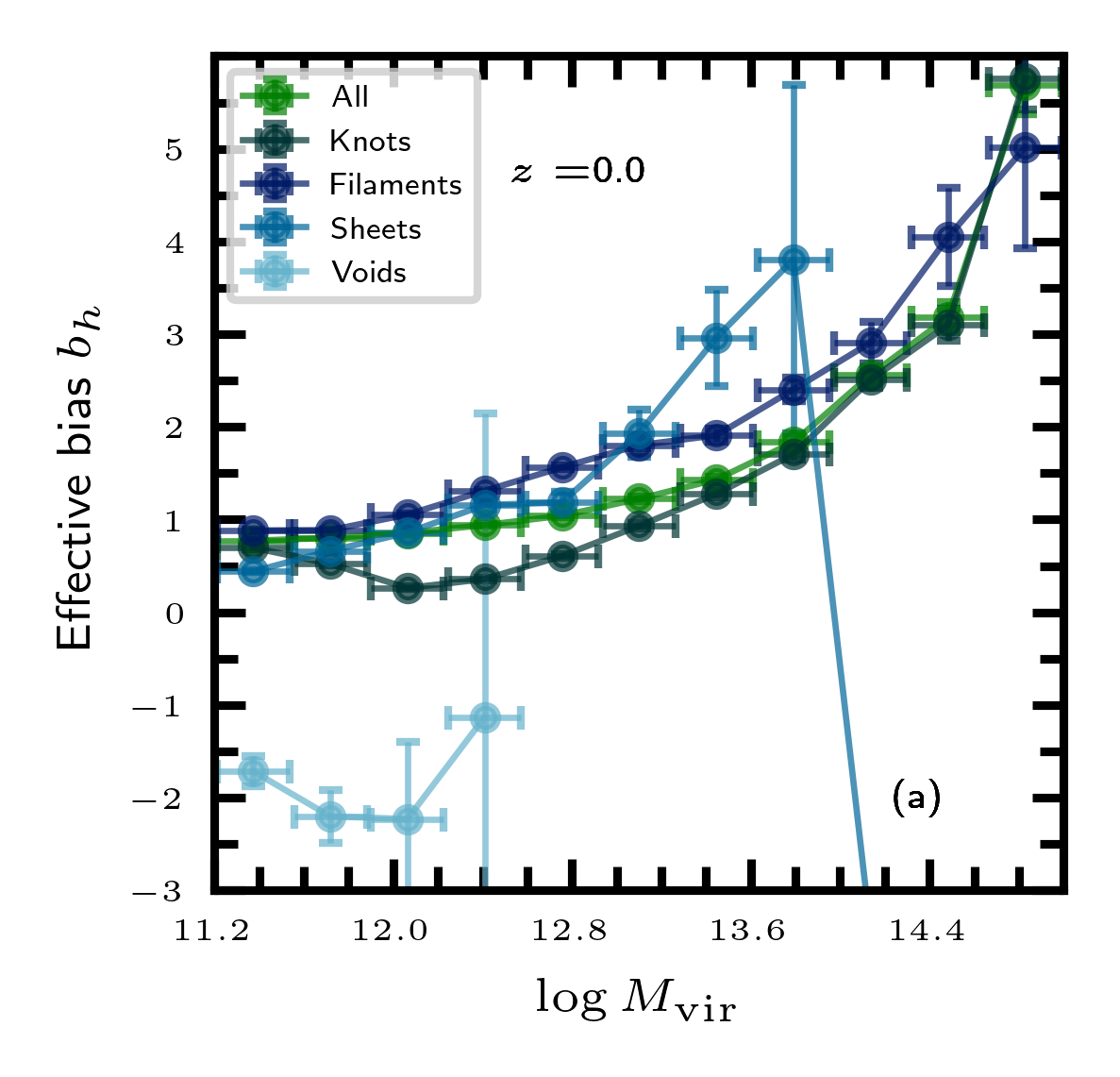} 
\includegraphics[trim = 0.3cm 0.2cm 0cm 0cm ,clip=true, width=0.32\textwidth]{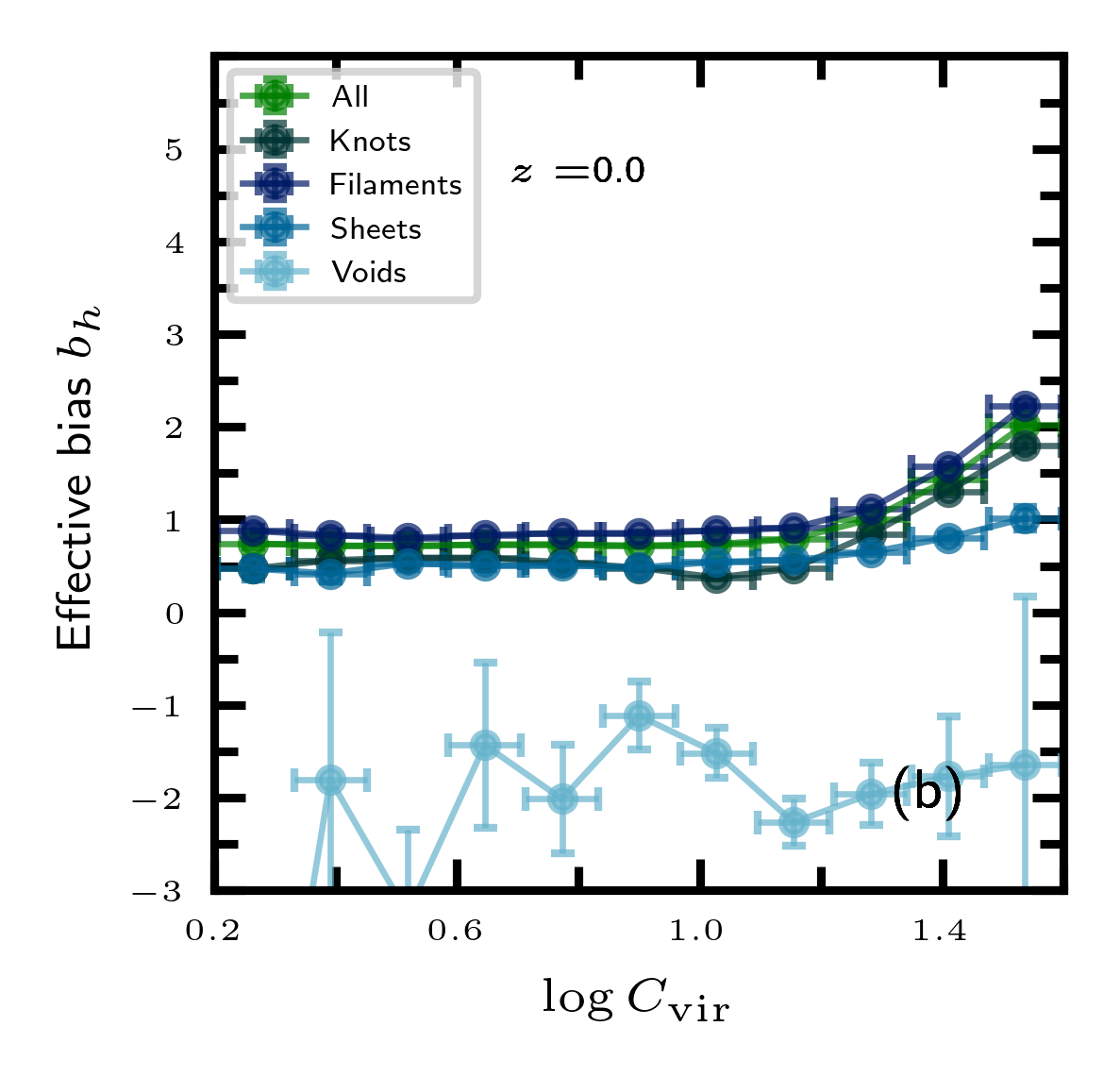} 
\includegraphics[trim = 0.25cm 0.2cm 0cm 0cm ,clip=true, width=0.32\textwidth]{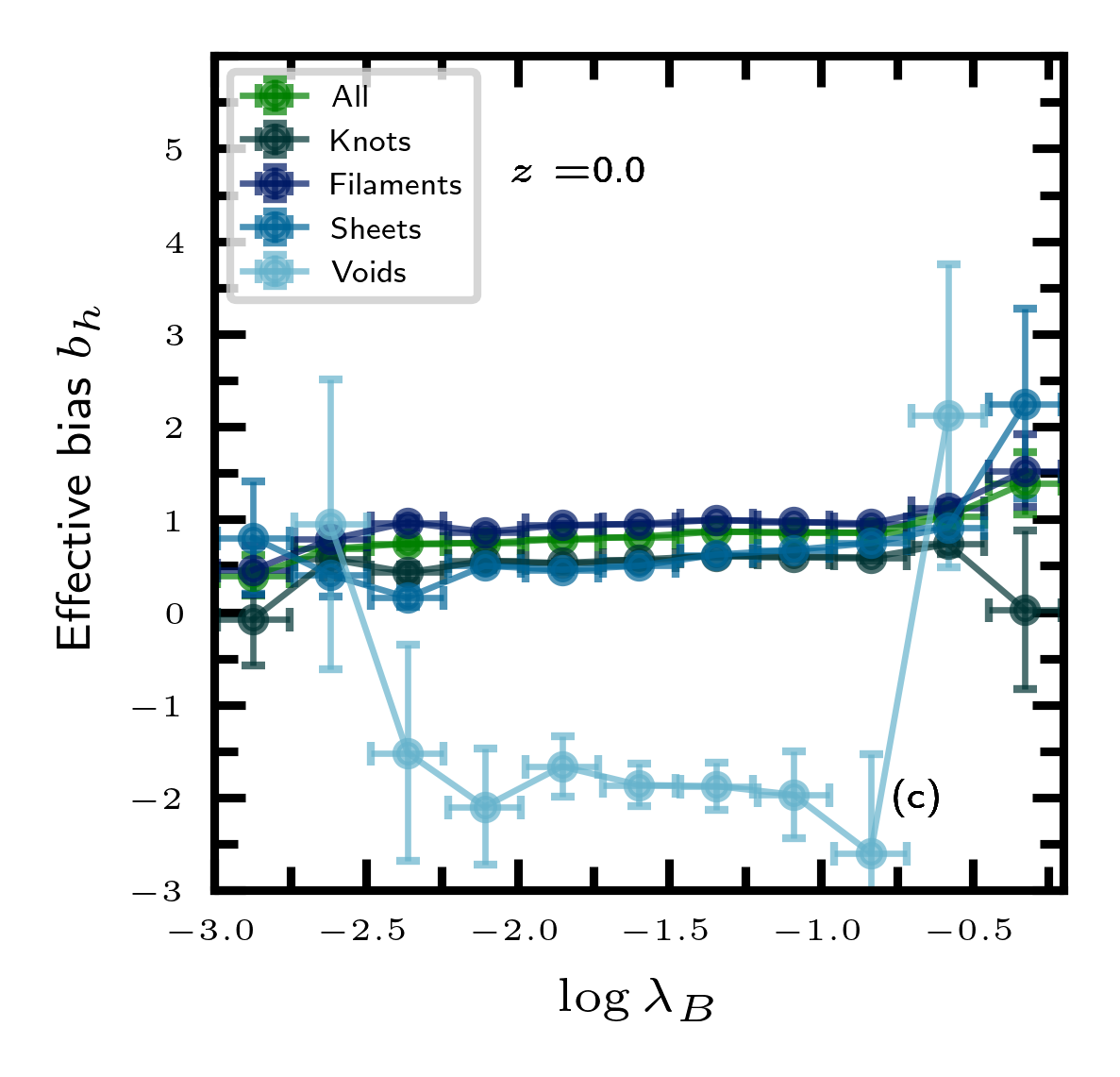} 
\includegraphics[trim = 0.25cm 0.2cm 0cm 0cm ,clip=true, width=0.32\textwidth]{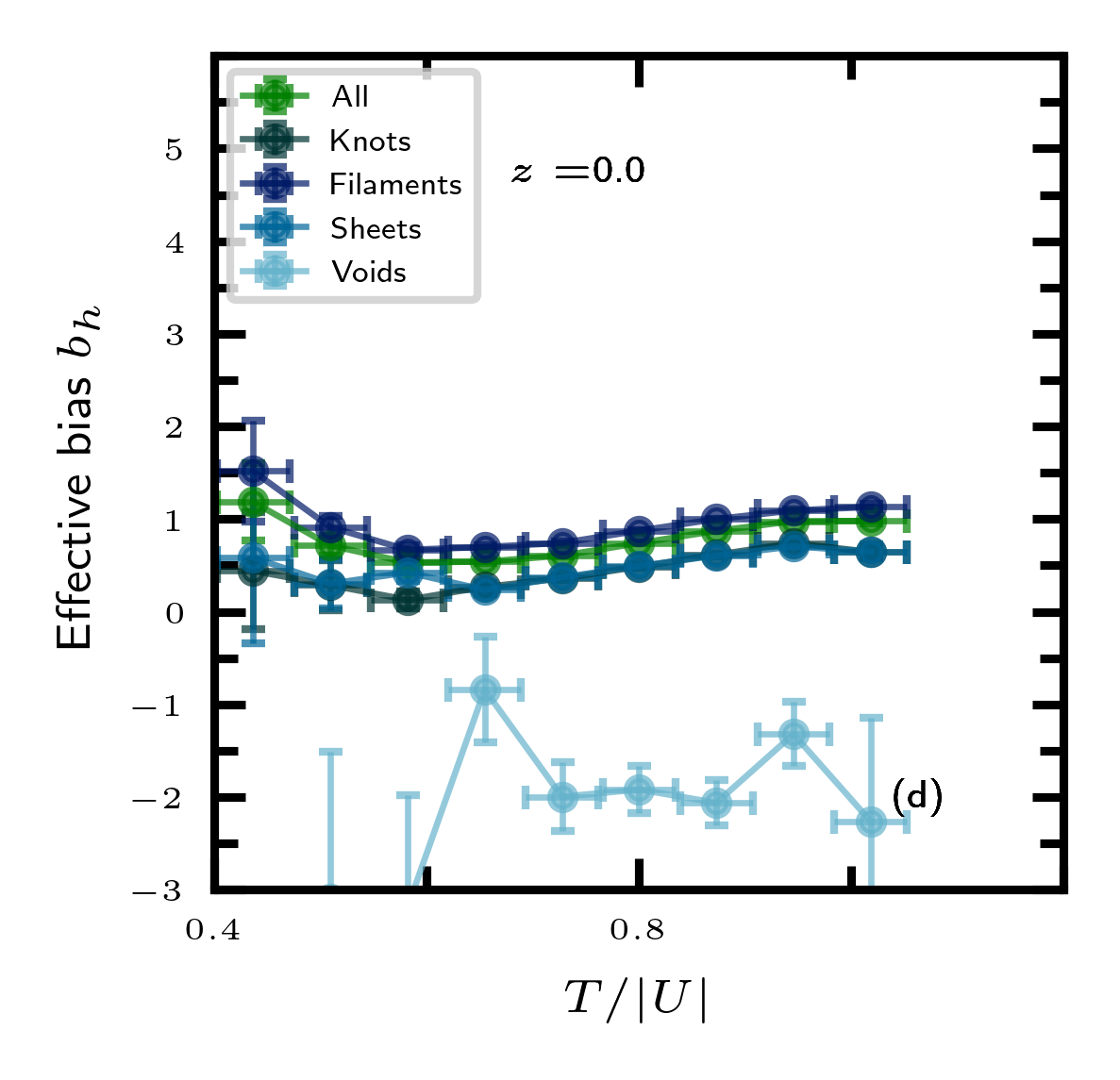} 
\includegraphics[trim = 0.25cm 0.2cm 0cm 0cm ,clip=true, width=0.32\textwidth]{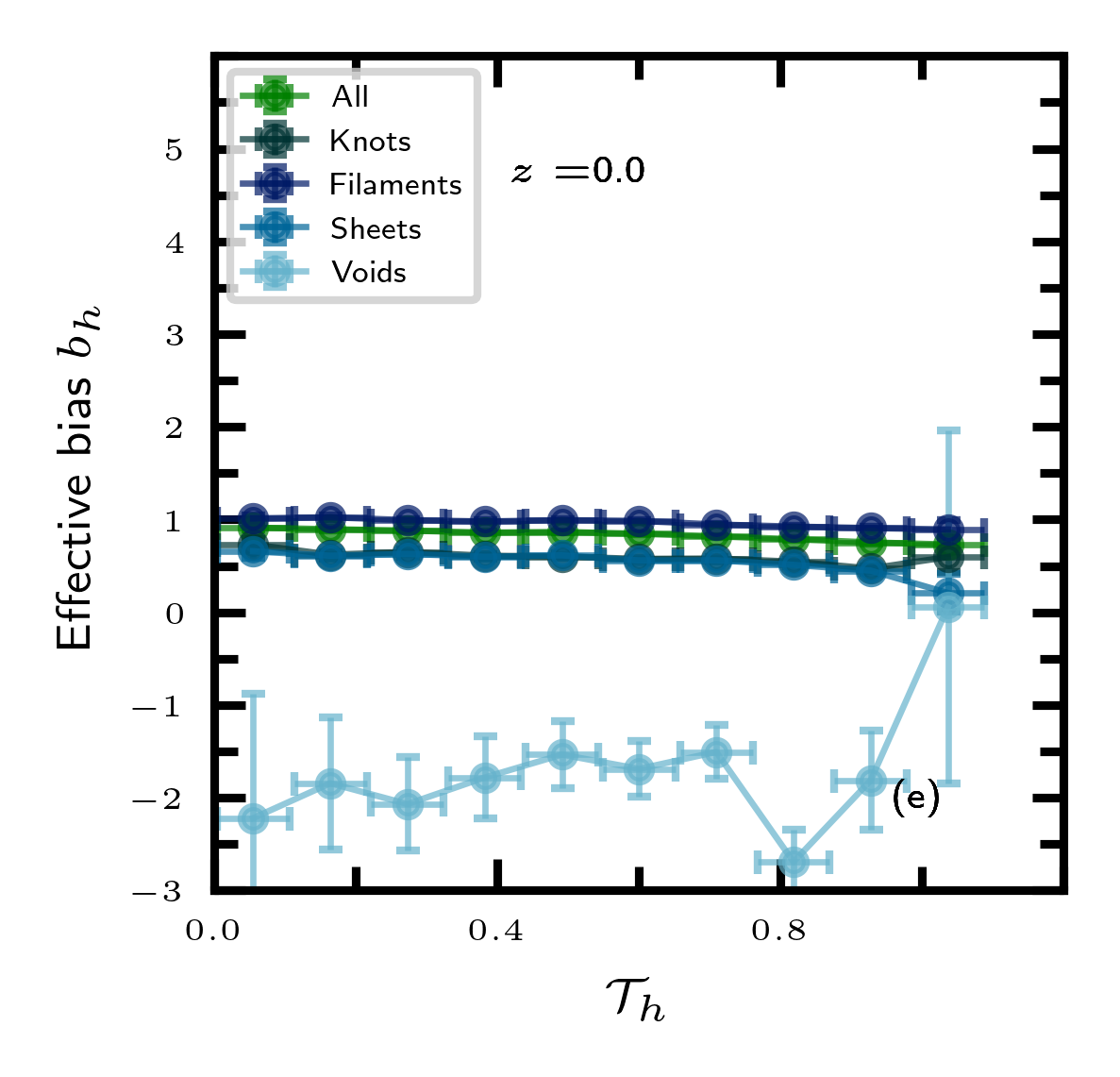} 
\includegraphics[trim = 0.25cm 0.2cm 0cm 0cm ,clip=true, width=0.32\textwidth]{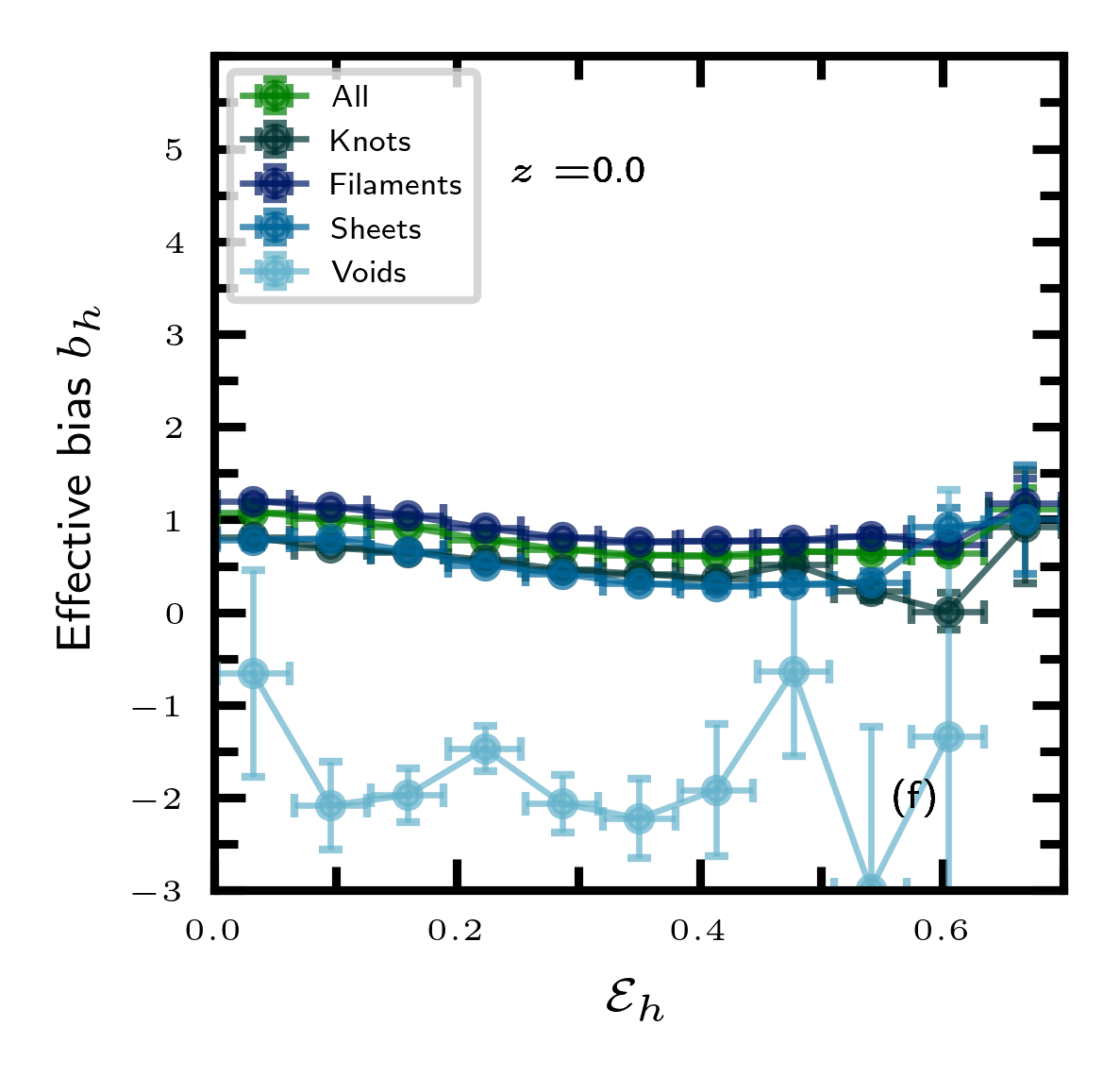} 
\includegraphics[trim = 0.25cm 0.2cm 0cm 0cm ,clip=true, width=0.32\textwidth]{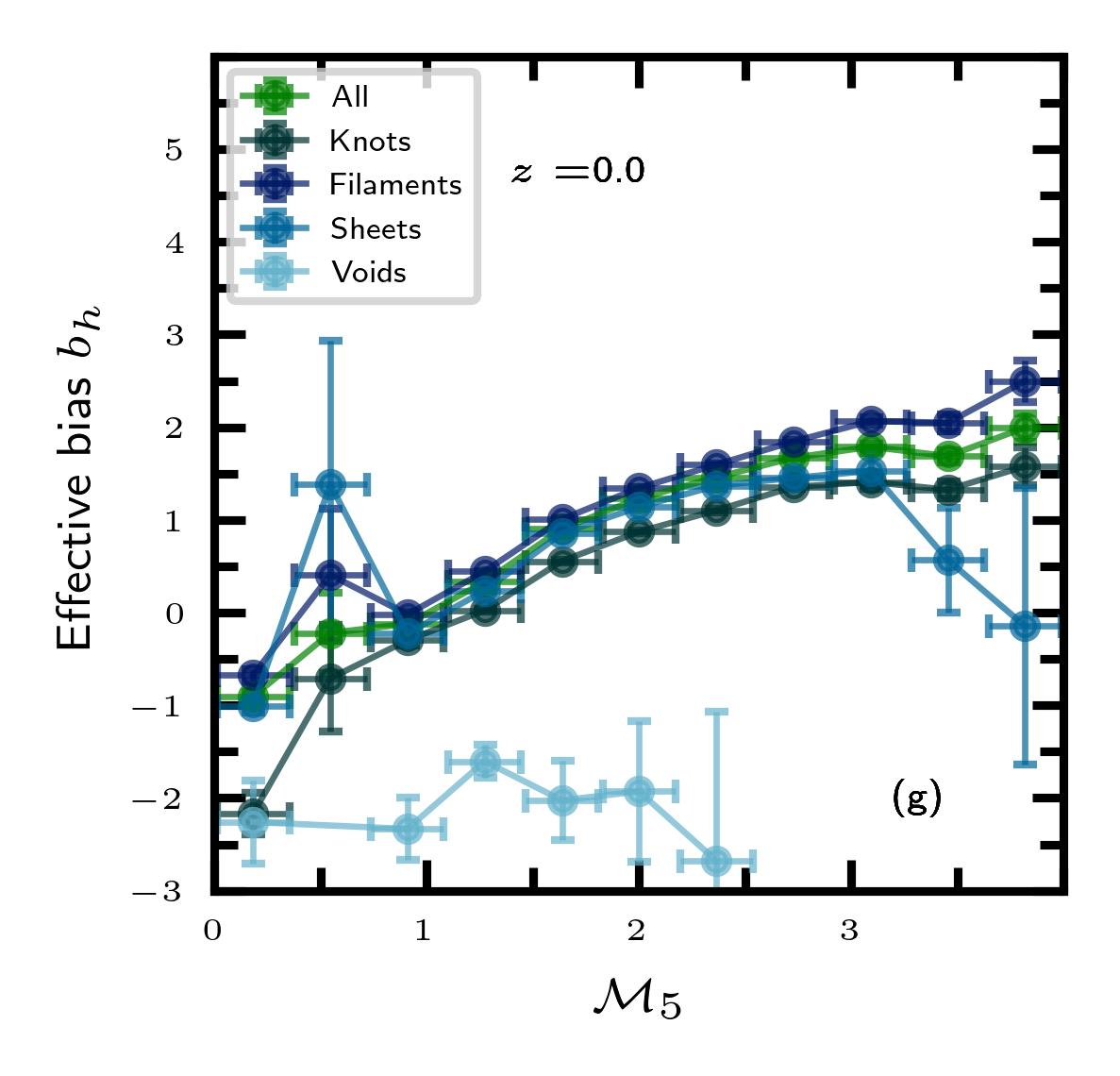} 
\includegraphics[trim = 0.25cm 0.2cm 0cm 0cm ,clip=true, width=0.32\textwidth]{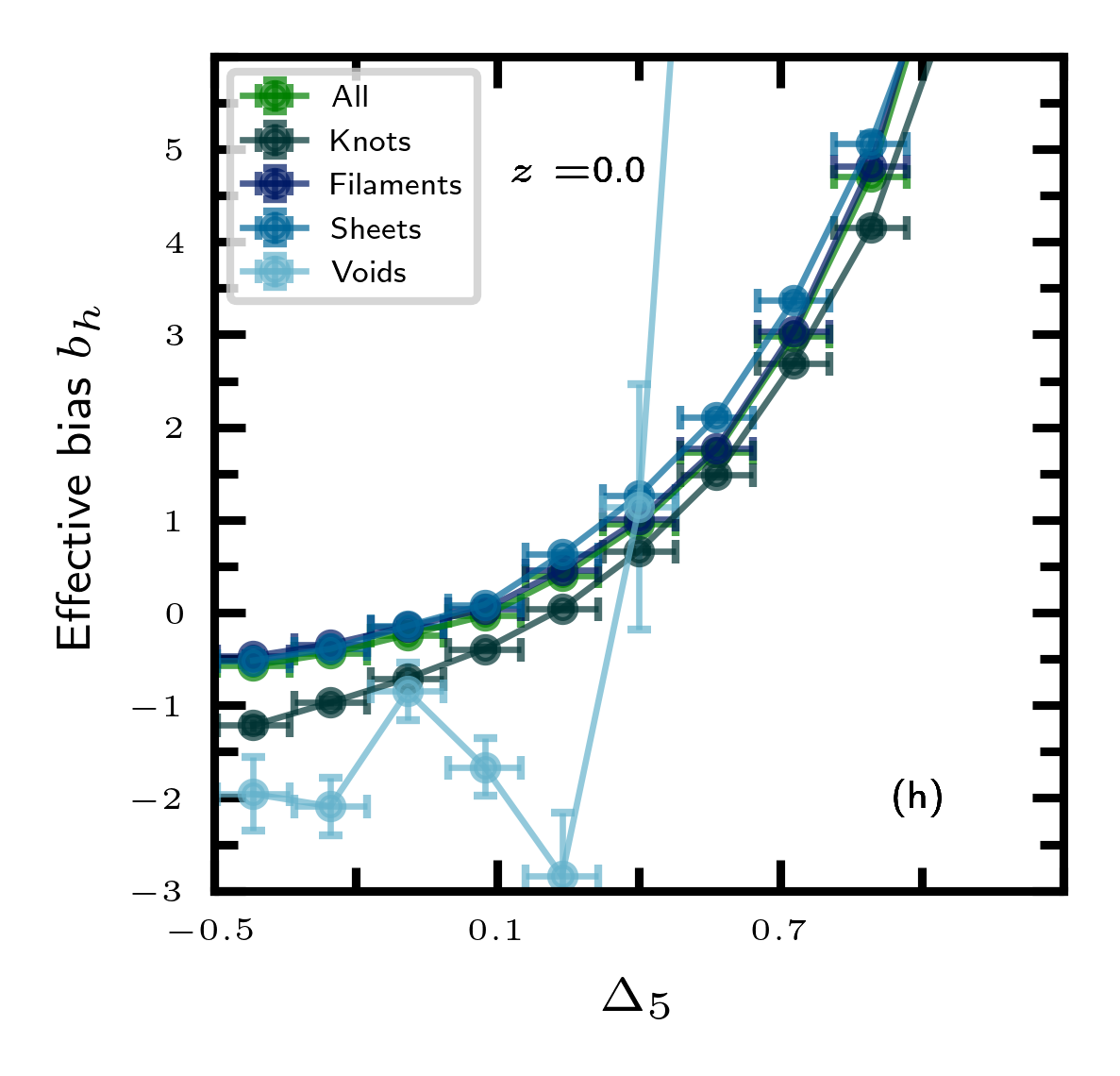} 
\includegraphics[trim = 0.25cm 0.2cm 0cm 0cm ,clip=true, width=0.32\textwidth]{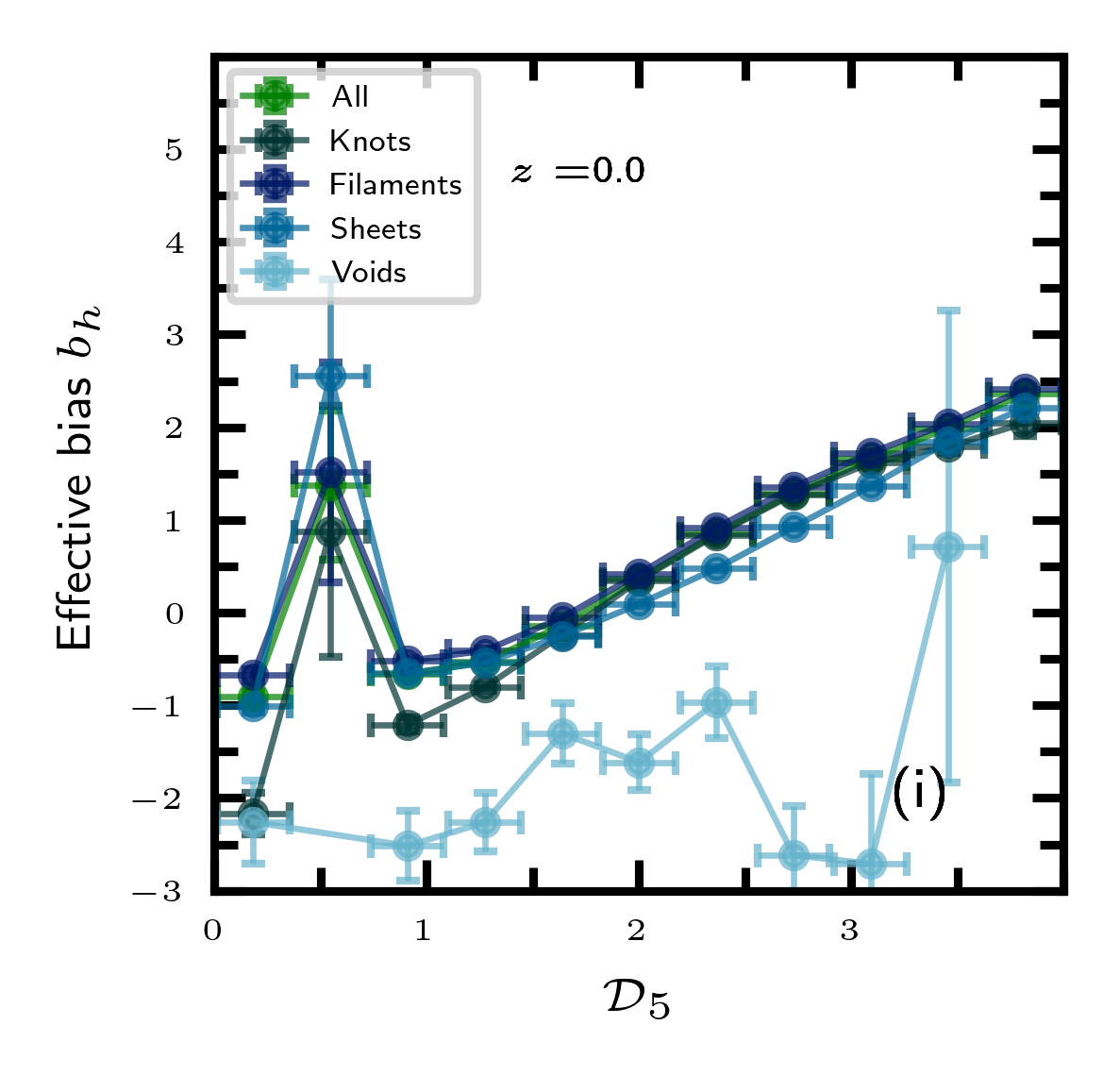} 
\includegraphics[trim = 0.25cm 0.2cm 0cm 0cm ,clip=true, width=0.32\textwidth]{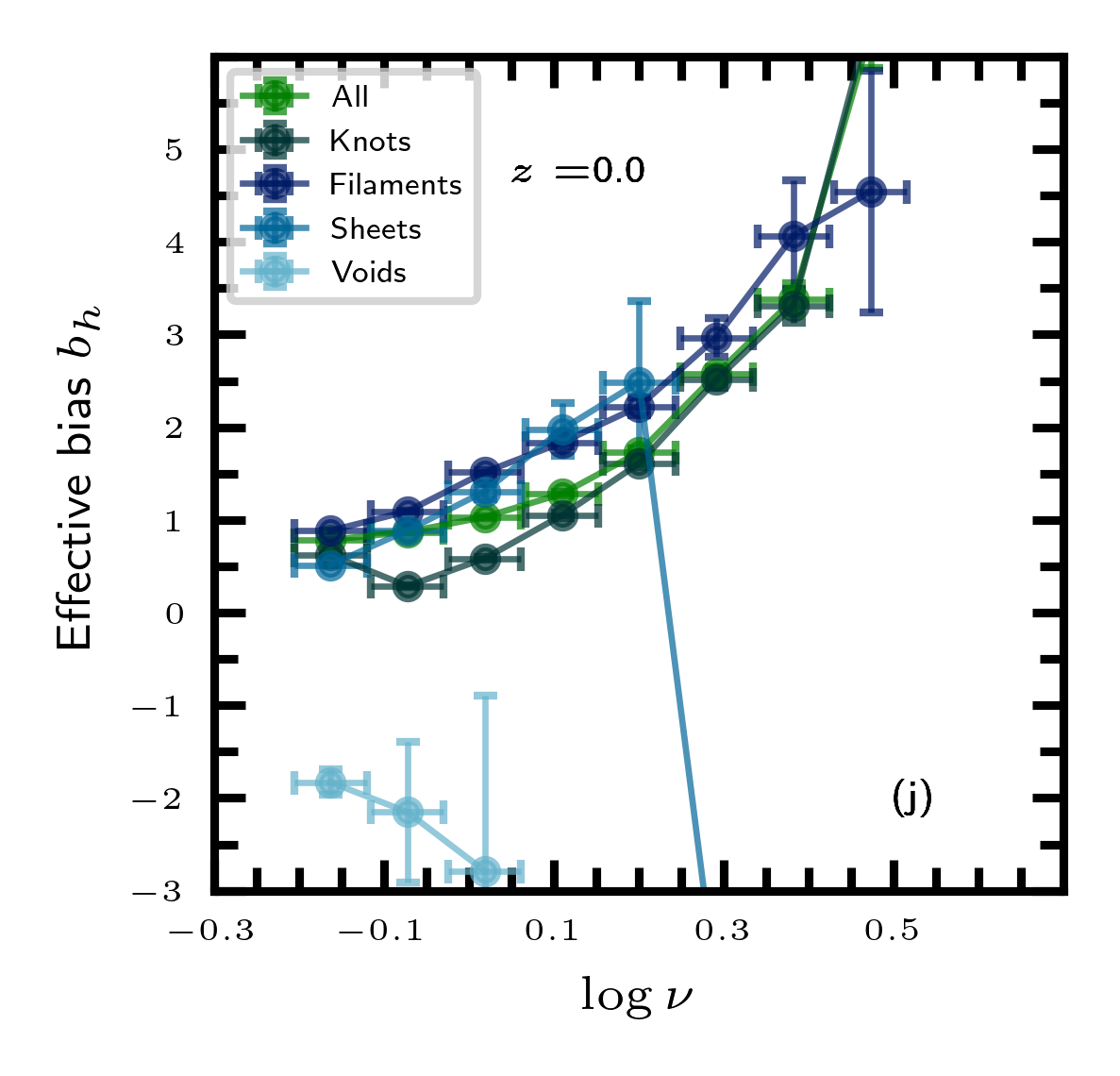} 
\includegraphics[trim = 0.25cm 0.2cm 0cm 0cm ,clip=true, width=0.32\textwidth]{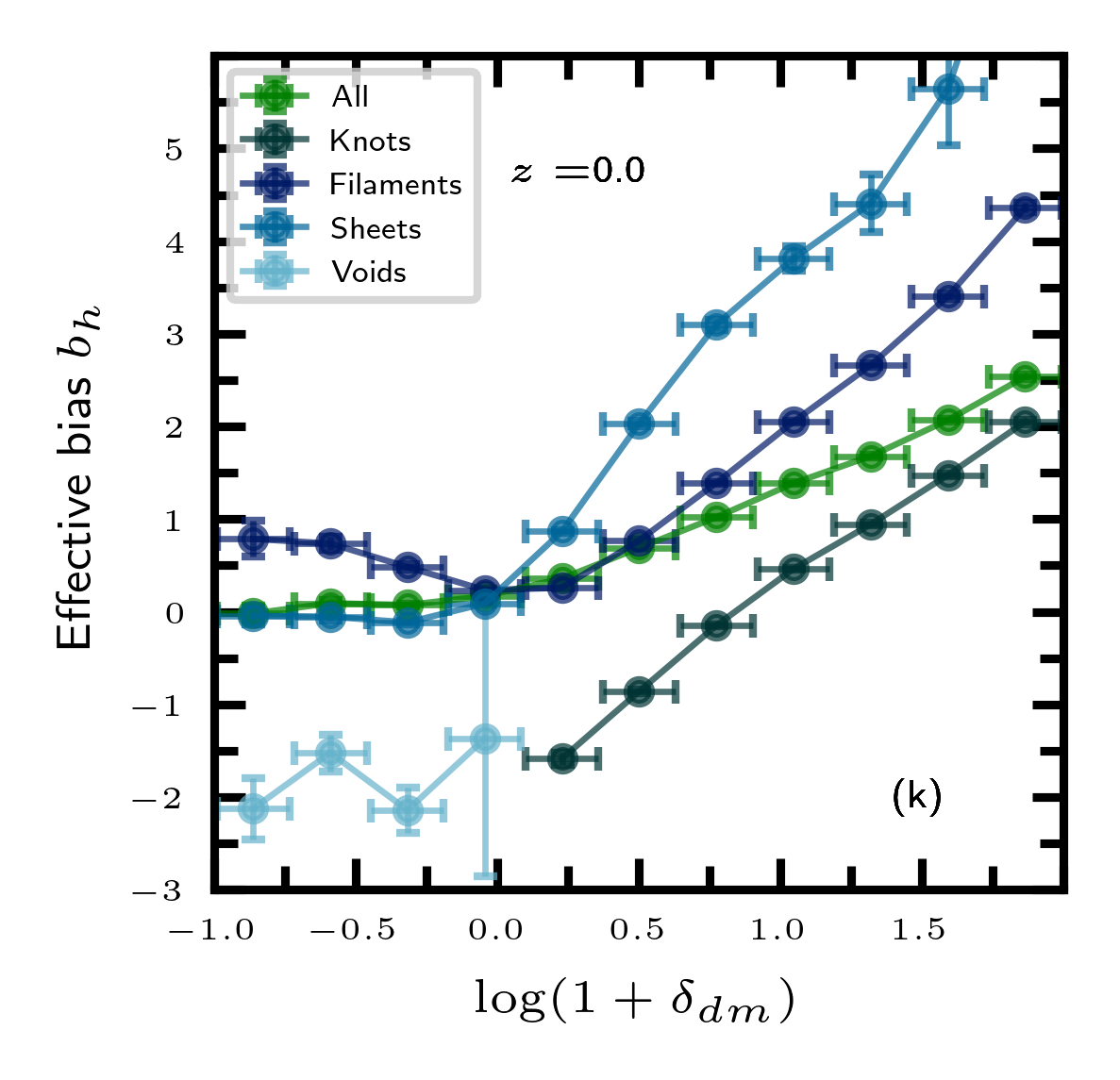} 
\includegraphics[trim = 0.25cm 0.2cm 0cm 0cm ,clip=true, width=0.32\textwidth]{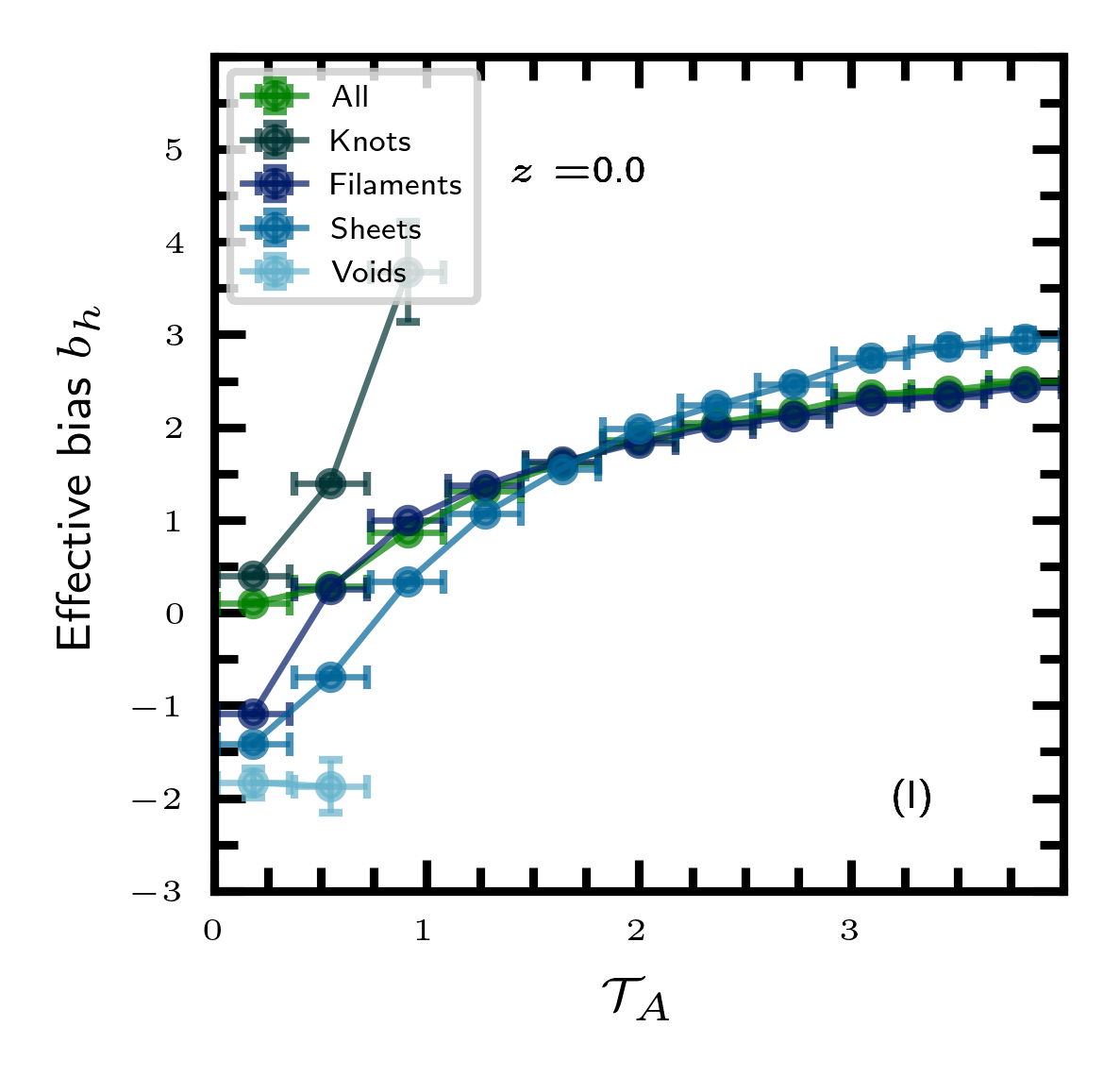}
\caption{\small{Mean effective bias $\langle b_{h}|\theta, w_{i}\rangle$ measured in bins of halo properties $\theta$ in the different cosmic-web $\omega_{i}$ at $z=0$. Intrinsic halo properties are: virial mass $M_{\rm vir}$ (panel a), halo concentration $C_{\rm vir}$ (panel b), halo spin $\lambda_{B}$ (panel c), the ratio $T/|U|$ (panel d), halo triaxiality $\mathcal{T}_{h}$ (panel d), halo ellipticity  $\mathcal{E}_{h}$ (panel g). Nonlocal properties: relative Mach number $\mathcal{M}_{5}$ (panel g), local overdensity $\Delta_{5}$ (panel h) and neighbor statistics $\mathcal{D}_{5}$ (panel i). Panel (j) shows the peak height-halo bias relation. Environmental properties: local dark matter density (panel k) and tidal anisotropy $\mathcal{T}_{A}$ (panel l).}}\label{fig:bias_prop_cwt}
\end{figure*}
%========================================================
%========================================================

%========================================================
%========================================================
\begin{figure*}[]
%bias_scaling_relation_assembly_twoz.py
%the name of files for spin and concentration are exchanged
%but the plots are correctly identified with the secondary property reported in the legend
\centering
\includegraphics[trim = .0cm .59cm 0cm 0cm ,clip=true, width=0.45\textwidth]{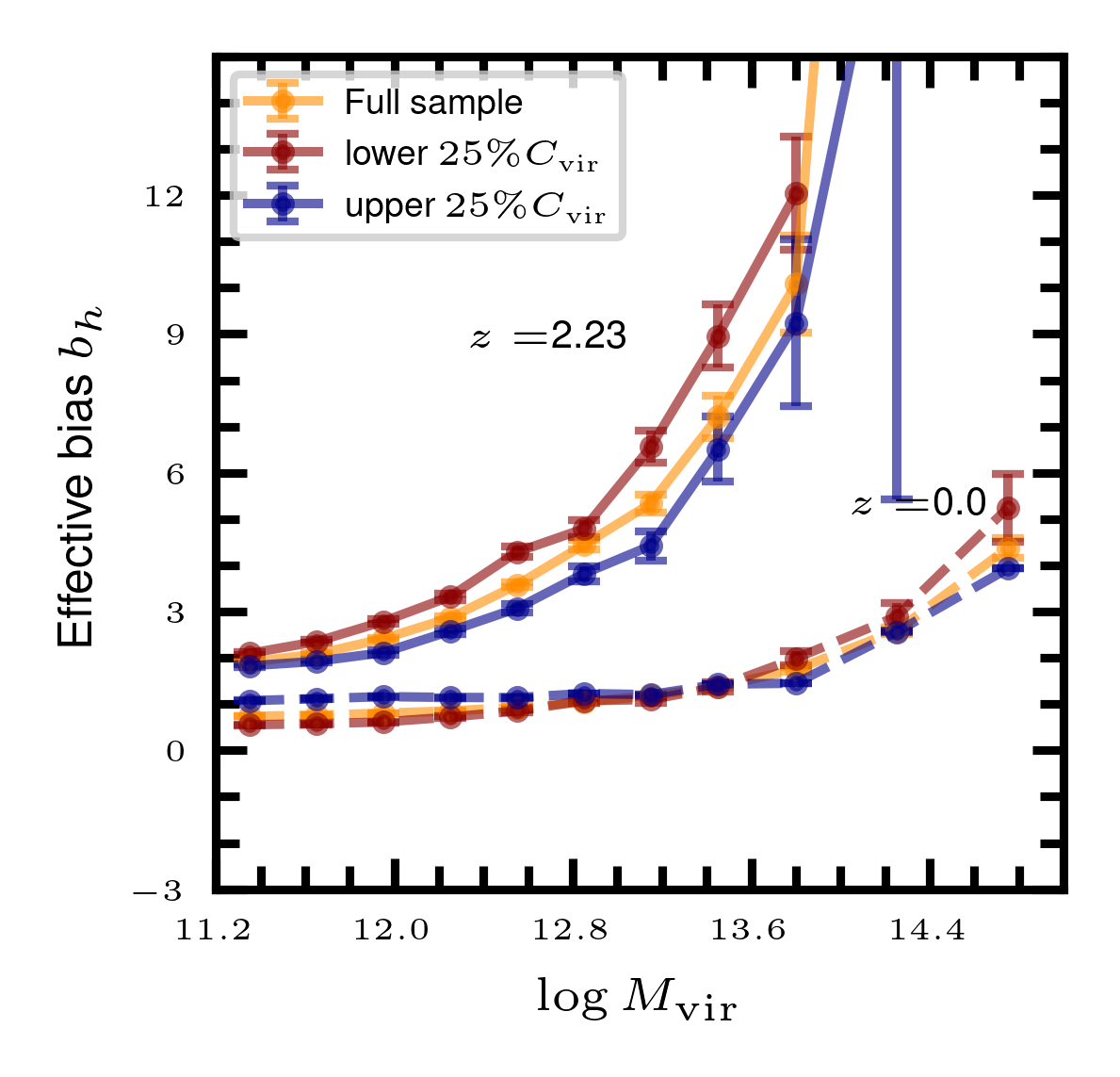}
\includegraphics[trim = .0cm .59cm 0cm 0cm ,clip=true, width=0.45\textwidth]{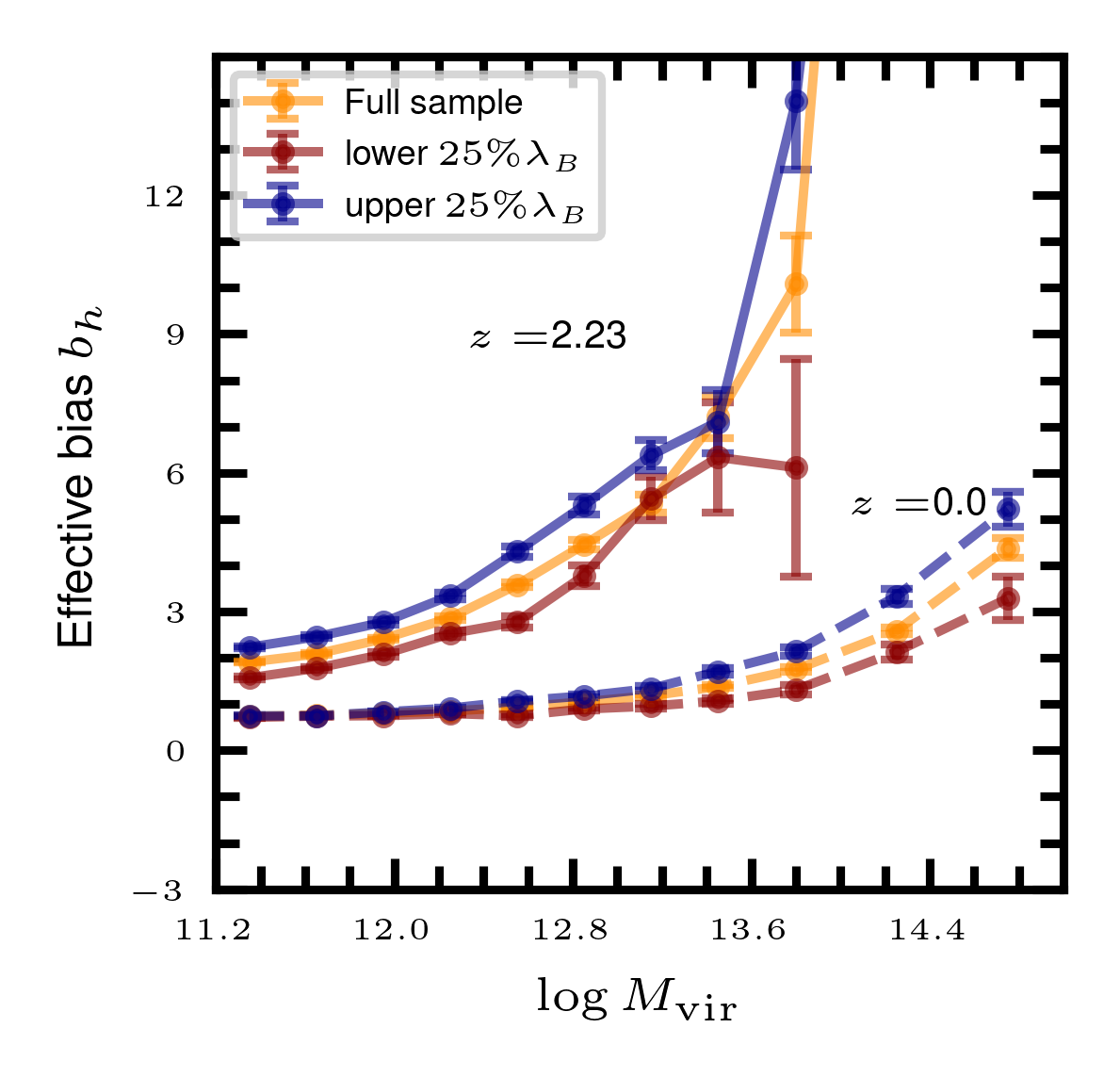}
\includegraphics[trim = .0cm .59cm 0cm 0cm ,clip=true, width=0.45\textwidth]{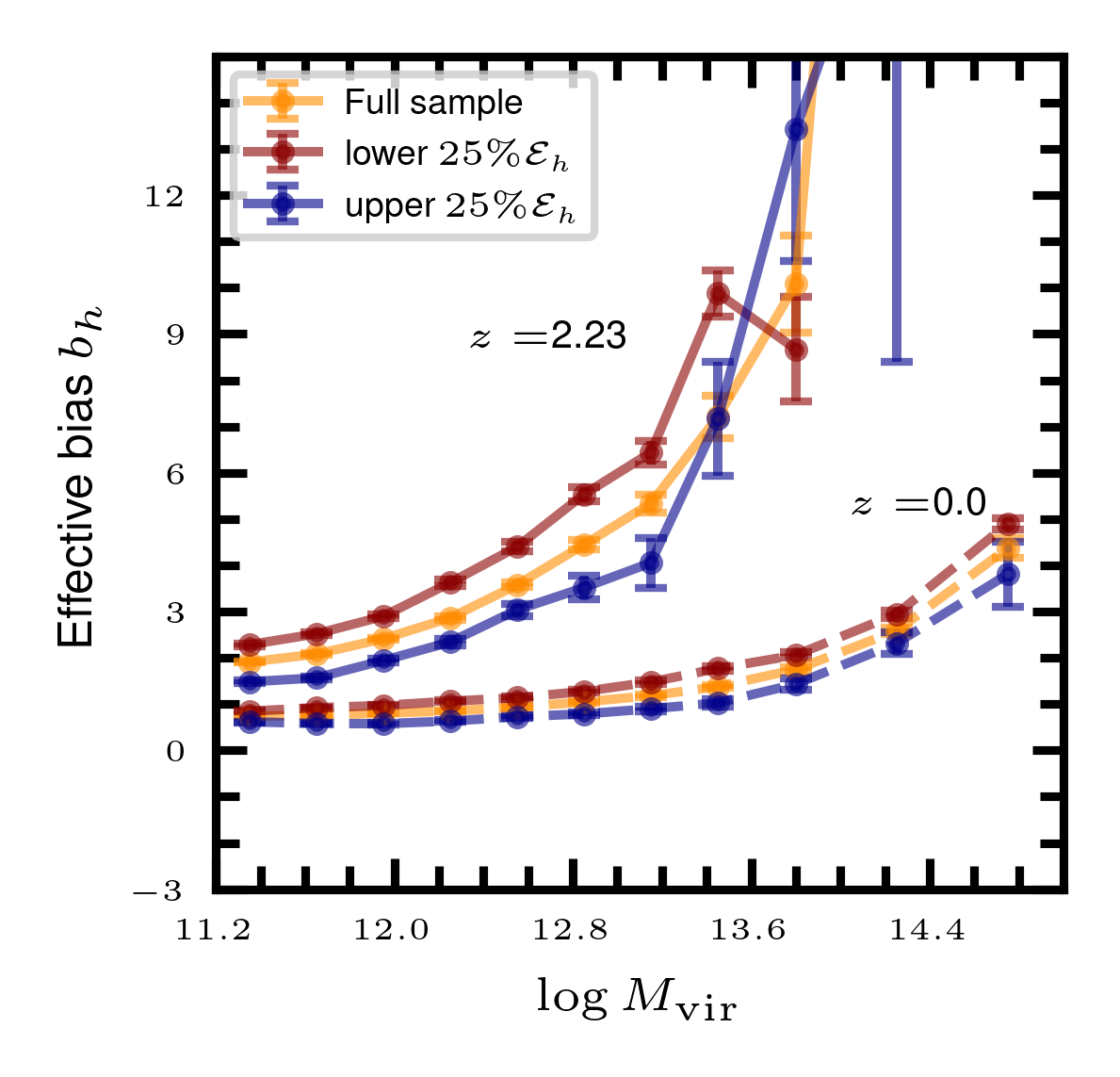}
\includegraphics[trim = .0cm .59cm 0cm 0cm ,clip=true, width=0.45\textwidth]{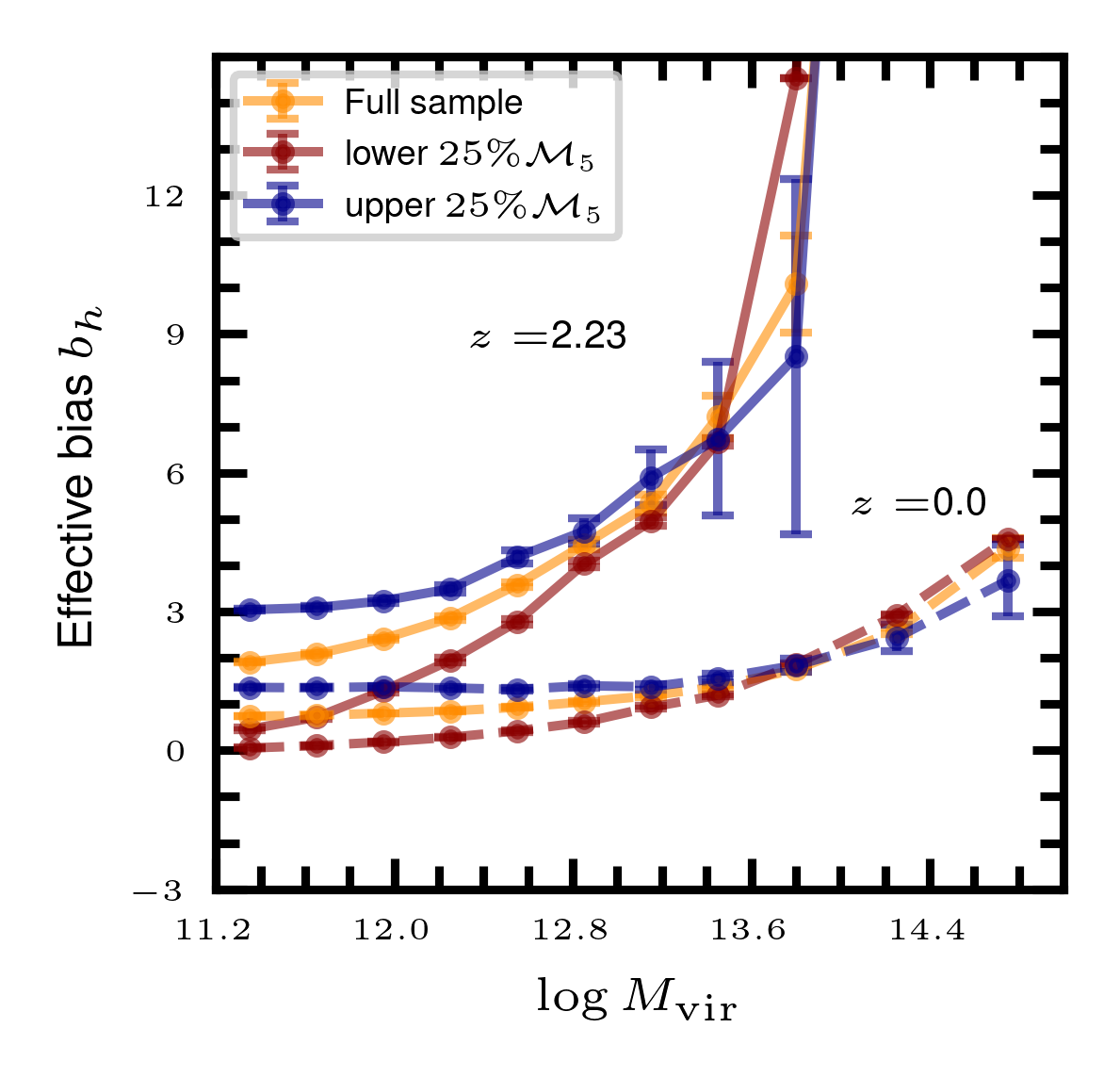}
\includegraphics[trim = .0cm .2cm 0cm 0cm ,clip=true, width=0.45\textwidth]{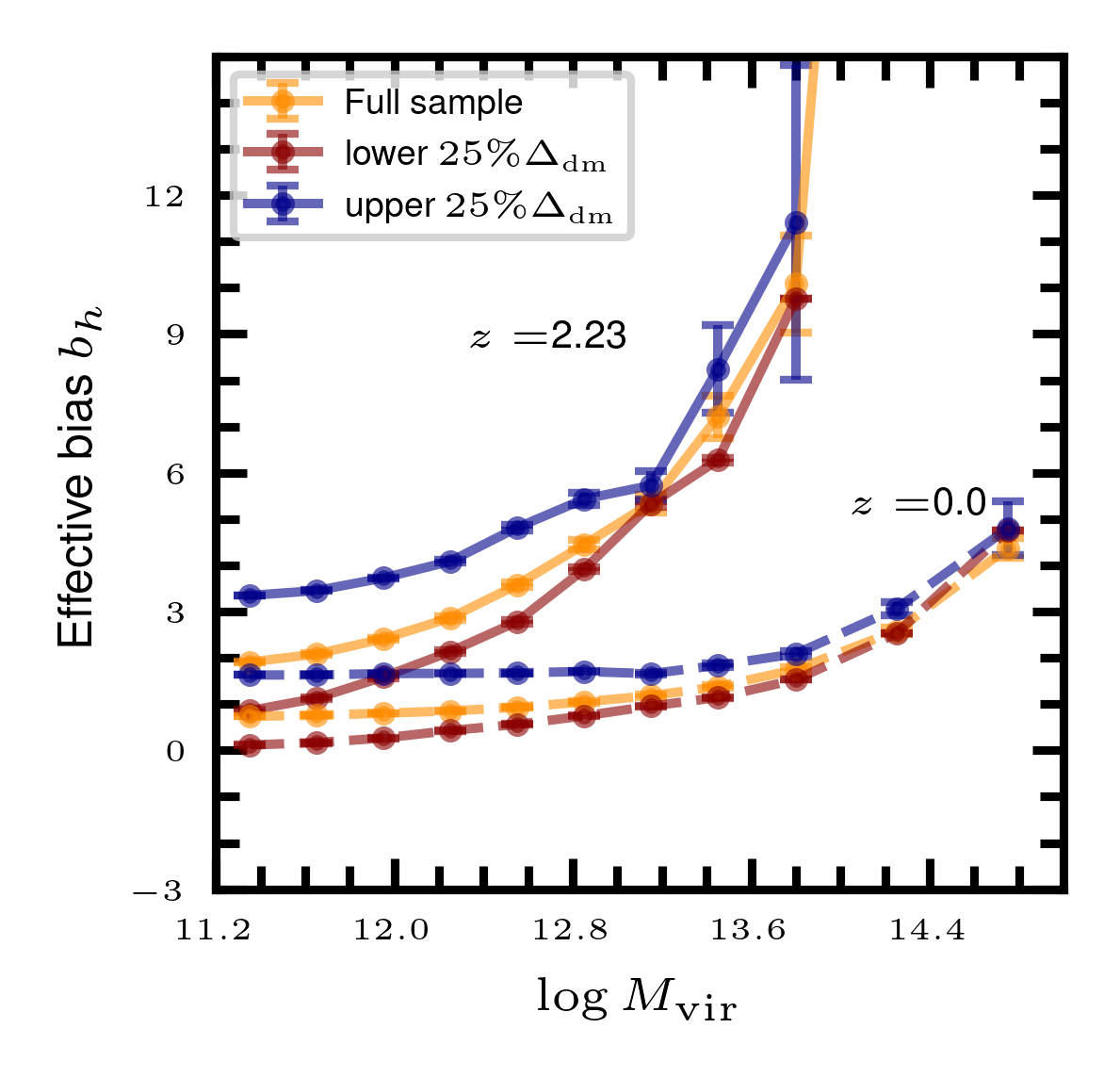}
\includegraphics[trim = .0cm .2cm 0cm 0cm ,clip=true, width=0.45\textwidth]{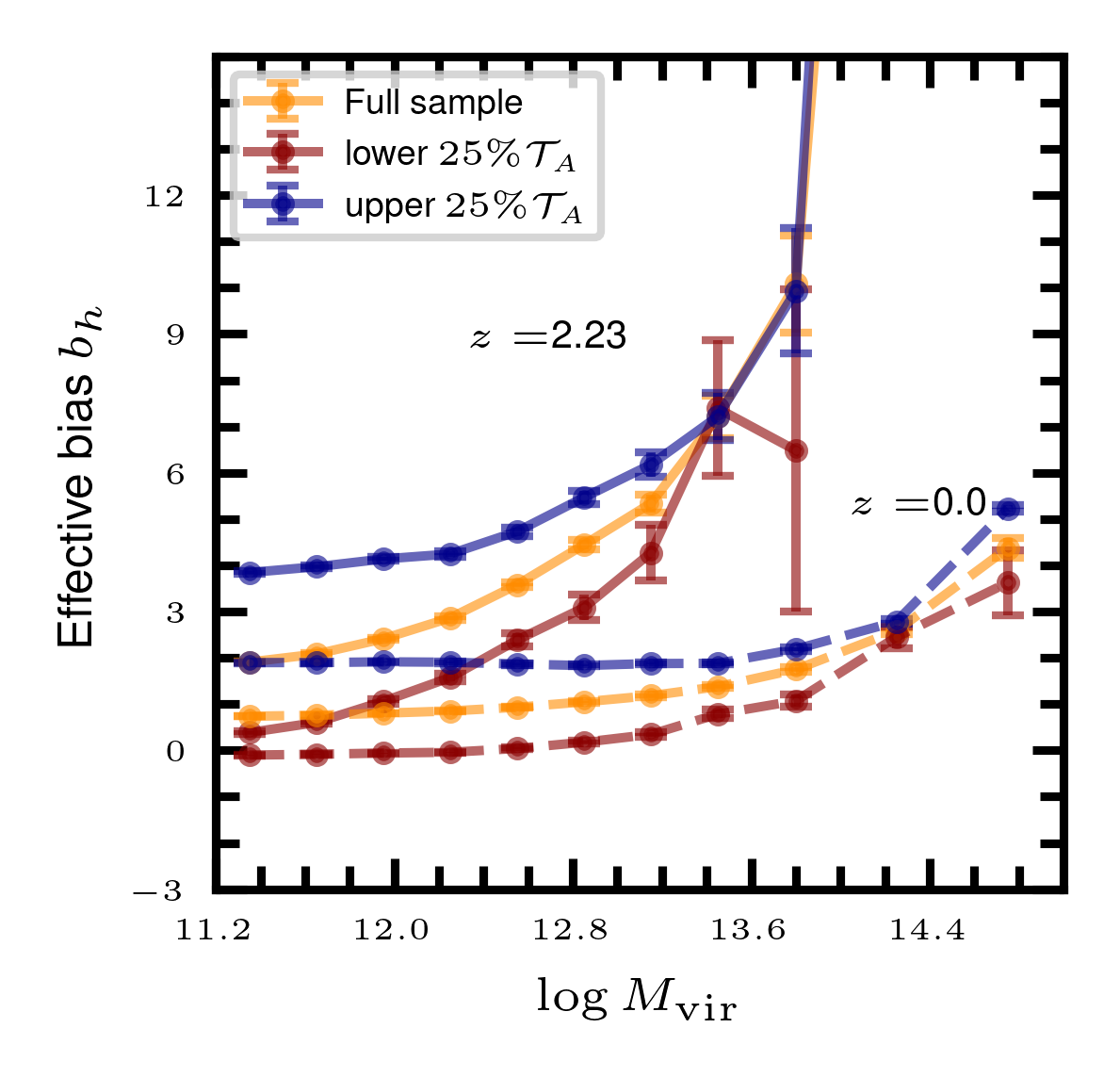}
\caption{\small{Example of secondary halo bias: mean halo effective bias $\langle b_{h}|M_{\rm vir} \rangle_{q}^{(\theta)}$ measured in bins of the halo (log) mass using a number of halo properties as secondary properties $\theta$, namely, concentration $C_{\rm vir}$, spin $\lambda_{B}$, ellipticity $\mathcal{E}_{h}$, Mach number $\mathcal{M}_{5}$, local dark matter overdensity $\Delta_{\rm dm}$, and tidal anisotropy $\mathcal{T}_{A}$. In each mass-bin, the sample has been divided in quartiles $q$ of the property $\theta$. We show results from the lower (first) quartile (red symbols) and the upper (fourth) quartile (blue error bars), along with the results from the full sample (in each mass bin, yellow error bars), at redshifts $z=0$ (dashed lines) and $z=2.2$ (solid lines).}}
\label{fig:sec_bias_ex1}
\end{figure*}
%==========================================
%==========================================

%==========================================
%==========================================
\begin{figure*}
%assembly_bias_delta.py
\includegraphics[trim = 0cm 0.8cm 0cm 0cm ,clip=true, width=.47\textwidth]{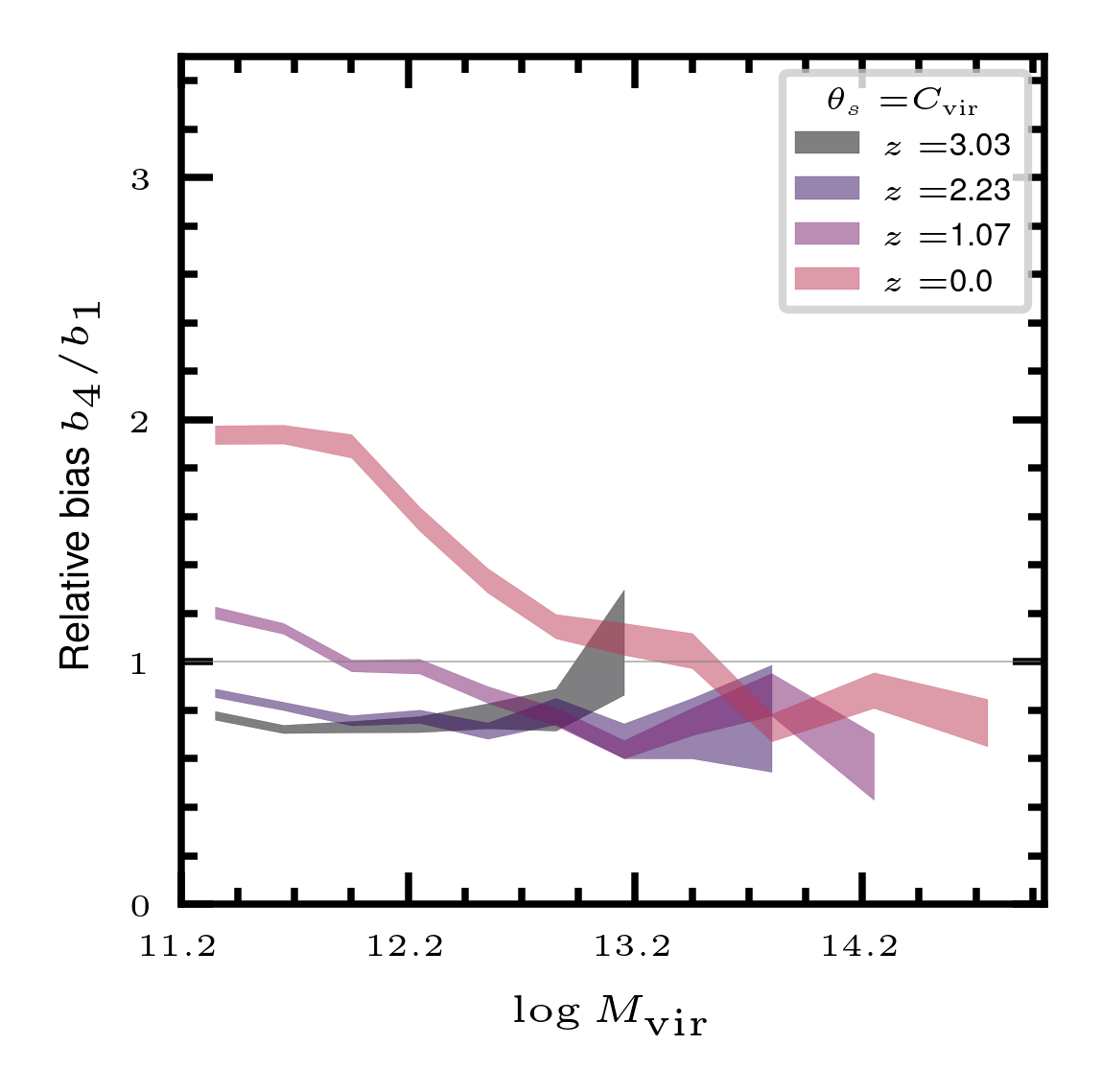}
\includegraphics[trim = 0.cm 0.8cm 0cm 0cm ,clip=true, width=.47\textwidth]{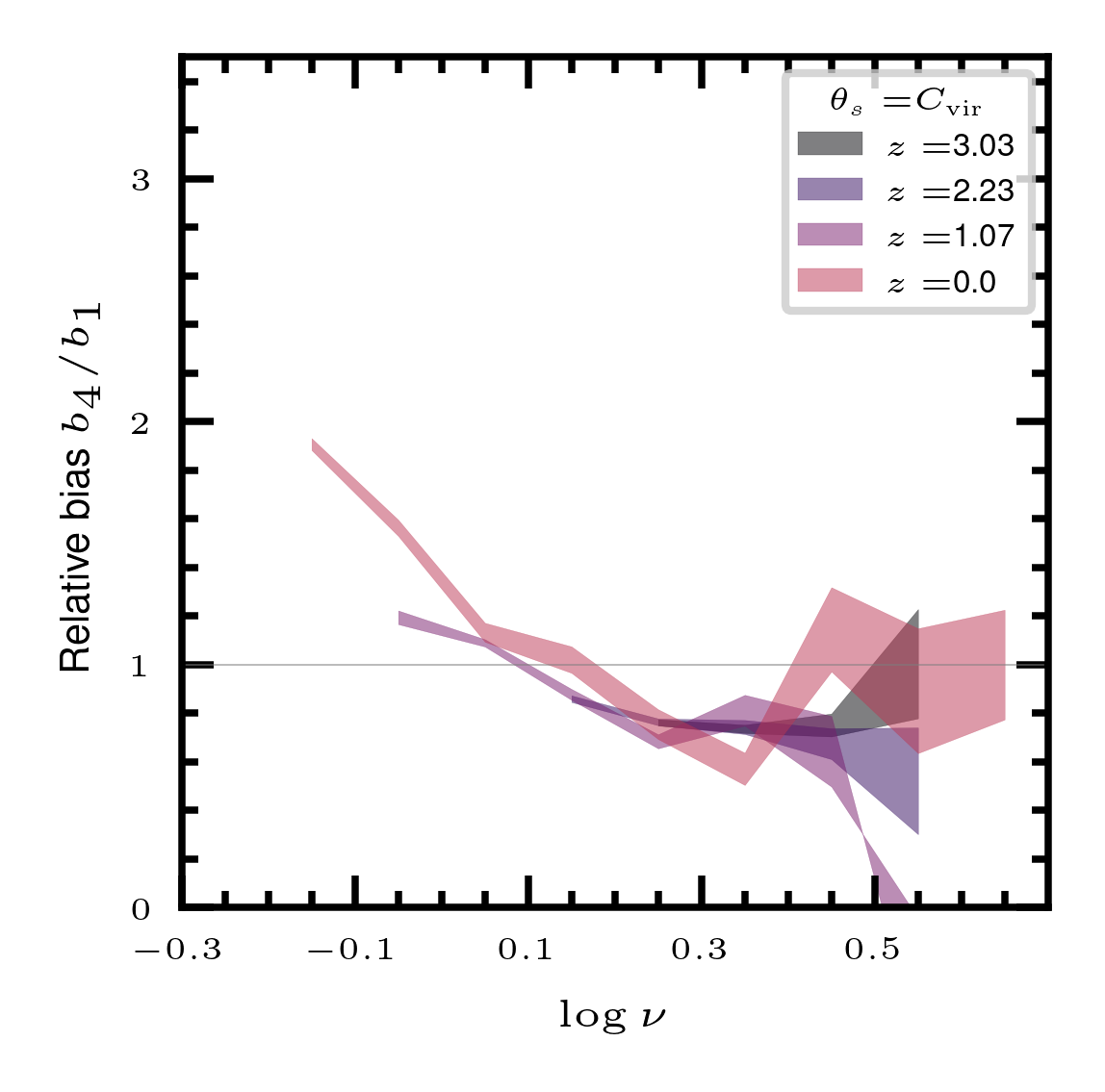}

\includegraphics[trim = 0cm 0.2cm 0cm 0cm ,clip=true, width=.47\textwidth]{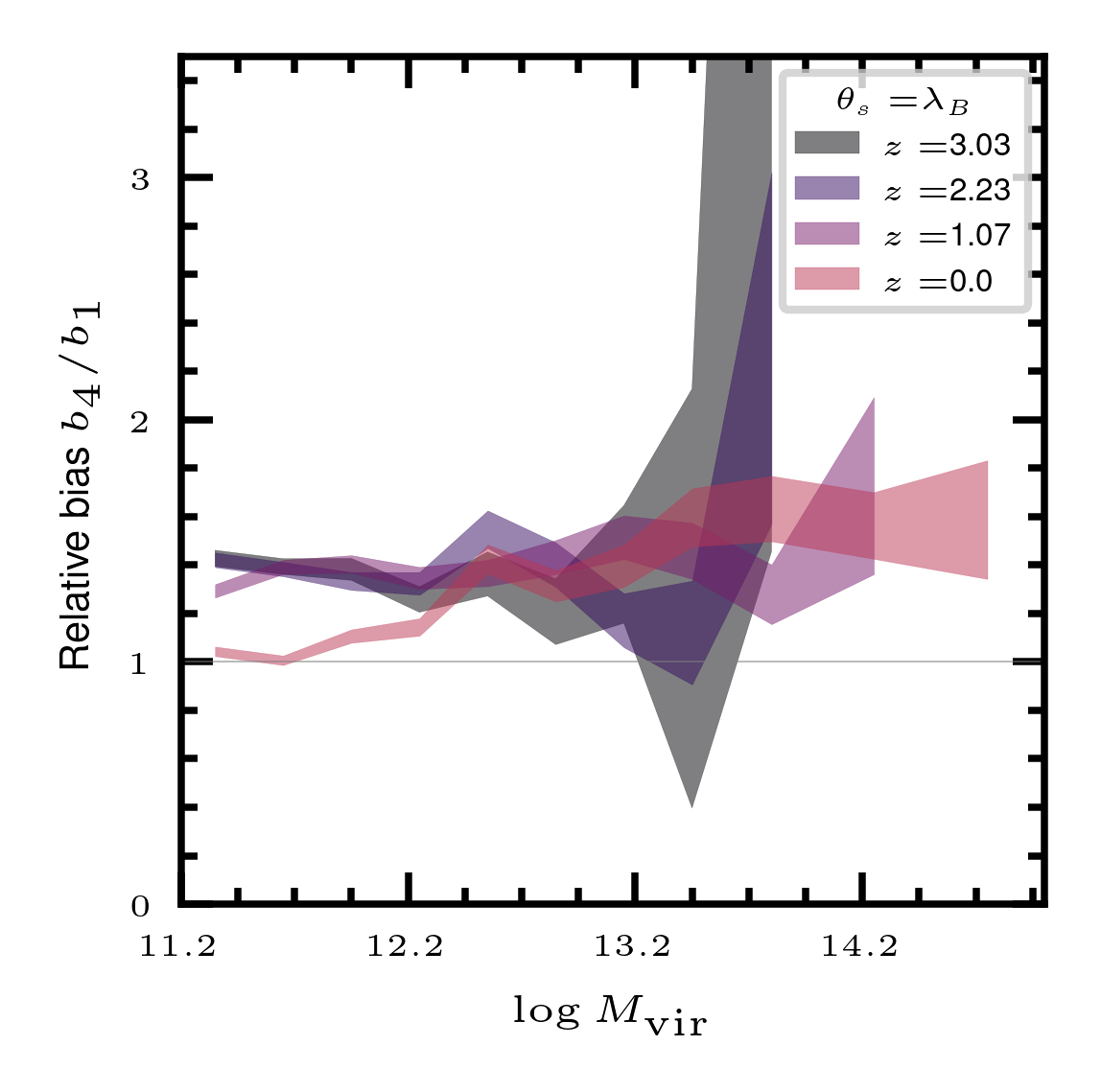}
\includegraphics[trim = 0cm 0.2cm 0cm 0cm ,clip=true, width=.47\textwidth]{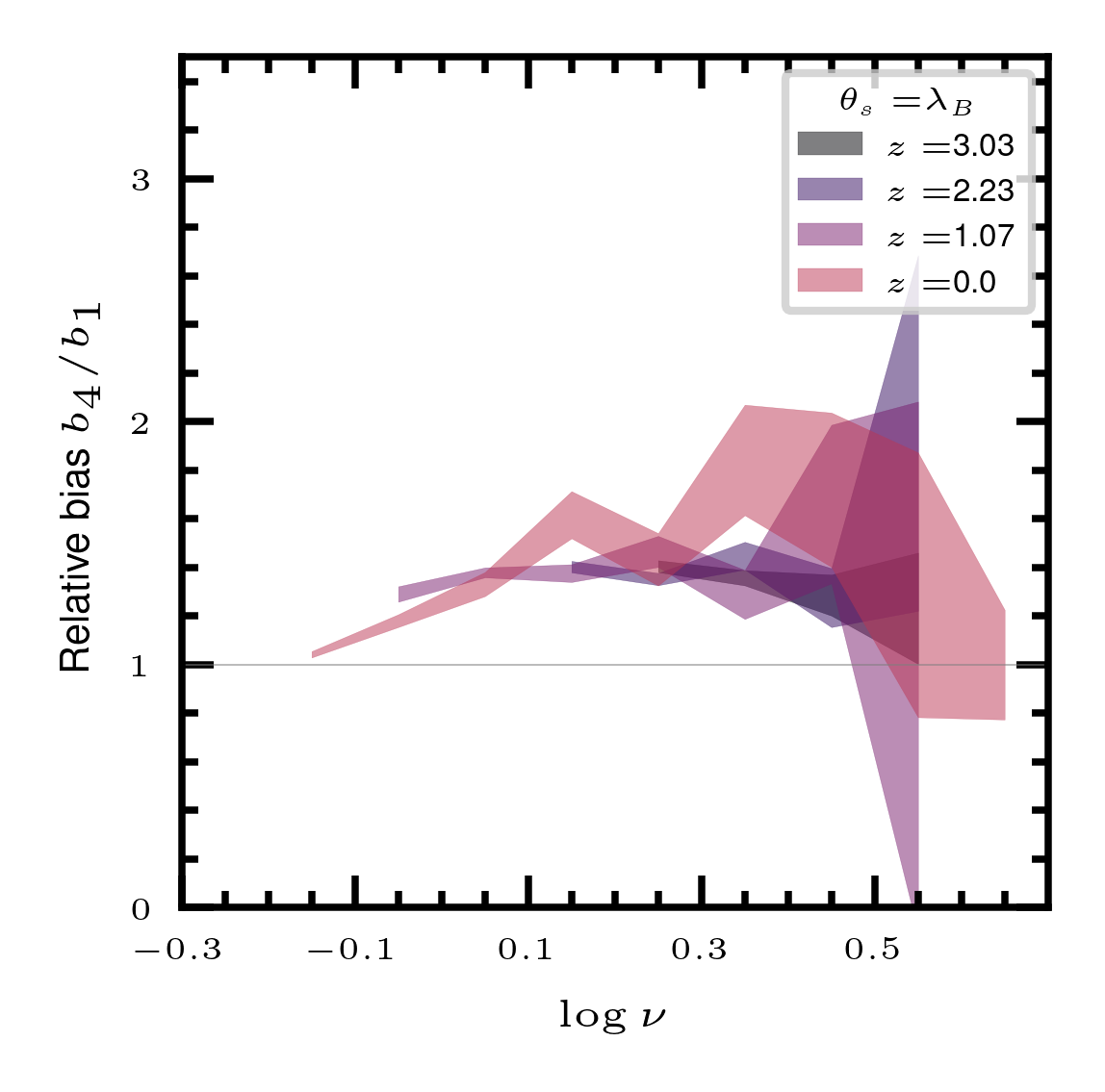}
\caption{\small{Ratio between the halo bias as a function of the halo mass (left column) and peak height (right column) obtained in two different quartiles of halo concentration $C_{\rm vir}$ (top) and halo spin $\lambda_{B}$ (bottom). The shaded area denotes the uncertainty of the ratio computed from the error bars associated with each effective bias.}}
\label{fig:sec_bias_rat}
\end{figure*}
%==========================================
%==========================================

A full characterization of the halo scaling relations is beyond the scope of this paper \citep[see e.g.,][for a detailed analysis on halo and cluster scaling relations]{2012A&A...539A.120B,2014MNRAS.438...49R,2014MNRAS.441.3562E}. We nevertheless describe some relevant statistical features, the first being the correlation among properties. In Fig.~\ref{fig:bias_spear} we show the Spearman's rank correlation coefficient \footnote{We have also computed the Pearson correlation coefficient, leading to similar results.} $\rho_{s}$ between the effective halo bias and a number of halo properties, as a function of redshift. In general, the correlation coefficient is small, meaning that for all properties, the link between effective halo bias and halo properties cannot be described through a monotonic relation (as expected, given the large scatter shown in e.g., Fig.~\ref{fig:bias_nu}). Among the intrinsic properties, halo mass and halos spin display the largest correlations, a trend that varies at low redshifts where halo concentration displays, interestingly, a larger correlation. The halo geometry, characterized by the ellipticity $\mathcal{E}_{h}$, shows a significant anticorrelation (with respect to intrinsic properties), similar in absolute value to that displayed by mass.

The main message from this plot is that nonlocal and environmental properties have higher-degree correlation with halo bias (compared to intrinsic properties). While this might be expected for the nonlocal halo properties (e.g., $\mathcal{D}_{5}$ and $\Delta_{5}$ are probes of small-scale clustering and hence can be coupled via nonlinear evolution with the large-scale modes used to measure the effective bias), the behavior with respect to the environmental properties (local density, tidal anisotropy) plots an scenario in which halo bias is mainly driven by the environment where halos reside. This will be key in order to explain the signal of secondary bias, as was previously addressed by \cite{2018MNRAS.476.3631P, 2019MNRAS.489.2977R}.

In Fig.~\ref{fig:bias_spear_cwt} we present the correlation coefficient $\rho_{s}$ for knots (high density) and sheets (intermediate to low density) as a function of redshift. The environmental properties display in general more correlation with bias in high density regions, while the intrinsic halo properties mildly change from one cosmic-environment to another. The behavior in voids is very noisy and oscillates around $\rho_ {s}=0$ for almost all properties thought the redshift range.

%==========================================
%==========================================
\begin{figure*}
%assembly_bias_delta.py
\centering
\includegraphics[trim = 0.1cm 0.52cm 0cm 0cm ,clip=true, width=.42\textwidth]{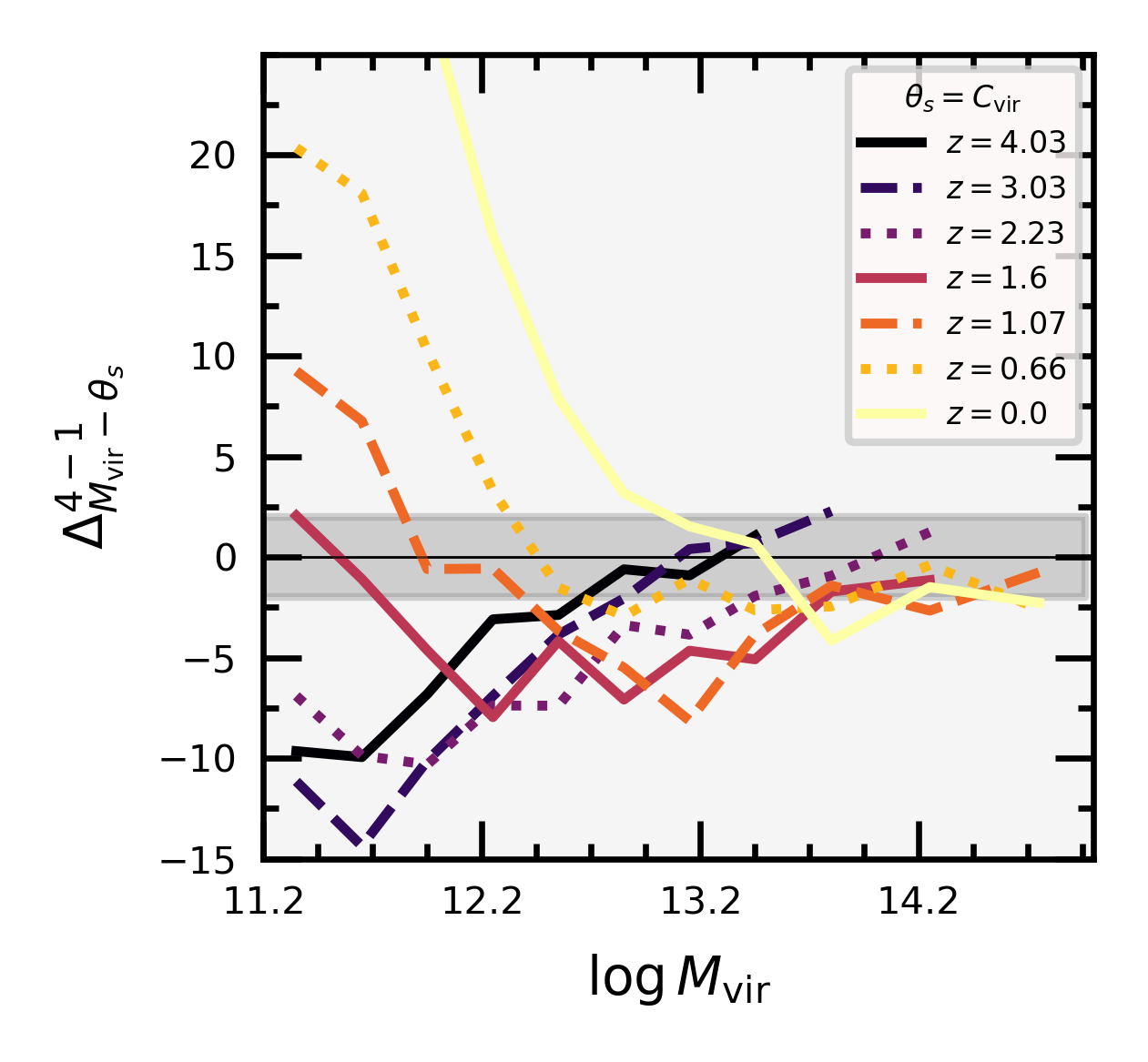}
\includegraphics[trim = 0.1cm 0.52cm 0cm 0cm ,clip=true, width=.42\textwidth]{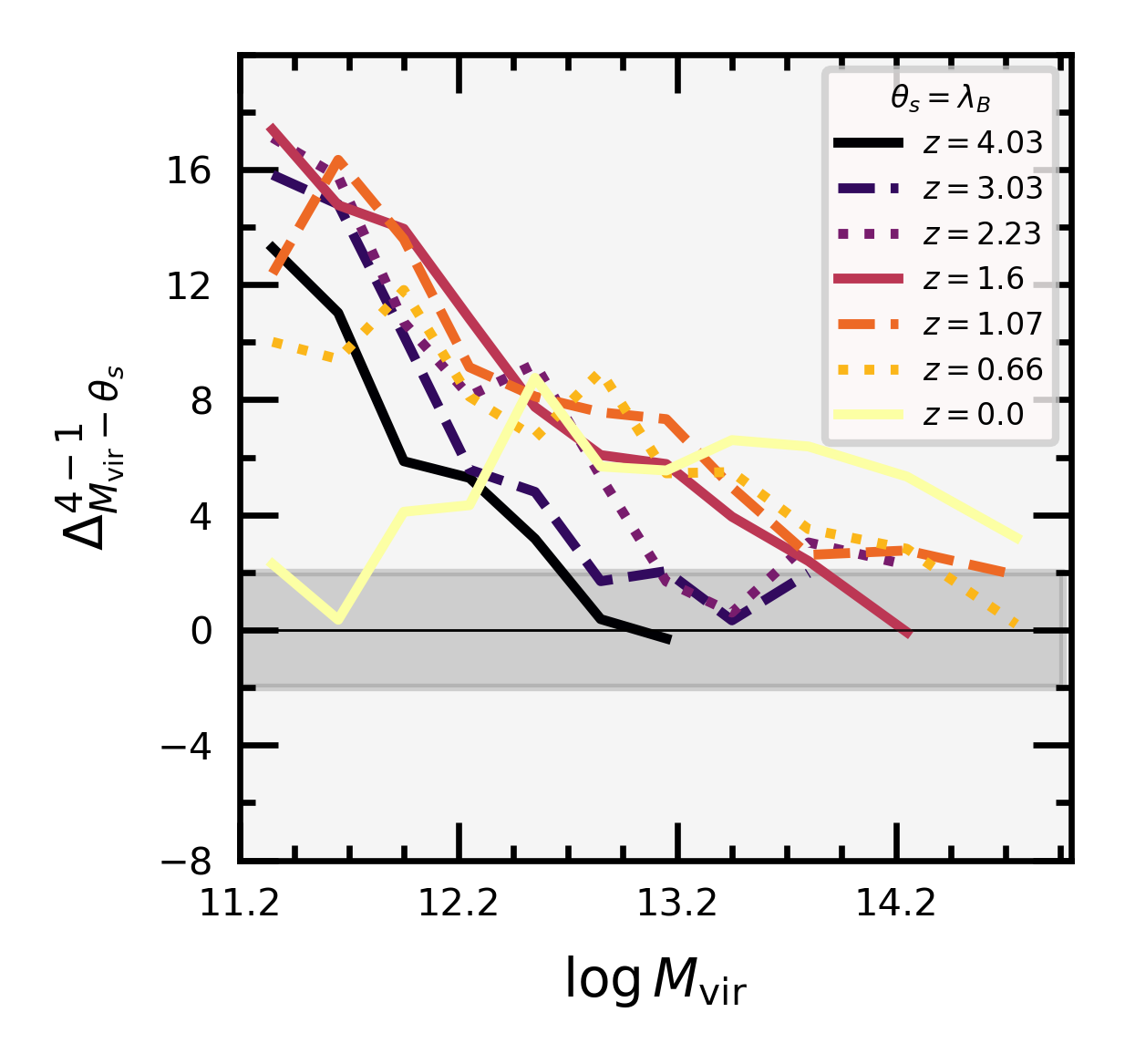}
\includegraphics[trim = 0.1cm 0.52cm 0cm 0cm ,clip=true, width=.42\textwidth]{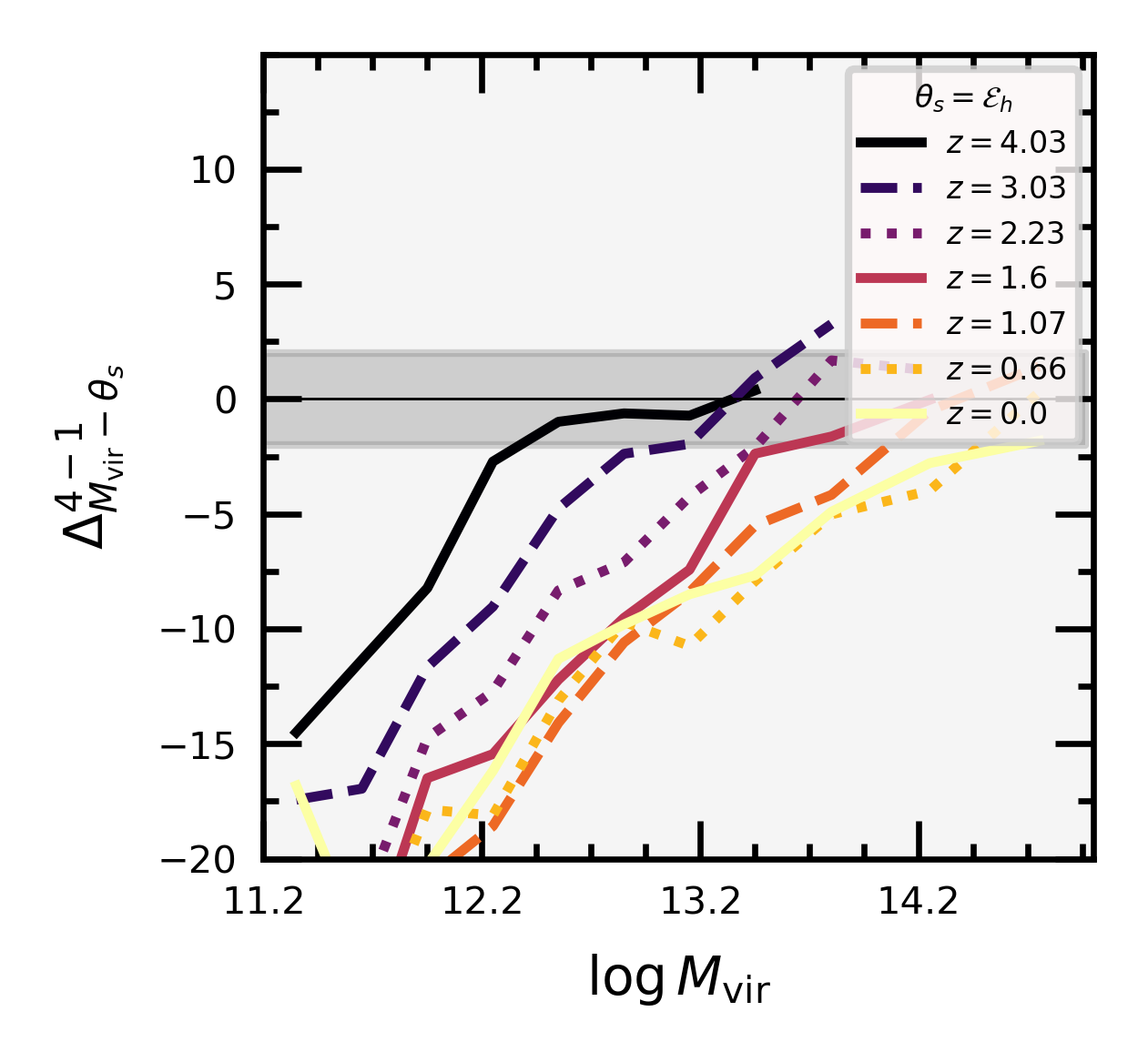}
\includegraphics[trim = 0.1cm 0.52cm 0cm 0cm ,clip=true, width=.42\textwidth]{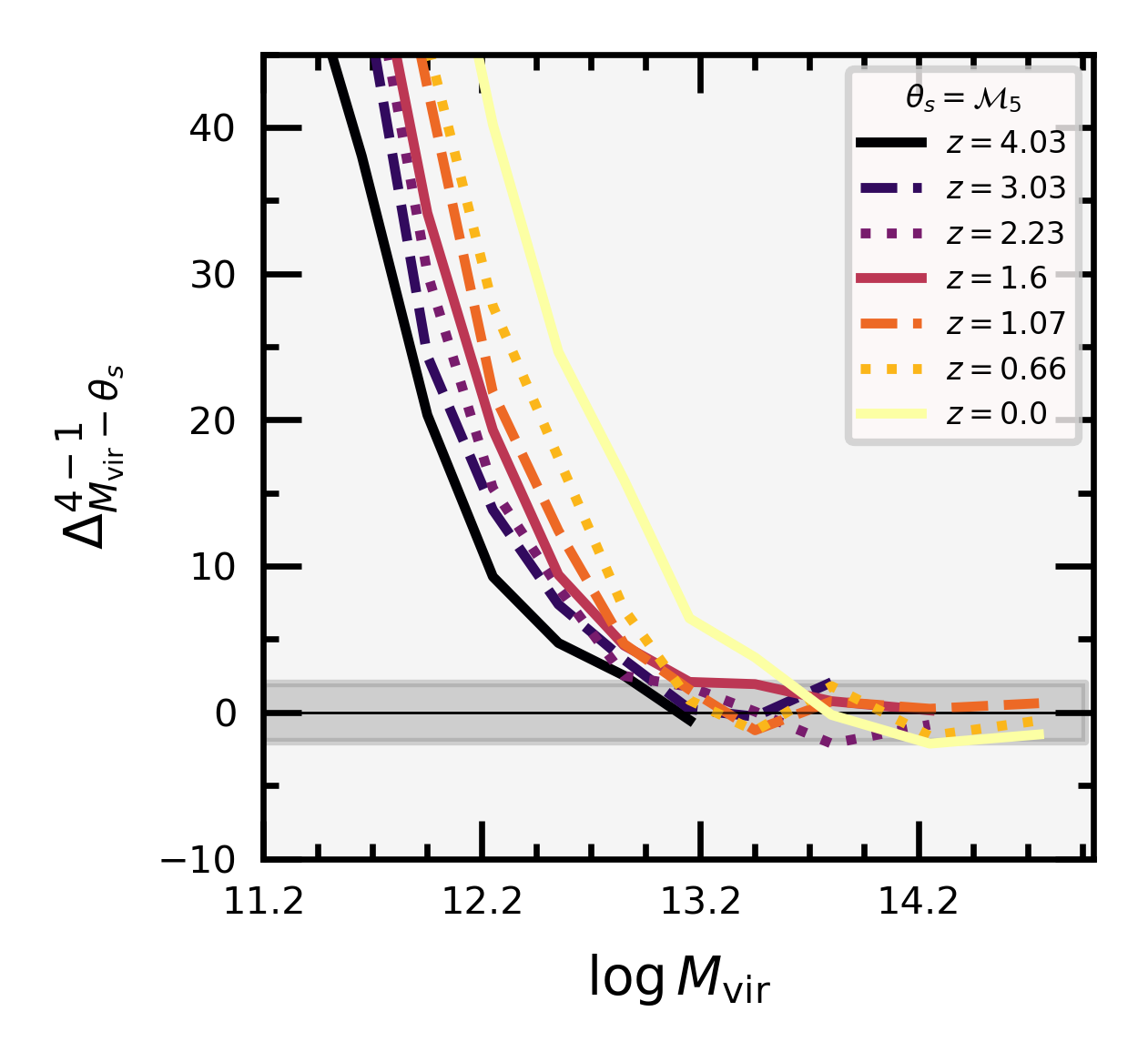}
\includegraphics[trim = 0.1cm 0.1cm 0cm 0cm ,clip=true, width=.42\textwidth]{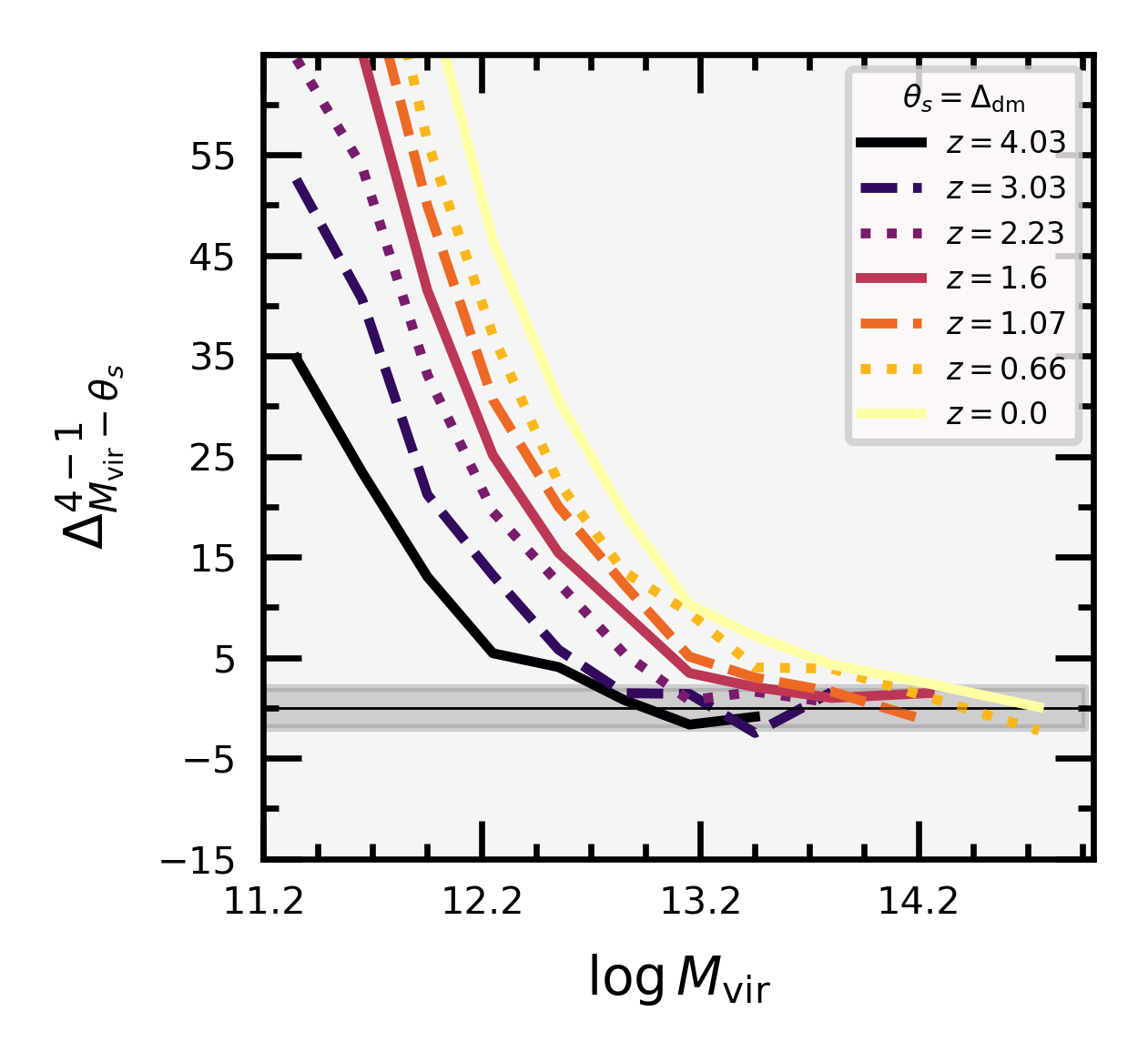}
\includegraphics[trim = 0.1cm 0.1cm 0cm 0cm ,clip=true, width=.42\textwidth]{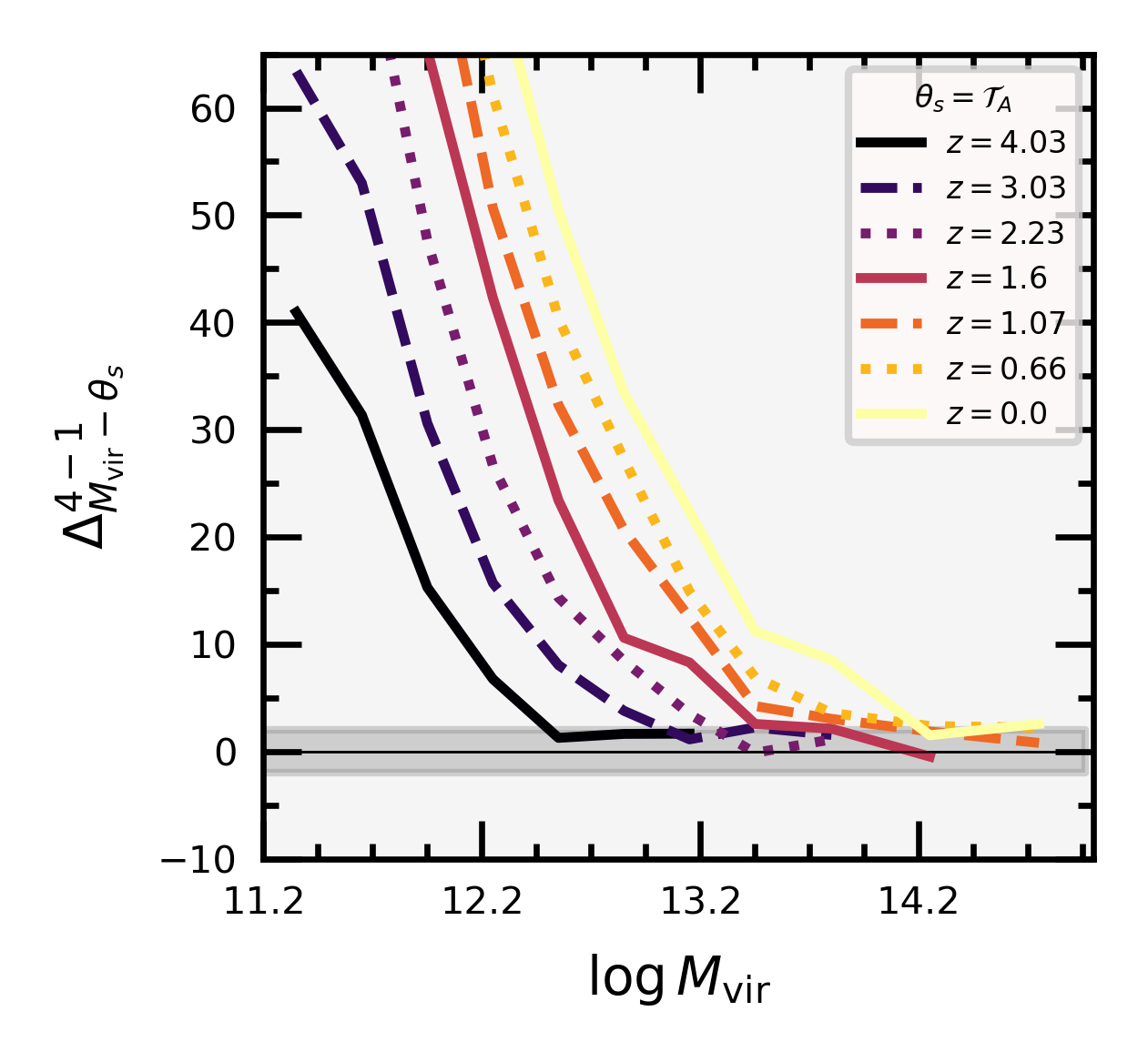}
\caption{\small{Statistical significance of the secondary halo bias as a function of halo mass, computed using Eq.~(\ref{eq:Delta}) for different redshifts and for a number of secondary halo properties: concentration $C_{\rm vir}$, spin $\lambda_{B}$, ellipticity $\mathcal{E}_{h}$, Mach number $\mathcal{M}_{5}$, local dark matter overdensity $\Delta_{\rm dm}$, and tidal anisotropy $\mathcal{T}_{A}$. The dark-shaded stripe denotes a strip of $\Delta^{4-1}_{p-s} \pm 2$ and the horizontal line marks $\Delta^{4-1}_{p-s}=0$. $\Delta^{4-1}_{p-s}>0$ ($<0$) implies that upper (lower) quartiles display higher clustering strength than lower (upper) quartiles.}}
\label{fig:sec_bias_allz1}
\end{figure*}

%==========================================
%==========================================
\begin{figure*}[htbp]
\centering
%bias_scaling_relation_assembly_cwt.py
\includegraphics[trim = .1cm 0.87cm 0cm 0cm ,clip=true, width=0.232\textwidth]{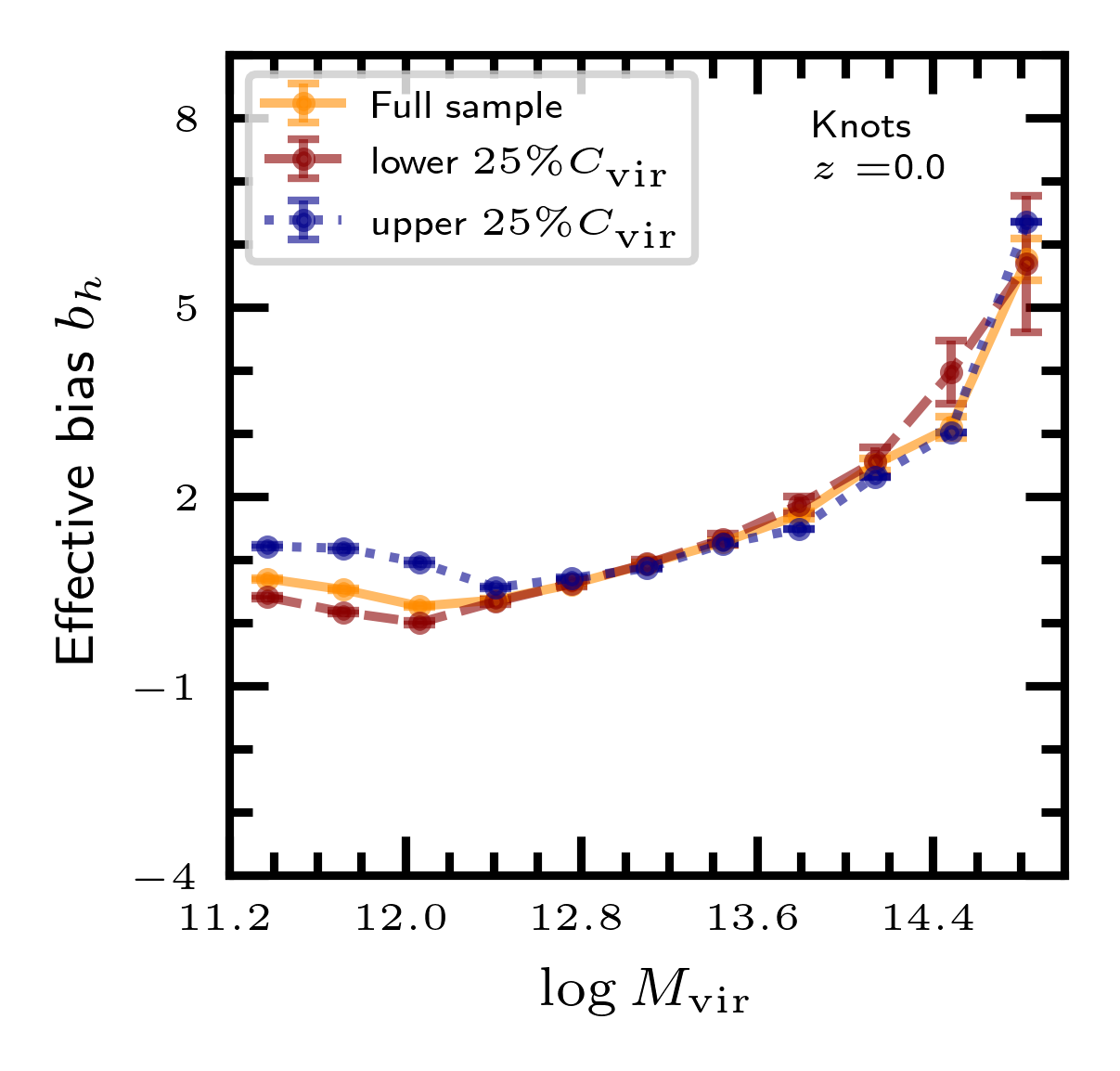}
\includegraphics[trim = .1cm .87cm 0cm 0cm ,clip=true, width=0.232\textwidth]{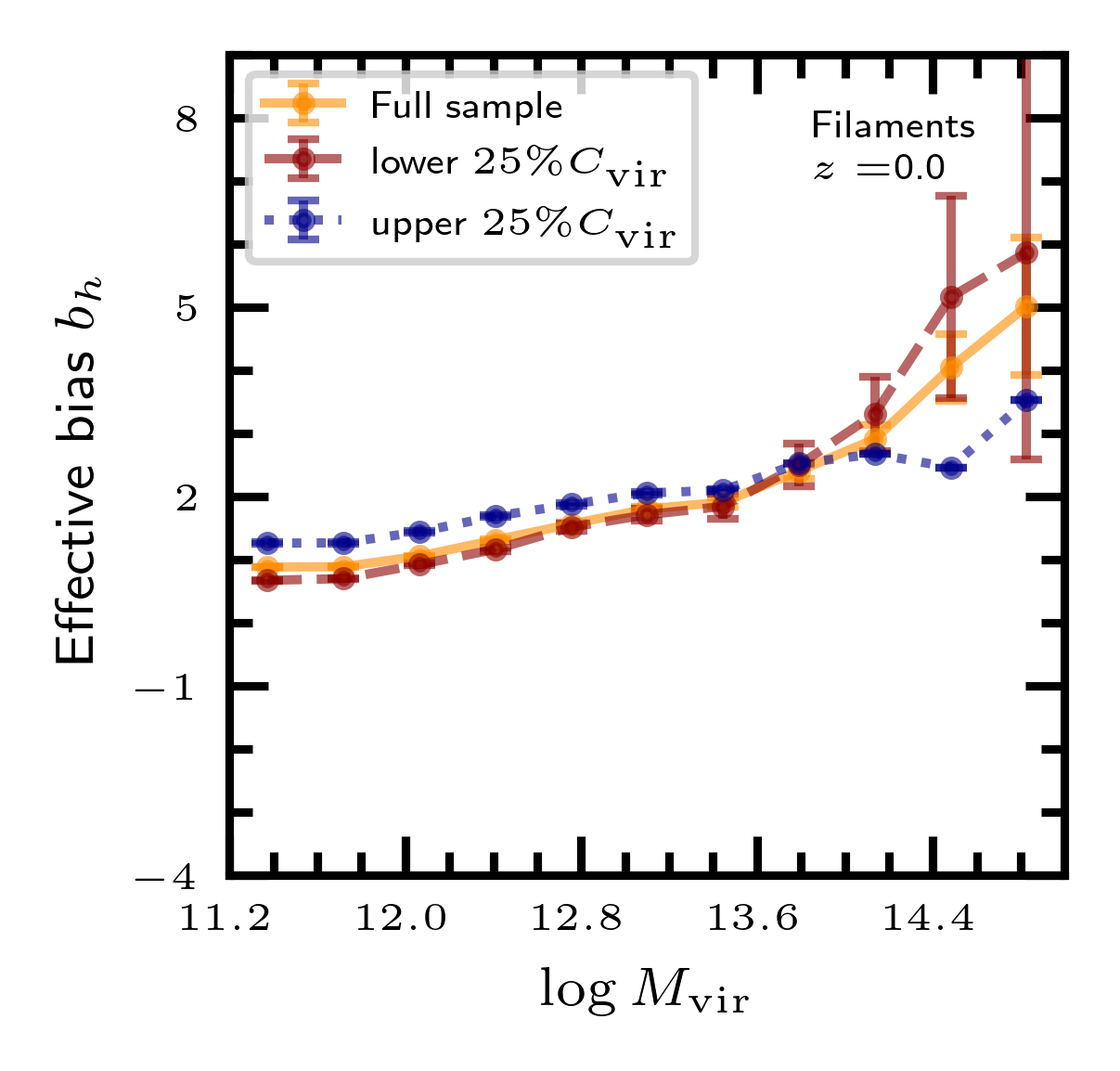}
\includegraphics[trim = .1cm .87cm 0cm 0cm ,clip=true, width=0.232\textwidth]{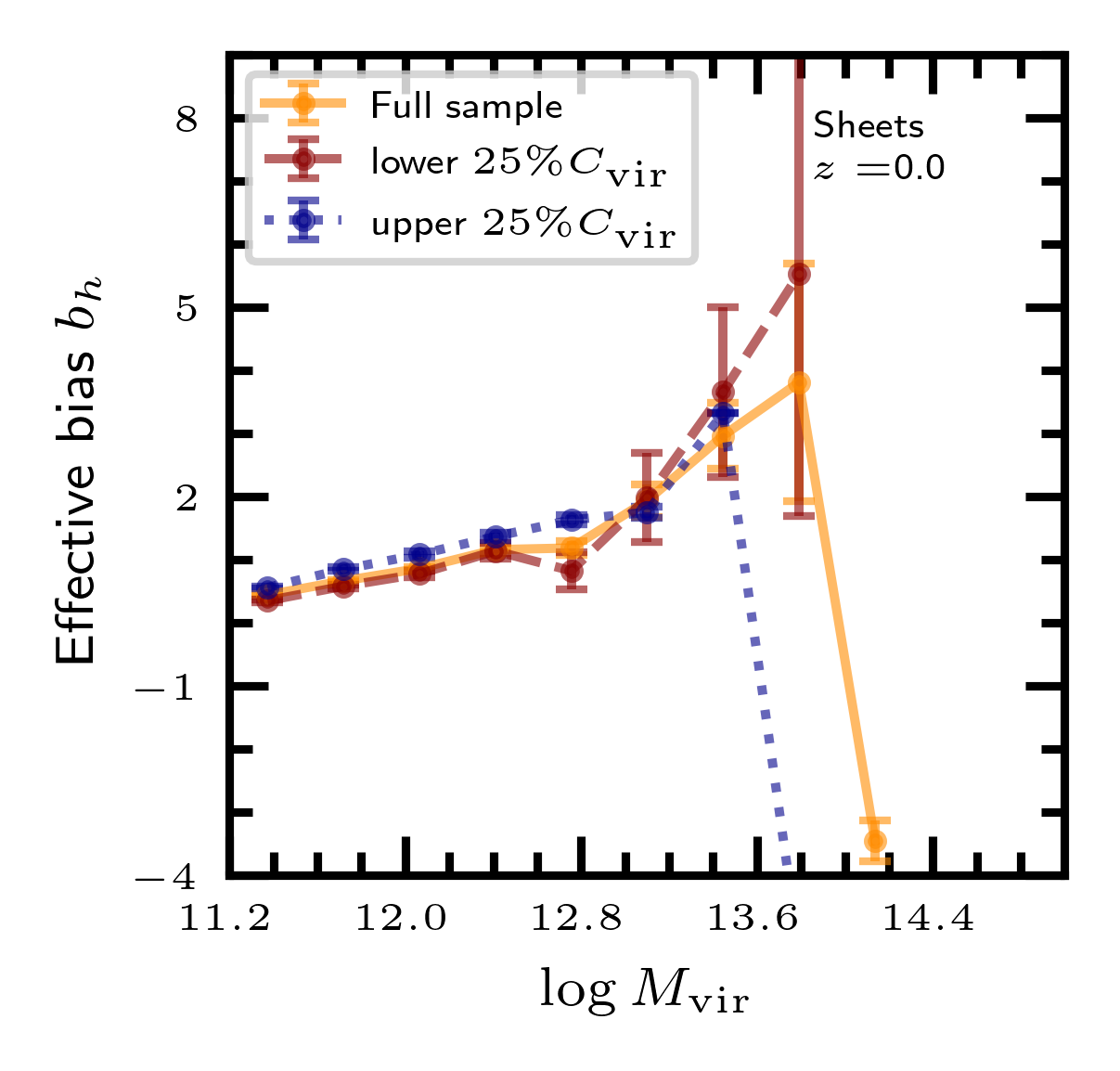}
\includegraphics[trim = .1cm .87cm 0cm 0cm ,clip=true, width=0.232\textwidth]{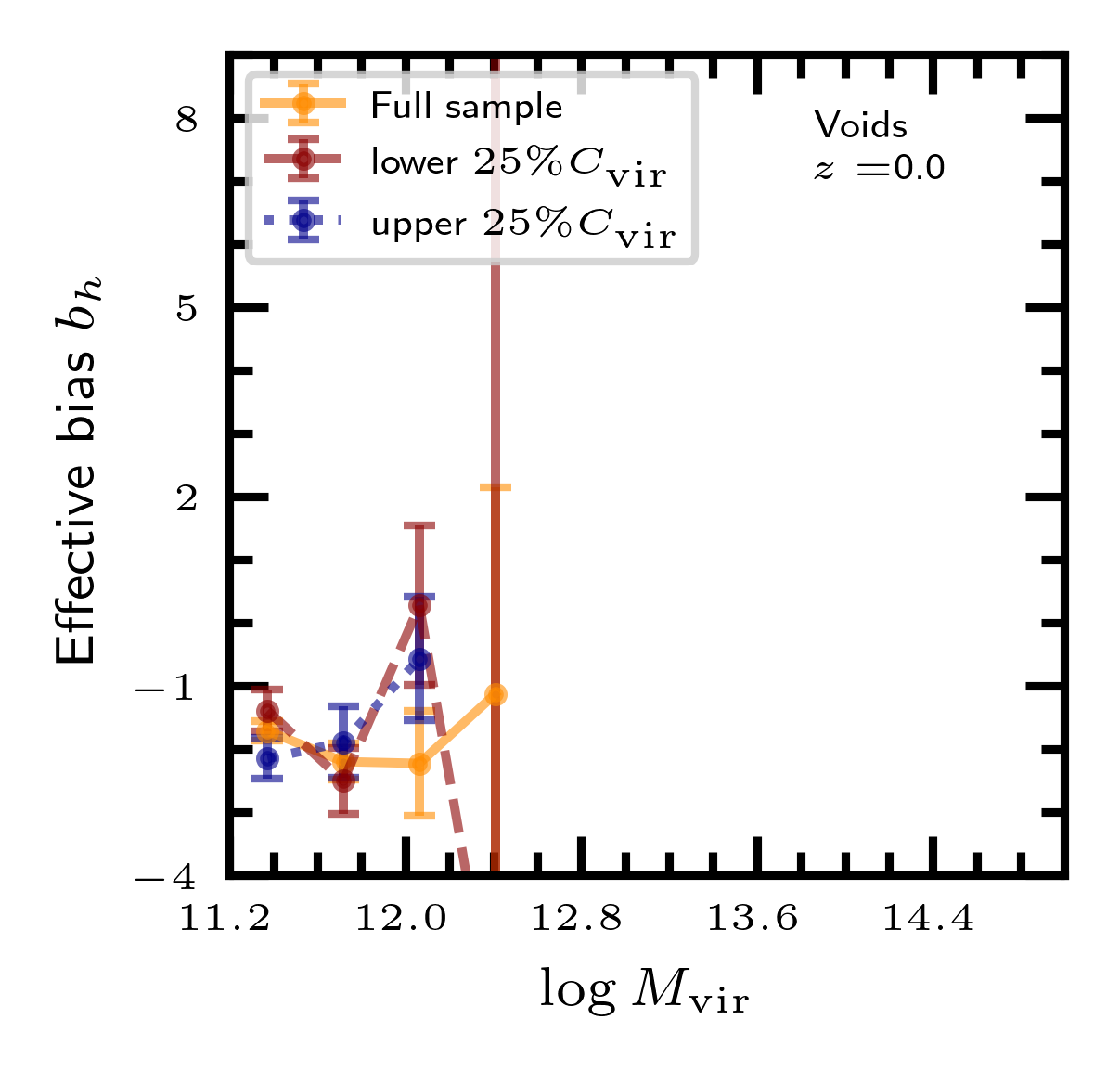}
\includegraphics[trim = .1cm 0.87cm 0cm 0cm ,clip=true, width=0.232\textwidth]{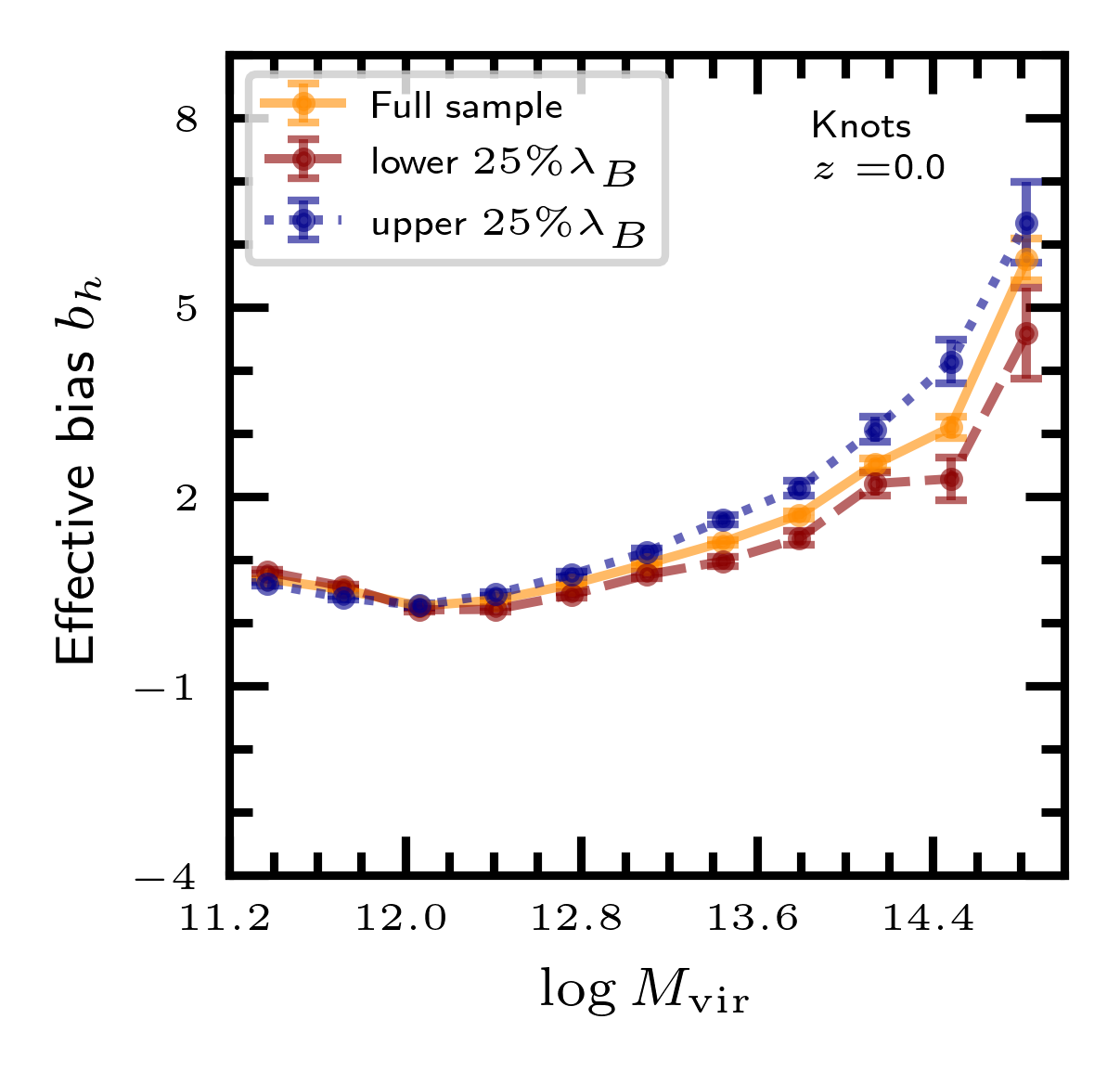}
\includegraphics[trim = .1cm .87cm 0cm 0cm ,clip=true, width=0.232\textwidth]{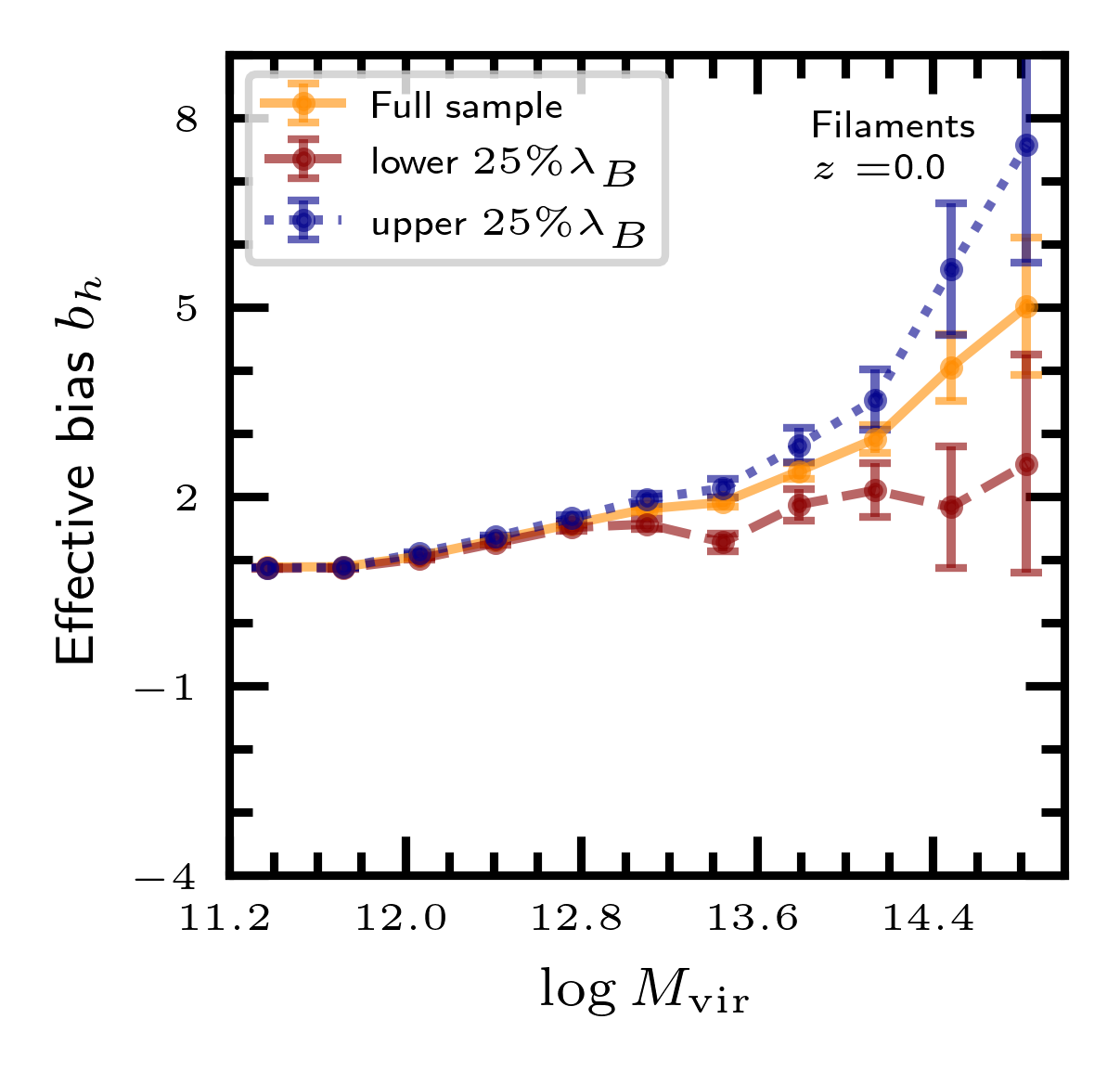}
\includegraphics[trim = .1cm .87cm 0cm 0cm ,clip=true, width=0.232\textwidth]{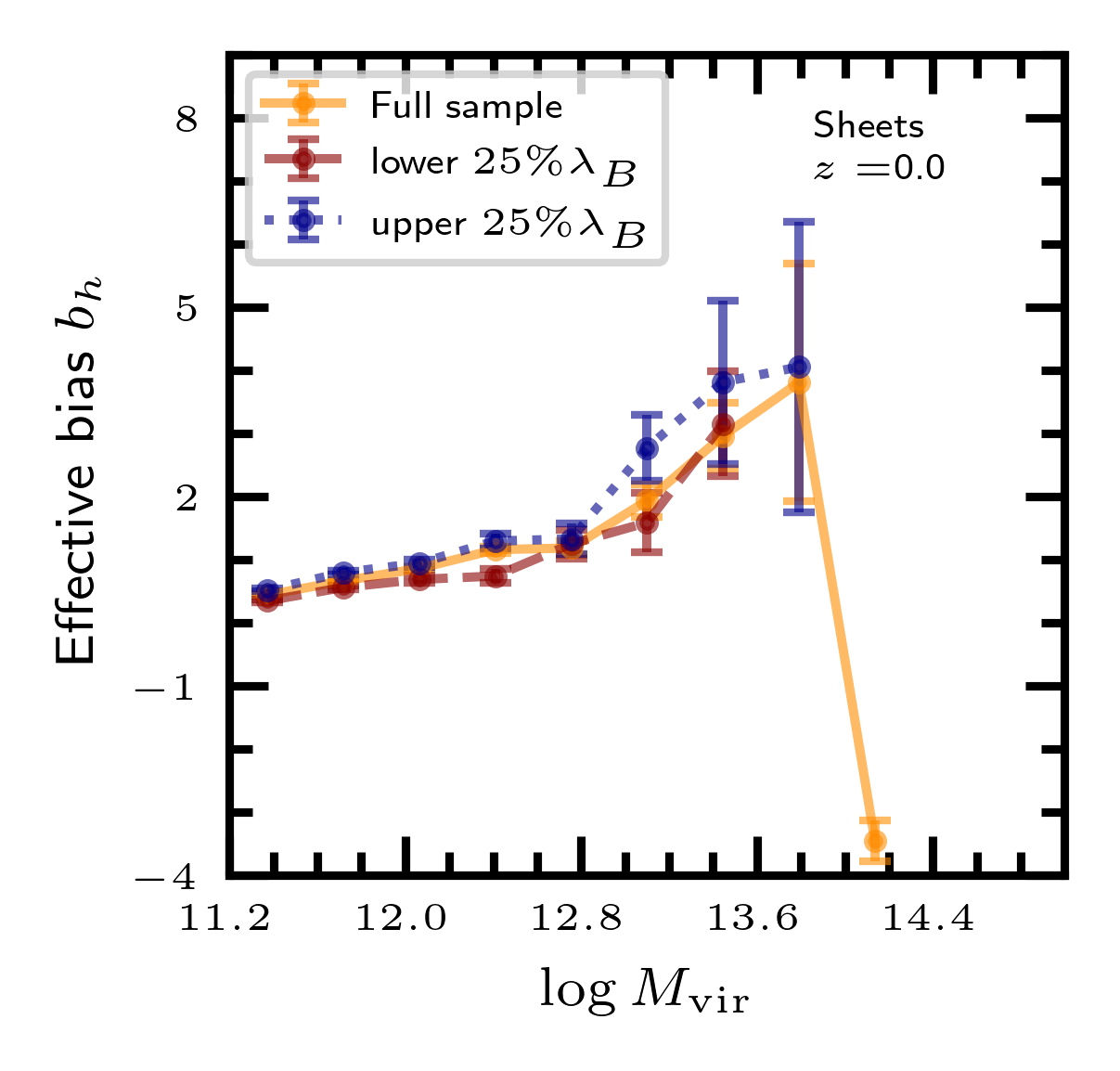}
\includegraphics[trim = .1cm .87cm 0cm 0cm ,clip=true, width=0.232\textwidth]{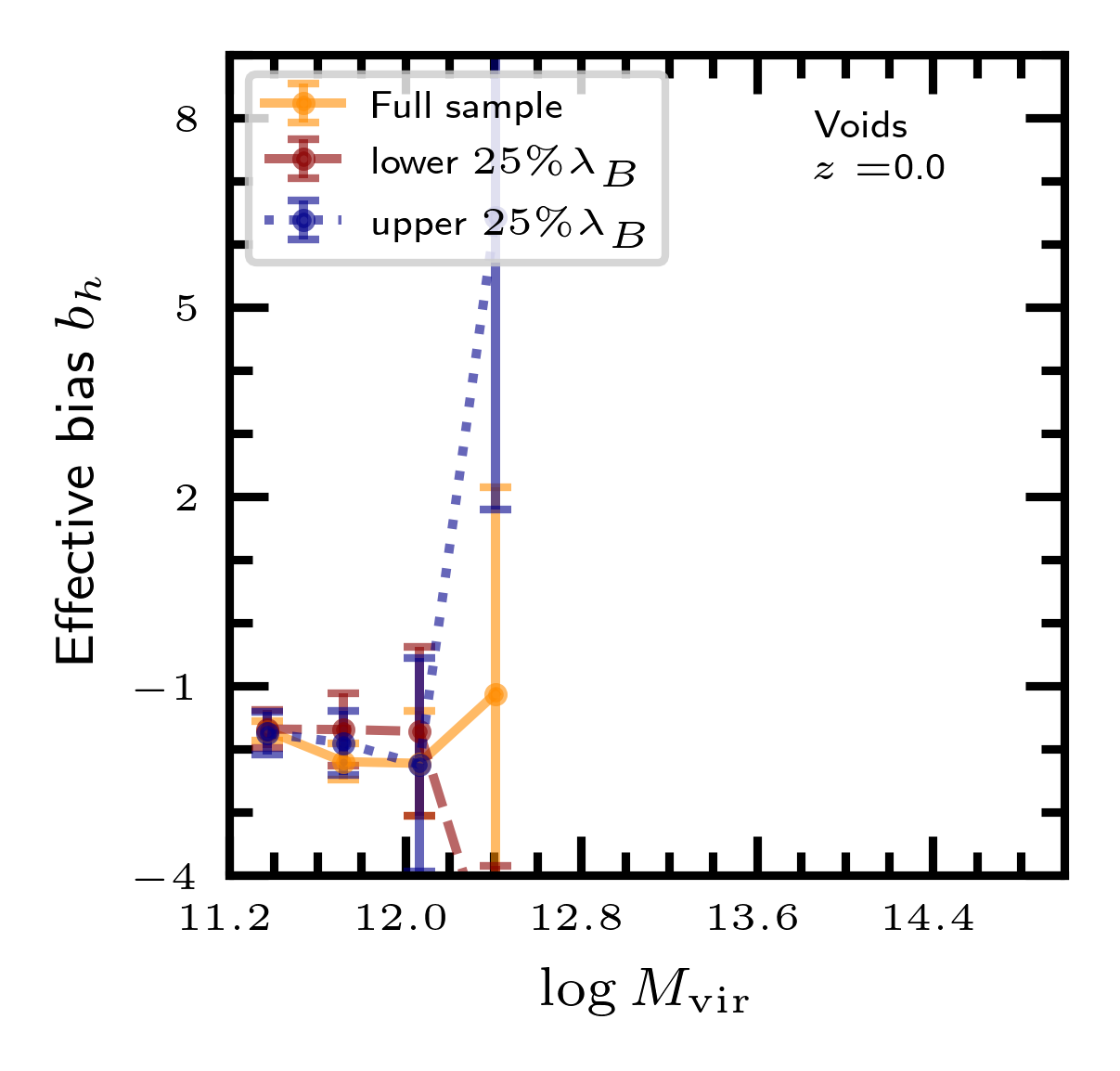}
\includegraphics[trim = .1cm 0.87cm 0cm 0cm ,clip=true, width=0.232\textwidth]{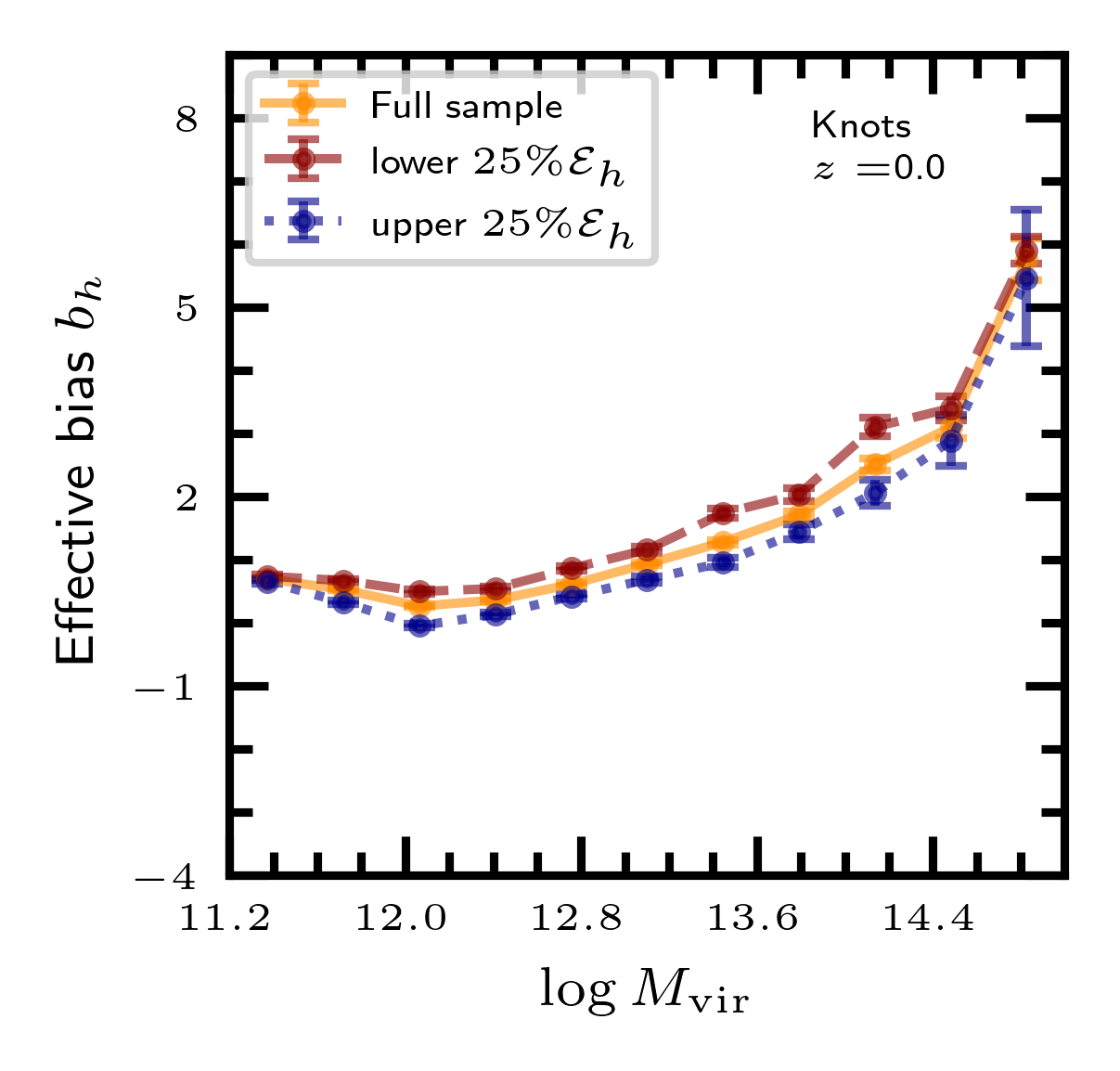}
\includegraphics[trim = .1cm .87cm 0cm 0cm ,clip=true, width=0.232\textwidth]{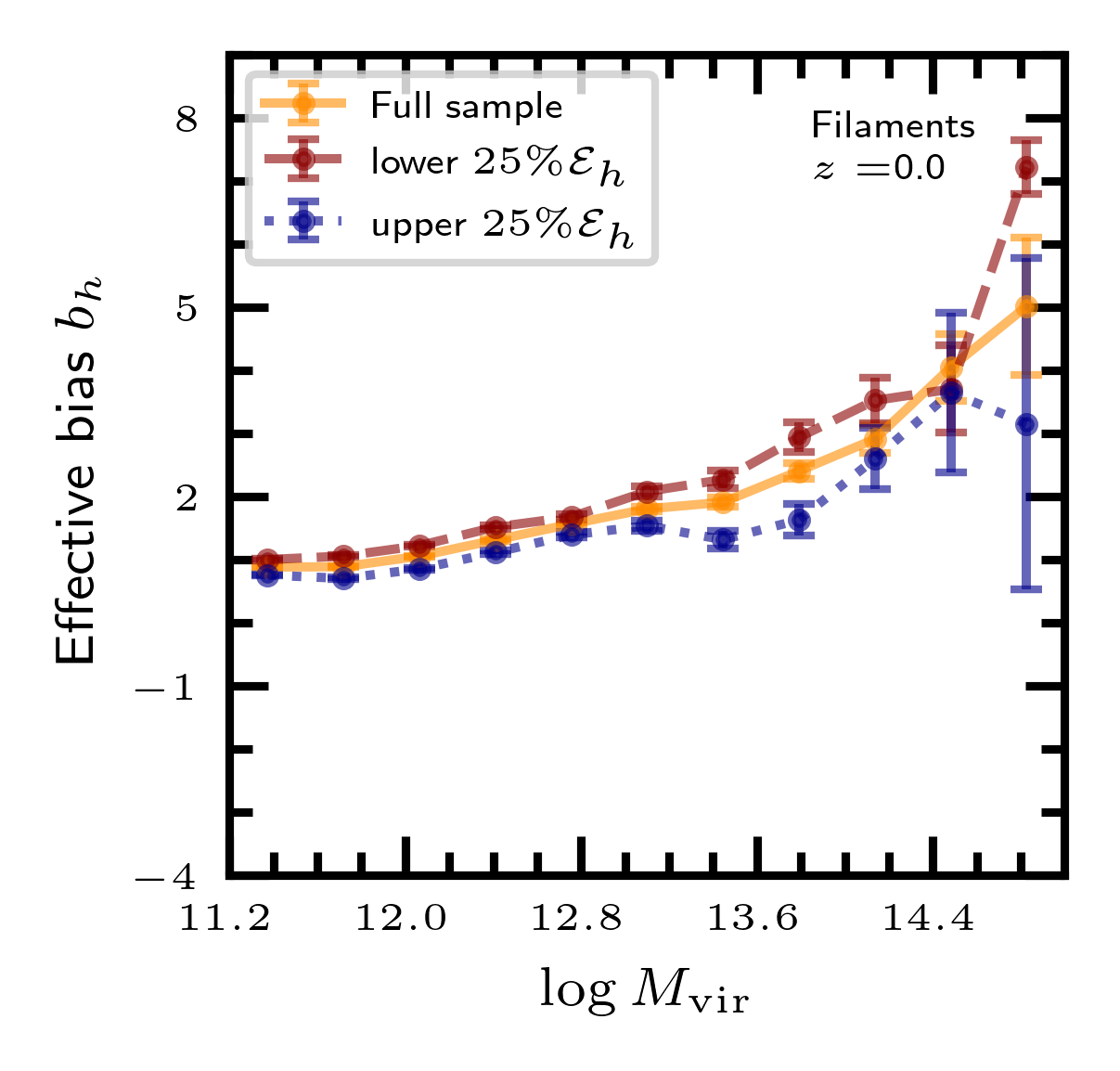}
\includegraphics[trim = .1cm .87cm 0cm 0cm ,clip=true, width=0.232\textwidth]{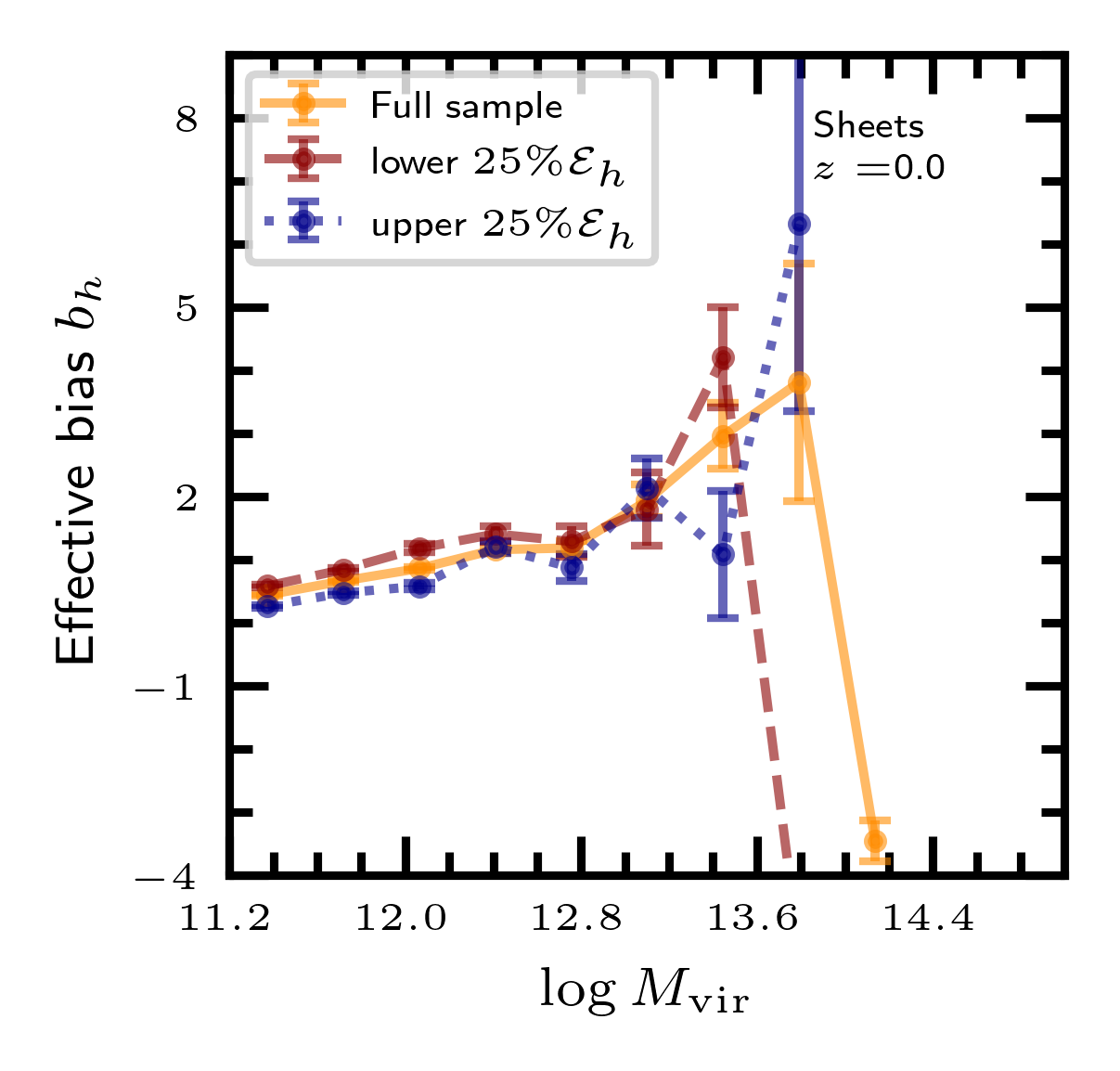}
\includegraphics[trim = .1cm .87cm 0cm 0cm ,clip=true, width=0.232\textwidth]{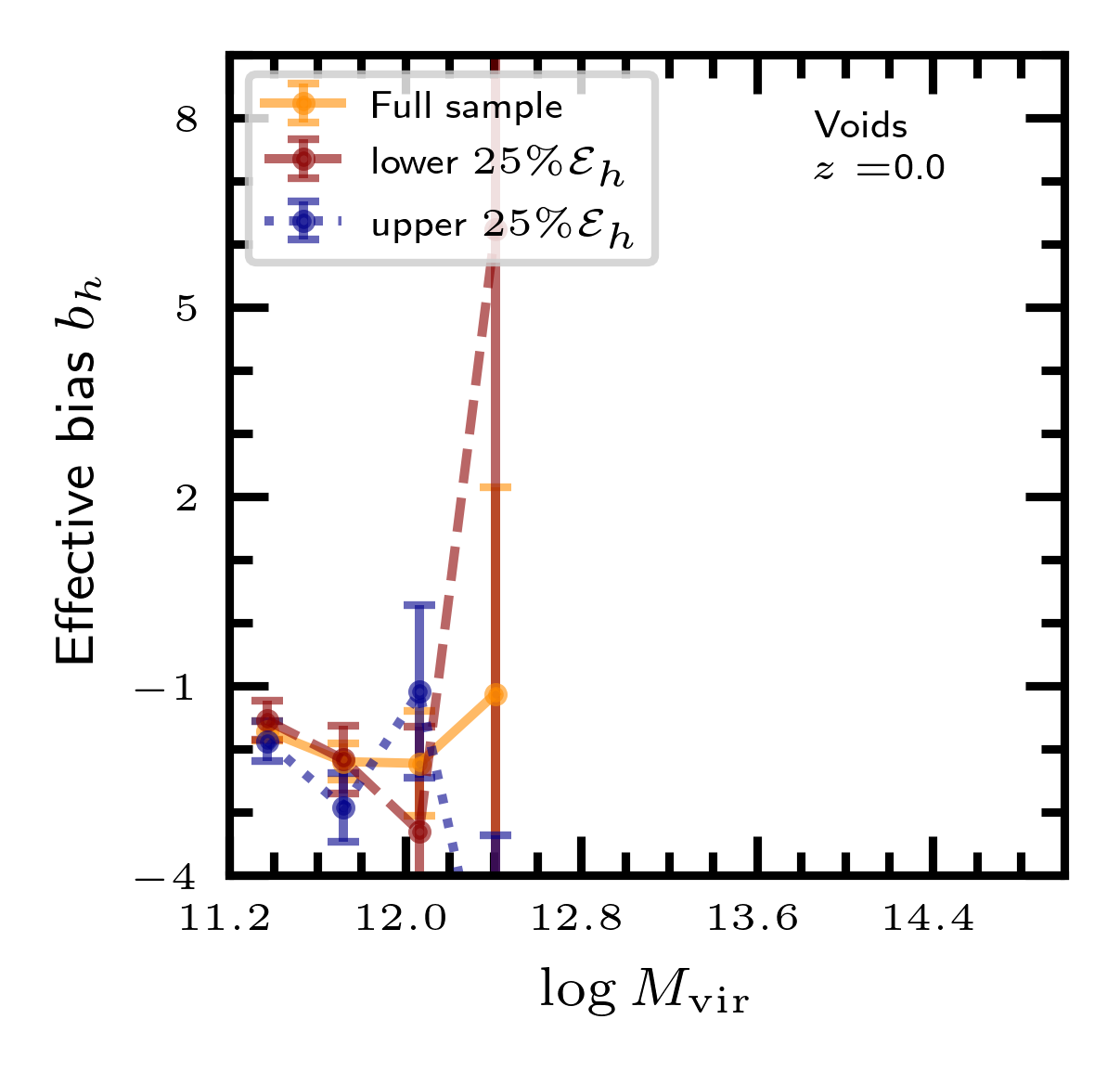}
\includegraphics[trim = .1cm 0.87cm 0cm 0cm ,clip=true, width=0.232\textwidth]{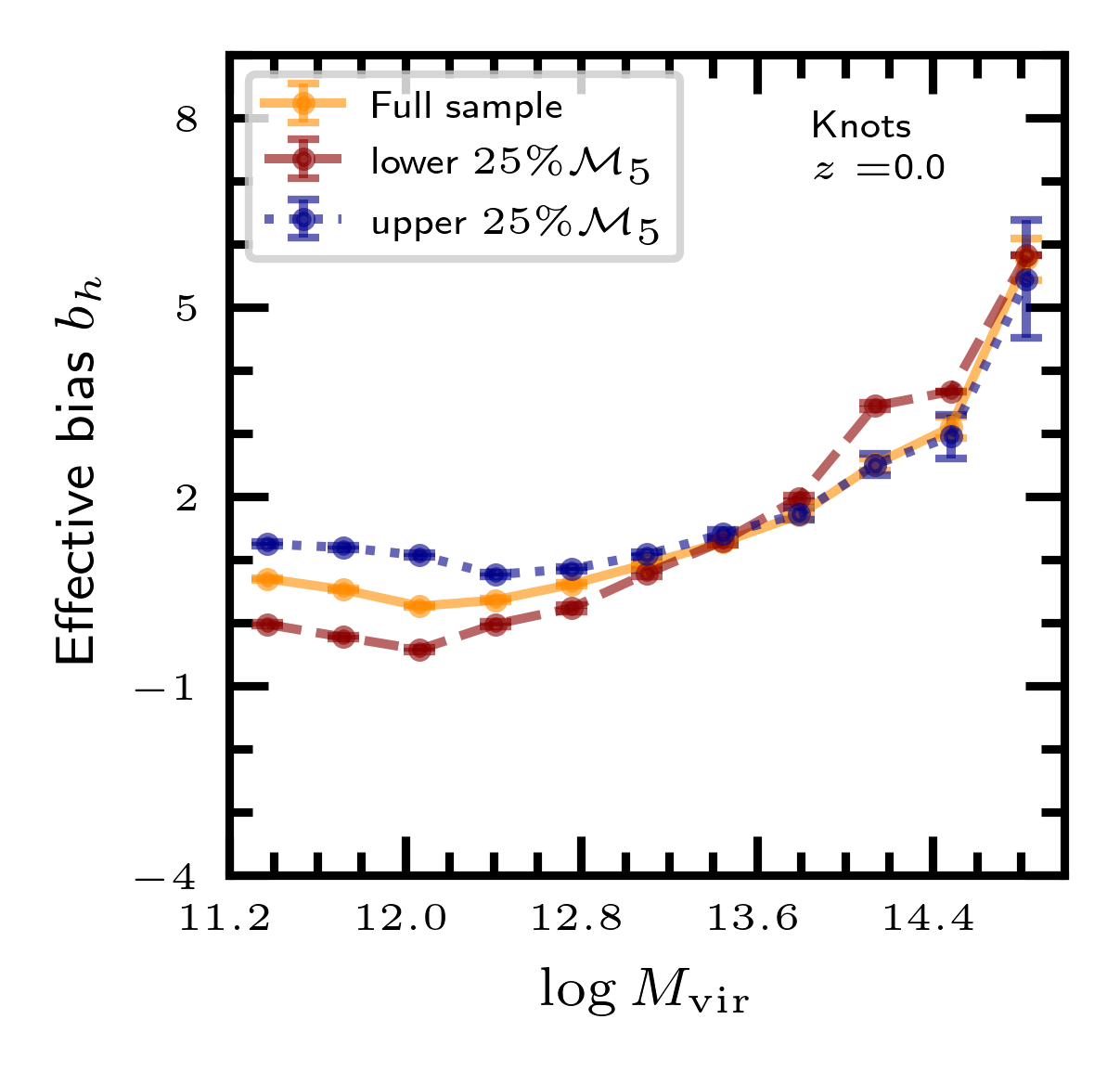}
\includegraphics[trim = .1cm .87cm 0cm 0cm ,clip=true, width=0.232\textwidth]{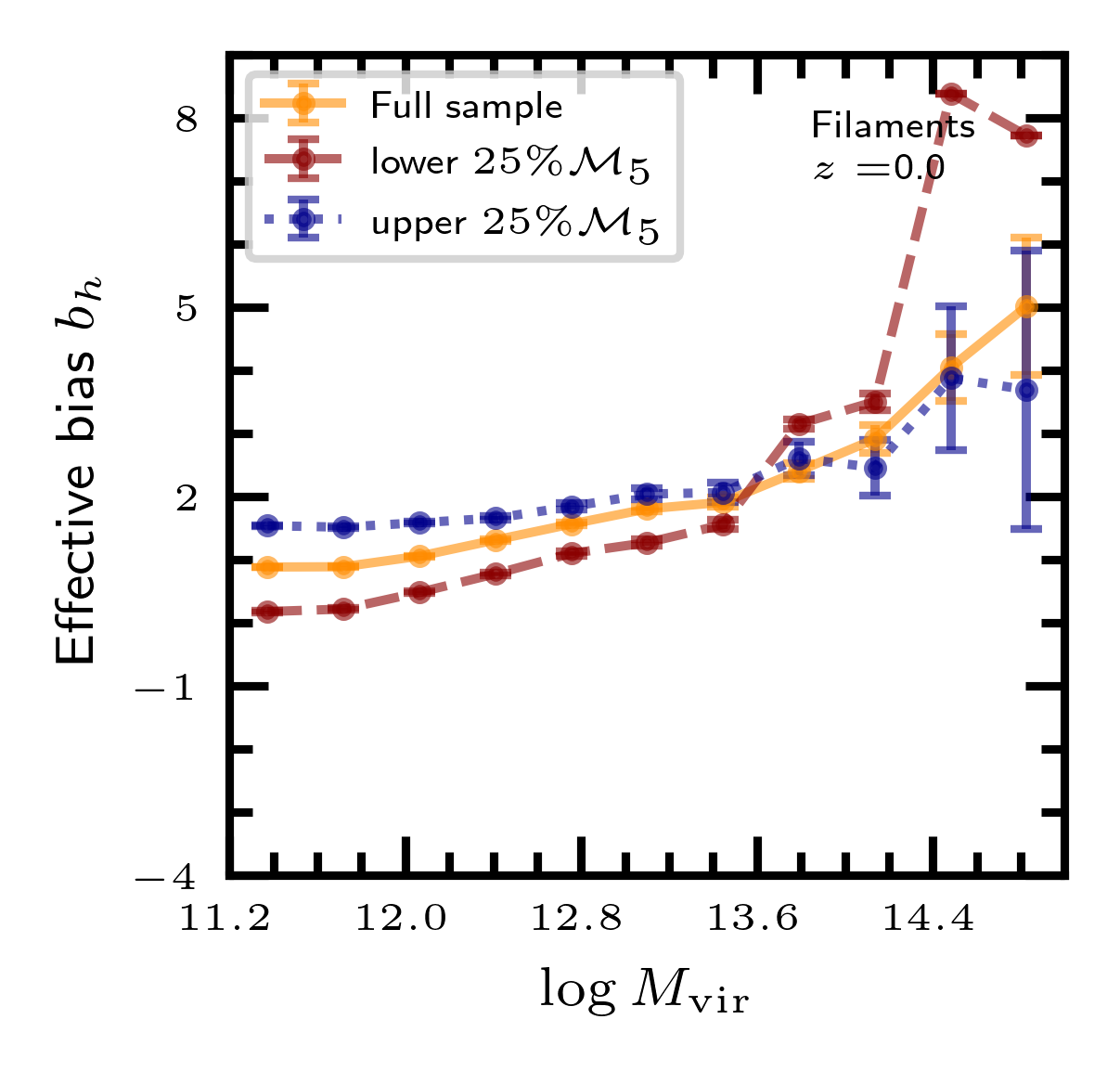}
\includegraphics[trim = .1cm .87cm 0cm 0cm ,clip=true, width=0.232\textwidth]{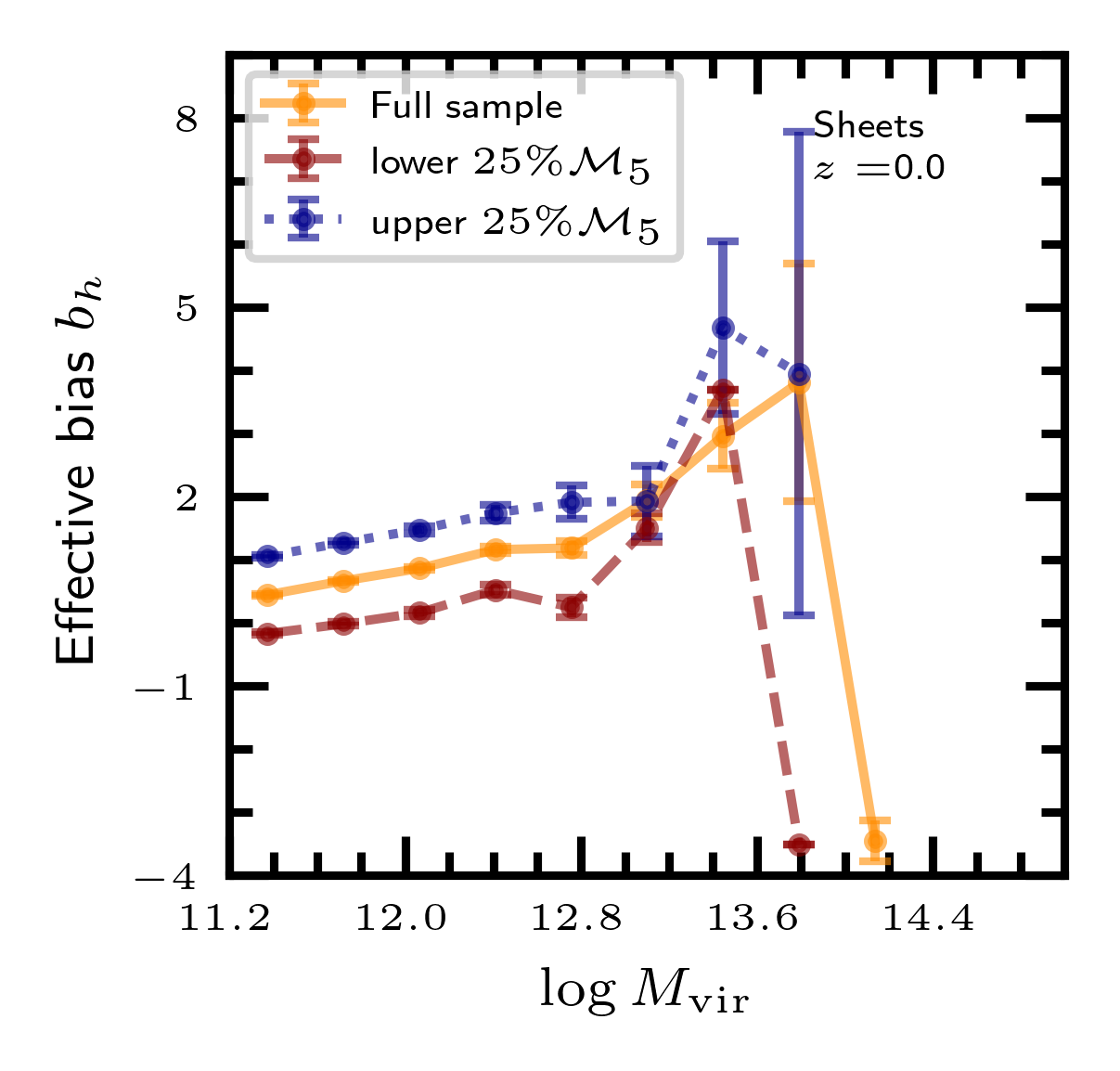}
\includegraphics[trim = .1cm .87cm 0cm 0cm ,clip=true, width=0.232\textwidth]{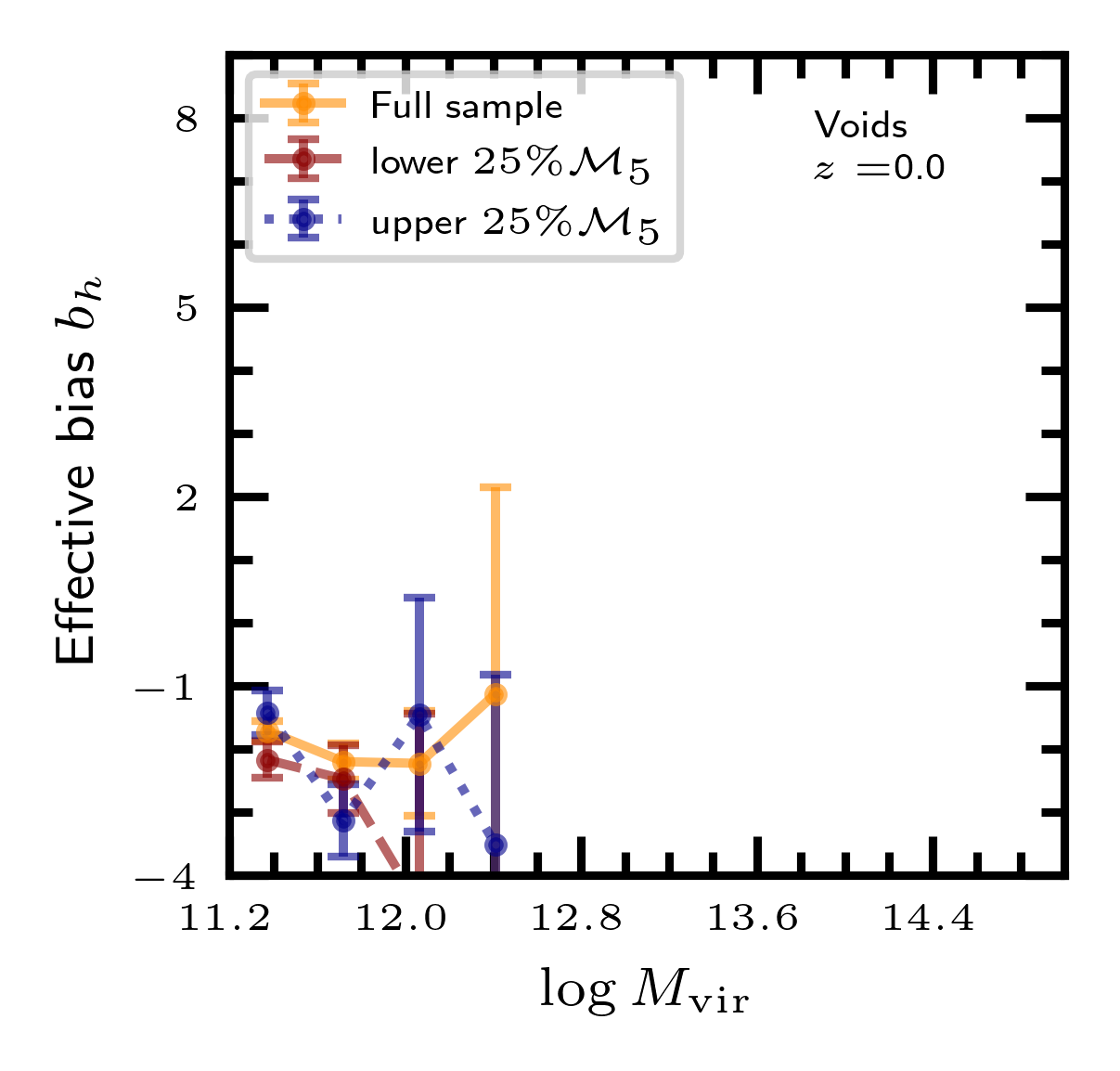}
\includegraphics[trim = .1cm 0.87cm 0cm 0cm ,clip=true, width=0.232\textwidth]{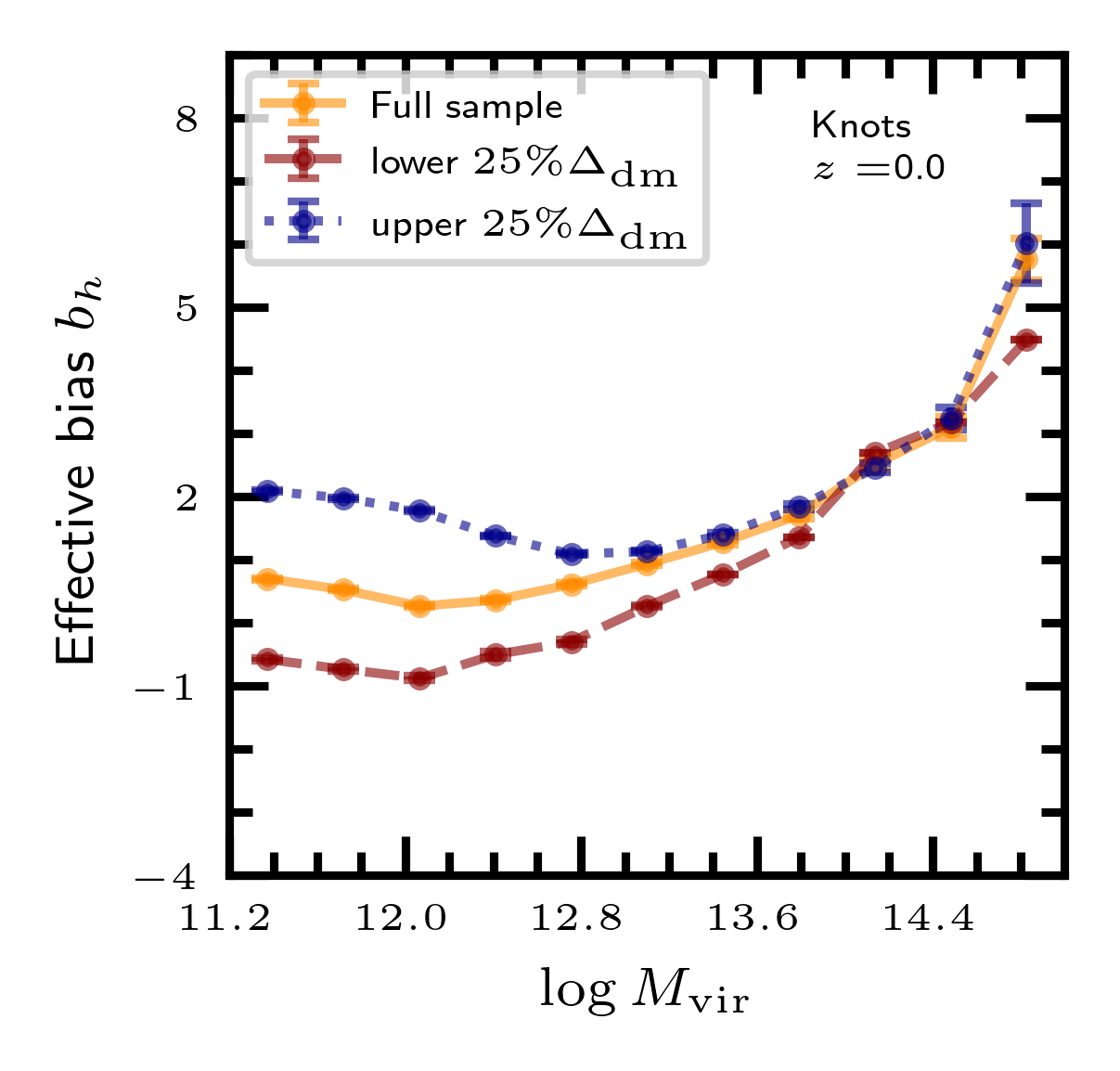}
\includegraphics[trim = .1cm .87cm 0cm 0cm ,clip=true, width=0.232\textwidth]{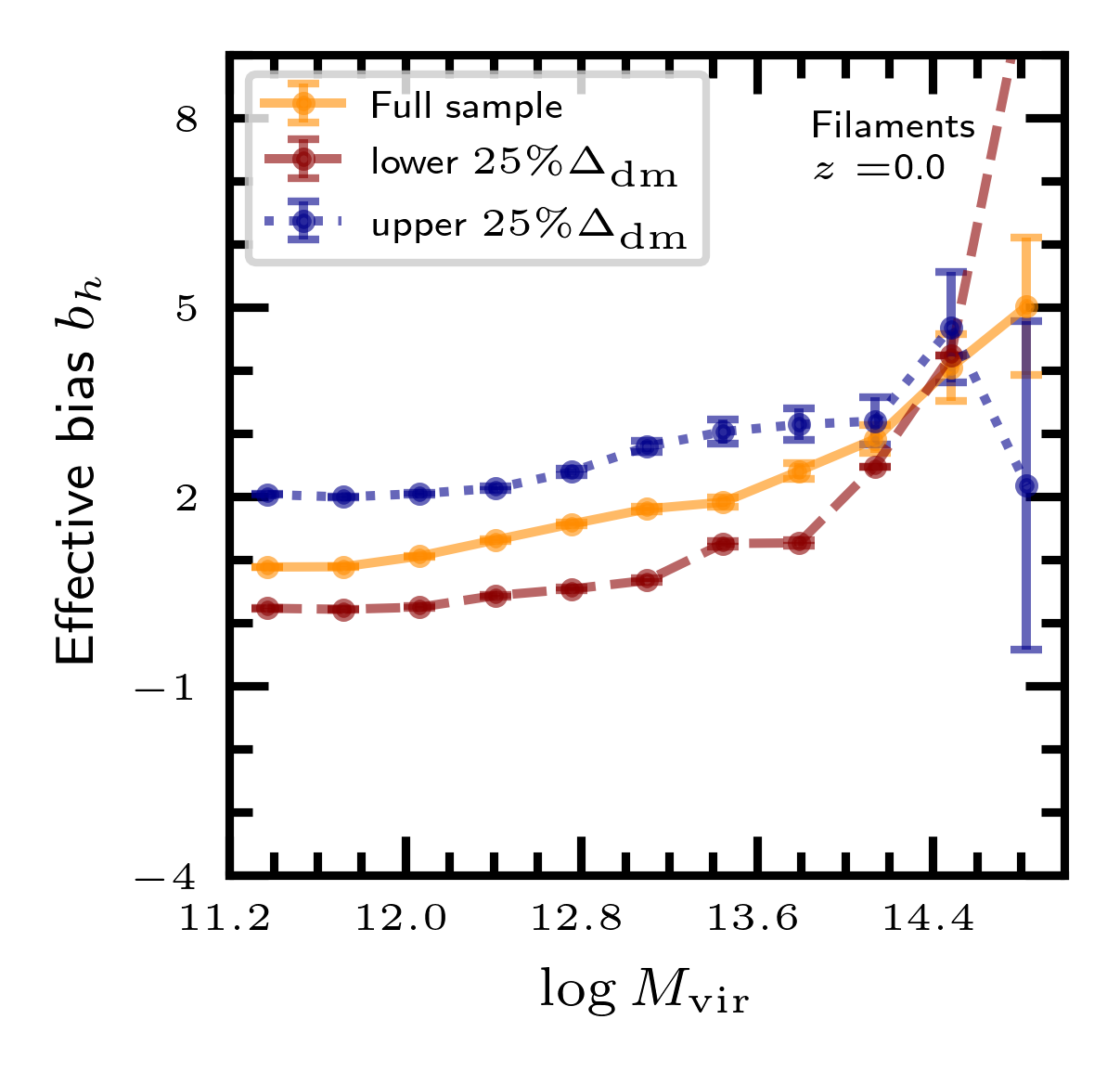}
\includegraphics[trim = .1cm .87cm 0cm 0cm ,clip=true, width=0.232\textwidth]{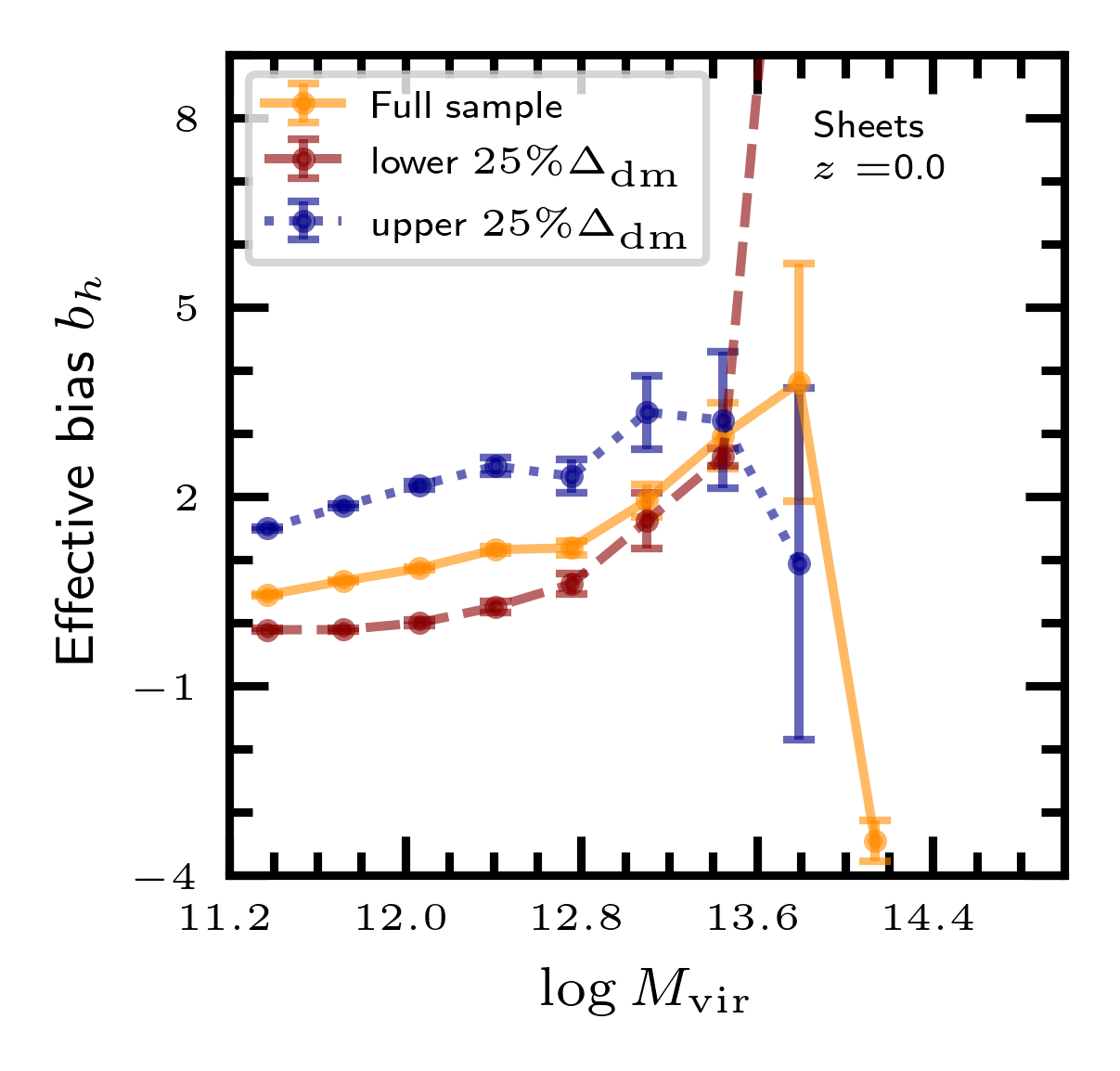}
\includegraphics[trim = .1cm .87cm 0cm 0cm ,clip=true, width=0.232\textwidth]{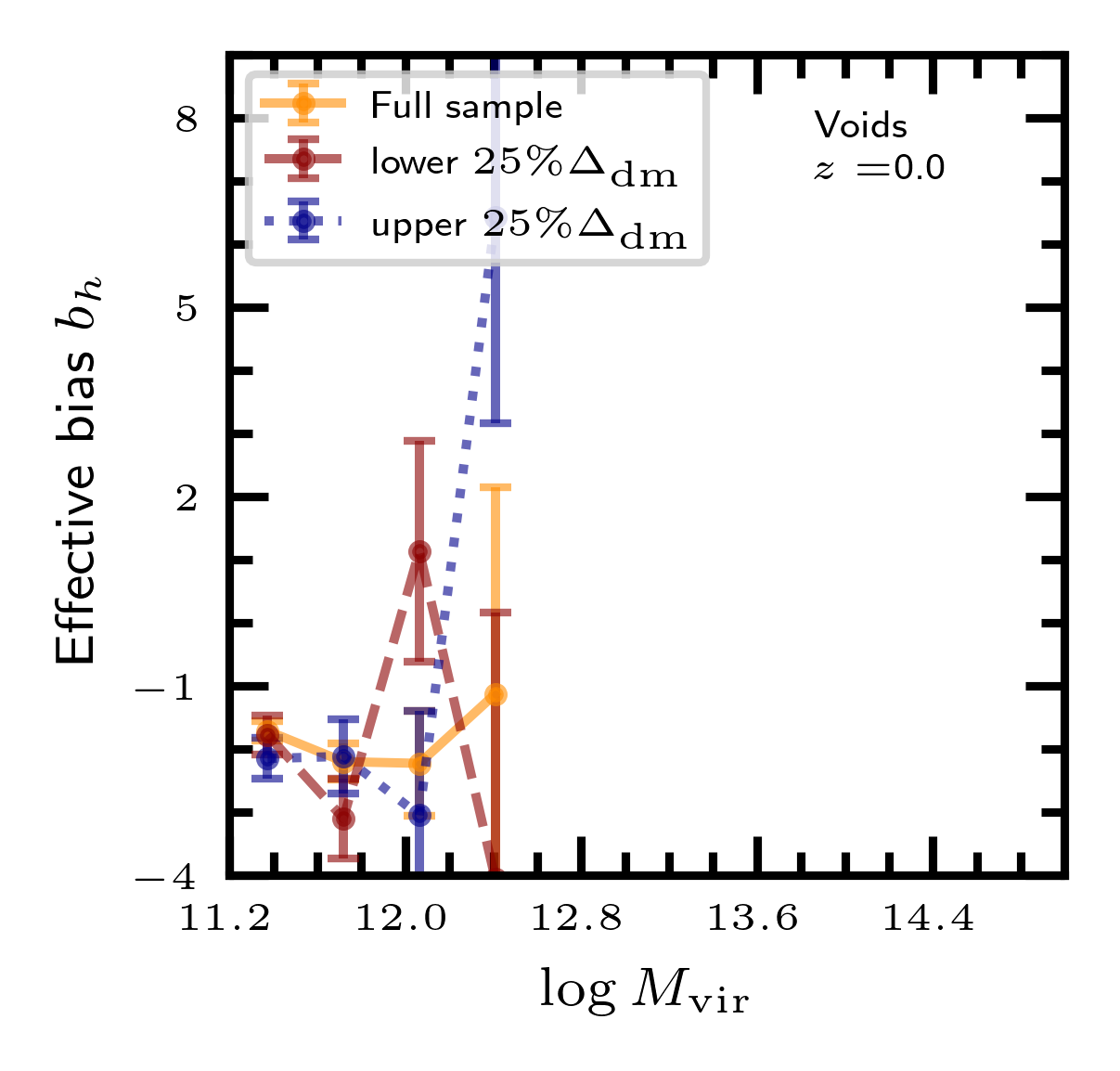}
\includegraphics[trim = .1cm 0.22cm 0cm 0cm ,clip=true, width=0.232\textwidth]{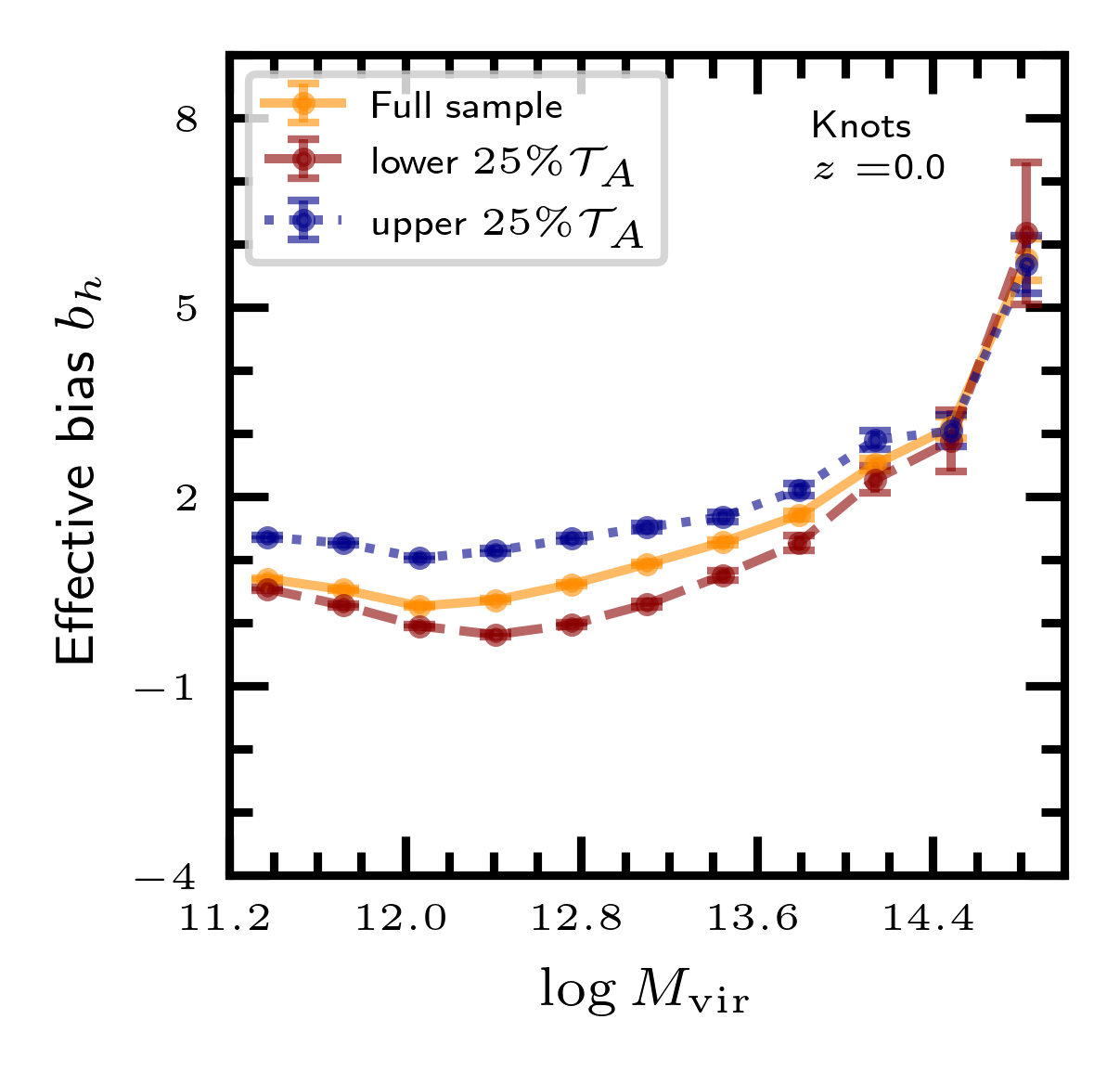}
\includegraphics[trim = .1cm .22cm 0cm 0cm ,clip=true, width=0.232\textwidth]{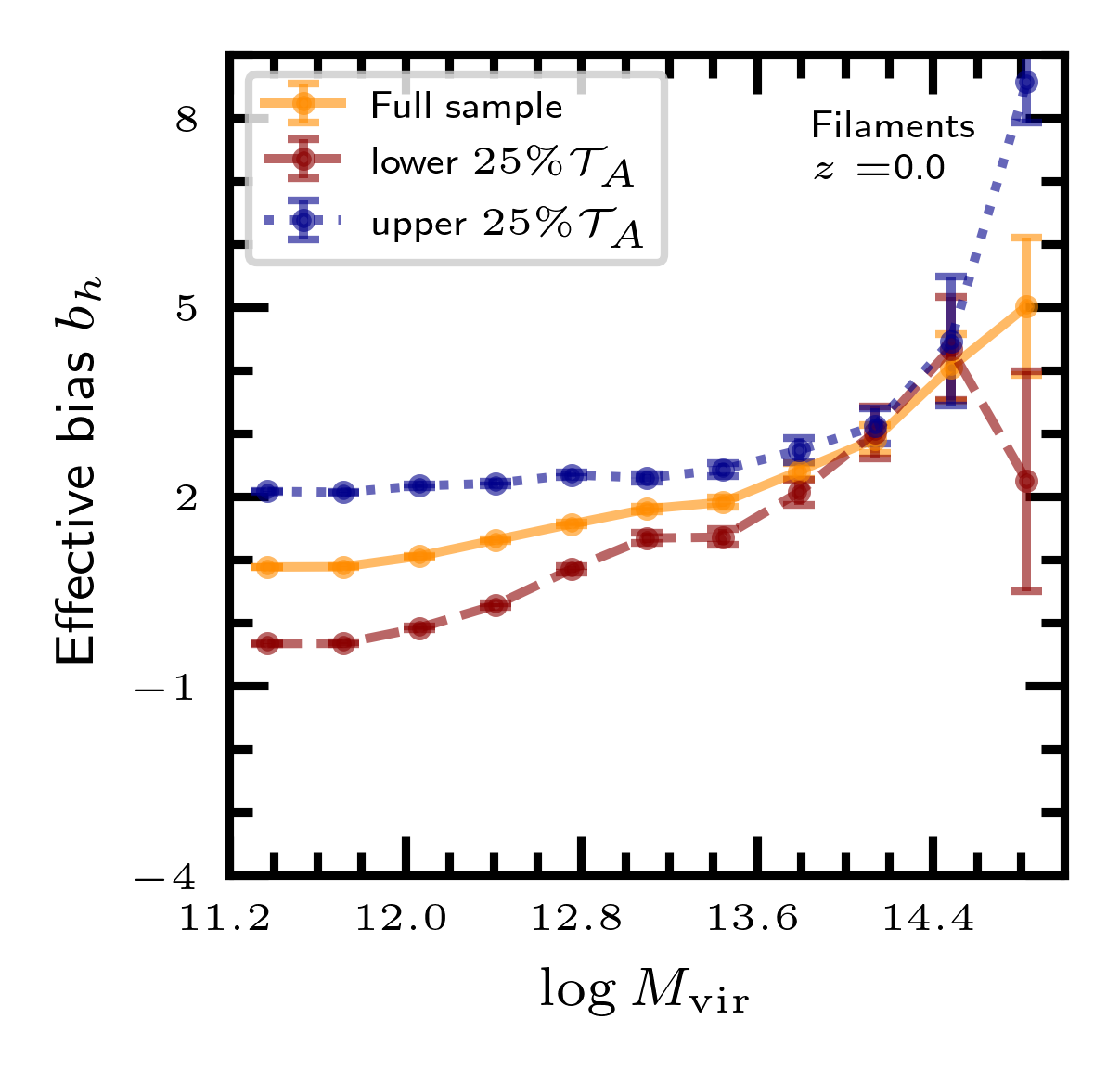}
\includegraphics[trim = .1cm .22cm 0cm 0cm ,clip=true, width=0.232\textwidth]{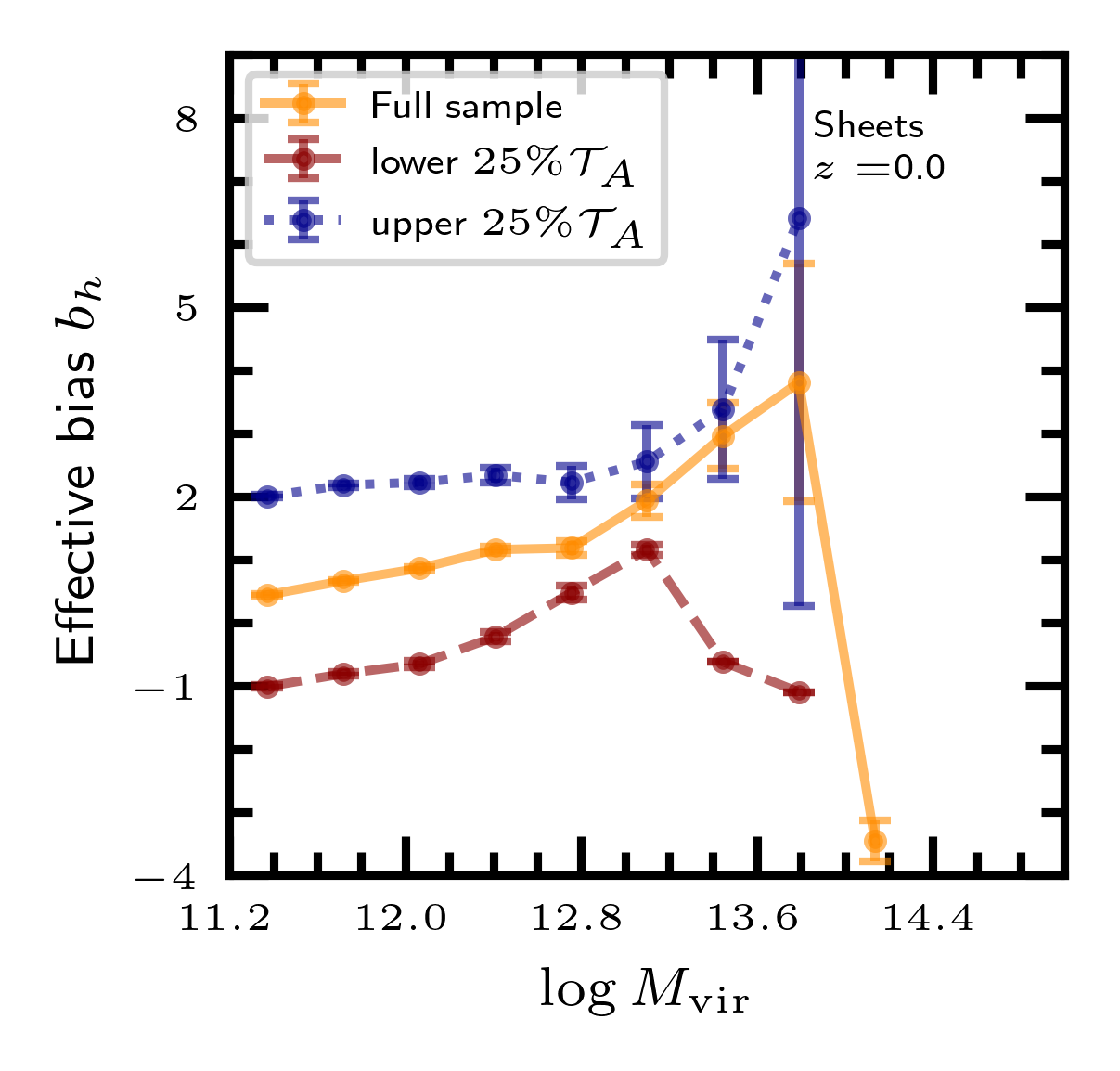}
\includegraphics[trim = .1cm .22cm 0cm 0cm ,clip=true, width=0.232\textwidth]{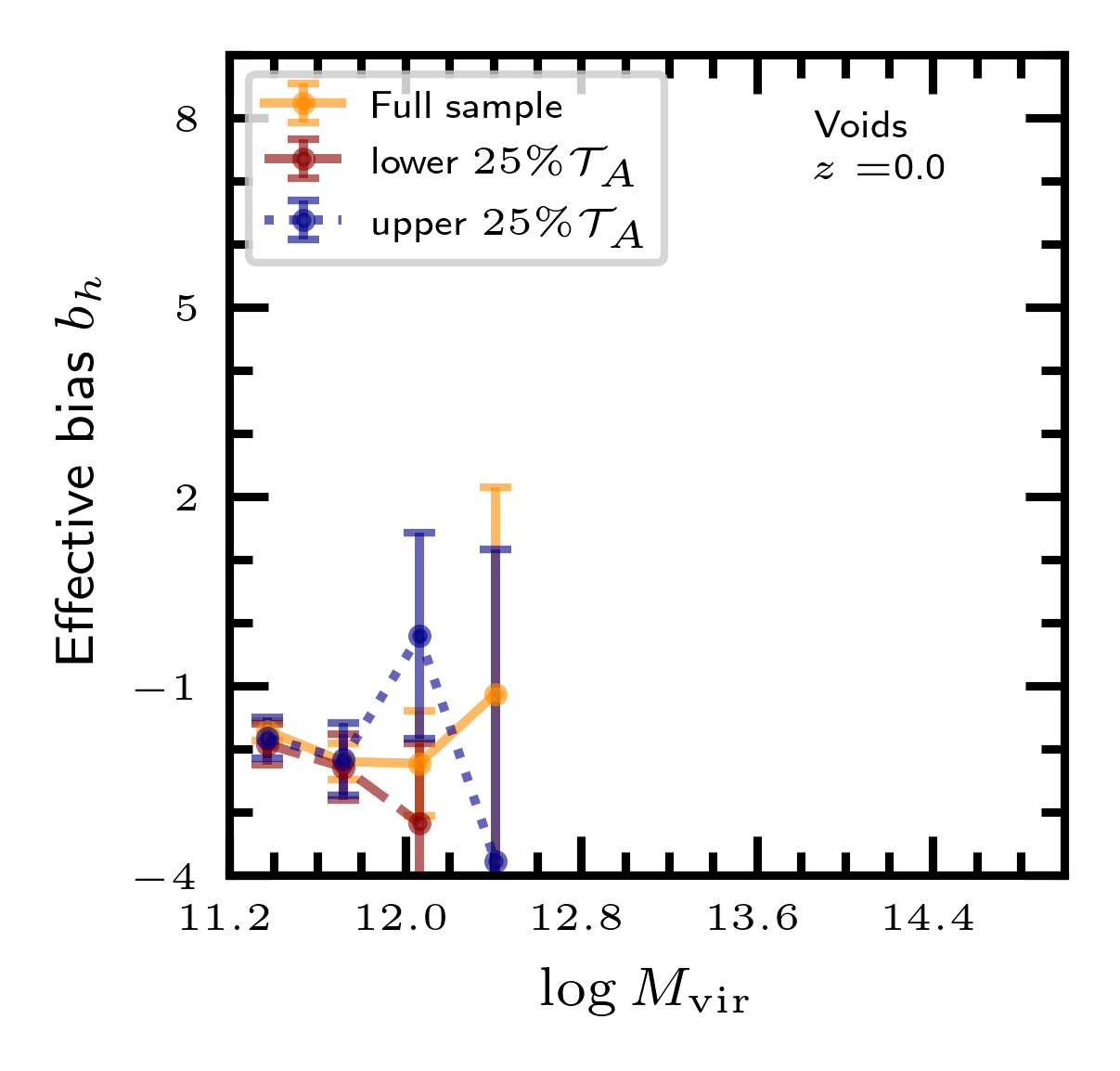}
\caption{\small{Secondary halo bias in different cosmic-web environments at $z=0$. From top to bottom: halo concentration $C_{\rm vir}$, spin $\lambda_{B}$, ellipticity $\mathcal{E}_{h}$, Mach number $\mathcal{M}_{5}$, local overdensity $\Delta_{\rm dm}$ and tidal anisotropy $\mathcal{T}_{A}$).}}
\label{fig:sec_bias_ex2}
\end{figure*}
%==========================================
%==========================================

\subsection{Effective halo bias and the cosmic web}\label{sec:eff_cwt}

In Fig.~\ref{fig:bias_prop_cwt} we show the measurements of mean effective bias as a function of halo mass and cosmic-web types $\langle b_{h}|\theta, w_{i}\rangle$ at $z=0$. 
Some conclusions based on these results are:

\begin{itemize}
    \item A statistically significant difference in the mean effective halo bias can be observed as a function of comic-web types, especially for environmental properties.
    \item The mean relation between halo bias and intrinsic properties is positive for all cosmic-web types, except for voids, as shown in panels (a) to (f) of Fig.~\ref{fig:bias_prop_cwt}.
    \item Halo effective bias displays a rather weak dependence with intrinsic properties such as concentration (panel b), spin (panel c) and triaxiality (panel e).
    \item  Nonlocal and environmental properties show negative bias not only in voids, but also in filaments and knots (see panels g-i). In general these set of properties display the largest differences in their links with effective bias as a function of the cosmic-web type.
    \item When explored in different cosmic-web types, the universality of the bias-peak height relation is waned, as shown in panel (j) of Fig.~\ref{fig:bias_prop_cwt}.
   \item The halo effective bias is mildly higher in environments probing intermediate densities (i.e, filaments) than in extreme environments (knots and voids), as shown in panel (k) of Fig.~\ref{fig:bias_prop_cwt}.
\end{itemize}

One key feature arising from these results is the fact that we can measure, with statistical significance, a negative halo bias. Notice that negative values of mean effective bias are not reachable from estimators such as Eq.(\ref{eq:bhh_k}), as these is based on positive definite quantities. On the other hand, such values should not be surprising, since it can arise from halo overdensities ($\delta_{h}>0$) in voids ($\delta_{\rm dm}<0$). This does not imply that all objects living in voids acquire a negative bias. In fact, we have verified that nearly $60\%$ of the halos living in voids (as defined by our cosmic-web classification) have a negative effective bias at all redshifts explored in this paper.

%==========================================
%==========================================
\begin{figure*}
\centering
%assembly_bias_delta.py
\includegraphics[trim = 0.1cm 0.87cm 0cm 0cm ,clip=true, width=.232\textwidth]{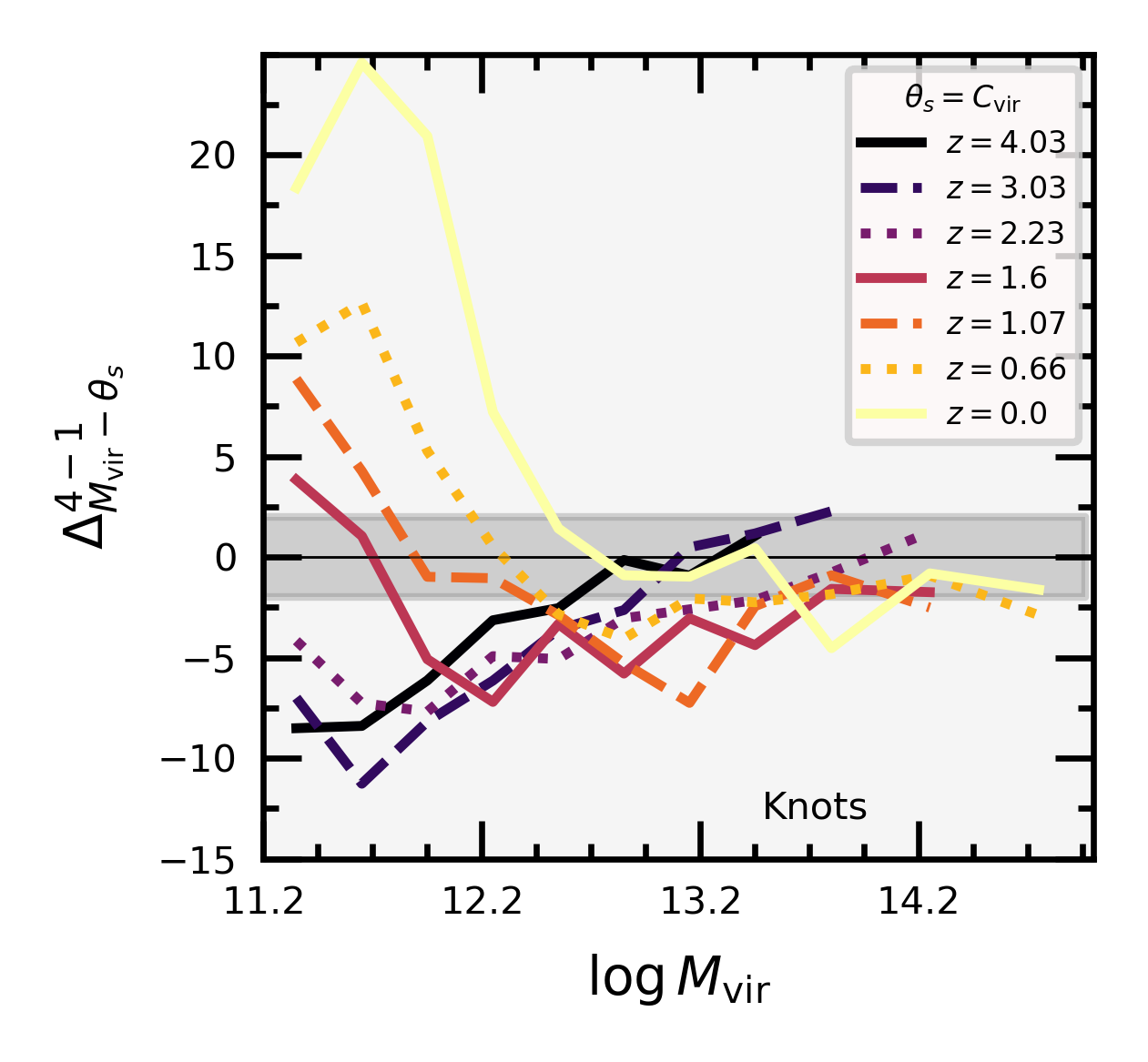}
\includegraphics[trim = 0.1cm 0.87cm 0cm 0cm ,clip=true, width=.232\textwidth]{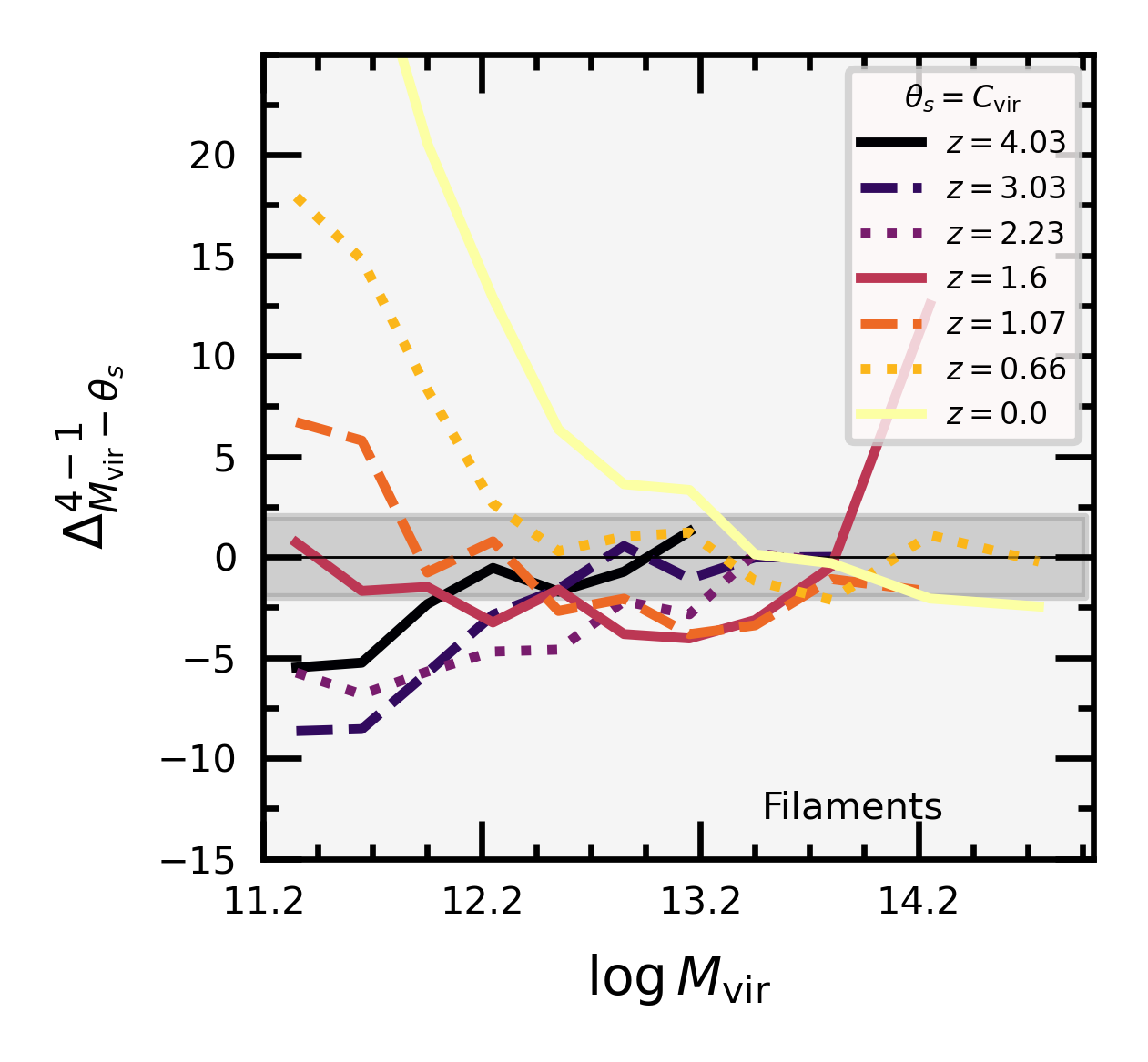}
\includegraphics[trim = 0.1cm 0.87cm 0cm 0cm ,clip=true, width=.232\textwidth]{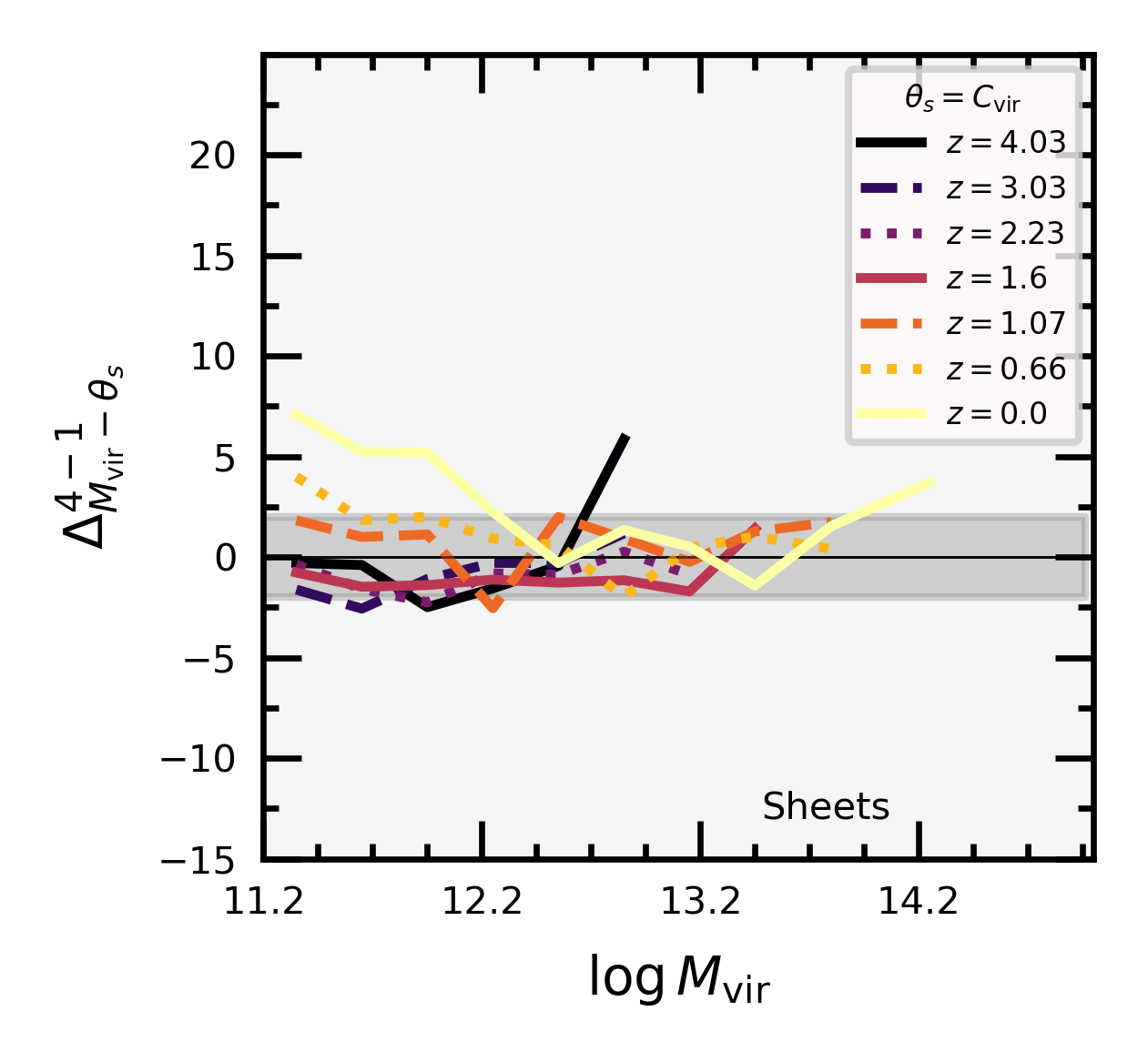}
\includegraphics[trim = 0.1cm 0.87cm 0cm 0cm ,clip=true, width=.232\textwidth]{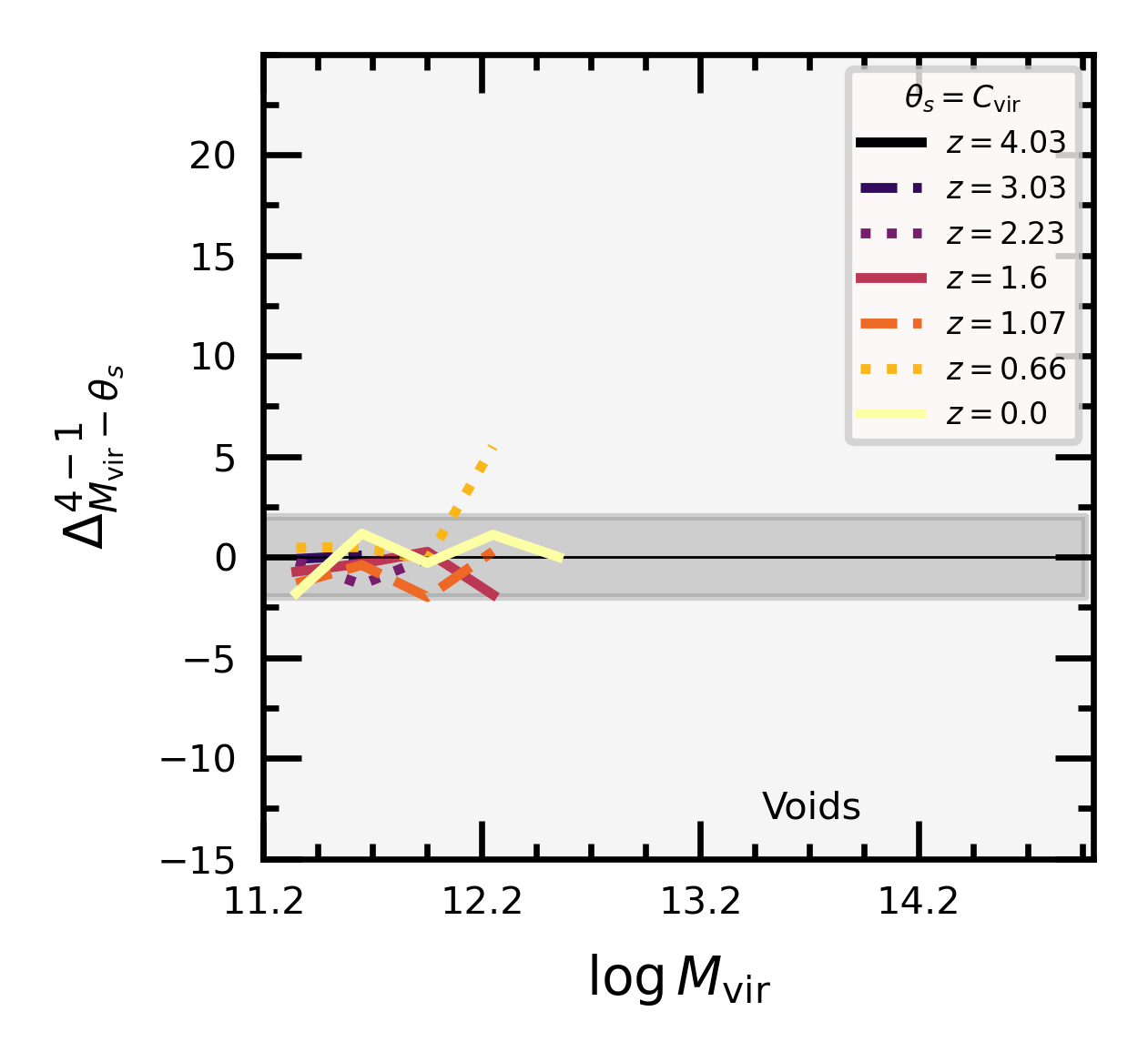}
\includegraphics[trim = 0.1cm 0.87cm 0cm 0cm ,clip=true, width=.232\textwidth]{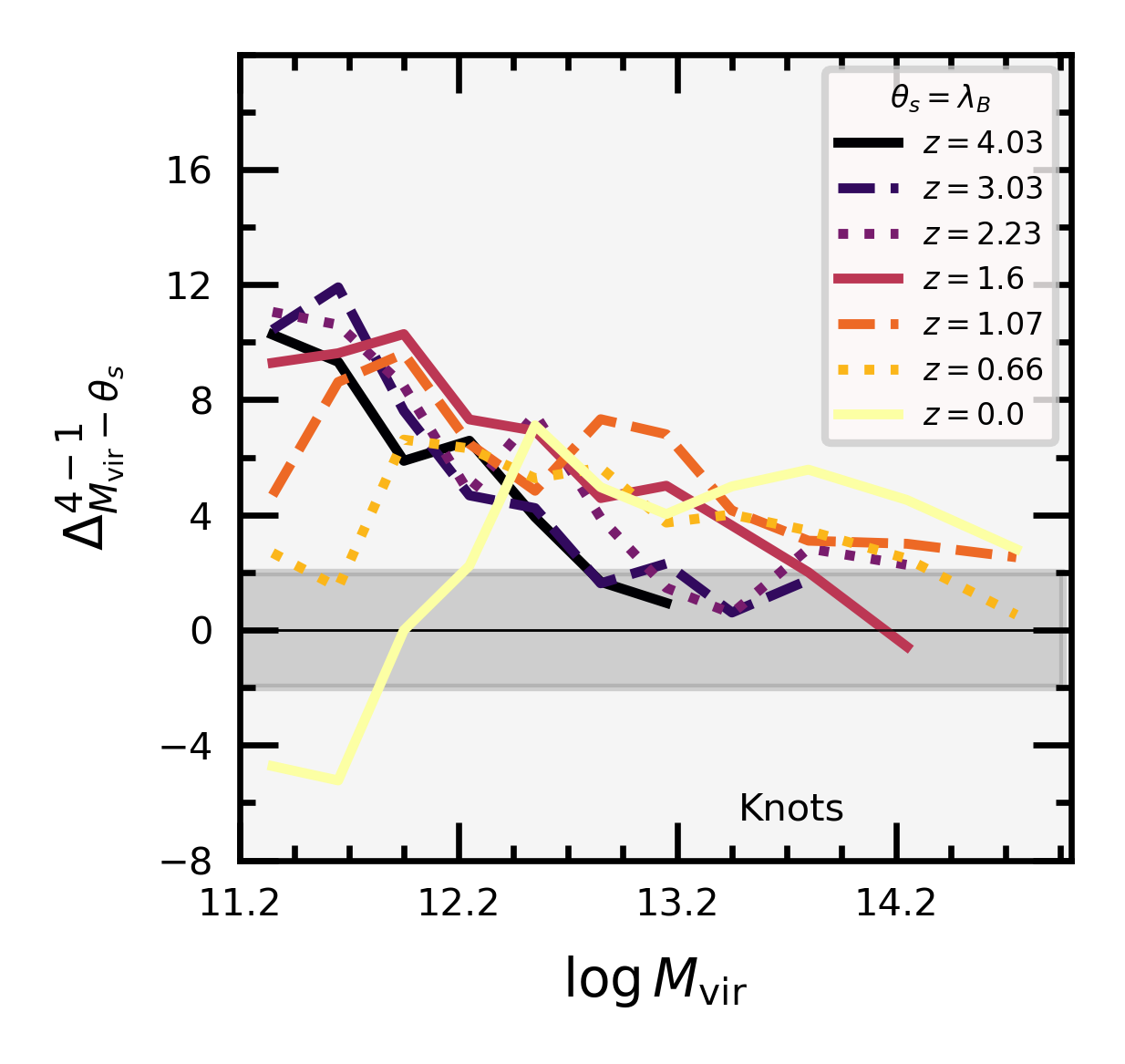}
\includegraphics[trim = 0.1cm 0.87cm 0cm 0cm ,clip=true, width=.232\textwidth]{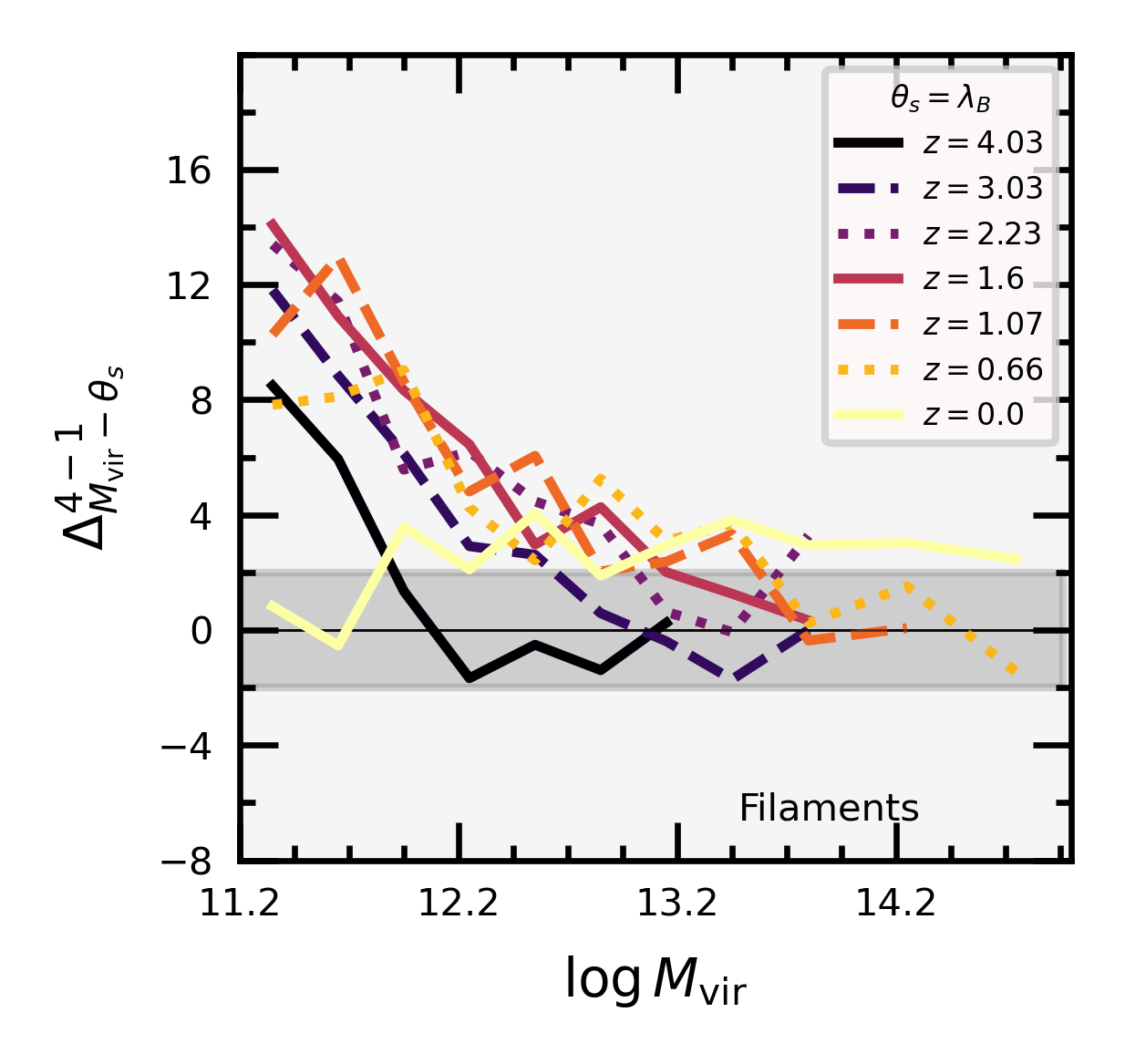}
\includegraphics[trim = 0.1cm 0.87cm 0cm 0cm ,clip=true, width=.232\textwidth]{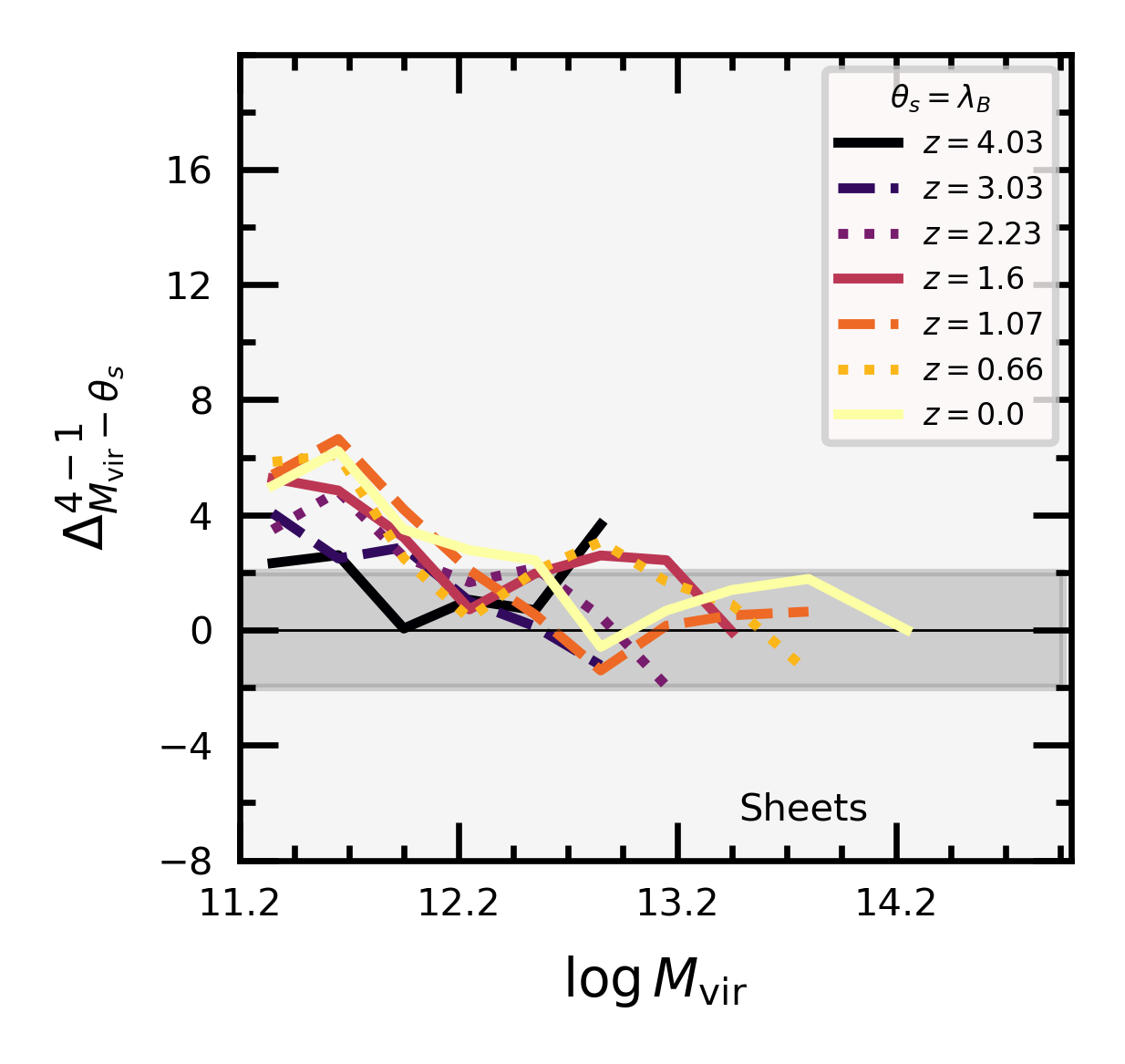}
\includegraphics[trim = 0.1cm 0.87cm 0cm 0cm ,clip=true, width=.232\textwidth]{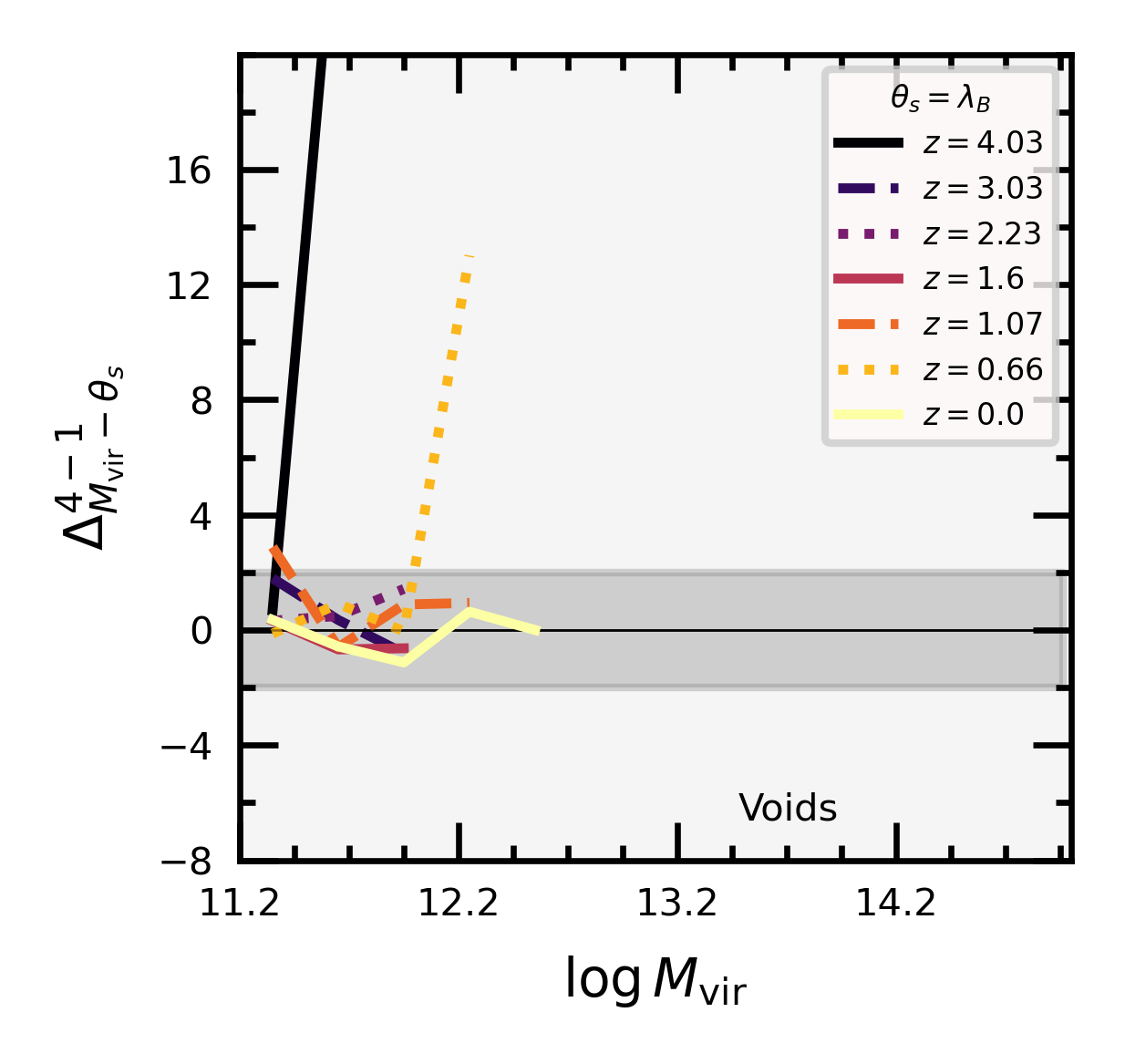}
\includegraphics[trim = 0.1cm 0.87cm 0cm 0cm ,clip=true, width=.232\textwidth]{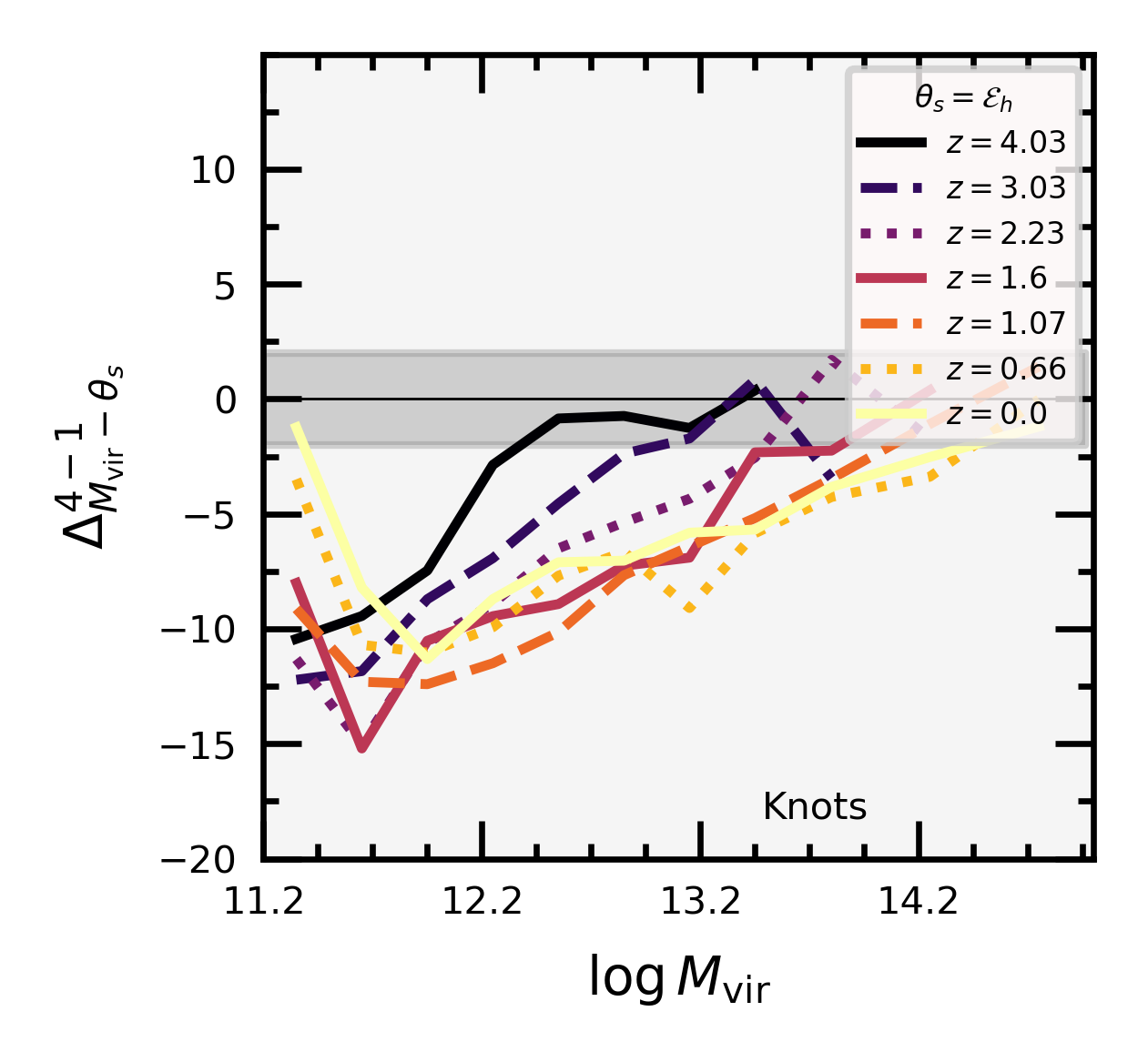}
\includegraphics[trim = 0.1cm 0.87cm 0cm 0cm ,clip=true, width=.232\textwidth]{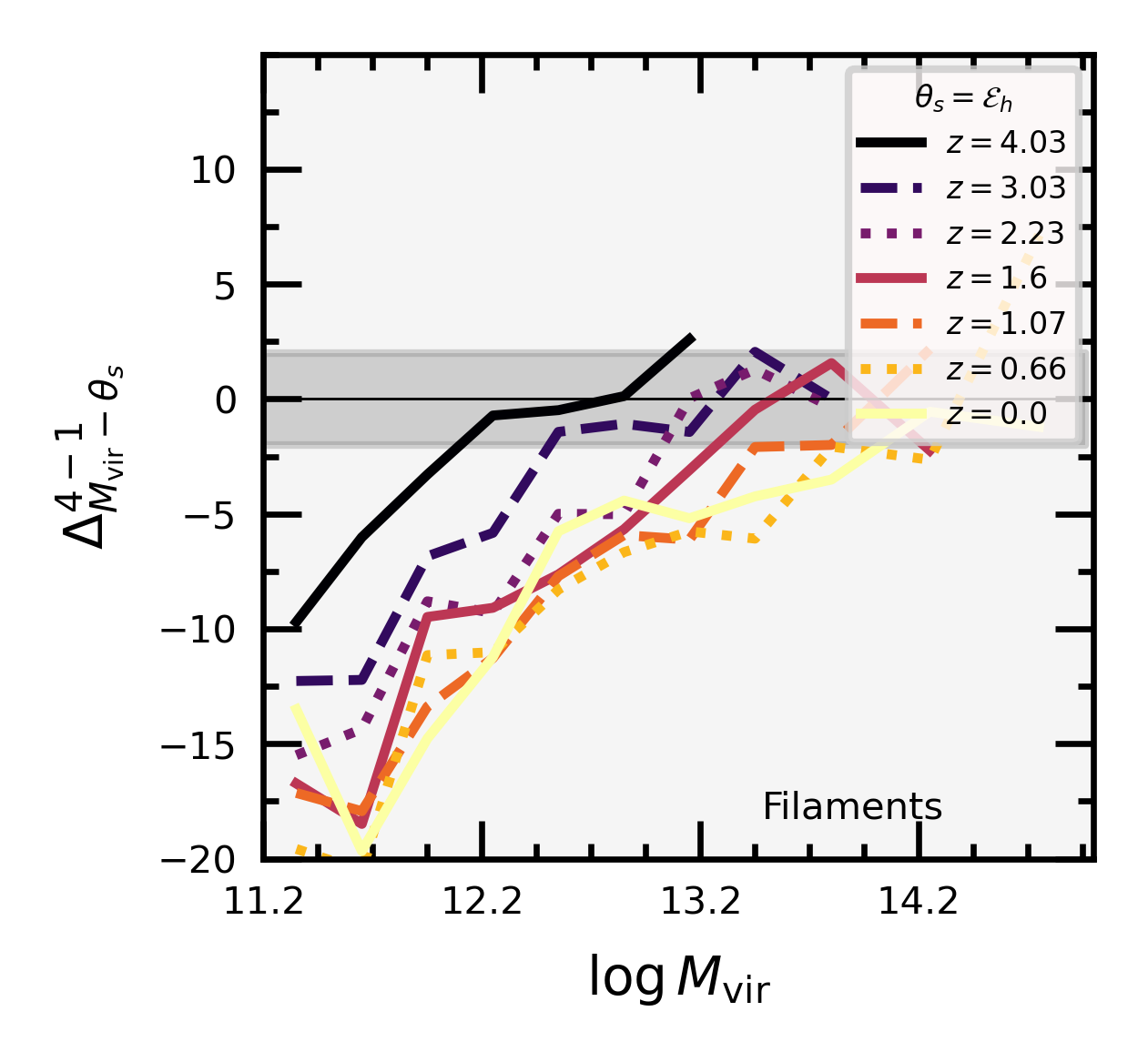}
\includegraphics[trim = 0.1cm 0.87cm 0cm 0cm ,clip=true, width=.232\textwidth]{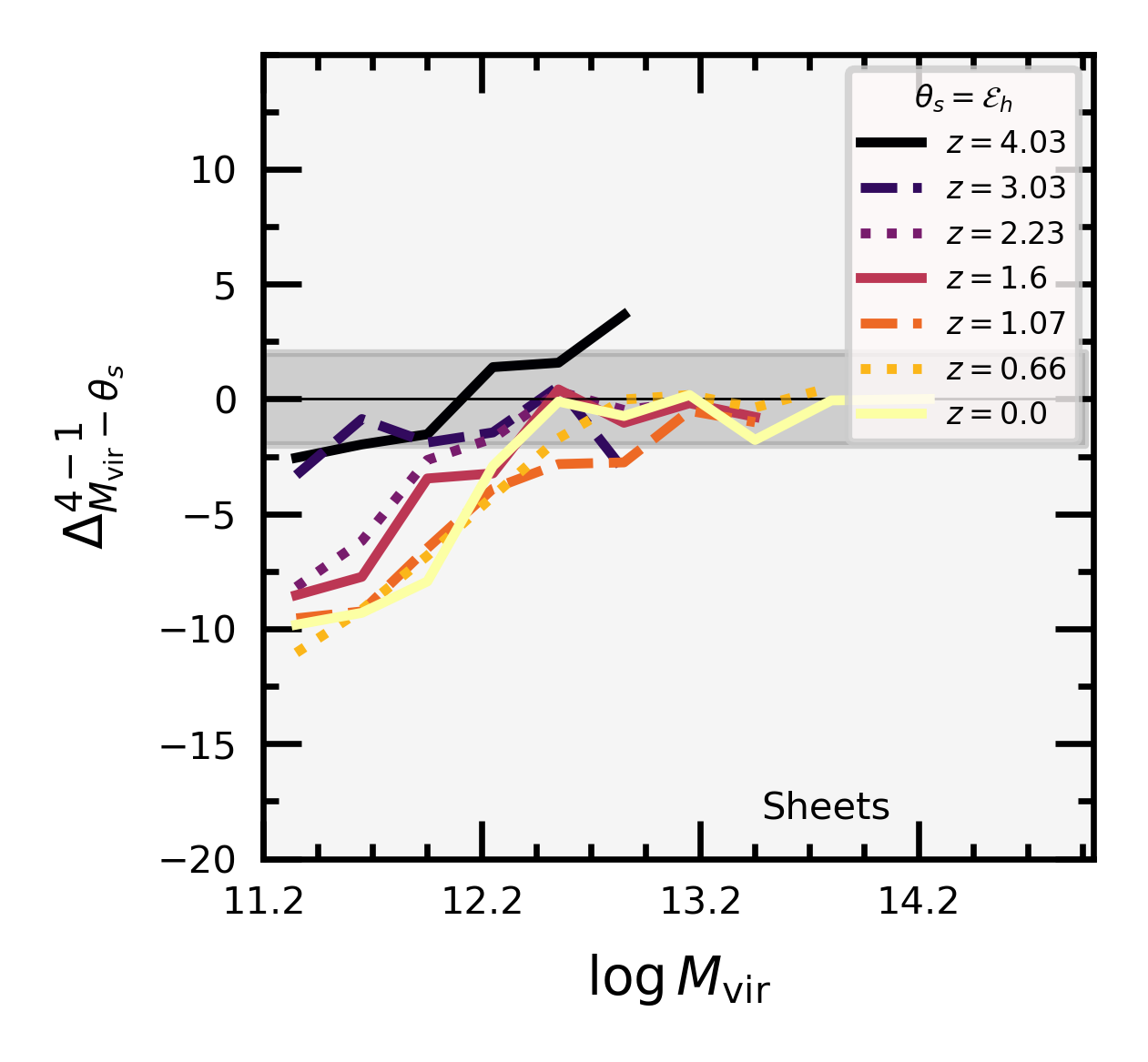}
\includegraphics[trim = 0.1cm 0.87cm 0cm 0cm ,clip=true, width=.232\textwidth]{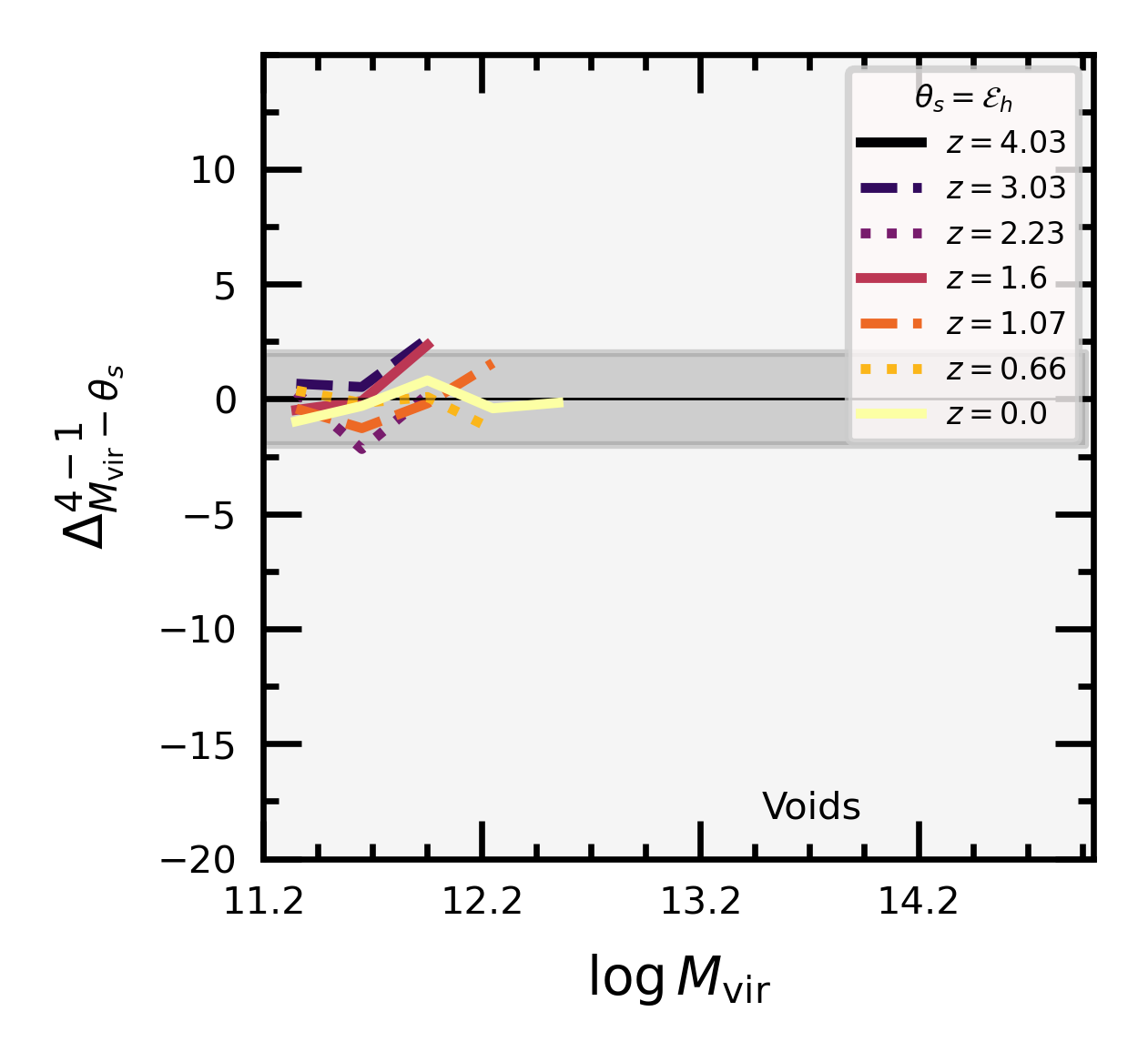}
\includegraphics[trim = 0.1cm 0.87cm 0cm 0cm ,clip=true, width=.232\textwidth]{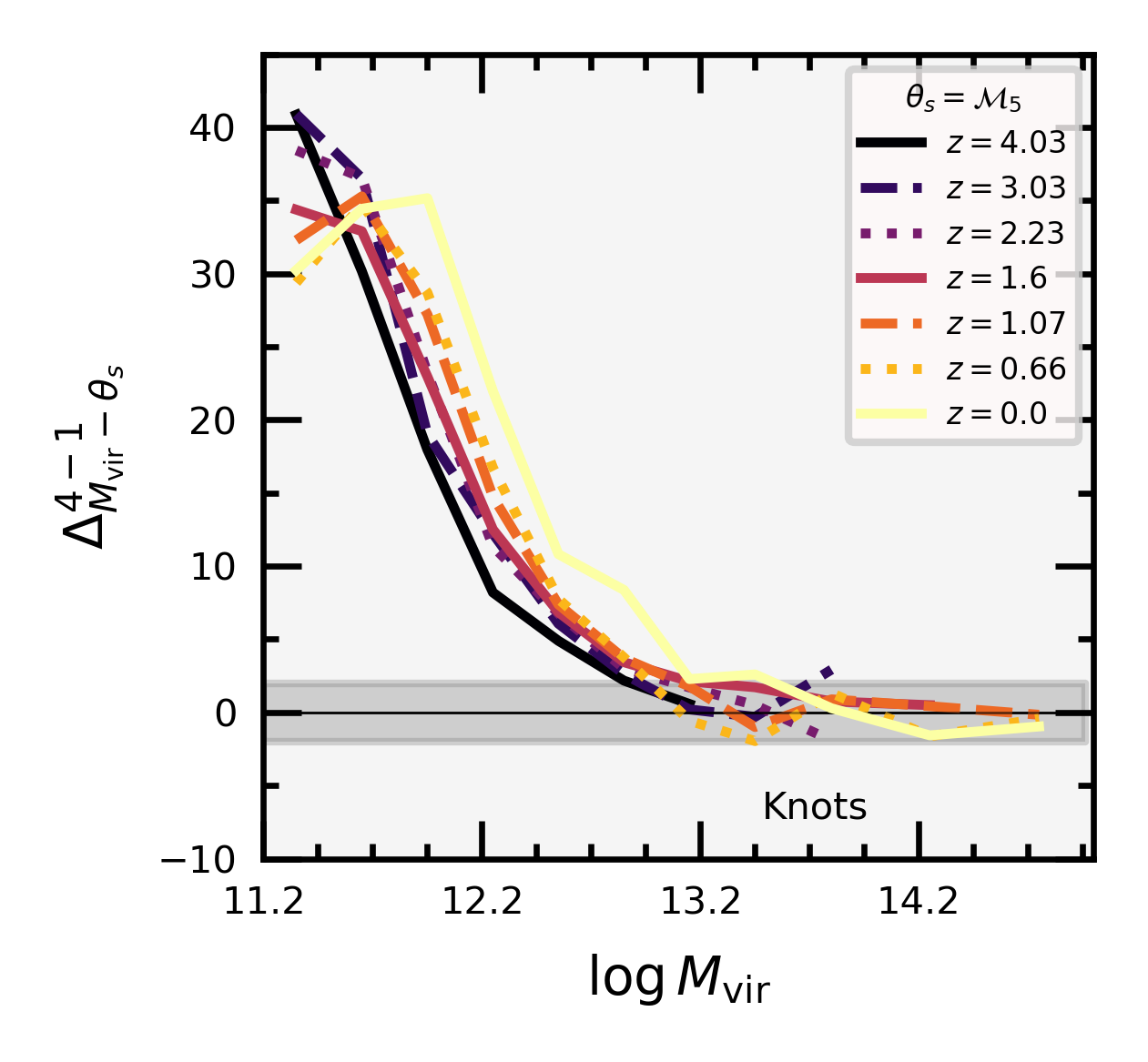}
\includegraphics[trim = 0.1cm 0.87cm 0cm 0cm ,clip=true, width=.232\textwidth]{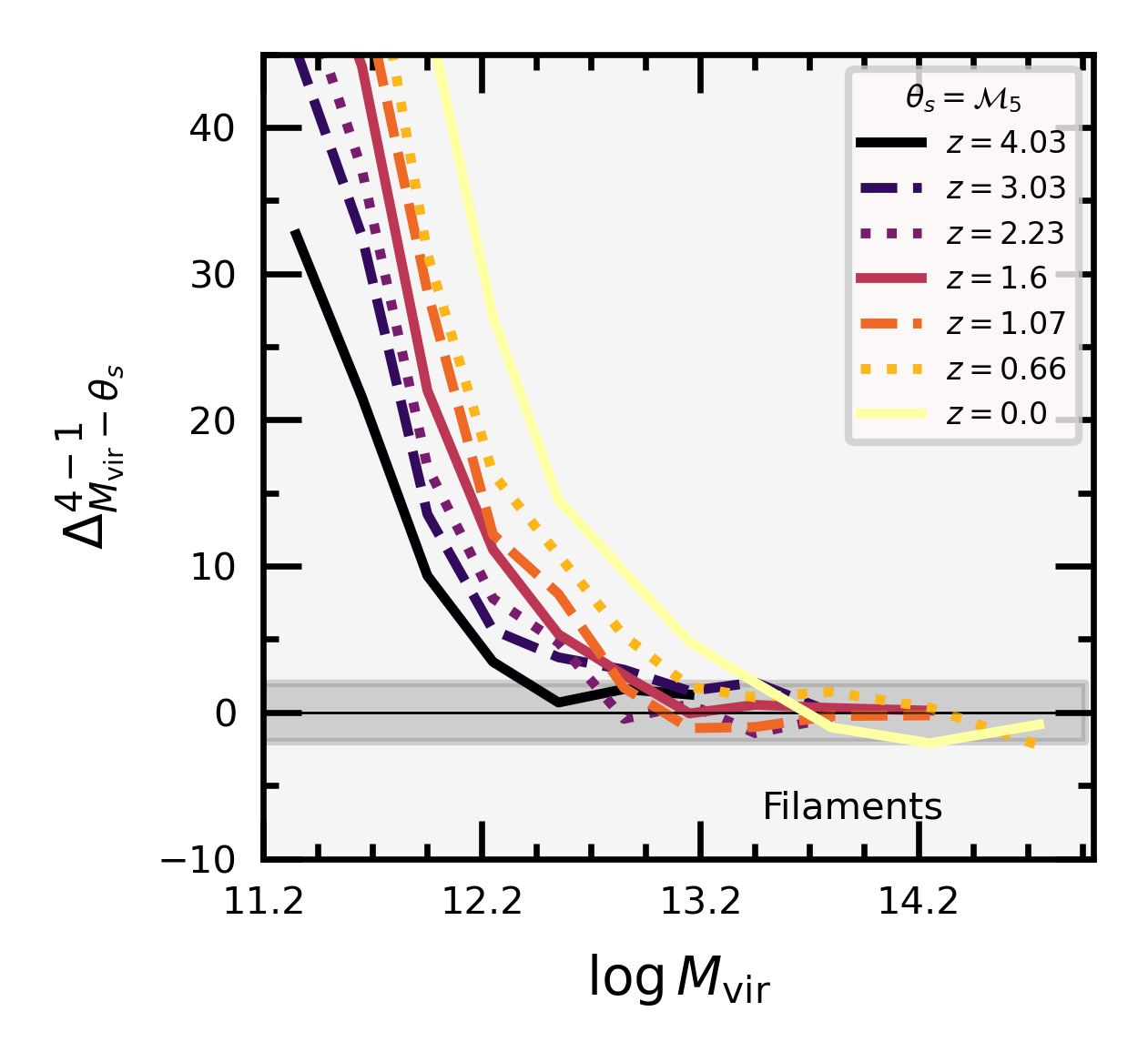}
\includegraphics[trim = 0.1cm 0.87cm 0cm 0cm ,clip=true, width=.232\textwidth]{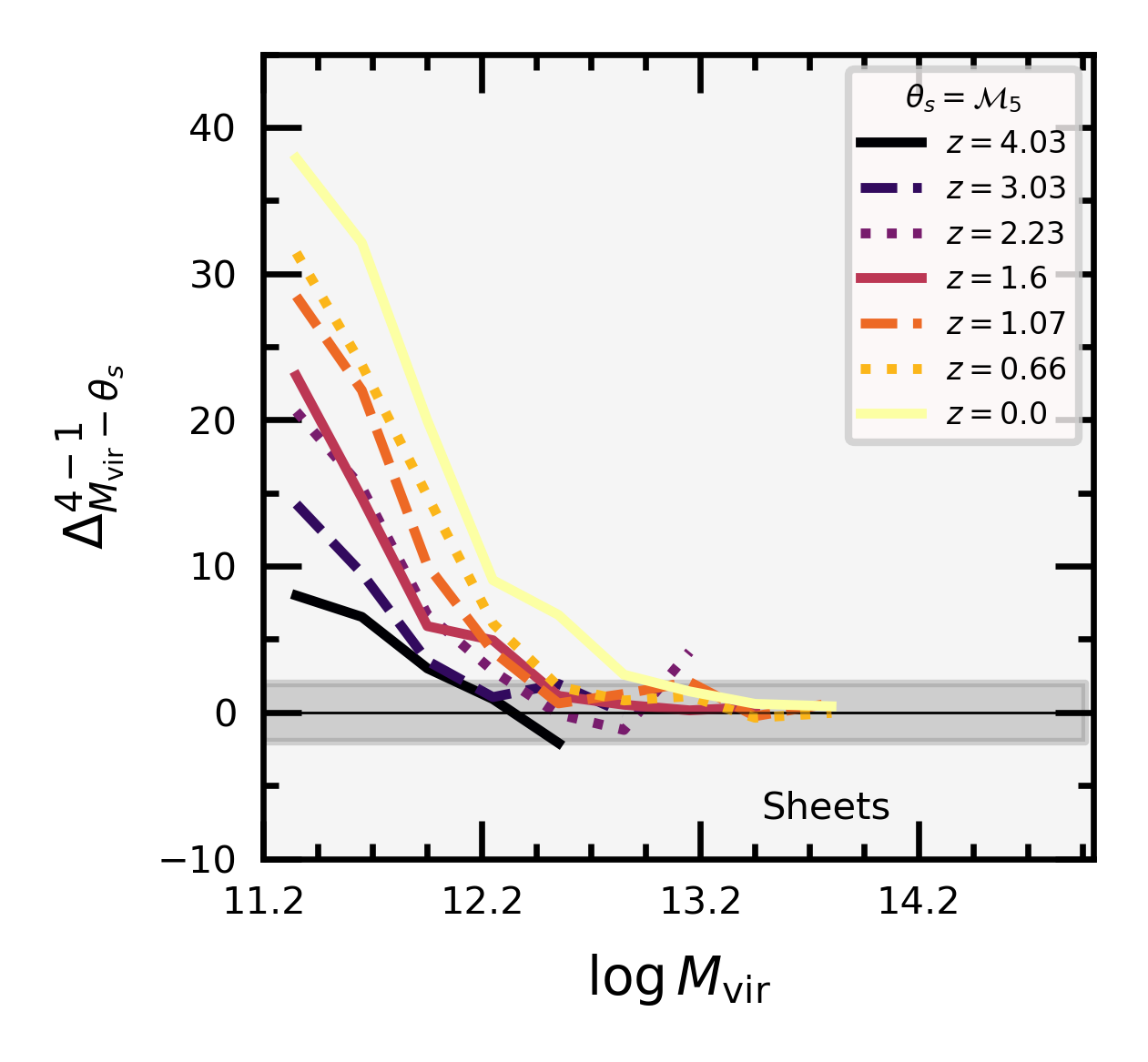}
\includegraphics[trim = 0.1cm 0.87cm 0cm 0cm ,clip=true, width=.232\textwidth]{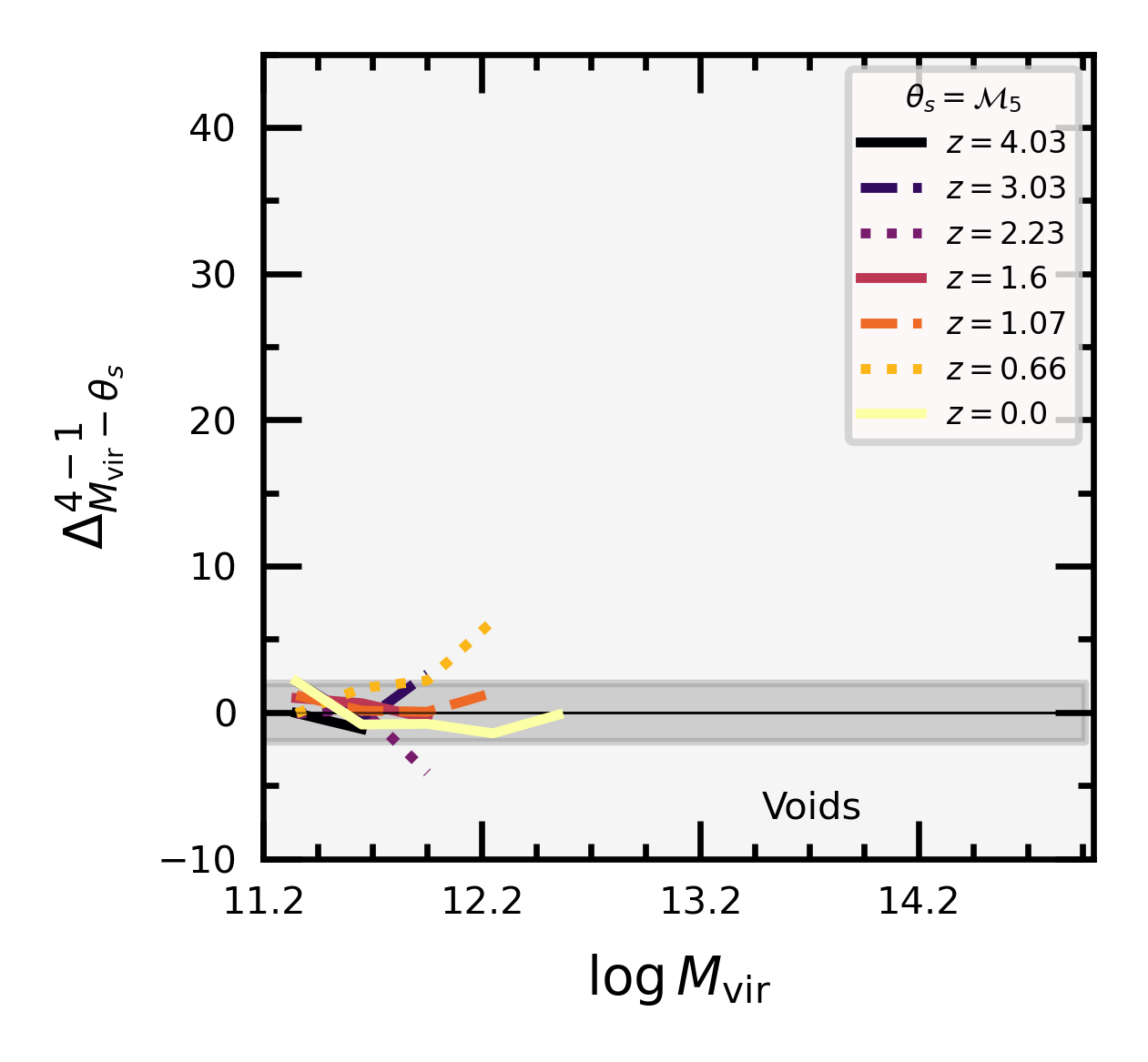}
\includegraphics[trim = 0.1cm 0.87cm 0cm 0cm ,clip=true, width=.232\textwidth]{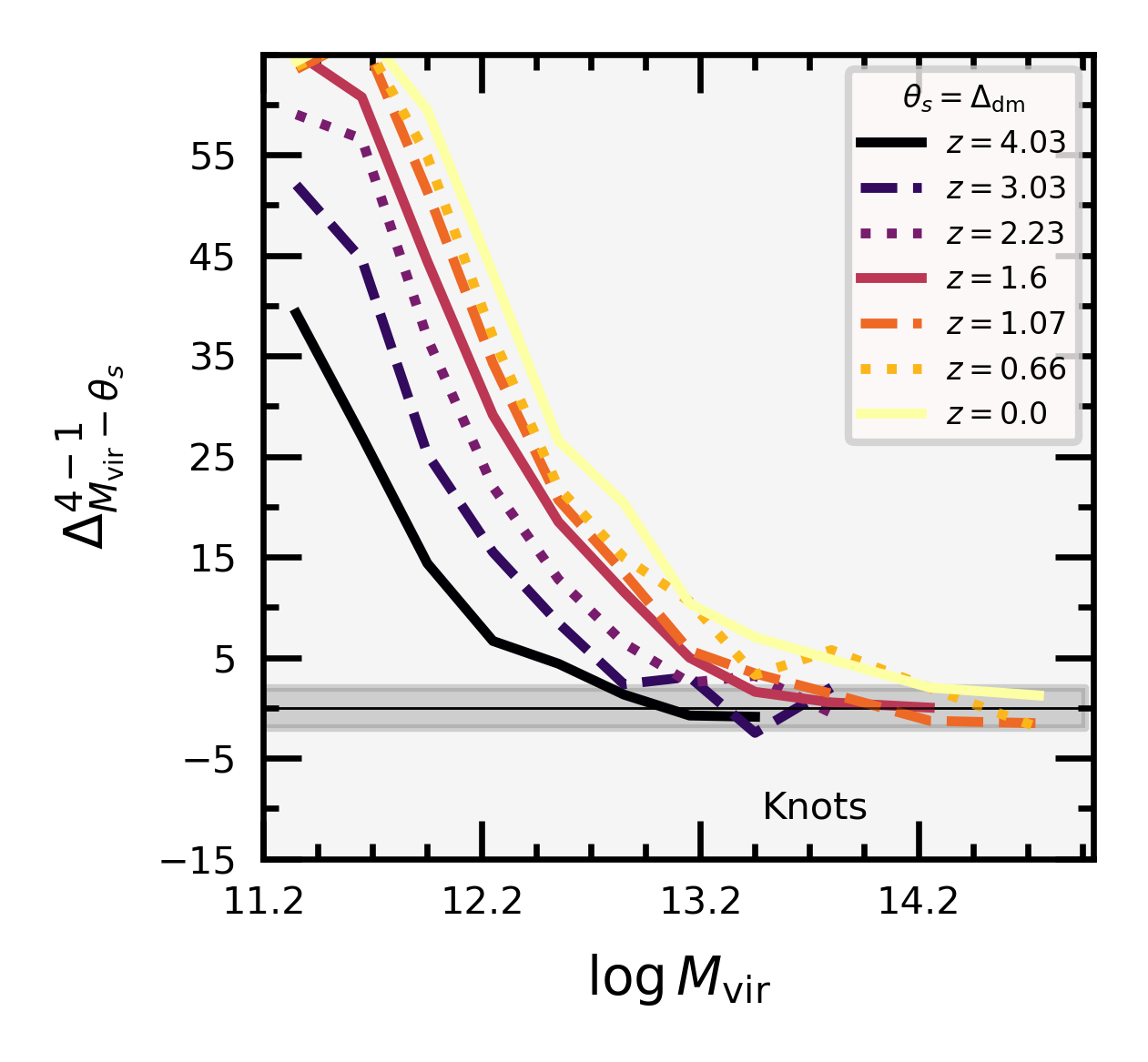}
\includegraphics[trim = 0.1cm 0.87cm 0cm 0cm ,clip=true, width=.232\textwidth]{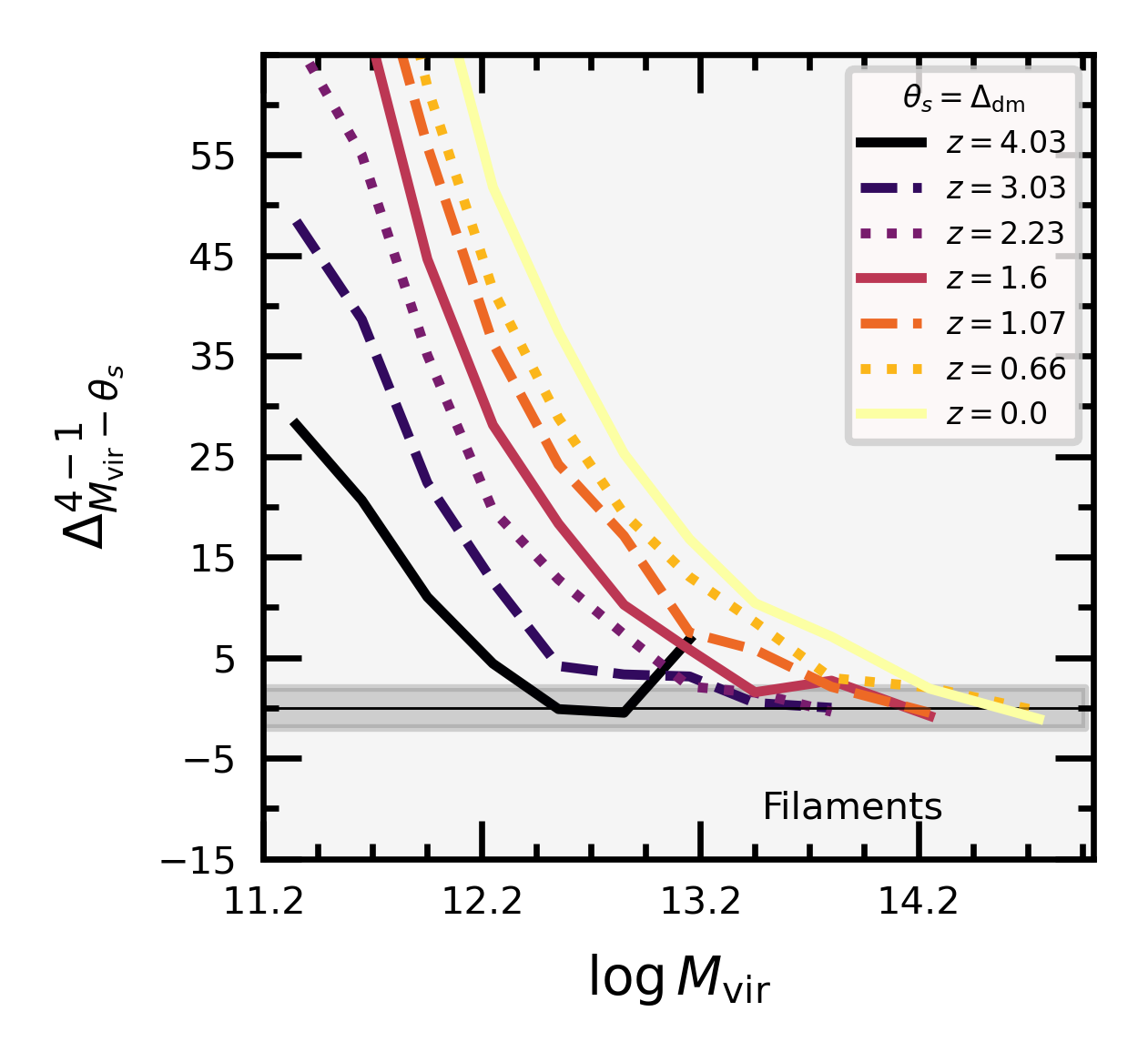}
\includegraphics[trim = 0.1cm 0.87cm 0cm 0cm ,clip=true, width=.232\textwidth]{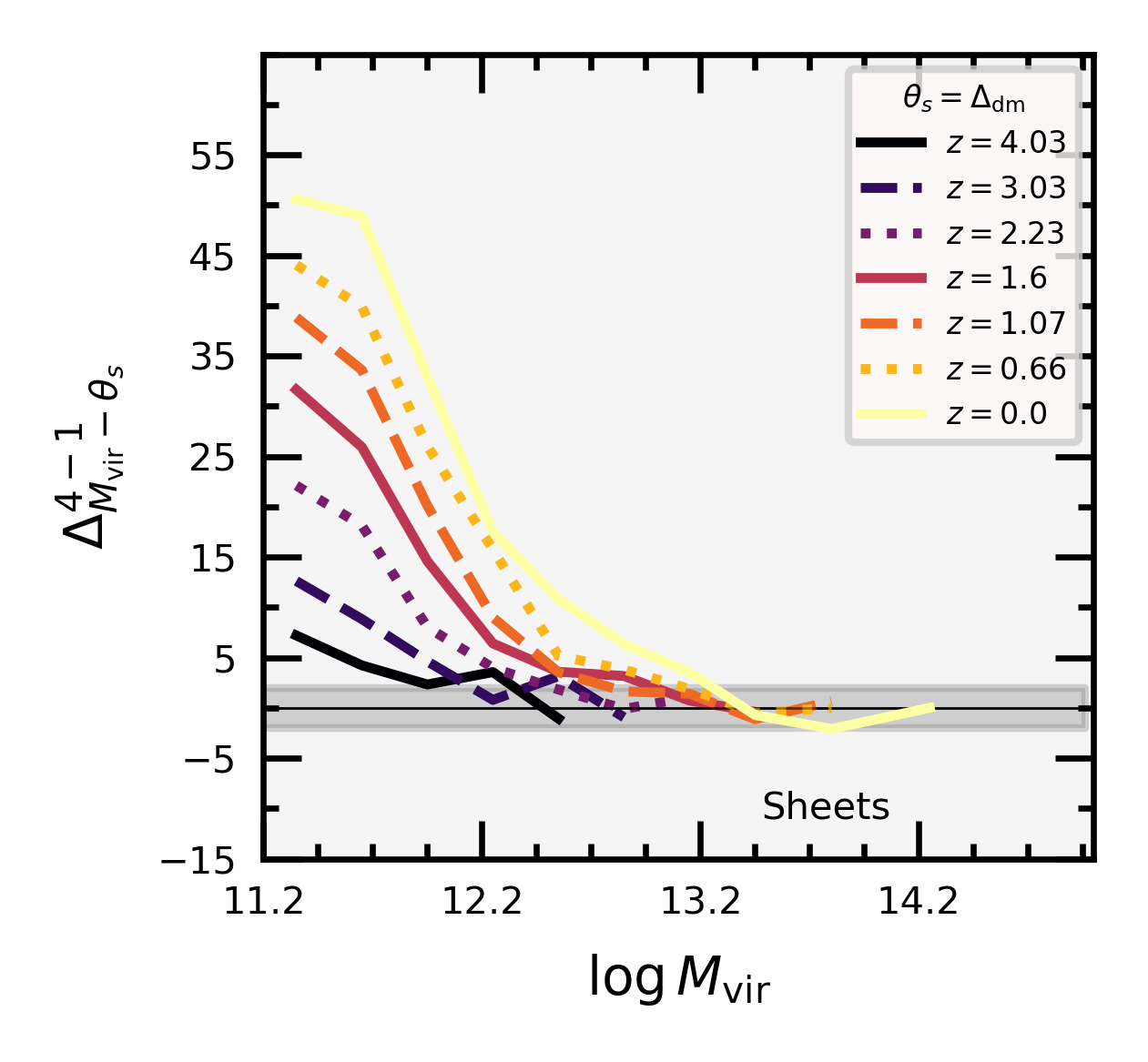}
\includegraphics[trim = 0.1cm 0.87cm 0cm 0cm ,clip=true, width=.232\textwidth]{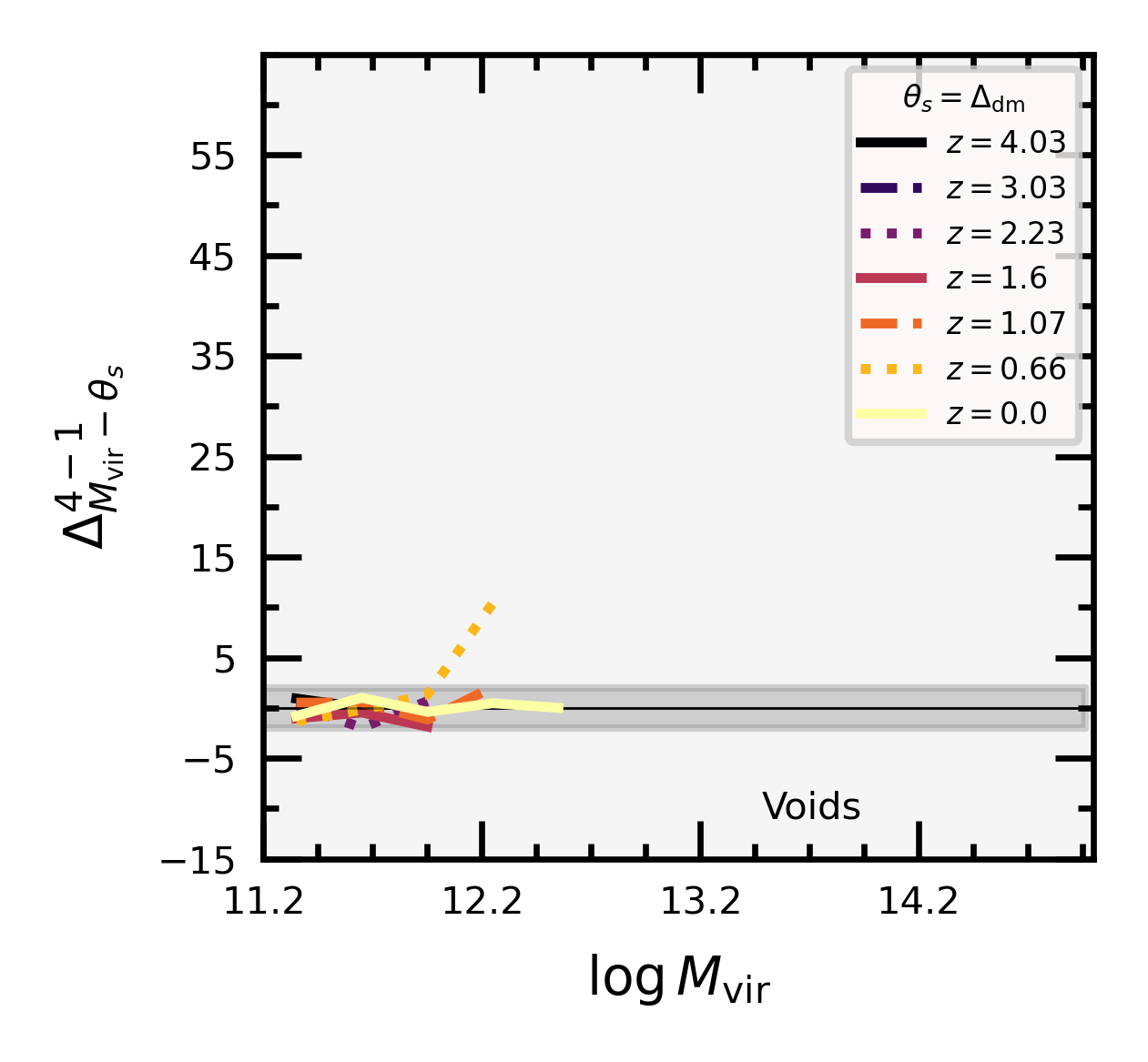}
\includegraphics[trim = 0.1cm 0.2cm 0cm 0cm ,clip=true, width=.232\textwidth]{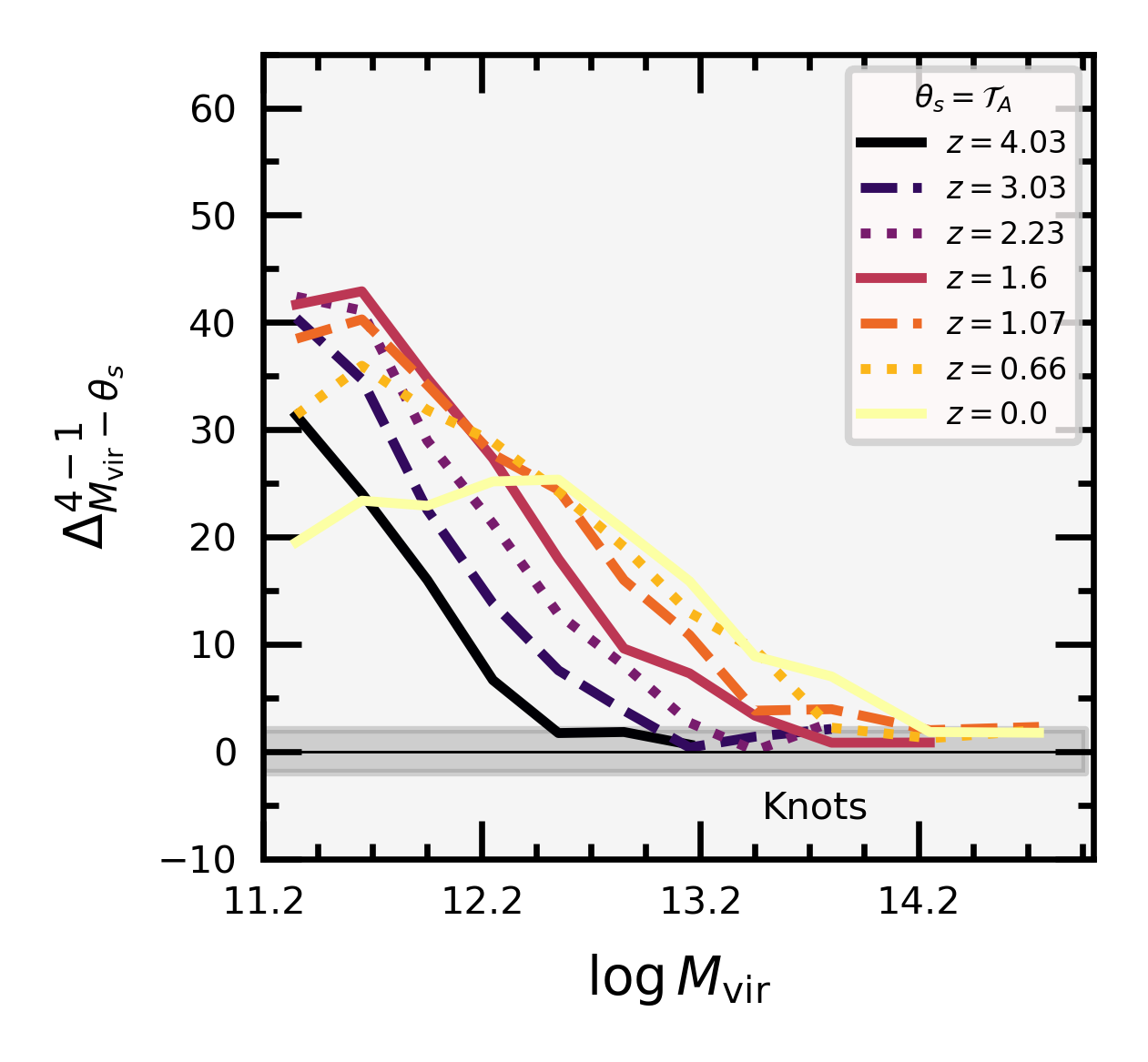}
\includegraphics[trim = 0.1cm 0.2cm 0cm 0cm ,clip=true, width=.232\textwidth]{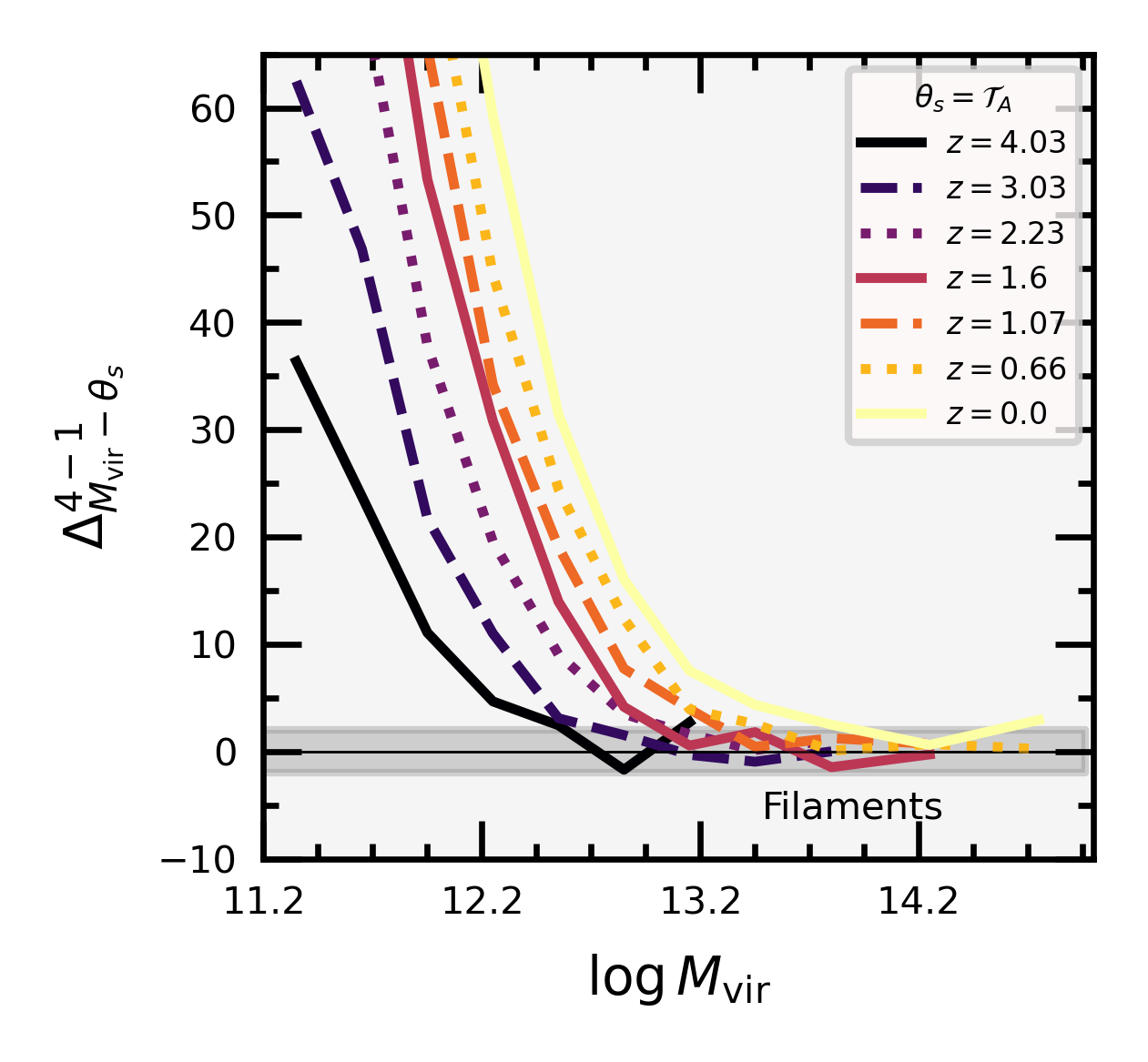}
\includegraphics[trim = 0.1cm 0.2cm 0cm 0cm ,clip=true, width=.232\textwidth]{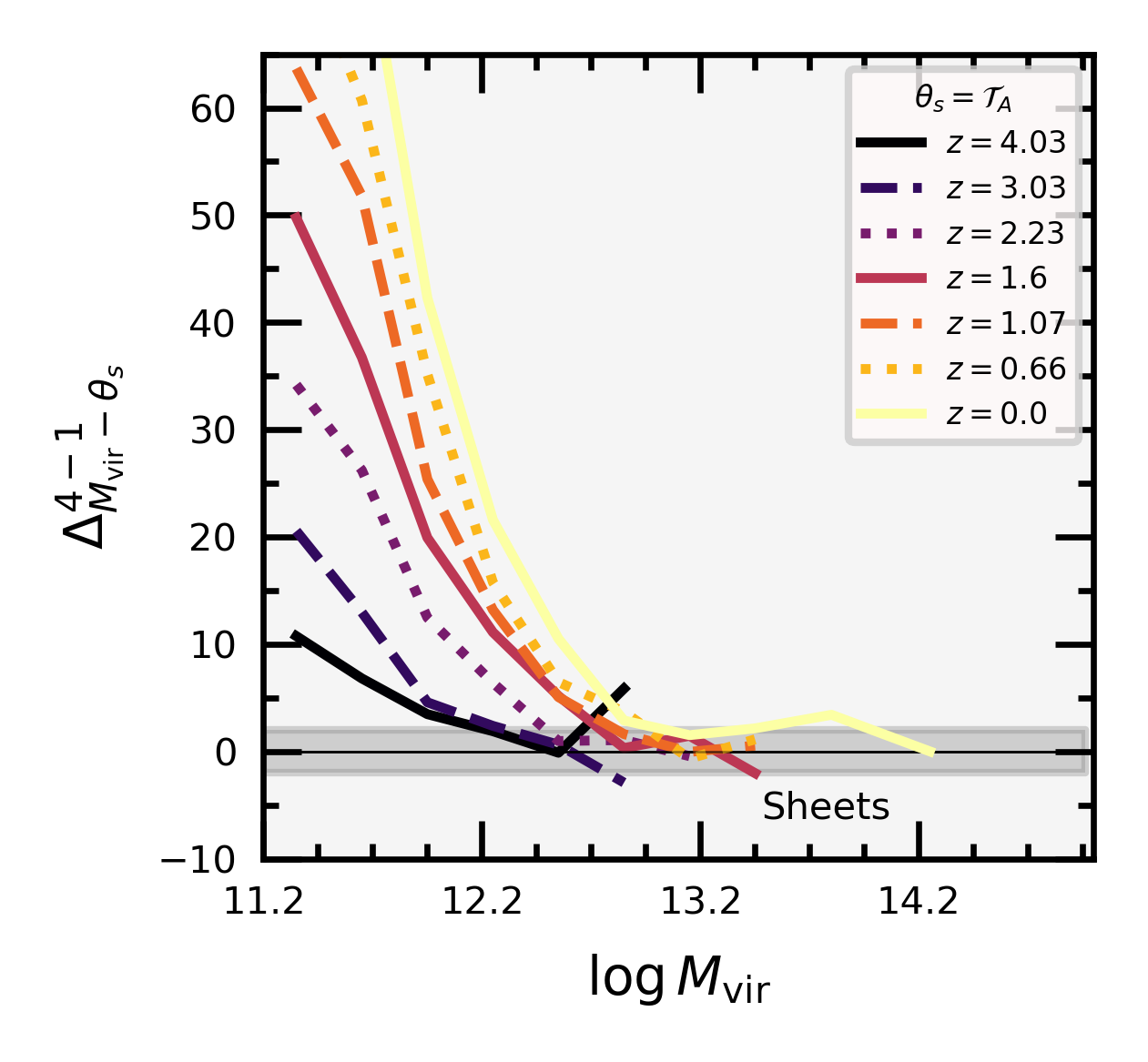}
\includegraphics[trim = 0.1cm 0.2cm 0cm 0cm ,clip=true, width=.232\textwidth]{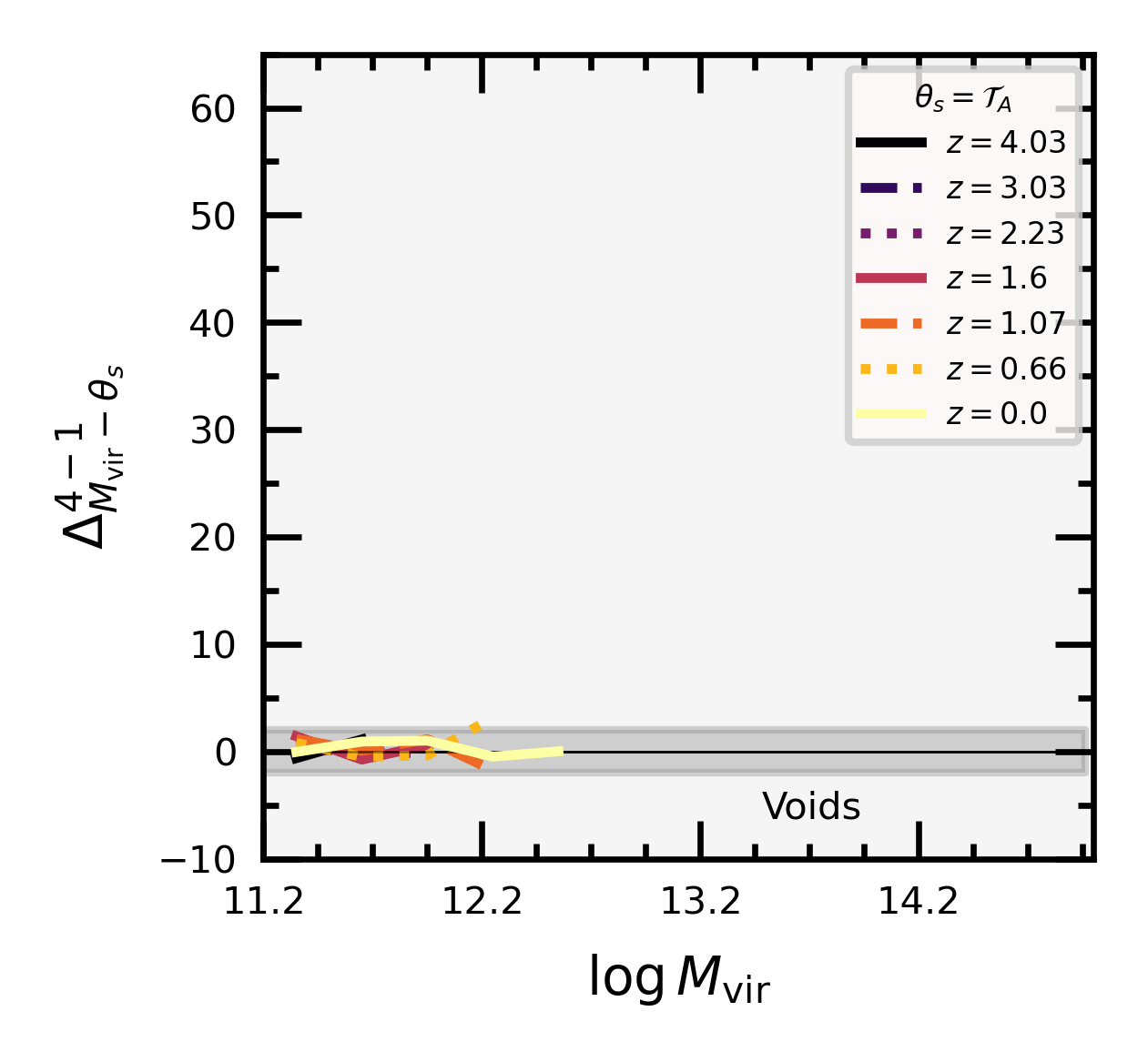}
\caption{\small{Significance of secondary bias (Eq.~\ref{eq:Delta}), evaluated in different cosmic-web types at $z=0$ for different secondary properties: concentration $C_{\rm vir}$, spin $\lambda_{B}$, ellipticity $\mathcal{E}_{h}$, Mach number $\mathcal{M}_{5}$, local overdensity $\Delta_{\rm dm}$, and tidal anisotropy $\mathcal{T}_{A}$. As in Fig.~\ref{fig:sec_bias_allz1}, the dark-shaded stripe denotes a strip of $\Delta^{4-1}_{p-s} \pm 2$ and the horizontal line marks $\Delta^{4-1}_{p-s}=0$.}}
\label{fig:sec_bias_allz2}
\end{figure*}
%=====================================================================
%=====================================================================

%=================================================================
%=================================================================
\begin{figure*}
\centering
\includegraphics[trim = .0cm 0.24cm 0cm 0cm ,clip=true, width=0.32\textwidth]{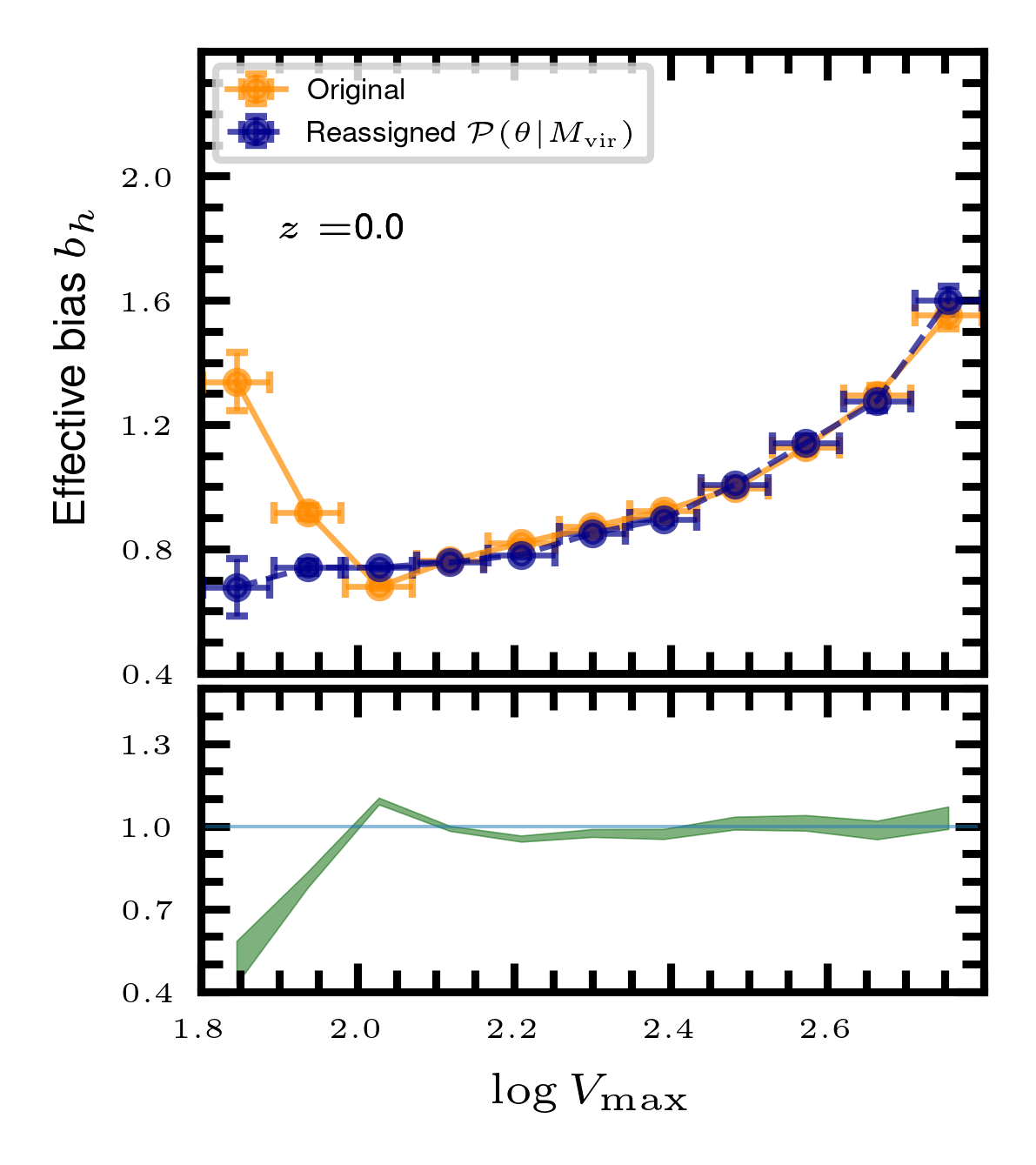}
\includegraphics[trim = .0cm 0.24cm 0cm 0cm ,clip=true, width=0.32\textwidth]{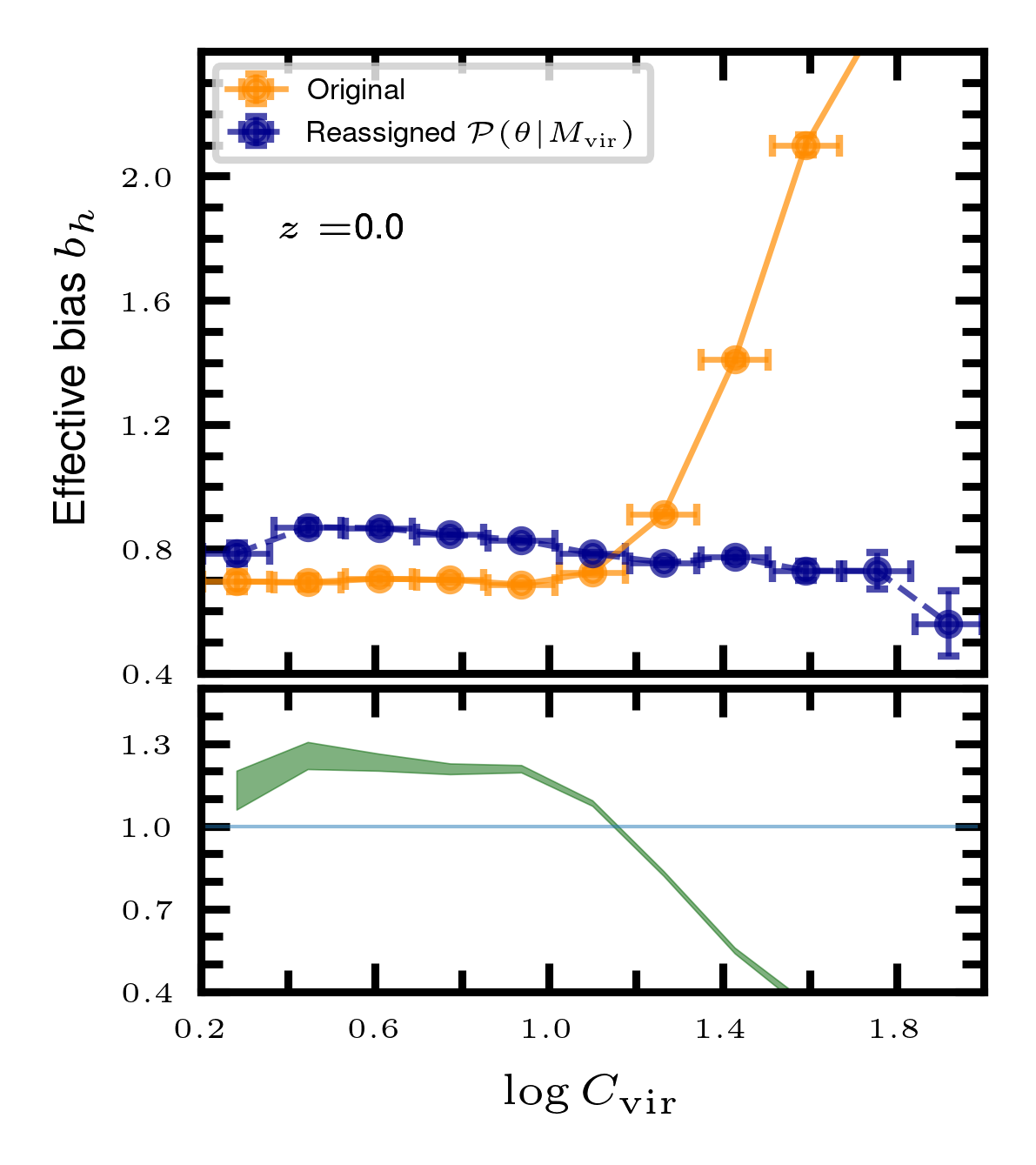}
\includegraphics[trim = .0cm 0.24cm 0cm 0cm ,clip=true, width=0.32\textwidth]{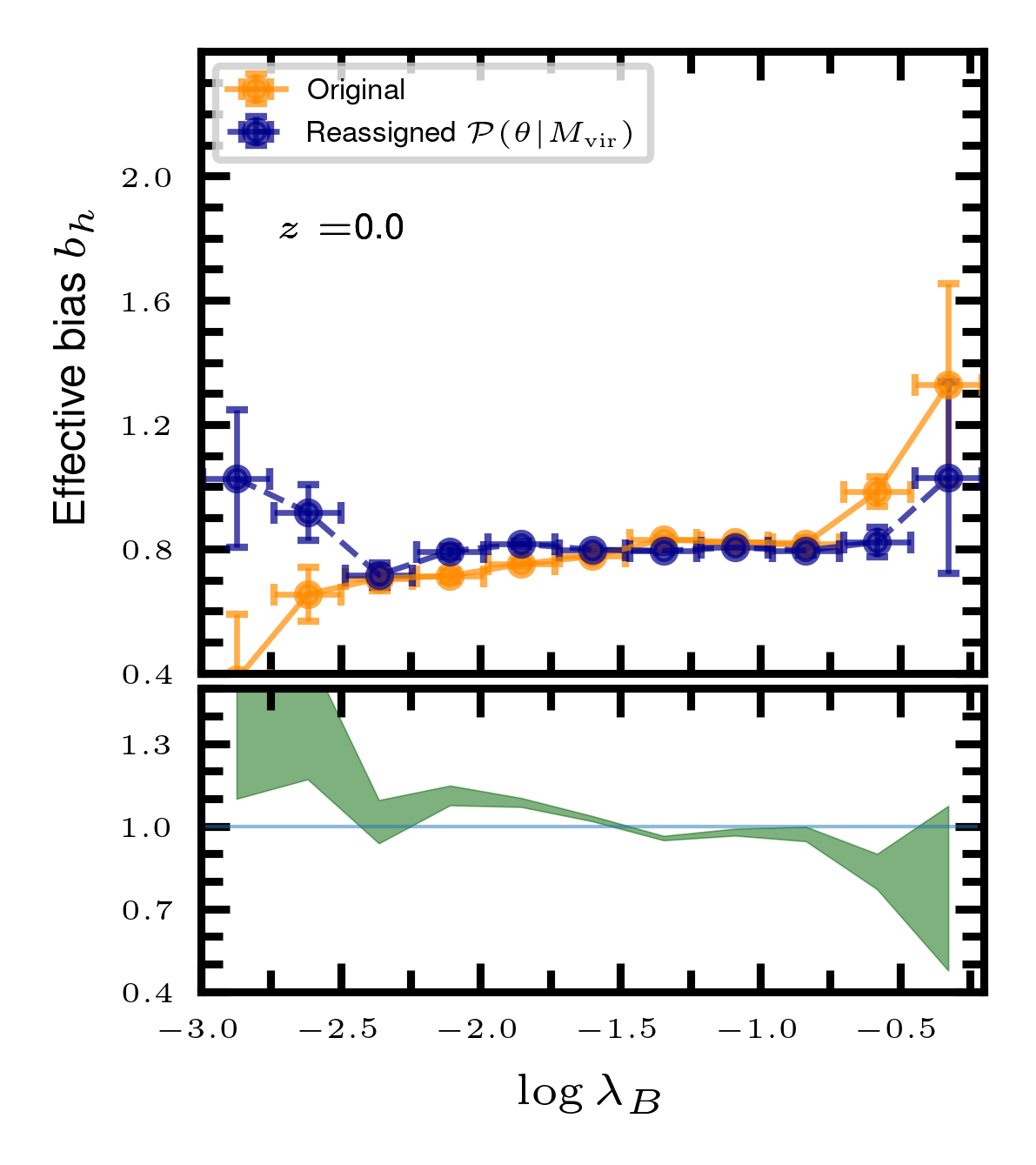}
\caption{\small{Comparison between the measurements of the effective halo bias as a function $V_{\rm max}$, $C_{\rm vir}$, and $\lambda_{B}$ as read from the original catalog against the signal measured used after reassigning the same set of properties using a scaling relation based solely on the virial mass. The bottom panels show the ratio between the bias in the bins of the ``reassigned" to that measured in bins of the ``original" property.}}
\label{fig:randtest0}
\end{figure*}

%=================================================================

\subsection{Secondary halo bias}\label{sec:secbias}

The signal of secondary halo bias is obtained by assessing the mean effective halo bias in bins of a primary property, in which the sample has been divided in quartiles of a secondary property \citep[see e.g.,][for a similar analysis, but using the correlation function computed from this type of subsets]{MonteroDorta2020B,SatoPolito2019,Salcedo2022}. For simplicity, we concentrate here in 6 representative secondary properties, namely, halo spin, concentration, ellipticity, Mach number, dark matter density, and tidal anisotropy. Figure ~\ref{fig:sec_bias_ex1} shows effective halo bias as a function of halo virial mass computed in the first (lower) and fourth (upper) quartiles of all these properties at redshifts $z\sim 2, 0$,  using the entire sample. All the properties shown in this figure display signatures of secondary bias, with a significance that can vary with halo mass and redshift. In particular, key features can be recognized from this type of plots such as inversions of trends for quartiles at different mass scales, and redshifts at which such inversions take place (when more redshifts are shown).  For example, notice  how tracers with low halo concentration at $z\sim 2$ are more biased than low concentrated halos at all masses, a trend that remains at $z=0$ but only for very high halo mass. In general, this plot illustrates a well-known result: \emph{virtually all intrinsic and external halo properties display a certain level of secondary halo bias in some mass ranges.} We have verified that the mean effective bias measured in quartiles of primary properties such as the maximum circular velocity or the velocity dispersion is statistically compatible with that from the full sample, i.e, these type (namely, primary) of properties do not display significant secondary bias. In other words, a halo property displays secondary bias as long as it has a weak connection with halo mass (or peak height).

As we aim to measure signals of secondary bias for a number of halo properties in different cosmic-web environments and cosmological redshifts, we need to condense the information shown in Fig.~\ref{fig:sec_bias_ex1} in a simple but still informative way. One possibility that has been employed in the literature is to compute the the {\it{relative bias}} between effective halo bias in the first and fourth quartiles of $\theta_{s}$. We present one example of this procedure in Fig.~\ref{fig:sec_bias_rat}, which displays the secondary halo bias with respect to concentration and spin, at three different redshifts. Here, some of the main features of the signals are clearly visible. For concentration, there is a well-known crossover of the signal at around $\log M_{\rm vir}[h^{-1}M_\odot] = 13$, so low-concentrated halos have higher (lower) bias than same-mass high-concentrated halos above (below) this characteristic mass. Higher spin halos are, nevertheless, always more tightly clustered than lower spin halos at fixed halo mass. The inversion of the signal in the secondary bias induced by halo spin that was previously measured (at $\log M_{\rm vir}[h^{-1}M_\odot] \simeq 11.5$, \citealt{SatoPolito2019,Johnson2019}) is not observed in UNITSim when the entire sample is analyzed (it is found only for certain cosmic-web environments, see below). This inversion was shown to be caused by the so-called ``splashback halos'' in \cite{Tucci2021}, at a mass scale very close to the minimum mass adopted for this work (see e.g., Figure $2$ in \citet{Tucci2021}). We notice however that the relative bias tends to unity for low-mass halos at $z=0$.   

Importantly, the right-hand panels of Fig.~\ref{fig:sec_bias_rat} display the secondary halo bias as a function of peak height. At least to a first approximation, these plots show that the redshift dependence of these particular signals is approximately captured by $\nu$, as previously shown by, e.g., \cite{Wechsler2006, Gao2007}.

We can also assess the significance of the signal of secondary bias by computing
\be\label{eq:Delta}
\Delta^{i-j}_{p-s} \equiv \frac{\langle b_{h}|\theta_{p}\rangle^{(s)}_{i} -  \langle b_{h}|\theta_{p}\rangle^{(s)}_{j} }{\sqrt{\sigma^{2}_{s,i}+\sigma^{2}_{s,j}}},
\ee
where $\langle b_{h}|\theta_{p}\rangle^{(s)}_{i}$ is the mean sample bias measured in bins of a primary property $\theta_{p}$ in the $i$-th quartile of secondary property $\theta_{s}$. Similarly, $\sigma_{s,i}$ denotes the standard error in the mean bias obtained in the $i$th quartile of the secondary property. For a pair of primary and secondary properties, a value $|\Delta^{i-j}_{p-s}|\gg 1$ can be used as an statistical signature of secondary halo bias. It is important not to mistake this statistic with the amplitude of the secondary bias signal itself, that is, a particular property might display a large signal at a given mass range, but at low statistical significance. 

In Fig.~\ref{fig:sec_bias_allz1}, we show the behavior of the variable defined by Eq.(\ref{eq:Delta}) as a function of halo mass, for the selected set of secondary properties and as a function of redshift. We include an uncertainty region of $\Delta^{i-j}_{p-s} = \pm 2$ to guide the reader. The information displayed in Figs.~\ref{fig:sec_bias_allz1} is expanded in Figs.~\ref{fig:sec_bias_ex2} and \ref{fig:sec_bias_allz2}, where we evaluate the dependence of secondary bias on the cosmic web. Also, the redshift evolution of secondary bias based on peak height can be found in the Appendix. This set of plots represents the measurement of secondary bias as a function of redshift and cosmic-web type, thus containing a vast amount of relevance information. Our main conclusions can be summarized as follows:

%=================================================================
%=================================================================
\begin{figure*}
\centering
\includegraphics[trim = .2cm 0.55cm 0cm 0cm ,clip=true, width=0.32\textwidth]{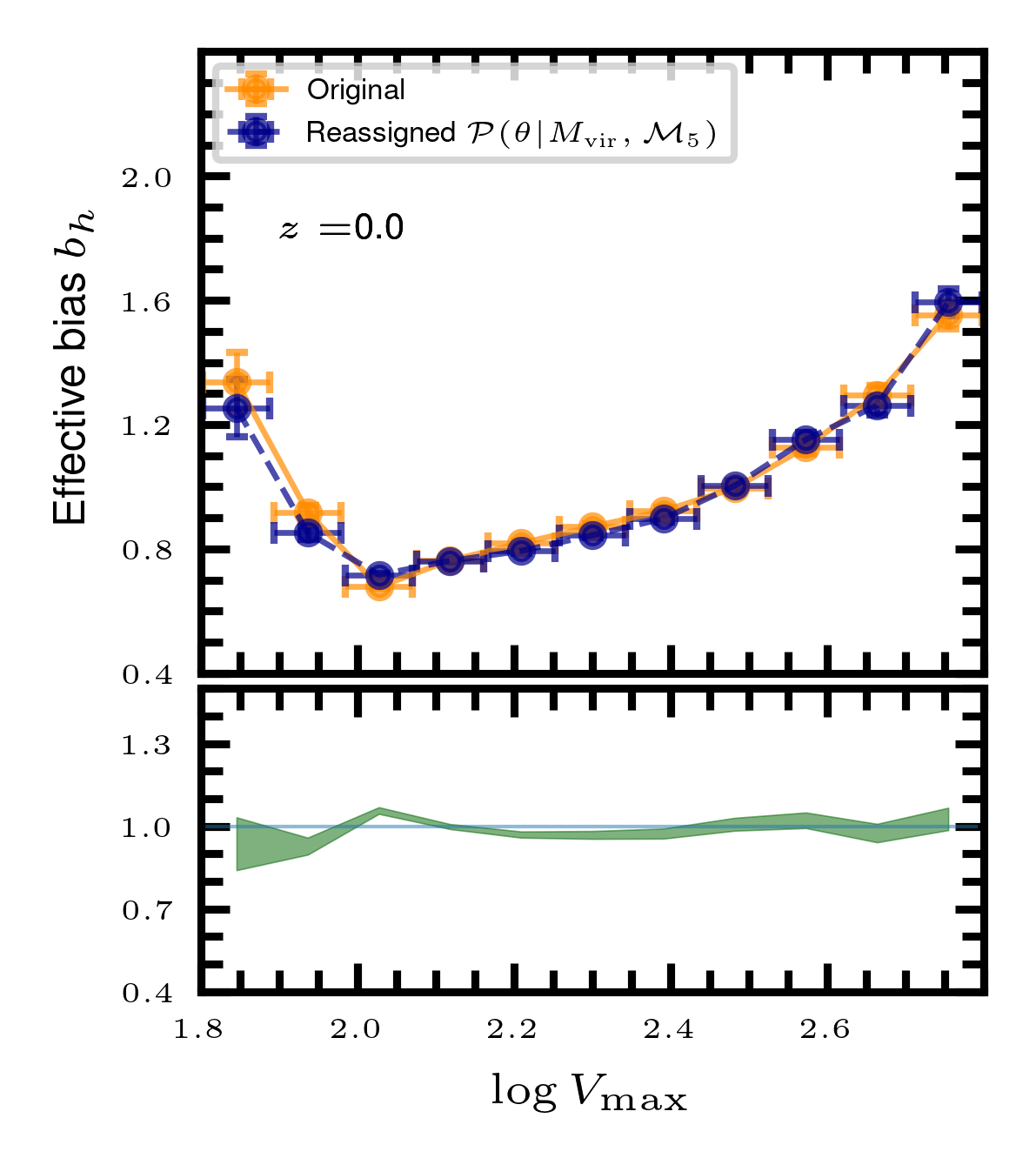}
\includegraphics[trim = .2cm 0.55cm 0cm 0cm ,clip=true, width=0.32\textwidth]{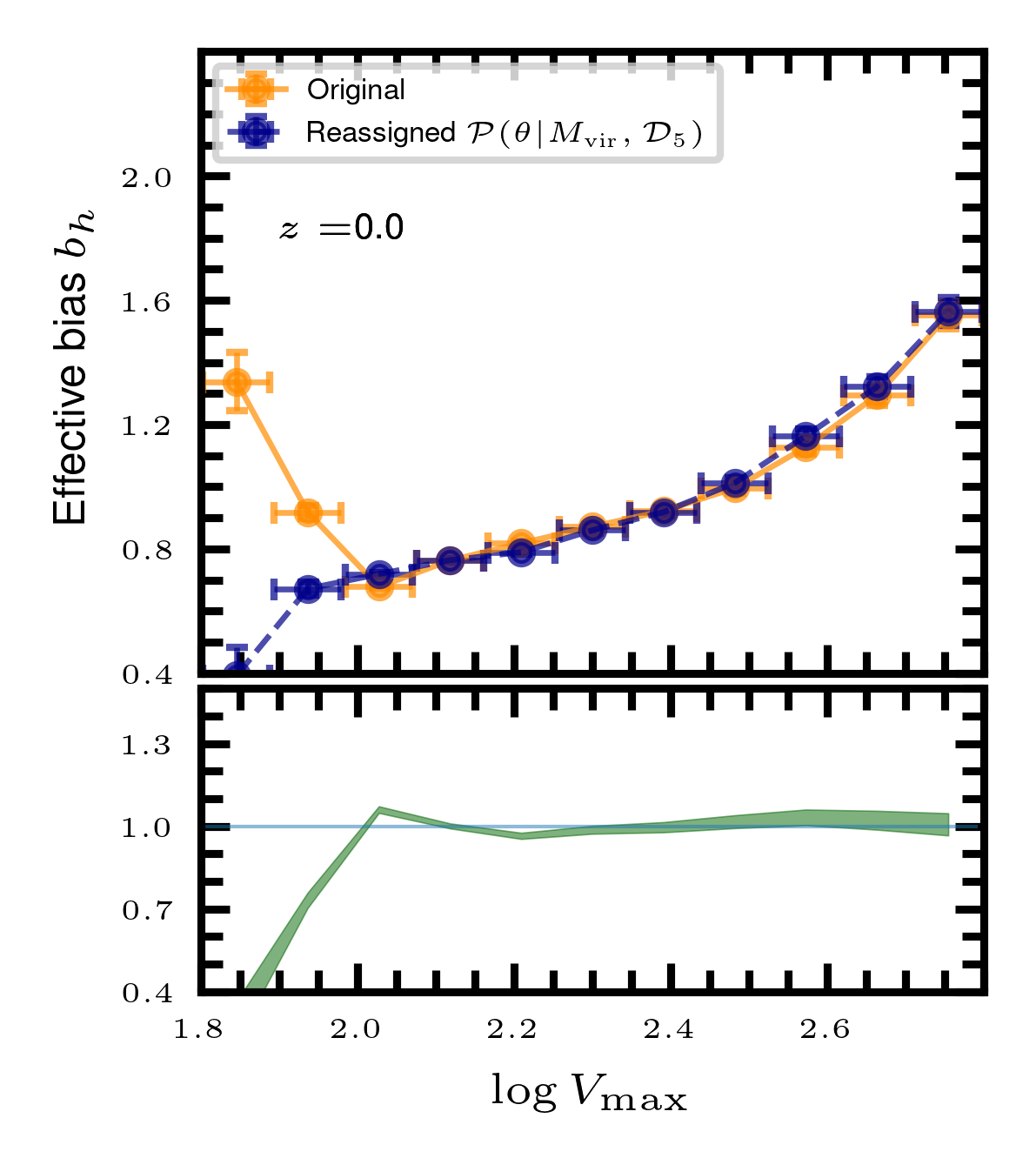}
\includegraphics[trim = .2cm 0.55cm 0cm 0cm ,clip=true, width=0.32\textwidth]{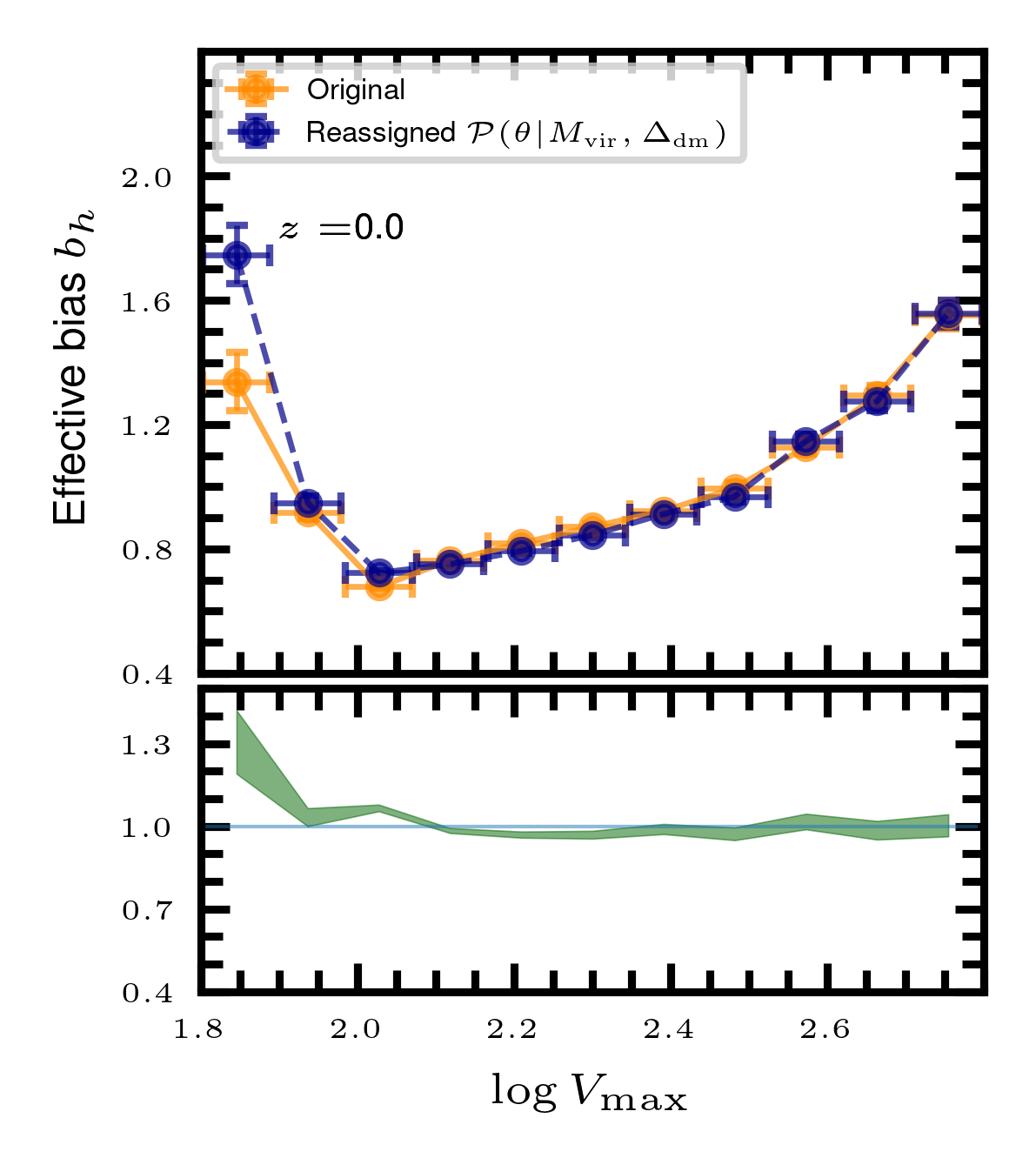}
\includegraphics[trim = .2cm 0.24cm 0cm 0cm ,clip=true, width=0.32\textwidth]{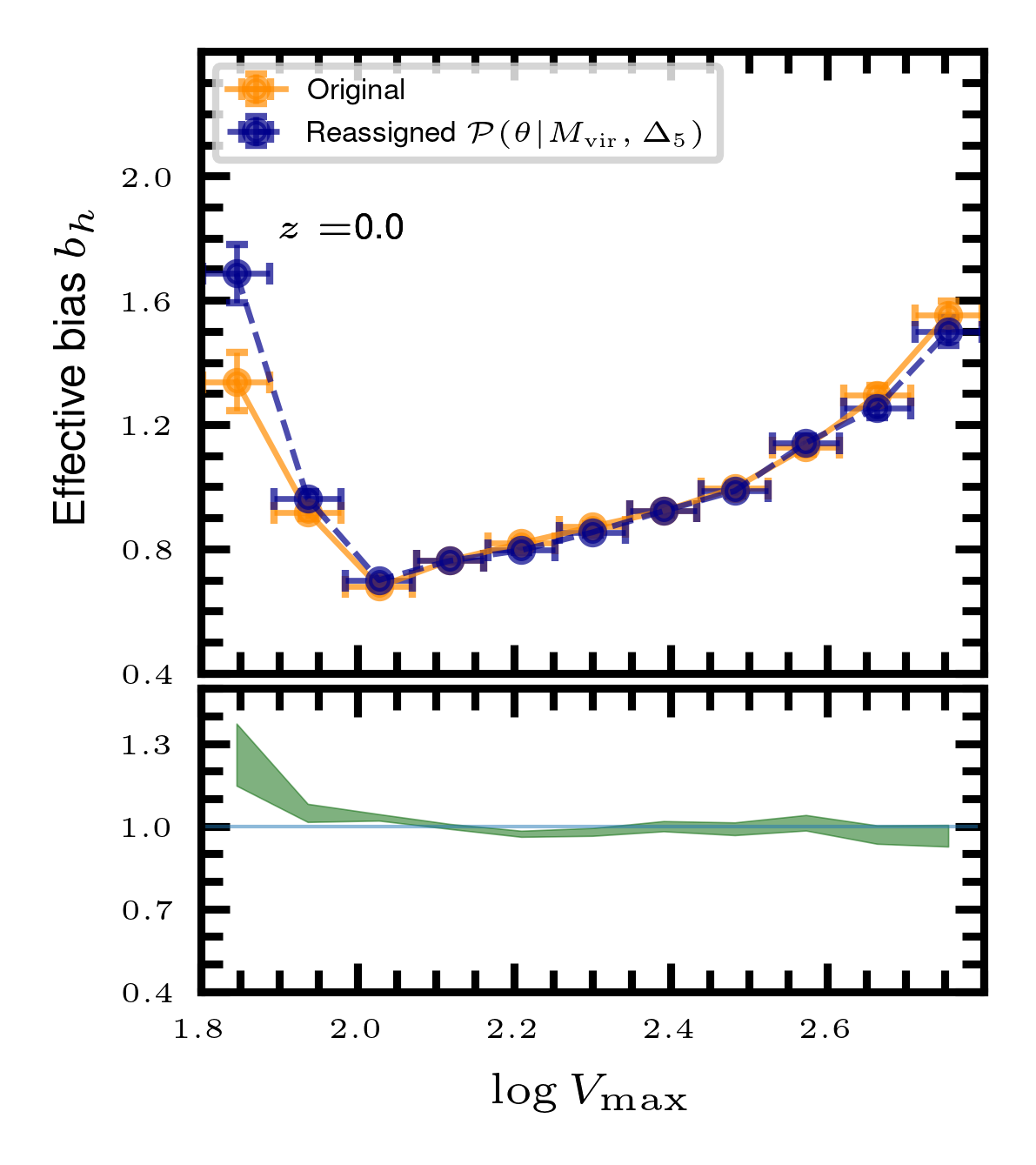}
\includegraphics[trim = .2cm 0.24cm 0cm 0cm ,clip=true, width=0.32\textwidth]{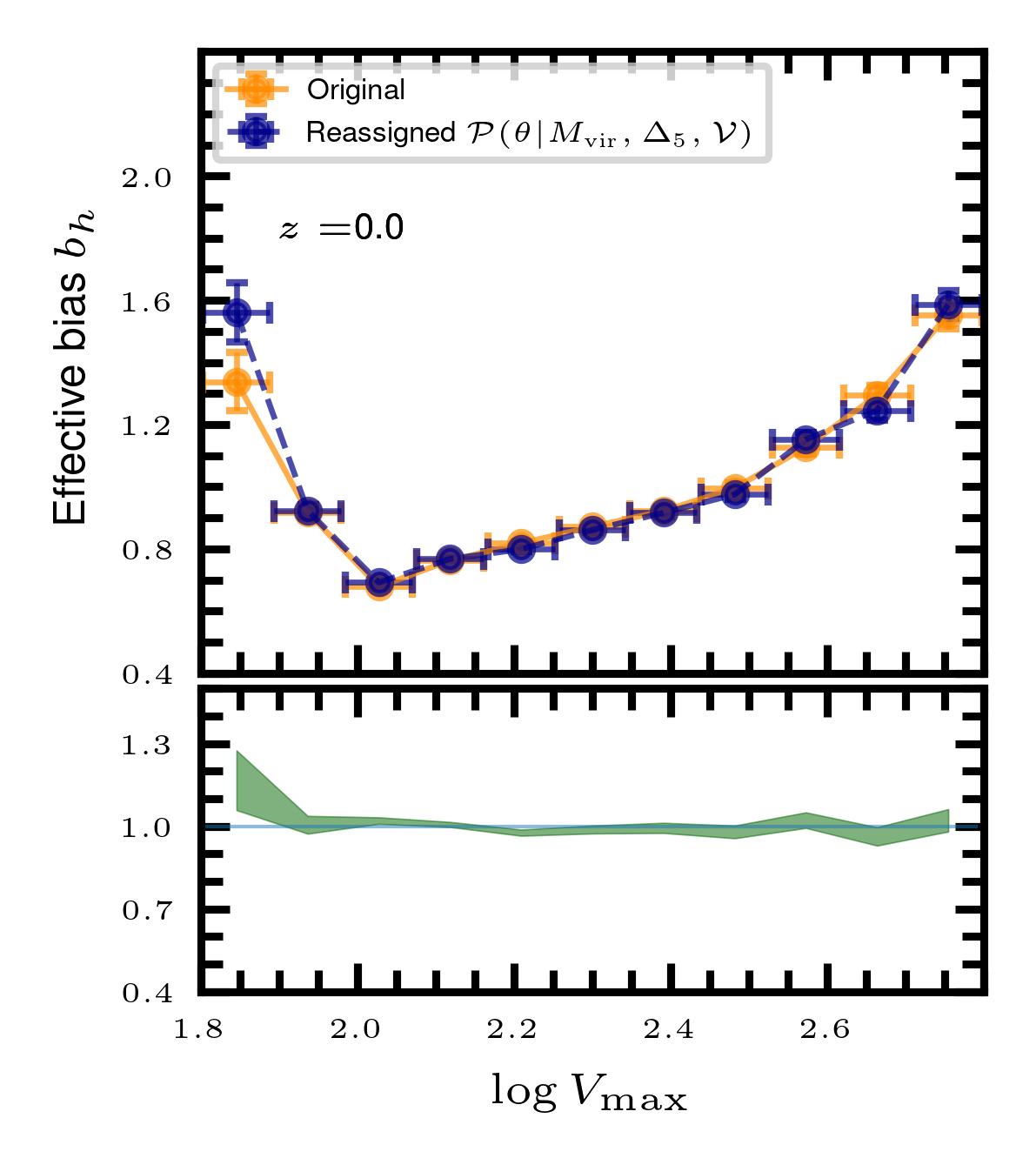}
\includegraphics[trim = .2cm 0.24cm 0cm 0cm ,clip=true, width=0.32\textwidth]{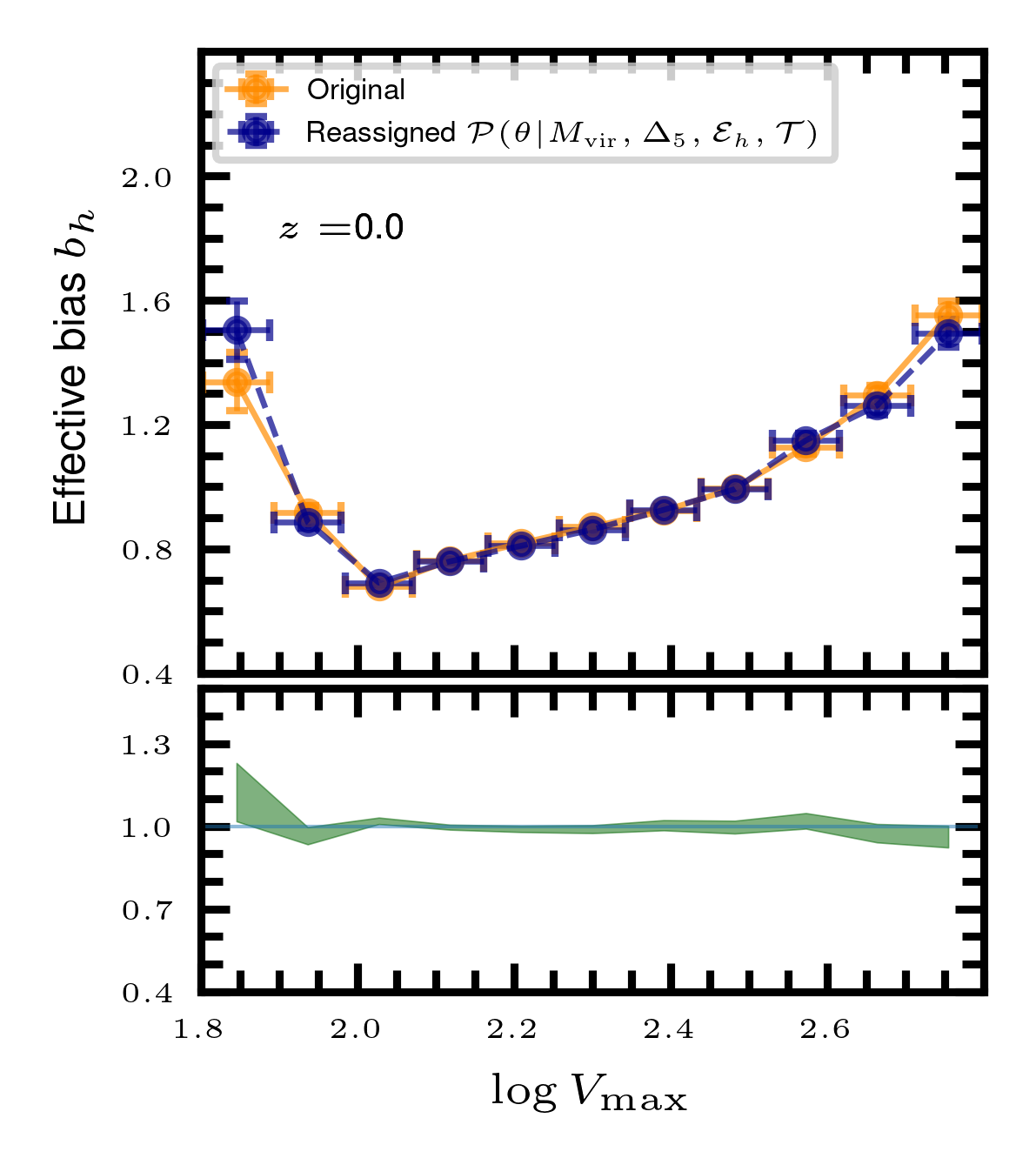}
\caption{\small{Same as Fig.~\ref{fig:randtest0} but focused on $V_{\rm max}$. In each panel, different dependencies (named in the legends) are used to reassign maximum circular velocity.}}
\label{fig:randtest_vmax}
\end{figure*}
%=================================================================
%=================================================================

\begin{itemize}
\item \emph{Halo concentration}.
The secondary bias based on concentration displays the well-known trend explained above, dominated by a inversion of the signal at high masses at $z=0$. Due to the intrinsic dependence of secondary bias on $\nu$, this crossover moves toward lower halo masses at higher redshift. Importantly, the significance of the concentration-based secondary bias depends on cosmic-web environment: it is quite significant for knots and filaments, but not so much for sheets and voids. Also, the "crossover mass" changes for different cosmic web types.

\item \emph{Halo spin}. As mentioned before,  {\it{spin bias}} conserves the link between clustering strength and quartiles, i.e, upper quartile has a larger effective bias at all mass scales in the full sample at $z=0$. It is only when the signal is analyzed as a function of cosmic-web environments that an inversion of the signal is observed for knots at low redshift (lower spin halos become more highly clustered below $\log M_{\rm vir}[h^{-1}M_\odot] = 12.2$ at $z=0$). This behavior might be due to the presence of splashback halos \citep[see e.g.,][]{Tucci2021} in the UNITSim.

\item \emph{Halo ellipticity}. At fixed halo mass, spherical halos (lower quartile in $\mathcal{E}_{h}$) are more clustered than nonspherical halos at $z=0$. This trend appears to be inverted at $z\gtrsim 2$ at a mass scale of $\log M_{\rm vir}\sim 13.2-13.4$. The $z=0$ result is in good agreement with \cite[][]{Faltenbacher2010}, where the redshift evolution was not measured. The secondary bias signal for this geometry indicator is more significant than that of halo triaxiality. We have checked that, at fixed halo mass, oblate halos are more clustered than prolate objects, also in agreement with \cite[][]{Faltenbacher2010}. Again, the secondary bias signals produced by these parameters are more significant in knots and filaments. 

\item \emph{Relative Mach number}. This property displays a significant positive secondary bias signal, at all redshifts, particularly toward the lower halo masses: at fixed halo mass, halos with higher $\mathcal{M}_{5}$ have higher bias. However, an inversion of the signal is detected, albeit with low significance, for high-mass halos at lower redshift (i.e, low-$\mathcal{M}_{5}$ halos are more clustered on large scales). In terms of cosmic-web types, again the presence of an inversion depends on environment (e.g., it is not observed clearly in sheets). 

\item \emph{Neighbor statistics, local density, local halo overdensity, tidal anisotropy}. All these halo external properties tend to produce very large and significant positive secondary bias signal (larger values at fixed halo mass are associated with higher bias). Again, the characteristic evolution dictated by $\nu$ is partially observed (with the signals shifting toward lower masses). There are particular redshifts and cosmic-web environments for which small and low (statistical) - significant inversions are detected. 

\item Among all properties considered, the one that produces the most statistically significant low-mass secondary bias signal are, in this order, tidal anisotropy, DM density, Mach number, and either ellipticity or concentration (at a reference mass of $\log M_{\rm vir}[M_{\odot}h^{-1}]\sim 12.2$).

\item Secondary halo properties tend to produce secondary bias signals that decrease in significance with halo mass (which could be partially due to the large uncertainty at the high-mas end of the mass function). At low redshift, spin remains the property that produces the largest signal in UNITSim.

\item The redshift evolution of secondary bias is approximately (but not completely) captured by the implicit redshift dependence of $\nu$ for internal halo properties (particularly concentration and ellipticity). That is not the case for the external halo properties, which still maintain significant evolution even when $\nu$ is chosen as a primary property (see also Fig. \ref{fig:sec_bias_rat2} from the Appendix). 

\end{itemize}

%=================================================================
%=================================================================
\begin{figure*}
\centering
\includegraphics[trim = .2cm 0.55cm 0cm 0cm ,clip=true, width=0.32\textwidth]{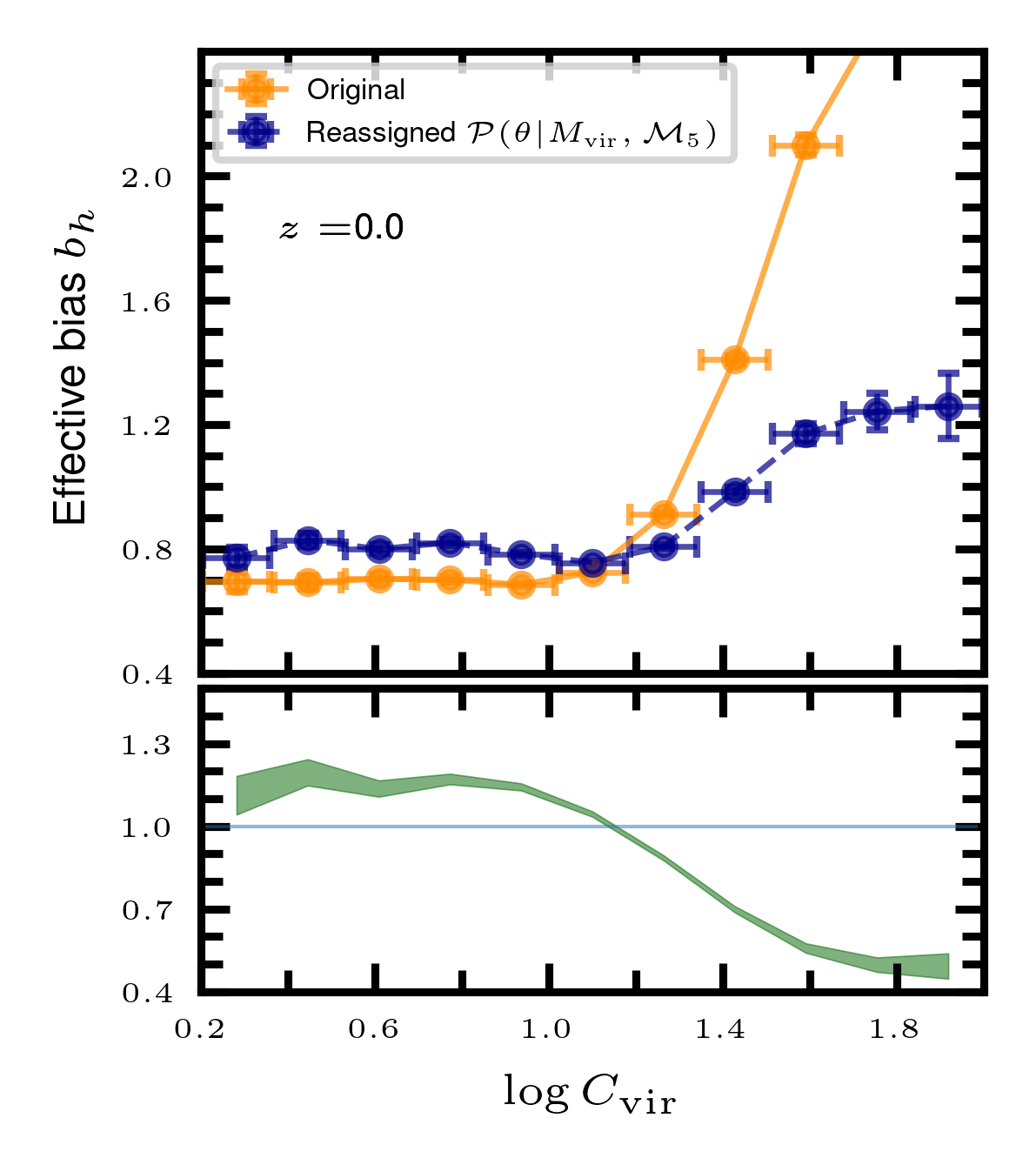}
\includegraphics[trim = .2cm 0.55cm 0cm 0cm ,clip=true, width=0.32\textwidth]{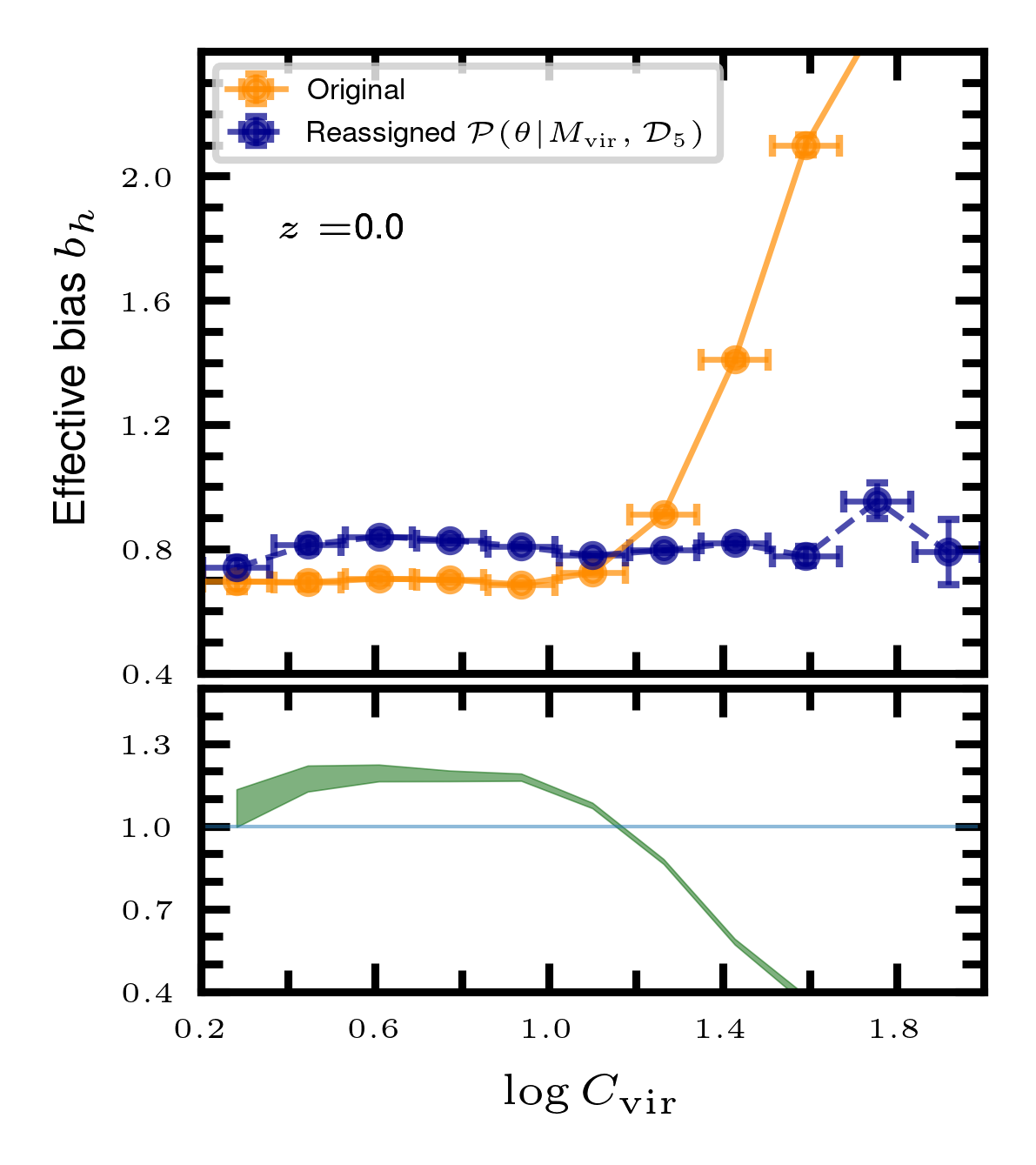}
\includegraphics[trim = .2cm 0.55cm 0cm 0cm ,clip=true, width=0.32\textwidth]{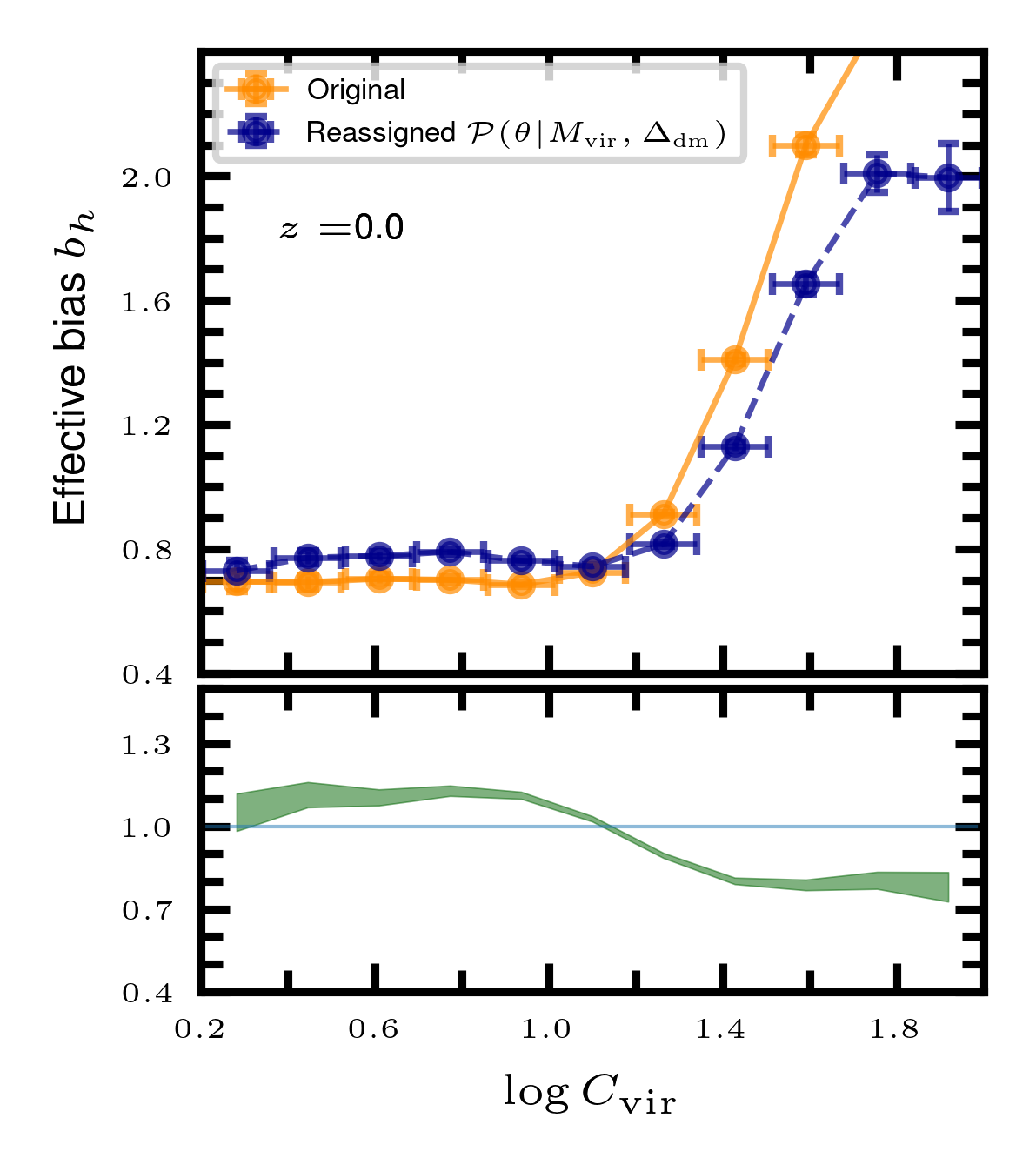}
\includegraphics[trim = .2cm 0.24cm 0cm 0cm ,clip=true, width=0.32\textwidth]{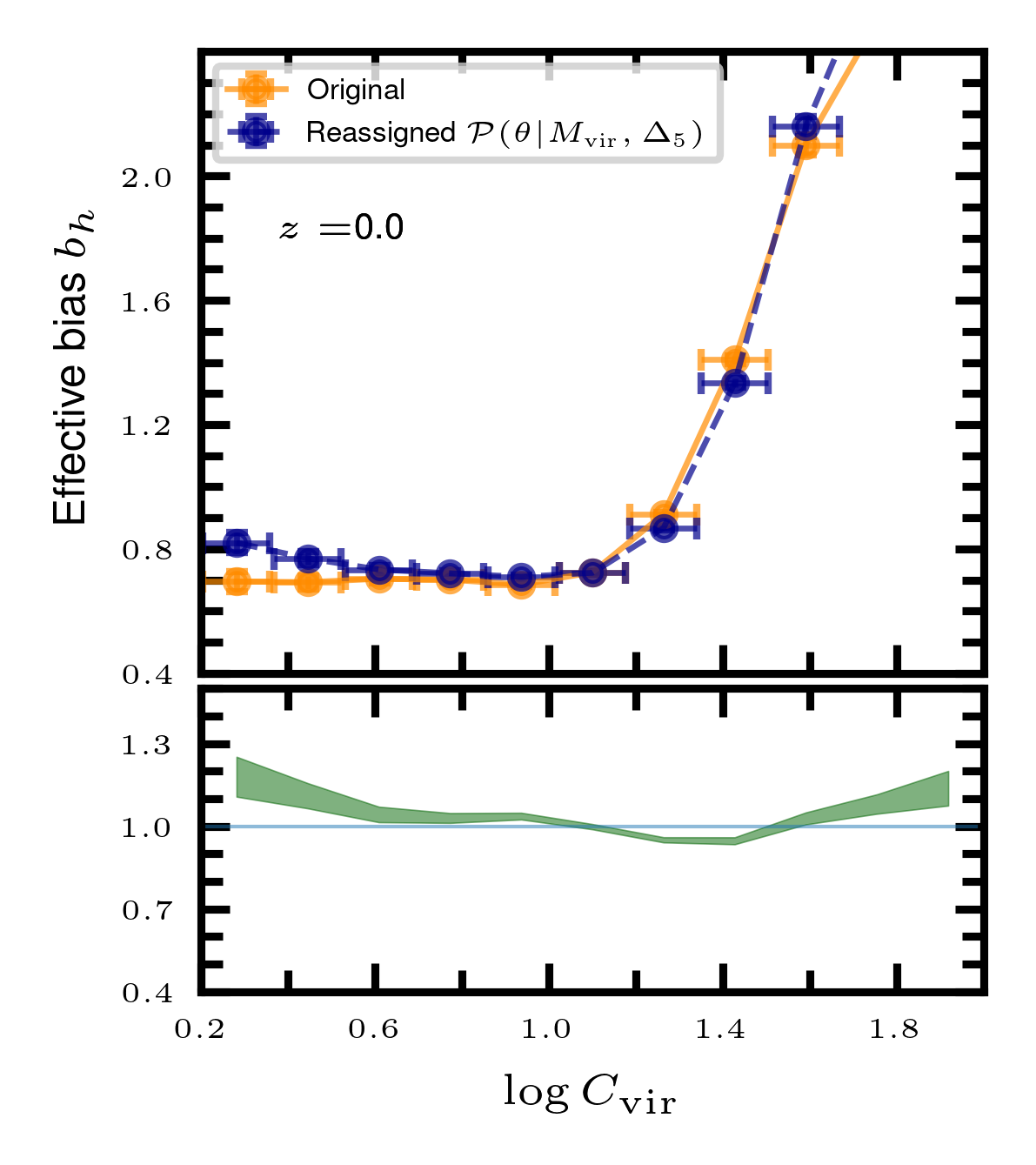}
\includegraphics[trim = .2cm 0.24cm 0cm 0cm ,clip=true, width=0.32\textwidth]{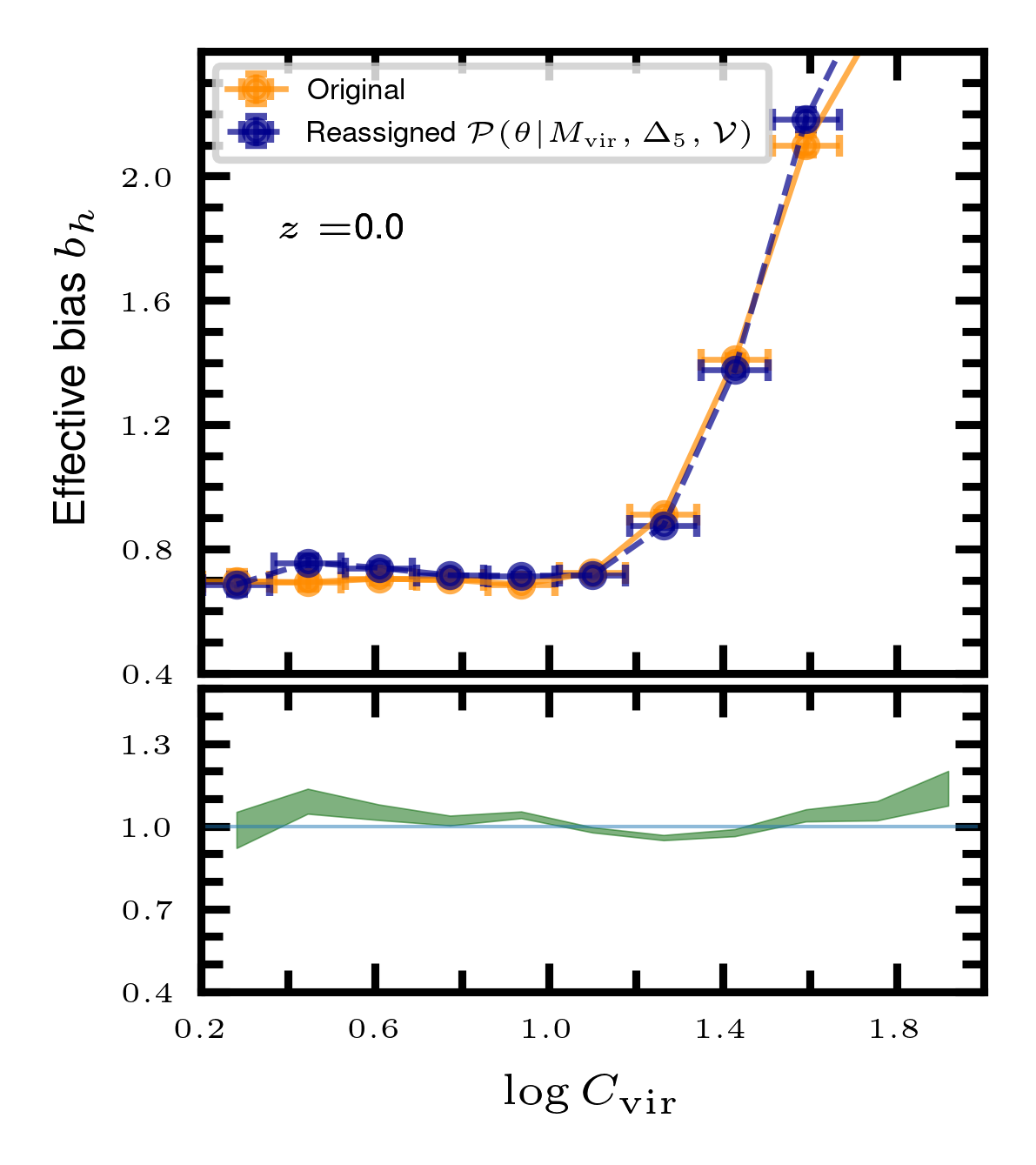}
\includegraphics[trim = .2cm 0.24cm 0cm 0cm ,clip=true, width=0.32\textwidth]{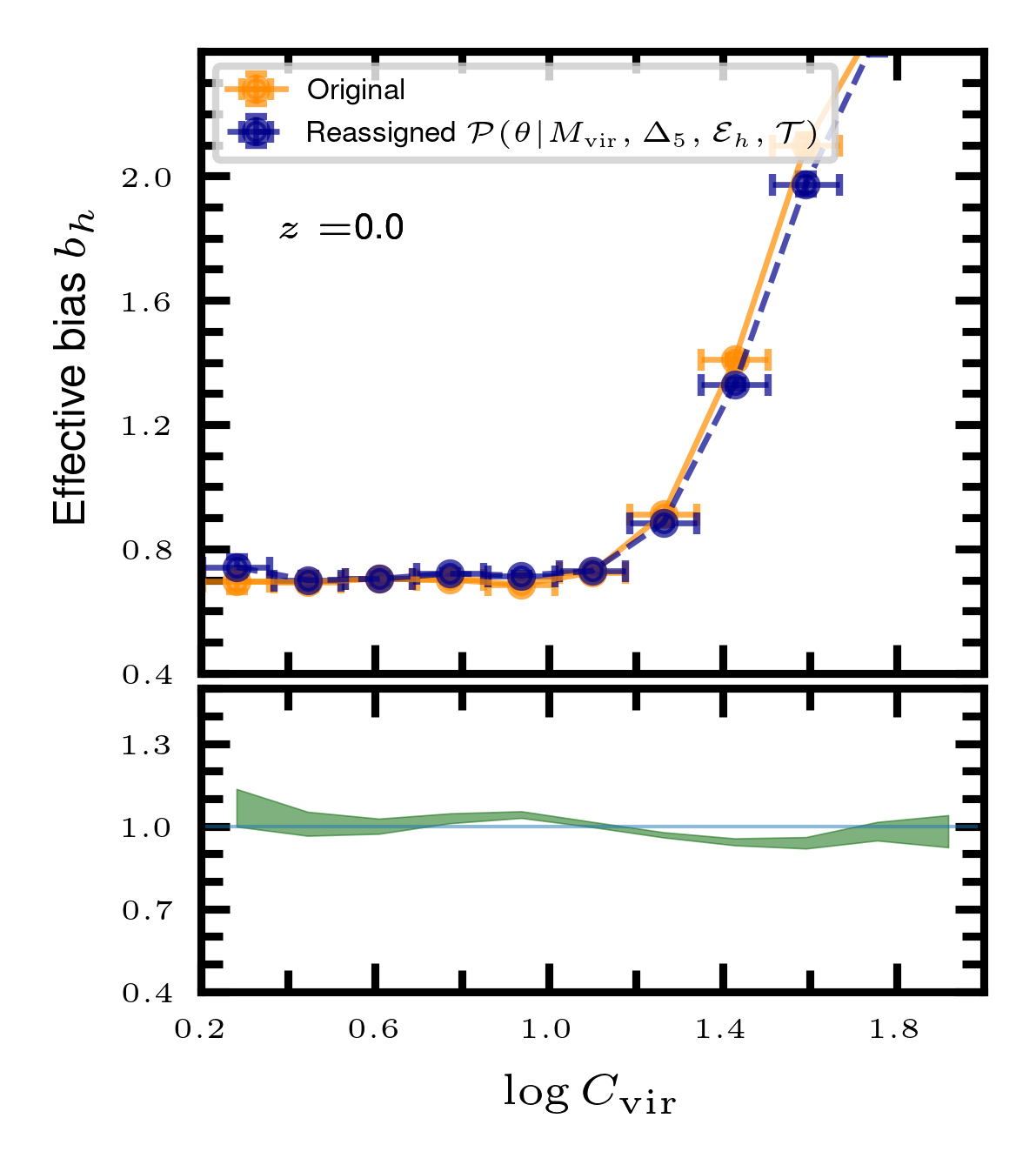}
\caption{\small{Same as Fig.~\ref{fig:randtest0} but using more information in the scaling relation to assign the property $\tilde{\theta}_{s}$ (halo concentration in this case) as shown in each panel.}}
\label{fig:randtest_con}
\end{figure*}
%=====================================================================
%=====================================================================

%=================================================================
%=================================================================
\begin{figure*}
\centering
\includegraphics[trim = .2cm 0.55cm 0cm 0cm ,clip=true, width=0.32\textwidth]{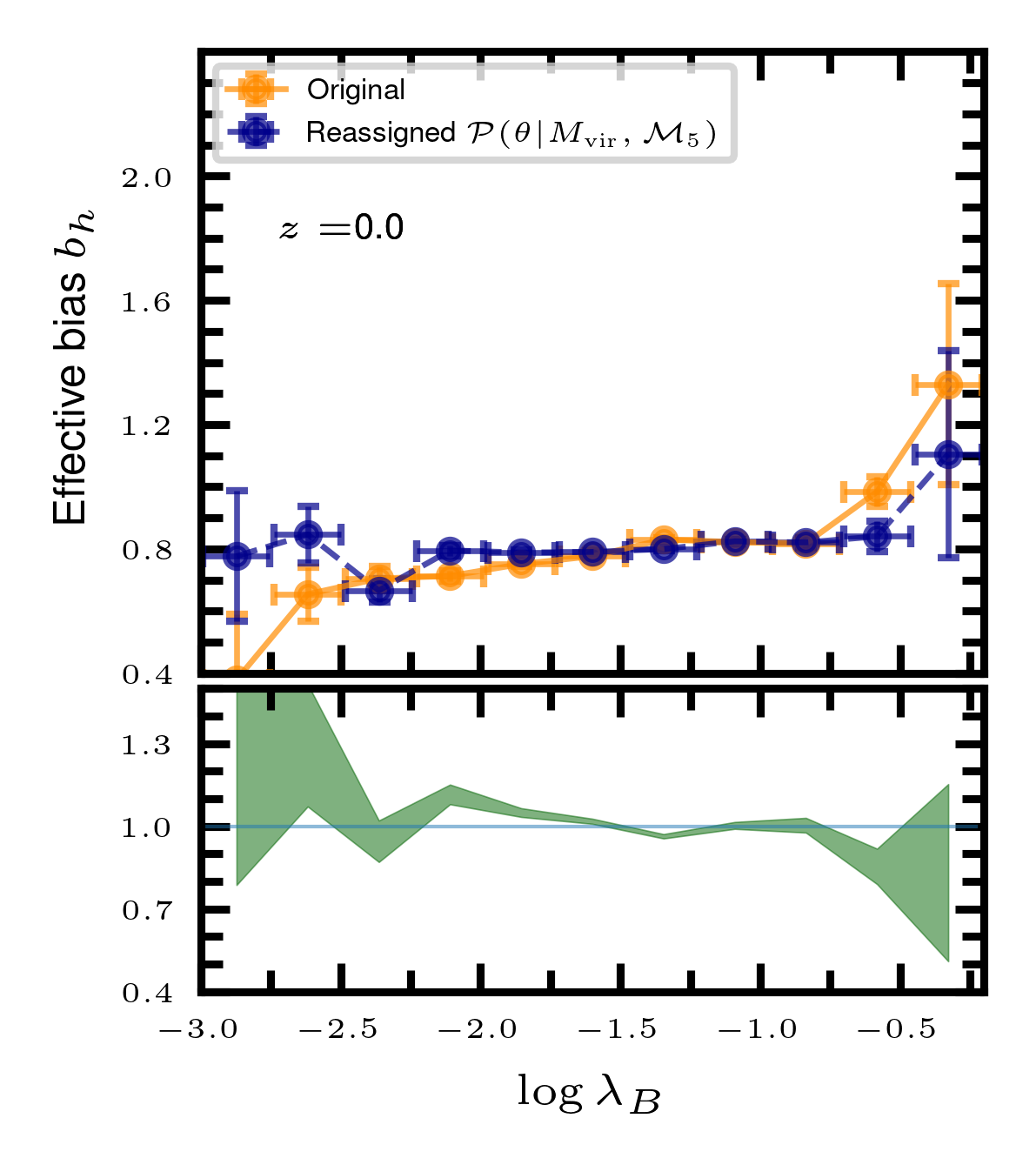}
\includegraphics[trim = .2cm 0.55cm 0cm 0cm ,clip=true, width=0.32\textwidth]{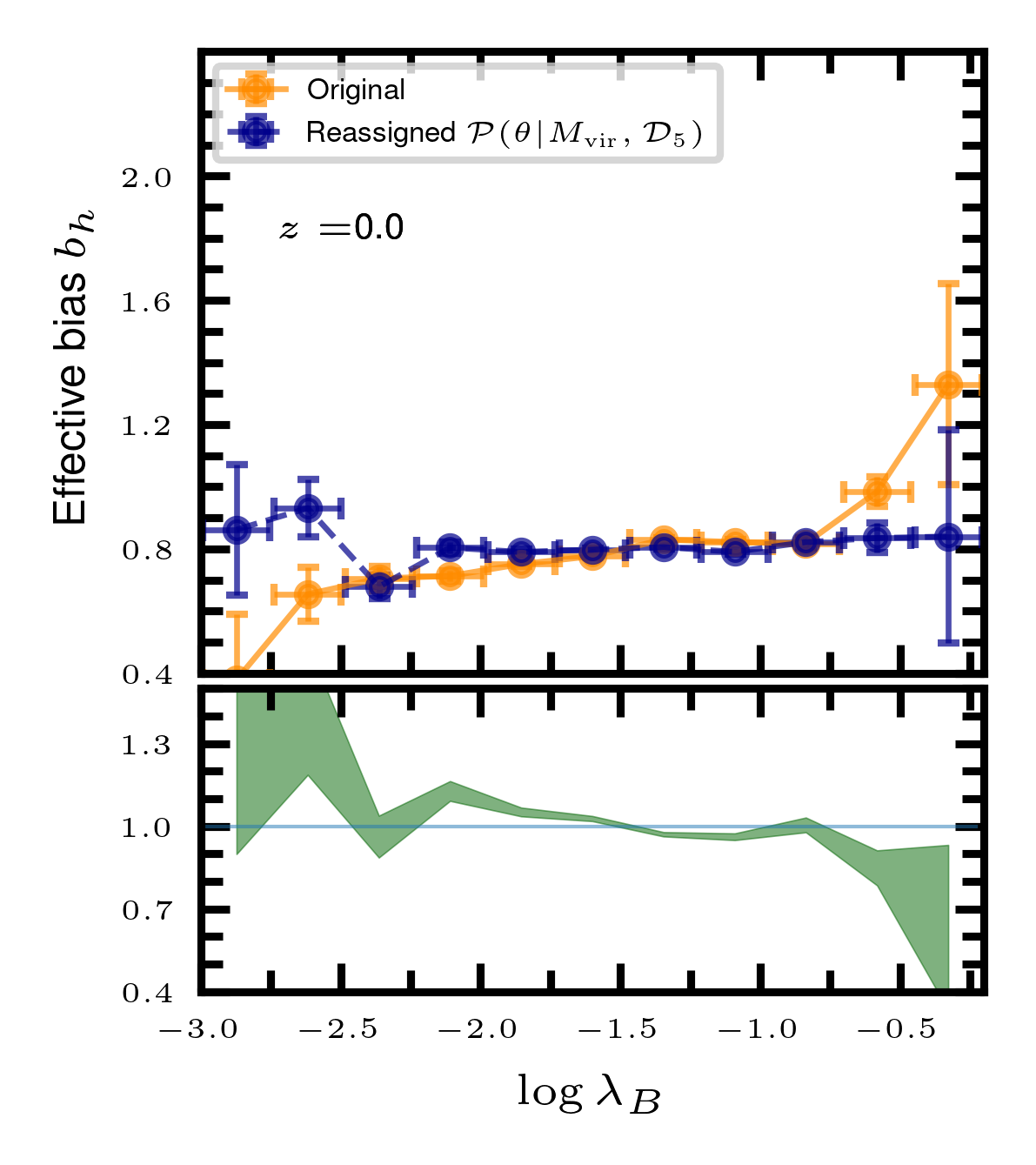}
\includegraphics[trim = .2cm 0.55cm 0cm 0cm ,clip=true, width=0.32\textwidth]{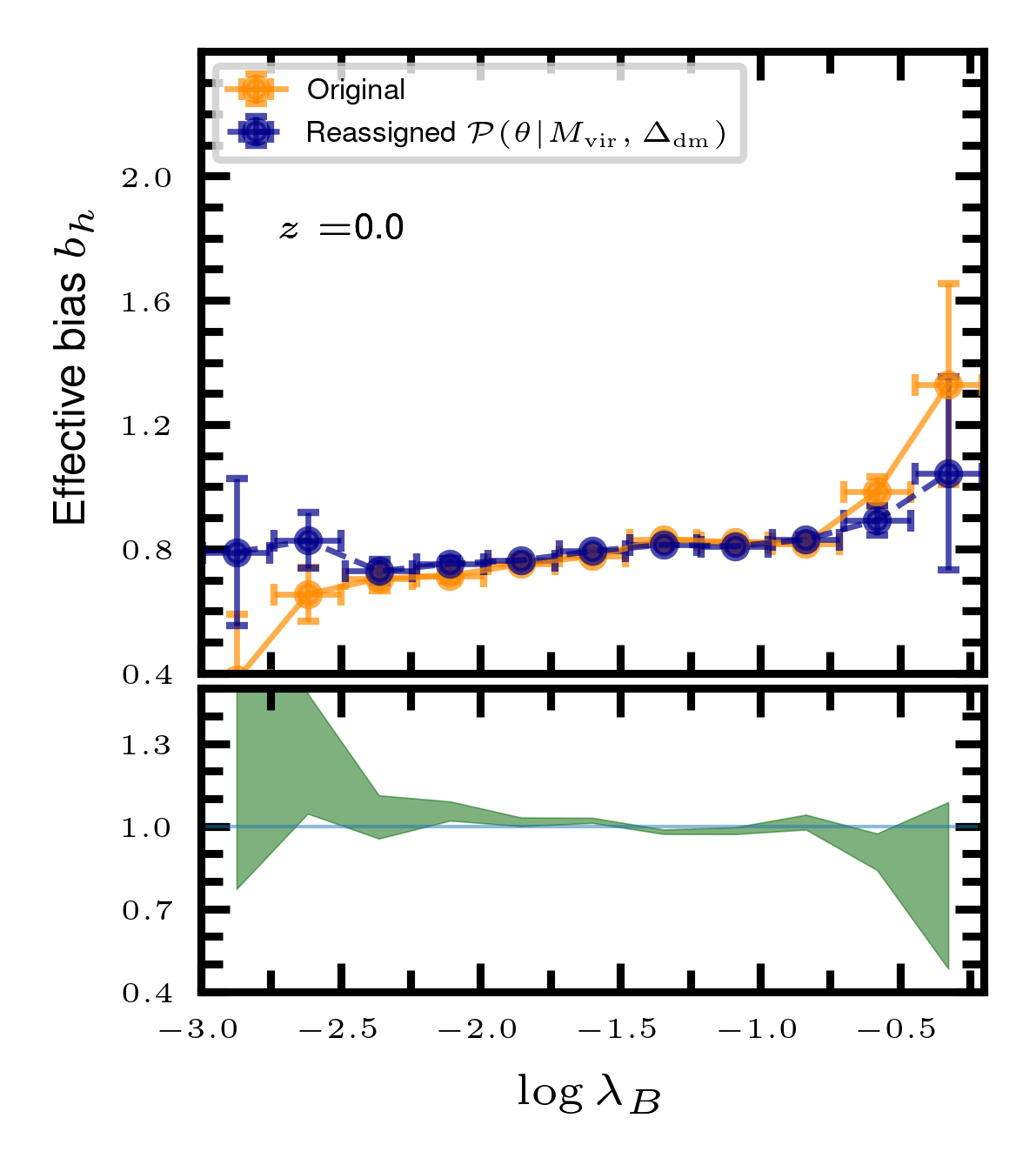}
\includegraphics[trim = .2cm 0.24cm 0cm 0cm ,clip=true, width=0.32\textwidth]{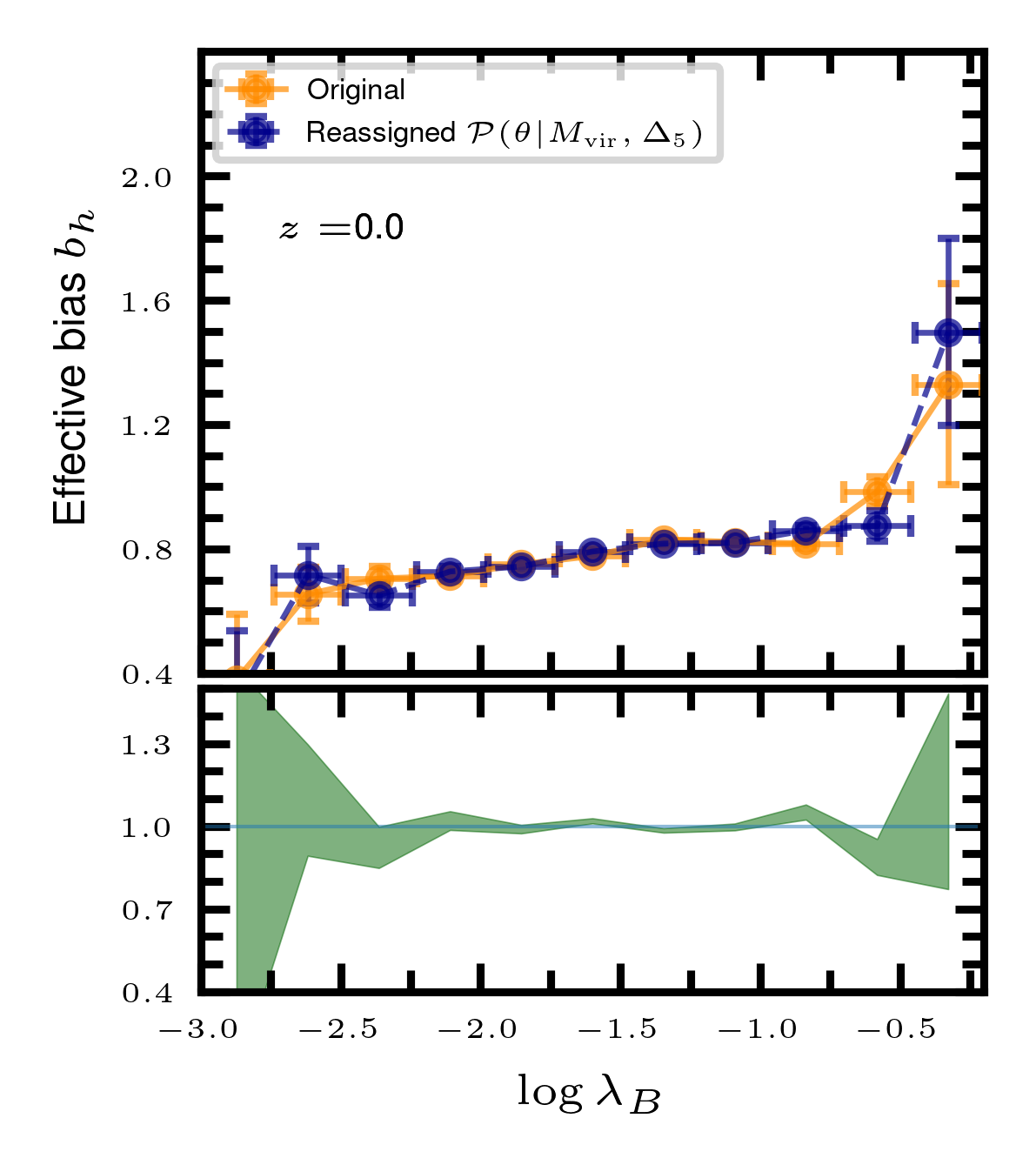}
\includegraphics[trim = .2cm 0.24cm 0cm 0cm ,clip=true, width=0.32\textwidth]{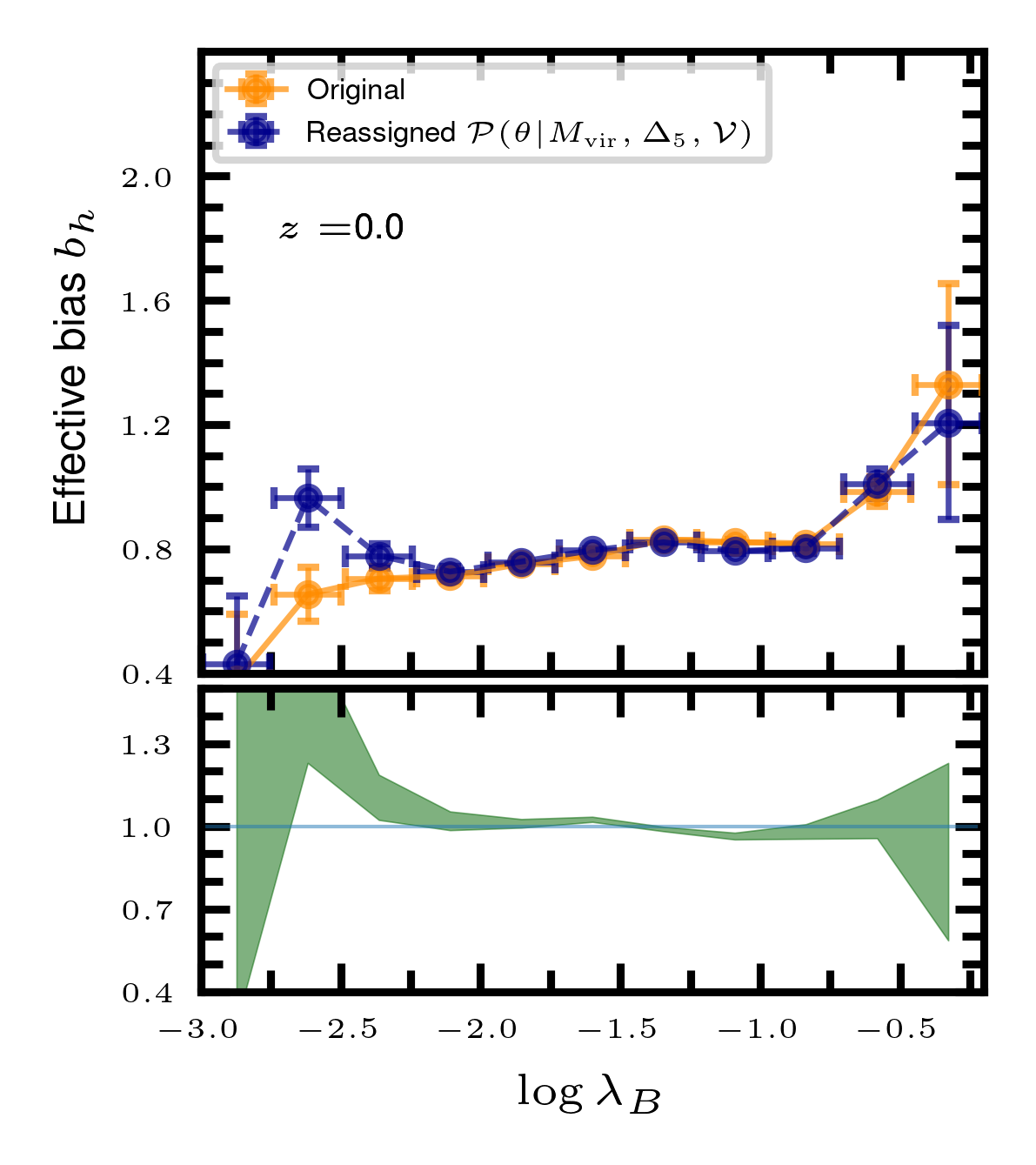}
\includegraphics[trim = .2cm 0.24cm 0cm 0cm ,clip=true, width=0.32\textwidth]{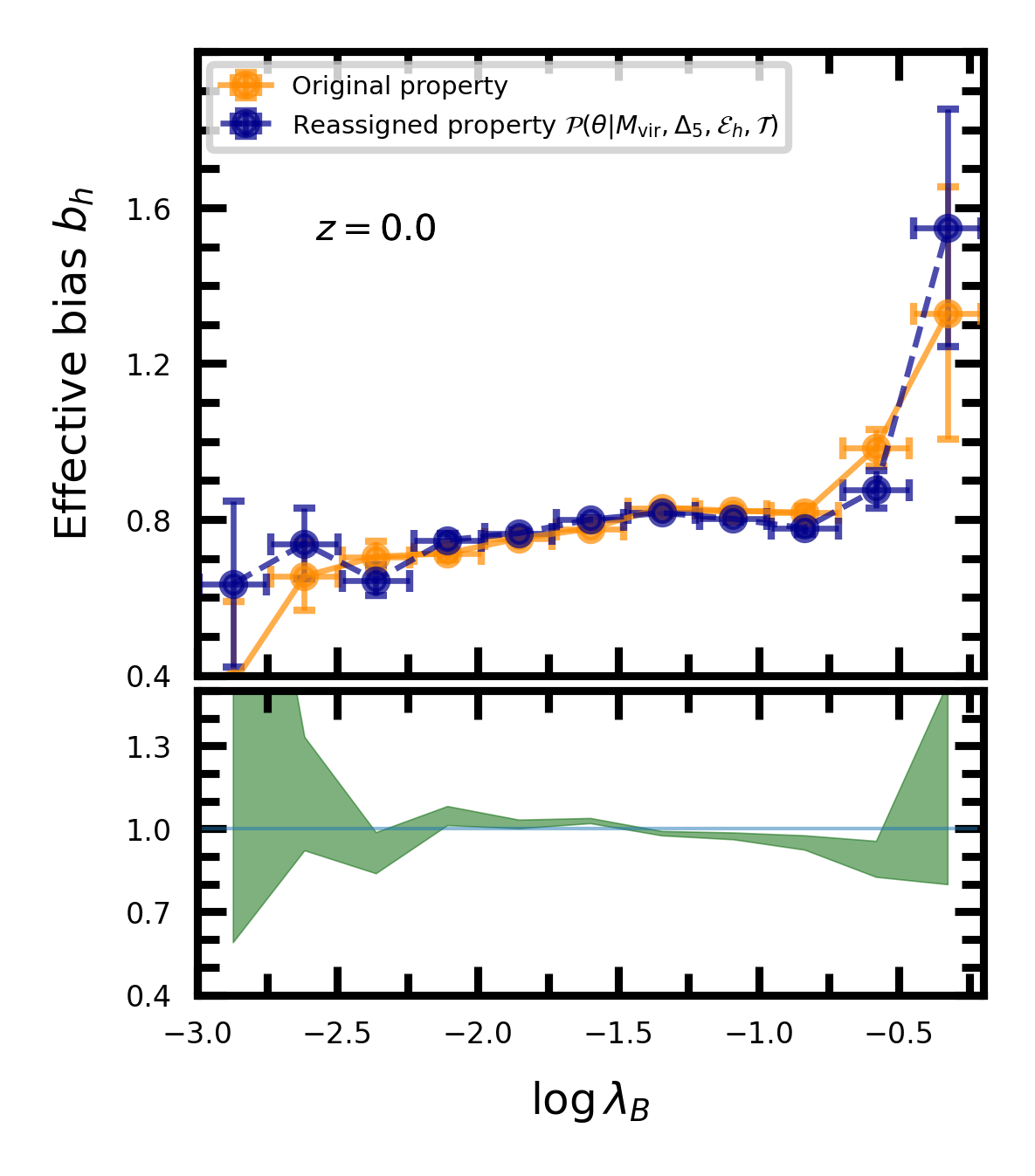}
\caption{\small{Same as Fig.~\ref{fig:randtest0} but using more information in the scaling relation to assign the property $\tilde{\theta}_{s}$ (halo spin in this case) as shown in each panel. }}
\label{fig:randtest_spin}
\end{figure*}
%=====================================================================
%=====================================================================

\section{Reconstructing the signal of secondary bias}\label{sec:ori}

In the previous section, we showed that secondary bias is a statistically significant effect in the context of halo clustering, as has been abundantly reported in the literature. In this section, we aim at a detailed statistical description of the signal. 

\subsection{The link between halo bias and halo properties}

Let us start by exploring the link between halo properties and halo bias. In \S~\ref{sec:bias}, we expressed halo effective bias as a function of a halo property $\theta_{a}$ (different from halo mass or peak height) as  
\be\label{eq:bias_assump2}
 \mathcal{P}(b_{h}|\theta_{a}) = \int_{0}^{\infty}  \mathcal{P}(b_{h}|M_{\rm vir}) \mathcal{P}(M_{\rm vir}|\theta_{a})\dd M_{\rm vir},
\ee
which implies that the statistical properties of halo bias as a function of $\theta_{a}$ are the result of the interplay between the secondary-primary property link and the bias-primary property connection (in our case, halo mass is taken as the primary property). As pointed out in \S\ref{sec:bias}, the latter can be readily obtained from a more fundamental property such as the peak-height as $ \mathcal{P}(b_{h}|M_{\rm vir})= \mathcal{P}(b_{h}|\nu=\nu_{0}(M_{\rm vir}))$, according to Eq.(\ref{eq:sigma}). 

Each of the conditional probability distributions under the integral in Eq.~(\ref{eq:bias_assump2}) is a marginalization over the halo properties; denoting with $\eta=\{\eta_{i}\}$ the remaining set of halo properties (i.e, not including mass or $\theta_{a}$), this is expressed as 
\be\label{eq:margi}
\mathcal{P}(M_{\rm vir}|\theta_{a}) = \int\, \mathcal{P}(M_{\rm vir}|\theta_{a}, \eta_{1}, \cdots,\eta_{i})\dd \eta_{1}\cdots \dd \eta_{i},
\ee
where $\mathcal{P}(M_{\rm vir}|\theta_{a}, \eta_{1}, \cdots,\eta_{i})$ represents a multivariate distribution of the halo mass conditional to the properties $\{\eta\}$ \citep[see e.g.,][]{2014MNRAS.441.3562E}. In general, the moments of the scaling relation $\mathcal{P}(M_{\rm vir}|\theta_{a})$ will be dependent of the correlation between viral mass and the set $\{\eta\}$. In particular, the mean and the scatter around it are a consequence of the propagation of the statistical properties of the links $\mathcal{P}(M_{\rm vir}|\eta_{i})$.

%=====================================================================
%=====================================================================
\begin{figure*}
\centering
\includegraphics[trim = .25cm 0.82cm 0cm 0cm ,clip=true, width=0.45\textwidth]{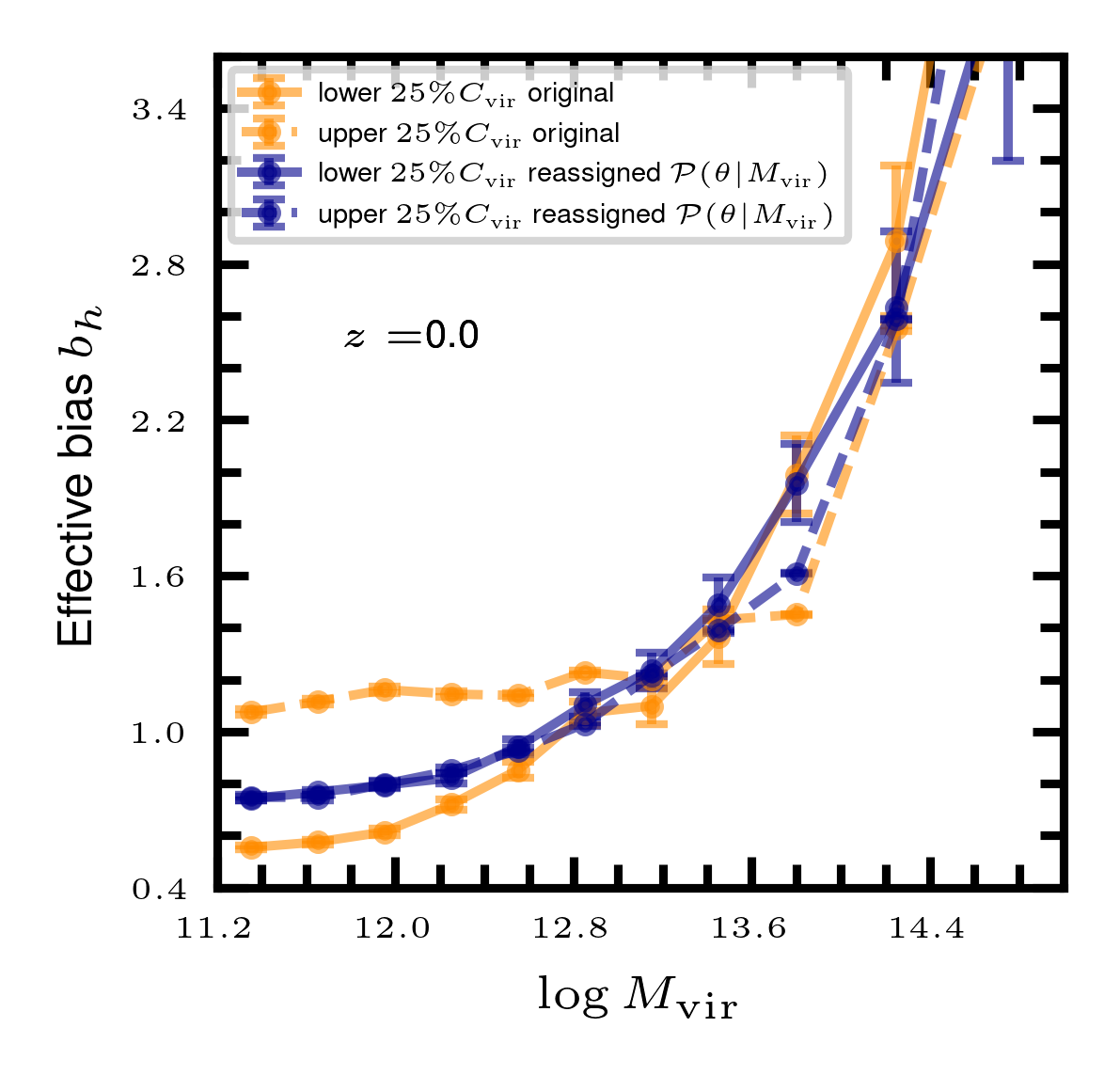}
\includegraphics[trim = .25cm 0.82cm 0cm 0cm ,clip=true, width=0.45\textwidth]{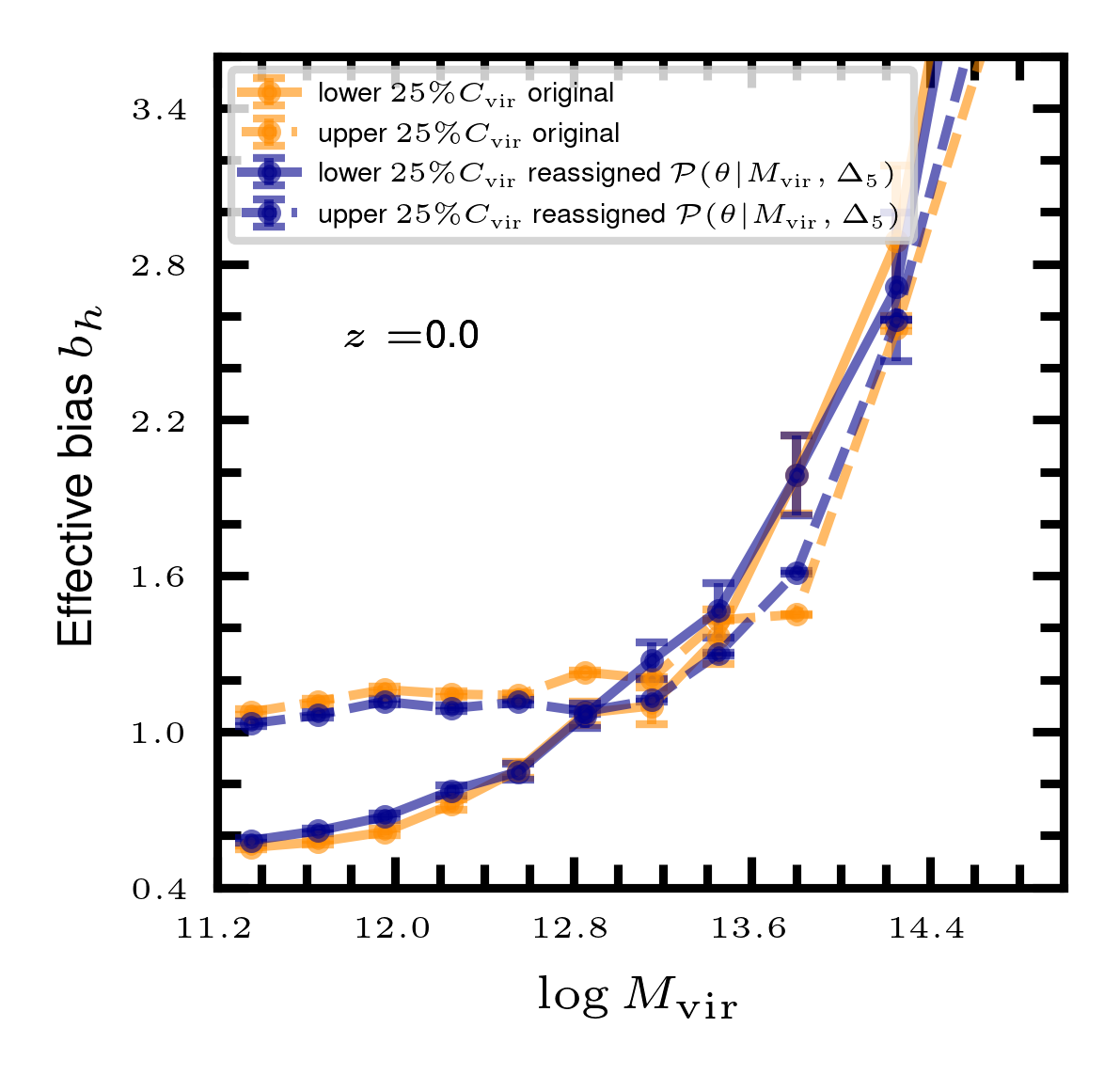}
\includegraphics[trim = .25cm 0.2cm 0cm 0cm ,clip=true, width=0.45\textwidth]{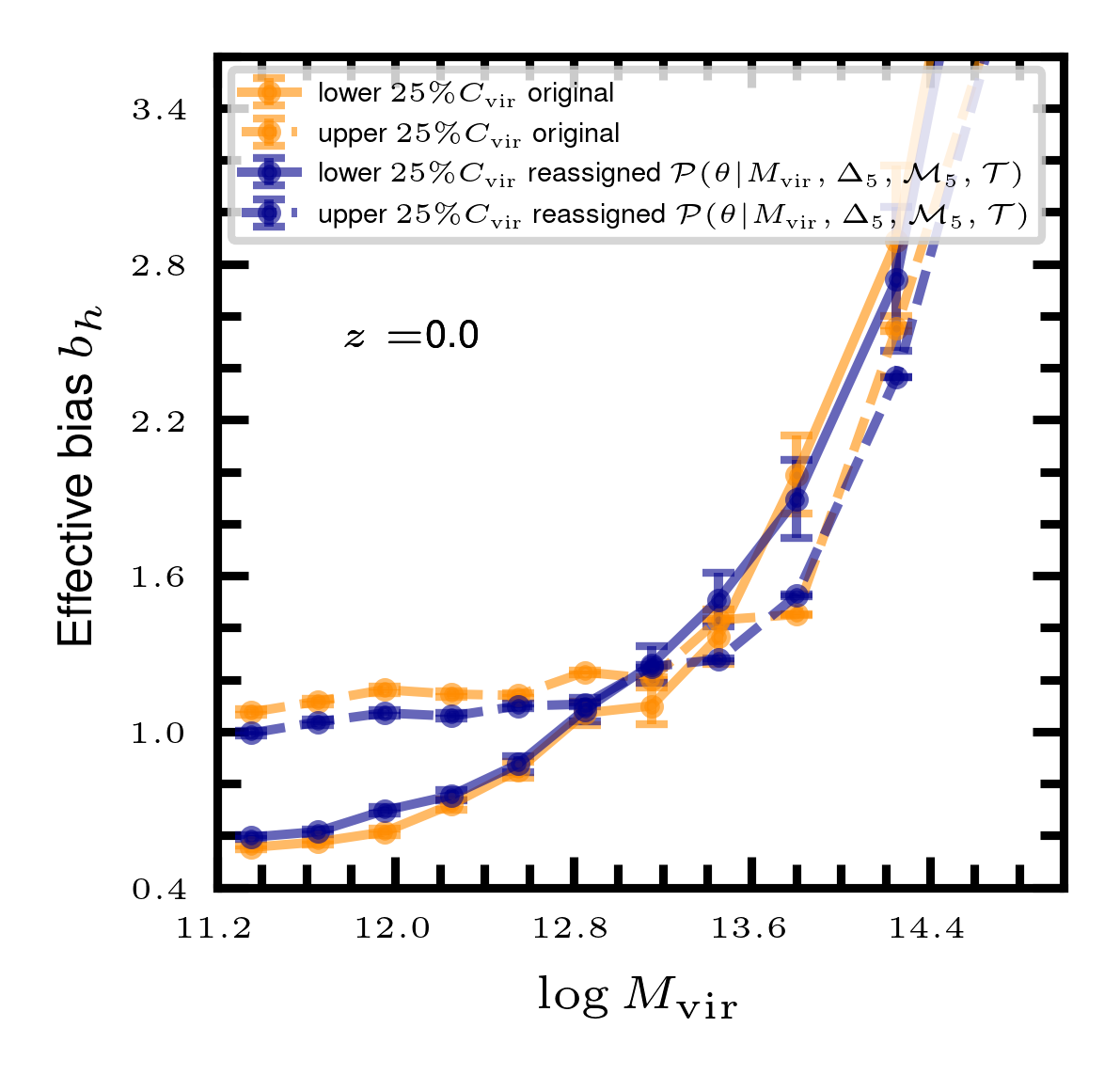}
\includegraphics[trim = .25cm 0.2cm 0cm 0cm ,clip=true, width=0.45\textwidth]{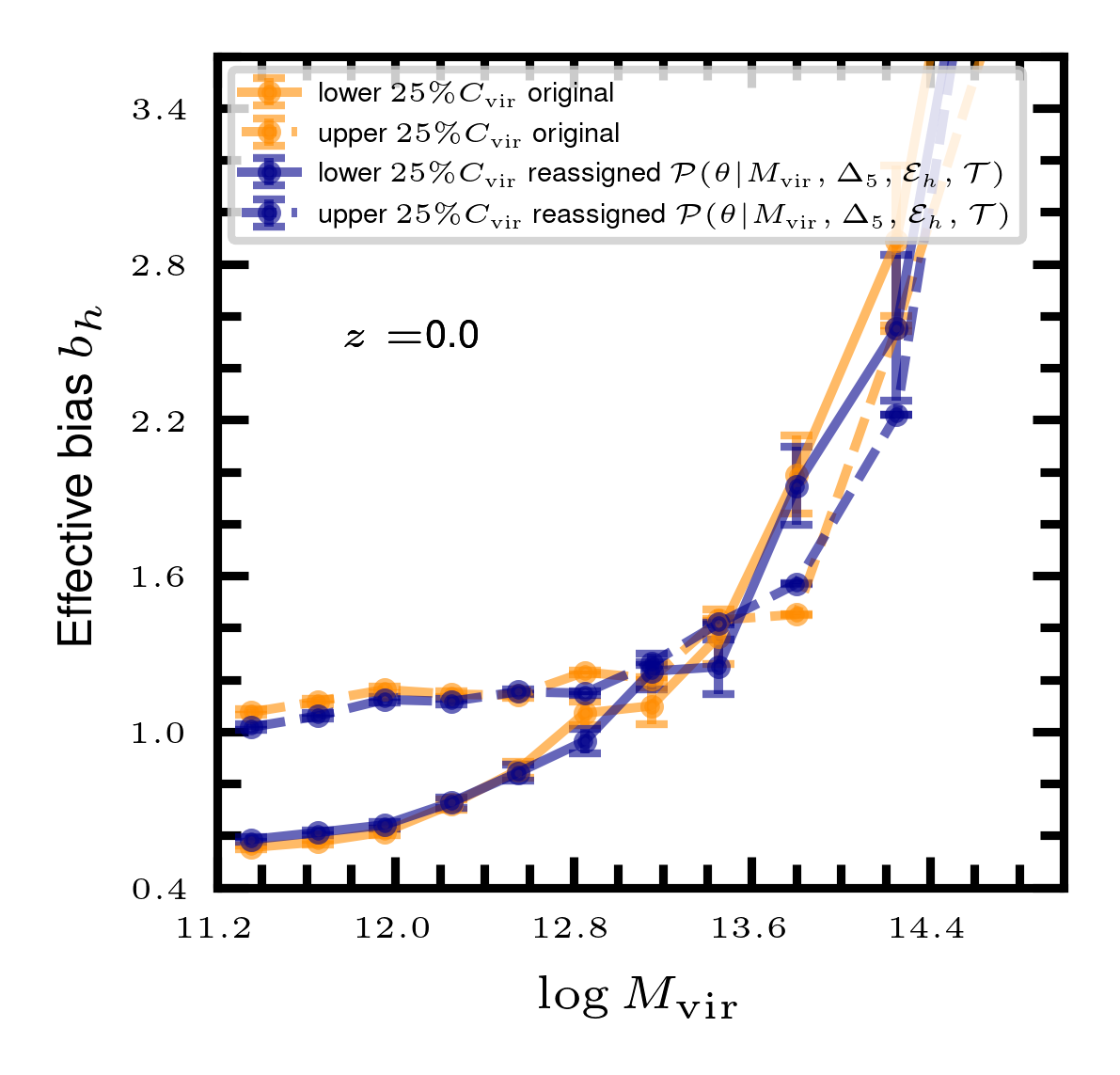}
\caption{\small{Reconstruction of the secondary bias: mean effective halo bias measured in bins of the halo mass using quartiles in halo concentration (upper panels) and spin (bottom panels). The signal is measured from the original set of halo properties and those reassigned using Eqs.(\ref{eq:newth}) with different combinations of nonlocal or environmental properties, including the information of tidal field $\mathcal{T}$ through the cosmic-web classification.}}
\label{fig:randtest_Q1}
\end{figure*}
%=====================================================================
\begin{figure*}
\centering
\includegraphics[trim = .25cm 0.82cm 0cm 0cm ,clip=true, width=0.45\textwidth]{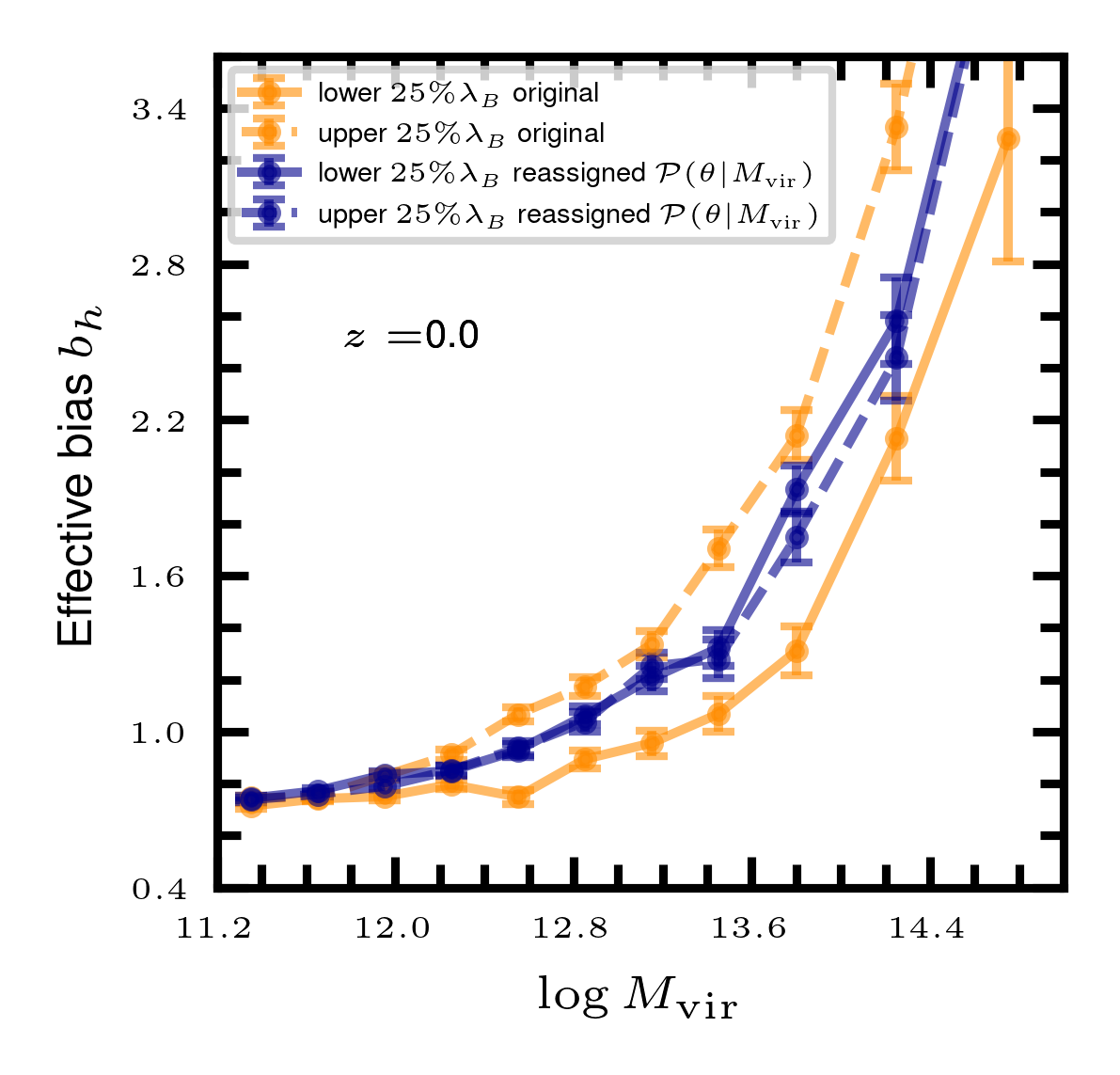}
\includegraphics[trim = .25cm 0.82cm 0cm 0cm ,clip=true, width=0.45\textwidth]{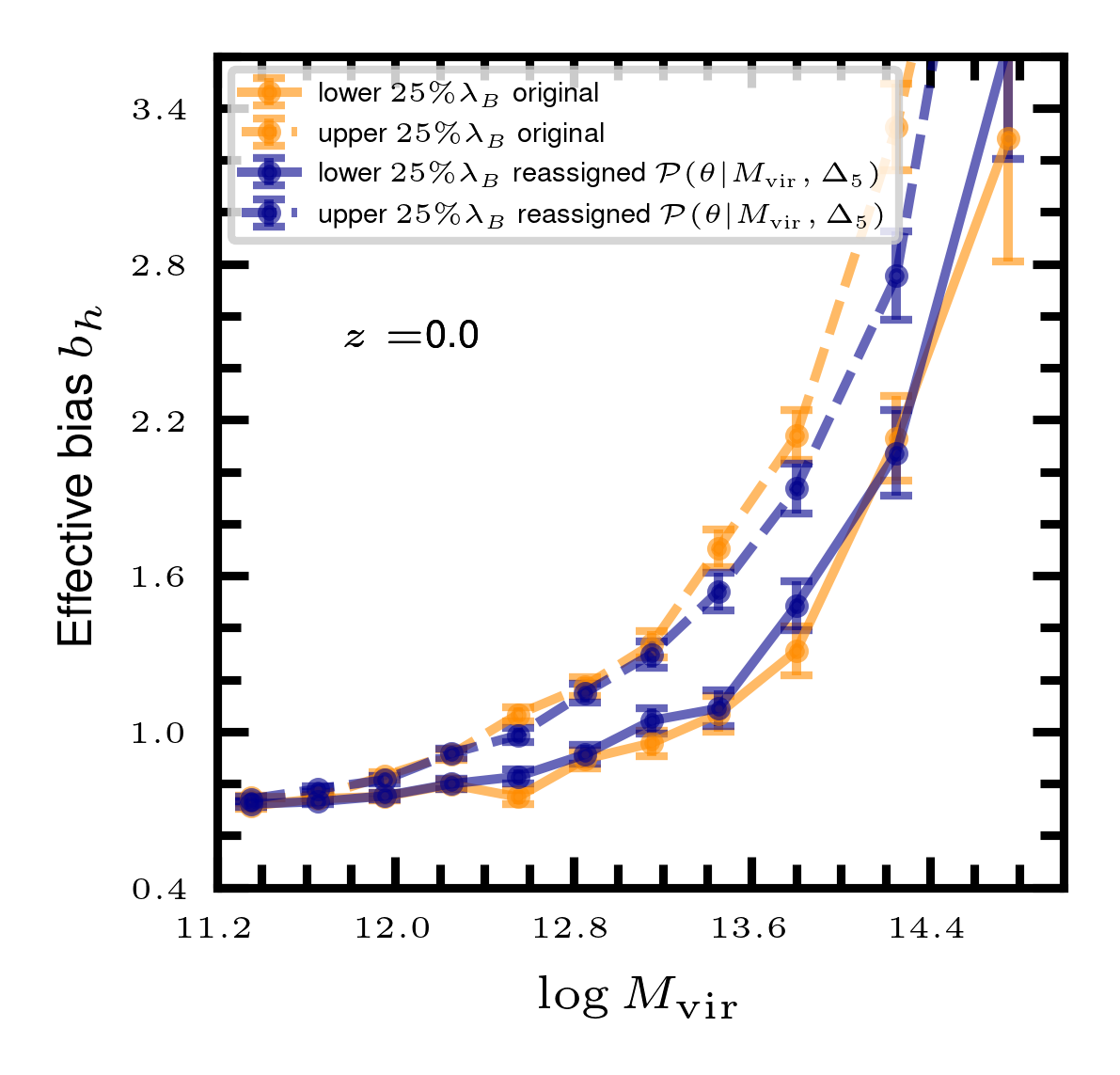}
\includegraphics[trim = .25cm 0.24cm 0cm 0cm ,clip=true, width=0.45\textwidth]{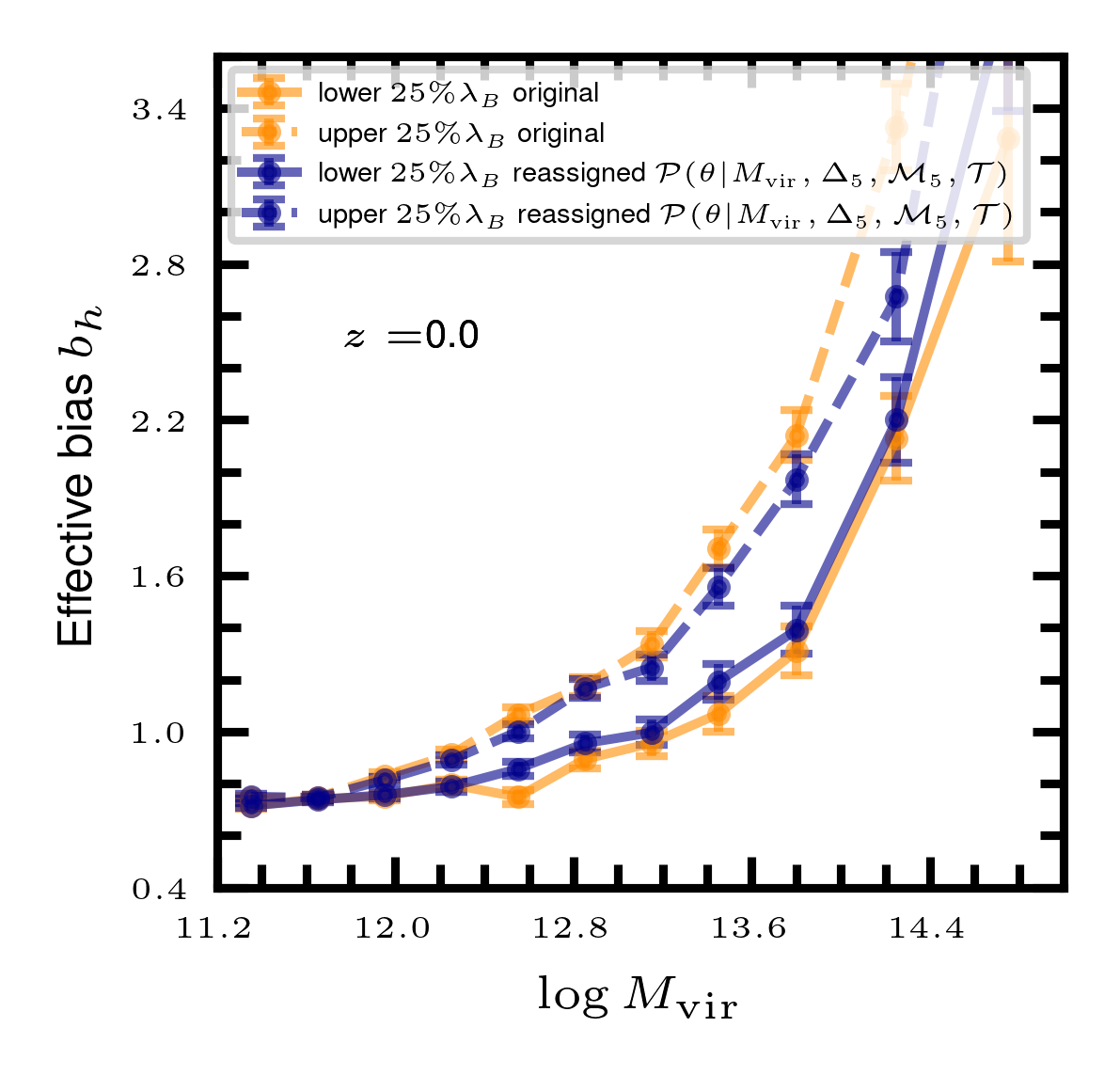}
\includegraphics[trim = .25cm 0.24cm 0cm 0cm ,clip=true, width=0.45\textwidth]{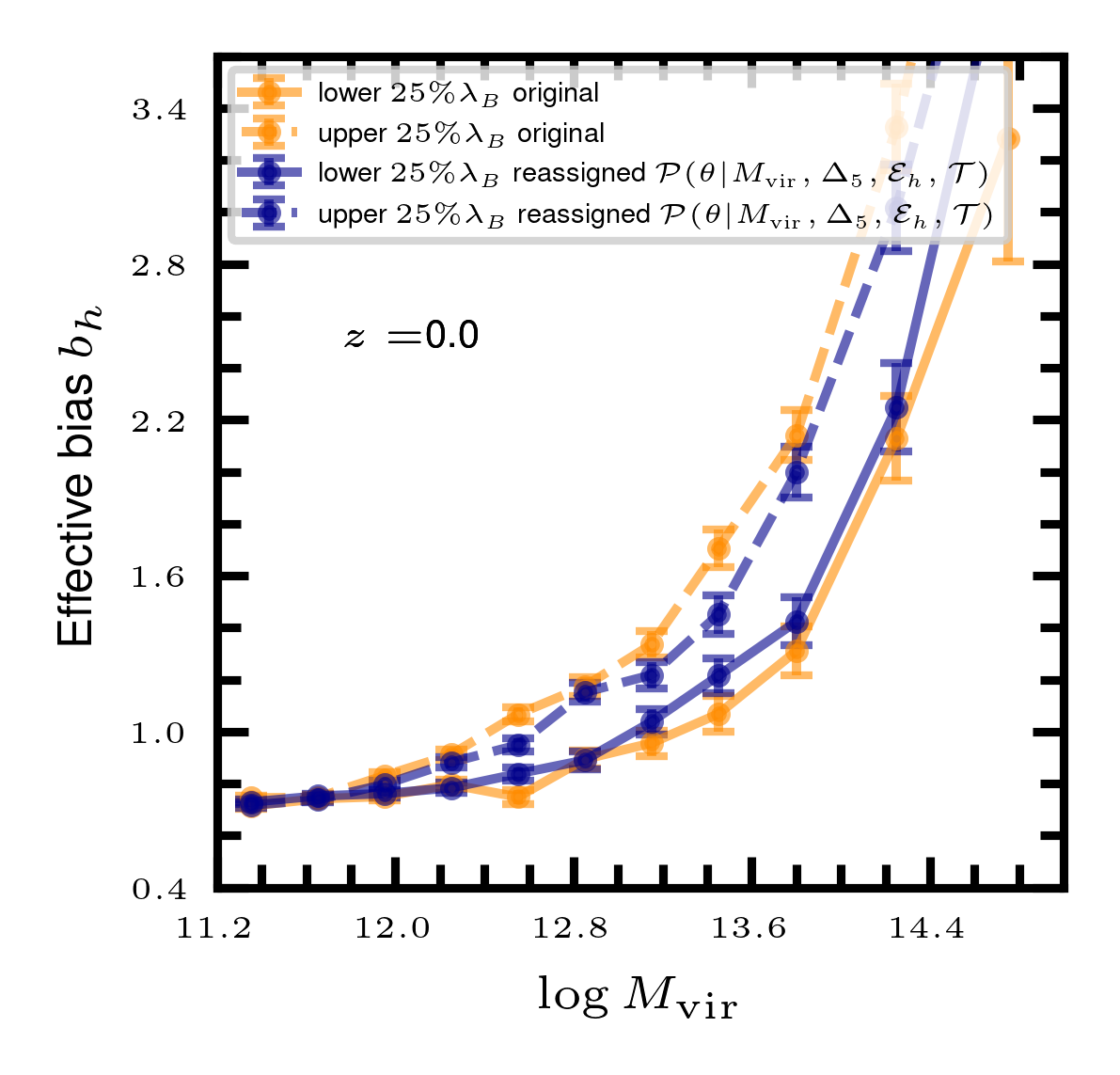}
\caption{\small{Same as Fig.~\ref{fig:randtest_Q1} but for halo spin as secondary property.}}
\label{fig:randtest_Q2}
\end{figure*}
%=====================================================================

We now evaluate the impact of partial marginalization over a reduced number of secondary properties in the scaling relations involved in Eq.(\ref{eq:bias_assump2}). To simplify the analysis, we use the results of Figs.~\ref{fig:spear1} and \ref{fig:bias_spear}, from which we have concluded that the correlation between effective halo bias and halo properties is in general smaller than the correlation between halo properties. Based on this, we fix the scaling relation $\mathcal{P}(b_{h}|M_{\rm vir})$ to the information provided by the $N$-body simulation, such that the impact of secondary properties (including environmental properties) in the $\mathcal{P}(b_{h}|\theta_{a})$ link will be addressed through the scaling relation $\mathcal{P}(M_{\rm vir}|\theta_{a})$.

We measure the scaling relation $\mathcal{P}(\theta_{a}|M_{\rm vir},\{\psi\})$, where $\{\psi\}\in \{\eta\}$ is a subset of secondary properties\footnote{We define a mesh of $N_{b}^{p}$ bins, where $p$ denotes the number of properties used to characterize the scaling relation (i.e, for a single mass dependency of a property $\theta$, $p=2$) and $N_{b}=100$, the same number of bins for each property. We count the tracers in each property bin and normalize it in bins of $\theta$.}. We use this measurement to reassign a new version of the property $\theta_{a}$ (labeled $\tilde{\theta}_{a}$) to each tracer of the simulated volume, i.e
\be\label{eq:newth}
\tilde{\theta}^{i}_{a}\curvearrowleft \mathcal{P}(\theta_{a}|M_{\rm vir}=M^{i}_{\rm vir}, \psi=\psi^{(i)}).
\ee
By construction, the new property $\tilde{\theta}_{a}$ obeys the same scaling relation with respect to the halo mass and the set $\{\psi\}$, while the correlations between $\tilde{\theta}_{a}$ and the rest of the halo properties (i.e, the set $\{\eta\}-\{\psi\}$ ) are partially inherited through the link between those properties and the halo mass.  Again, it is key to notice that \emph{even if the scatter is the same, it does not contain the full information of the scaling relations marginalized over in Eq.(\ref{eq:margi}).}

We compare the properties of  $\mathcal{P}(b_{h}|\tilde{\theta}_{a})$ (in particular, the mean) against those from the original scaling relation $\mathcal{P}(b_{h}|\theta_{a})$  \footnote{This procedure is similar to the ``galaxy shuffling'' performed in the context of galaxy assembly bias \citep[see e.g.,][]{2007MNRAS.374.1303C,2019MNRAS.484.1133C}; it aims at removing secondary correlations in order to assess their impact}. Figure \ref{fig:randtest0} shows the result of reassigning the each of the properties $\theta_{s}=\{V_{\rm max}, C_{\rm vir}, \lambda_{B}\}$ using only the virial mass in Eq.~(\ref{eq:newth}). The resulting mean effective bias displays large deviations with respect to the true one, especially for halo concentration and spin. Interestingly, $\sim 20\%$ deviations in the bias measured in bins of $V_{\rm max}$ with respect to the original behavior leads to the conclusion that $V_{\rm max}$, despite of being a probe of the potential well of dark matter halos, (and a relatively small scatter against halo mass compared with other properties) has secondary non-negligible correlations with other properties, specially at is low-end (low masses). This can be due to the high degree of anisotropy of the environment (see e.g., Fig.~\ref{fig:prop_mass}) at those mass-scales, which can be represented by tidal forces disturbing the gravitational potential and thereby the ``virial'' predictions for $V_{\rm max}$.  On the other hand, the results for the halo concentration indicate that virial mass is far from being its main primary property, while halo spin is more receptive to a single-mass dependence, but still with large variations with respect to the true halo bias.

We can extend the set $\{\psi\}$ to other properties beyond halo mass. In that regard, the correlation analysis of \S\ref{sec:corr} and the principal component analysis described in appendix \S\ref{sec:pca} indicate that environmental properties display large variability in the parameter space defined by the halo properties, and as such, these are likely to be the main driver for halo bias (primary and secondary). Inspired by this, we have generated a number of new halo catalogs with newly assigned properties using variations of the set $\psi=\{\mathcal{M}_{5},\Delta_{\rm dm},\Delta_{5},T/U, \mathcal{E}_{h},\omega_{i}\}$ (again, keeping halo effective bias and halo virial mass as their original values). The resulting mean halo bias measured as a function of $\{V_{\rm max}, C_{\rm vir},\lambda_{B}\}$ reassigned under different configurations (sets of $\{\psi\}$) are shown in Figs.\ref{fig:randtest_vmax},  \ref{fig:randtest_con} and  \ref{fig:randtest_spin}.

In general, these tests indicate that the signal of halo effective bias, measured as a function of halo properties different of halo mass (e.g., spin, concentration), can be obtained without systematic bias when nonlocal or environmental properties are included in the link between the observable and a primary property. Such extra-properties depend on the halo property used as observable. For example, the test shown in these figures reveals that the inclusion of the viral $\mathcal{V}$ and halo ellipticity $\mathcal{E}_{h}$ can be key to retrieve the effective bias as a function of $V_{\rm max}$, while for halo spin, the local halo overdensity is more relevant to obtain unbiased results.

In Figs.\ref{fig:randtest_Q1} and \ref{fig:randtest_Q2} we show the measurements of secondary bias (using halo spin and concentration as secondary properties) based on reassignments of halo properties previously mentioned, again at $z=0$. Left panels (top and bottom) in this plot show that \emph{the single mass-dependency used to sample the secondary property following Eq.~(\ref{eq:newth}) cannot account for the signal of secondary bias, even if partial information is present in the form of scatter, as suggested by Eq.~(\ref{eq:margi})}. Explicitly including more information such as local halo overdensity $\Delta_{5}$ or the cosmic-web classification can help to bring the set of reconstructed properties to the true signal of secondary bias, as shown in the different panels of the same plots. Again, the extra-secondary properties that help to improve this signal depends on the correlation between these and the secondary property under scrutiny, i.e, the set can vary e.g., between spin and concentration.

\begin{figure}
\centering
\includegraphics[trim = 0cm 0.2cm 0cm 0cm ,clip=true, width=0.49\textwidth]{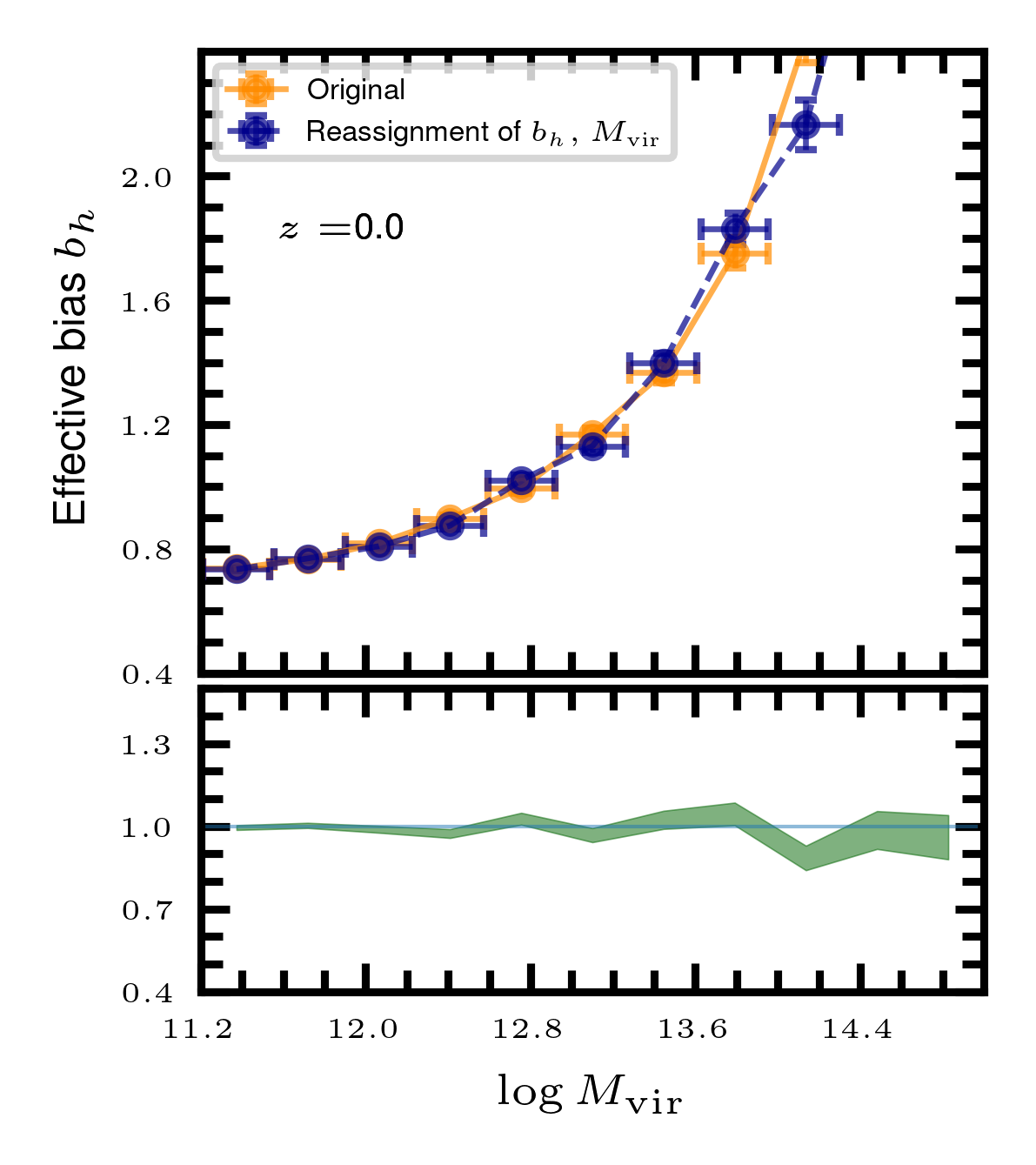}
\caption{\small{Effective halo bias measured in a synthetic halo catalog with bias and mass reassigned according to the learn-and-map approach (\S\ref{sec:forth_mocks}). The assignment of halo bias used the set of environmental properties $\{\Delta_{5}, \Delta_{\rm dm},\mathcal{M}_{5},\mathcal{D}_{5},\,\mathcal{T}_{A},\omega_{i}\}$, while halo masses are drawn from the scaling relation $\mathcal{P}(M_{\rm vir}| b_{h}, \Delta_{\rm dm},\Delta_{5},\omega_{i})$.}}
\label{fig:mocktest}
\end{figure}

\subsection{Toward applications to the construction of halo mock catalogs}\label{sec:forth_mocks}
One application of the possibility to have individual halo bias assigned to each tracers finds place in the generation of halo mock catalogs based on calibrated methods \citep[see e.g.,][]{1996ApJS..103....1B, 2015MNRAS.450.1856A, 2015MNRAS.446.2621C, 2016MNRAS.456.4156K}. In particular, the so-called ``bias assignment method'' \citep[][]{2019MNRAS.483L..58B,2020MNRAS.491.2565B,2020MNRAS.493..586P} has been designed to generate precise halo catalogs (in terms of two and three point statistics) while preserving the large-scale clustering signal as a function of halo properties. While different approaches can be followed to assign halo properties based on the features of the underlying dark matter \citep[][]{2015MNRAS.451.4266Z, 2023A&A...673A.130B}, the assignment of halo properties constraint to respect to a reference large-scale bias remains a difficult task, as such procedure in principle can be done mapping intrinsic scaling relations which do not carry clustering information. In that regard, the inclusion of small scale proxies for halo clustering such as $\Delta_{5}$ or the relative Mach number $\mathcal{M}_{5}$ can help improve the accuracy of the reconstruction of halo fields.

With the possibility to assign individual bias to tracers, we can select a number of environmental properties which, according e.g., to a correlation analysis of \S~\ref{sec:corr} (of Fig.~\ref{fig:bias_spear}), are most correlated with the object-by-object halo bias. Let us give a brief description of how it would work.
Labeling this set as $\{\eta\}$, we can measure (learn) the scaling relation $\mathcal{P}(b_{h}|\{\eta\})$ and reassign to each tracer a new bias  $\tilde{b}\curvearrowleft  \mathcal{P}(b_{h}|\{\eta\}=\{\eta\}_{new})$.
For practical applications, the set $\{\eta\}_{new}$ will represent nonlocal properties evaluated in independent realization of dark matter filed on top of which halos are to be mapped \citep[][]{2019MNRAS.483L..58B}. 
Once bias is assigned, halo properties would then be assigned in a learn-and-map procedure. For example, virial mass can be assigned as
\be \nonumber
\tilde{M}^{(i)}_{\rm vir}\curvearrowleft  \mathcal{P}(M_{\rm vir}| b_{h}=\tilde{b}_{h}^{i},\{\eta\}=\{\eta\}_{new}^{(i)}),
\ee
after which $V_{\rm max}$ can be assigned using
\be \nonumber
\tilde{V}^{(i)}_{\rm max} \curvearrowleft  \mathcal{P}(V_{\rm max}|\tilde{M}^{(i)}_{\rm vir}=M_{\rm vir}, \tilde{b}_{h}^{i}=b_{h},\{\eta\}_{new}^{(i)}=\{\eta\}),
\ee
and so forth for other halo properties, following the order established by e.g., the importance according to a principal component analysis (see appendix \S\ref{sec:pca}) of the halo properties. As an example, in Fig.~\ref{fig:mocktest} we show the halo effective bias as a function of mass after having reassigned both halos and masses following the learn-and-map prescription, with $\{\eta\}=\{\eta_{new}\}=\{\Delta_{5}, \Delta_{\rm dm},\mathcal{M}_{5},\mathcal{D}_{5},\,\mathcal{T}_{A},\omega_{i}\}$.
Contrary to previous approaches \citep[see e.g.,][]{2015MNRAS.451.4266Z, 2023A&A...673A.130B}, this methodology can guarantee that the large-scale clustering of a mock catalog, as a function of main halo properties, is in good agreement with respect to the expected results (taken from a reference $N$-body simulation). 

Halo catalogs with precise properties 
provide high flexibility at the time of producing galaxy mock catalogs using e.g., a halo occupation distribution (HOD) framework \citep[e.g.,][]{2002ApJ...576L.105C,2002PhR...372....1C,2002ApJ...575..587B,2004ApJ...609...35K} the sub-halo abundance matching (SHAM) \cite[see e.g.,][]{2004MNRAS.353..189V,2004ApJ...609...35K,2006ApJ...647..201C,2016MNRAS.461.3421F}
This is optimal for multi-tracer analyses \cite[see e.g.,][]{2012PhRvD..86j3513H,2013MNRAS.432..318A,2016MNRAS.455.3871A,2020RAA....20..158W,2021MNRAS.503.1149Z} as expected to be performed in many experiments \cite[][]{DESI}, EUCLID \cite[][]{Euclid},  J-PAS \citep{2014arXiv1403.5237B}, and the Nancy Grace Roman Space telescope \cite[][]{2015arXiv150303757S}. The approach presented in this paper opens the possibility to assign halo properties along with halo bias, printing the signal of secondary bias in halo mock catalogs. This will be explored in a forthcoming paper (Balaguera-Antolínez et al., in preparation).

%========================================================================================
%========================================================================================
\section{Discussion and conclusions}\label{sec:conclusions}

For this work, we measured the signal of the secondary halo bias for the first time in the \texttt{UNIT} simulation (UNITSim), a fixed-amplitude and paired set of $N$-body simulations following the evolution of dark matter and dark-matter halos
across cosmic time, since $z\sim 6$ to $z=0$, in a comoving box of a $1$ Gpc$h^{-1}$ side. The primary and secondary dependencies of the bias have been obtained using a new approach for the halo bias, in which every tracer can be assigned a value of effective large-scale bias, based on the estimator introduced by \citet[][]{2018MNRAS.476.3631P}.  We have verified the robustness of this estimator against the results obtained using  estimators with a large-scale bias based on the measurements of two point-statistics, in particular the auto- and cross-power spectra. The object-by-object assignment of the halo bias opens the possibility to analyze the statistical behavior of the halo effective bias as a function of a number of halo properties classified as {intrinsic}, {nonlocal}, and {environmental}.

Intrinsic properties (mass, velocity dispersion, and halo spin) are derived from the halo finder producing the catalogs of the UNITSim. These can be divided into a set of {primary} and {secondary} halo properties. This classification is based, on the one hand, on physical grounds (e.g., primary properties are those directly probing the depth of the halo potential well, that is, the mass, velocity dispersion, and maximum circular velocity) and, on the other hand, on statistical grounds (a principal component analysis, shown in Appendix \ref{sec:pca}, shows that these contain a higher level of variability, and therefore information, across time). 

{Nonlocal} halo properties refer to small-scale statistical diagnostics aimed at condensing information on halo clustering on small scales. Among these, we have introduced the {relative Mach number}, which is a measure of the kinetic {temperature} ``observed" by each tracer in spheres of $5$ Mpc$h^{-1}$. Similarly, we have used the {neighbor statistics}, which condenses information of the distribution of pair separations with respect to each tracer in the same spheres.

Environmental properties are defined in this paper as those related to the underlying dark-matter density field. This can be local (e.g., local density) and nonlocal (i.e., tidal field). The computation of the tidal field of the dark-matter distribution allows for the exploration of the halo bias as a function of key properties such as its invariants or functions generated with its eigenvalues. In particular, in this paper we have explored the behavior of the halo properties and effective bias as a function of cosmic-web environments and tidal anisotropy. 

In the first part of this paper, we have reported the measurements of halo scaling relations in different cosmological environments. The conclusions can be summarized as follows:
\begin{itemize}
    \item The halo scaling relations do not change in the same way through cosmic time. While links between the so-called {primary} properties (halo mass and maximum circular velocity) vary approximately along with expectations from, for example, the virial theorem, the scaling relations between the mass and {secondary properties} are more complex (e.g., these are not simple power laws) and, in general, they display a larger scatter around the mean relations. One particular example is the mass-concentration relation, whose slope varies with mass and redshift, ranging from positive values at a high redshift (i..e, more massive halos are more concentrated, with a Spearman's rank correlation of  $\rho_{s}\sim 0.1$, see Fig.~\ref{fig:spear1}), passing through a nearly mass-independent relation at $z\sim 2$ (with $\rho_{s}\sim 0.04$) and ending with a negative correlation (i.e., more massive halos are less concentrated) toward a lower redshift (with $\rho_{s}\sim -0.3$). This in agreement with a number of findings in the literature \citep[see e.g.,][]{2007MNRAS.378...55M, 2008ApJ...678..621K}.
    
    \item The variation in the halo spin with the halo mass at different redshifts is weak for low-mass halos ($\log M_{\rm vir} \sim 11.5$). In particular, at $z=0$ (with $\rho_{s}\sim -0.01$), only the very massive halos ($\log M_{\rm vir}>14$) display a relevant (negative) correlation with the halo mass. The dependency is slightly stronger at a higher redshift (e.g., at $z\sim 5$, $\rho_{s}\sim -0.02$).

    \item The halo-scaling relations involving secondary properties vary across the cosmic-web environment. Although the definition adopted based on the Hessian of the gravitational potential is arbitrary (as it depends on the threshold $\lambda_{\rm th}$ fixed to zero in this paper), the clear distinction of the scaling relations in over- and under-densities indicates how the evolution and growth of halos is heavily influenced by its environment (see Figs. \ref{fig:spear_cwt} and Fig.~\ref{fig:prop_mass_cwt}). 
\end{itemize}

The second (and main) part of this paper has been devoted to the assessment of the statistical properties of the large-scale halo bias as a function of the properties described in the first section of this work. The conclusions can be summarized as follows:
\begin{itemize}
    \item The object-by-object halo assignment proposed by \citet[][]{2018MNRAS.476.3631P} yields estimates for the large-scale halo bias in agreement with estimates based on the halo power spectrum (see Fig.~\ref{fig:bias_nu}). This approach has the great advantage of no explicit need to compute two-point statistics for subsamples divided into bins of primary and/or secondary properties.
    \item The mean effective bias expressed as a function of the peak height retains a degree of universality not only in this mean, but also in higher moments (see Fig.~\ref{fig:bias_nu_allz}). The probability distribution of the halo effective bias as a function of the peak height can be expressed as a normal distribution with a mean value, in good agreement with predictions from the conjunction of $N$-body simulations and the ellipsoidal collapse \citep[][]{2010ApJ...724..878T}, and a standard deviation that scales only with redshift.  The universality of the mean-effective bias is no longer conserved when the halo sample is split into terms of cosmic-web environments. 
    \item In general, the effective halo bias expressed as a function of halo properties is quite sensitive to the cosmic-web environment (see Fig.~\ref{fig:bias_prop_cwt}).
    \item The signal of  the secondary bias is present when exploring the bias as a function of the halo mass in quartiles of secondary properties (see Figs.\ref{fig:sec_bias_ex1}, \ref{fig:sec_bias_rat} and \ref{fig:sec_bias_allz1}), as widely reported in the literature. 
    \item Most of the secondary halo properties induce secondary bias signals that preserve a hierarchy across cosmic time, that is, the difference between the bias in the higher quartile and that in the lower quartile does not change sign with redshift, though the trend might differ from property to property. For instance, high spin halos have a larger bias than low-spin halos for all masses and redshifts explored (\S\ref{sec:secbias}).

    \item Among the secondary properties, the signal of the secondary bias due to the halo concentration displays a well-known behavior, inverting the hierarchy at a different mass scale in a different snapshot, a fact seen in previous works \citep[see e.g.,][]{Wechsler2006,SatoPolito2019}. Such a behavior can be seen as a direct consequence of the evolution of the mass-concentration scaling relation which in turn is inherited from that of the scaling radius $R_{s}$ (see Fig.~\ref{fig:prop_mass}).
    \item Given the mass resolution used in this work, we cannot detect the inversion of the clustering trend in the secondary bias induced by the halo spin, as previously reported \cite[e.g.,][]{SatoPolito2019, Tucci2021}. However, this type of crossover is observed when the analysis is conditioned to halos located in high density regions (knots).

    \item Nonlocal and environmental properties induce the most significant signals of secondary bias. In particular, the tidal anisotropy, the relative Mach number, and the local dark matter density 
    display the highest statistical significance (see Fig.~\ref{fig:sec_bias_allz1}).

    \item The redshift evolution of the secondary bias is approximately (but not completely) captured by the implicit redshift dependence of $\nu$ for internal halo properties (particularly concentration and ellipticity). That is not the case for the external halo properties, which still maintain significant evolution even when $\nu$ is chosen as a primary property.

\end{itemize}

We have demonstrated that the signal of the secondary bias based on halo properties is highly dependent on the environment where these halos reside. As we have used the cosmic-web classification based on the tidal field, such environments are associated with high (knots), intermediate (filaments, sheets), or low (voids) density regions. Extending the cosmic-web classification to other dynamical properties of the dark-matter field (such as the sheer of the velocity field or the local curvature) can provide more hints as to the physical origin of the secondary bias.
Our results conceptually agree with those derived from the analysis of \citet[][]{2021A&A...654A..67W} and \citep[][]{2023arXiv230915306W}, which implemented different estimators for the large-sale halo (or galaxy) bias, definitions of halo and galaxy properties, and a slightly different set-up for cosmological simulations, including the halo finder.

Evidently, more effort is required to fully understand the physical origin of the secondary bias. In that regard, we shall finish by providing some information on forthcoming research which extends some of the results presented here. These include the following:
\begin{itemize}
    \item The object-by-object estimator of Eq.(\ref{eq:bias_object}) can be generalized to provide estimates of individual redshift-space effective halo bias. Such a generalization can be used to test different subjects linked to redshift distortions, such as the {velocity bias}, measurement of the growth of structures, and marked statistics in redshift space.
    \item Extensions to redshift surveys can also be complemented by assigning relative (to the full galaxy population) effective bias or absolute effective bias to galaxies \citep[as done e.g., by][]{2021MNRAS.504.5205C}, provided that the underlying dark matter density field has been reconstructed using a number of techniques \citep[see e.g.,][]{10.1093/mnras/sty1203,2019A&A...625A..64J,2021MNRAS.502.3456K}. Assigning a relative bias can be important in the context of galaxy clusters, as the relative bias of clusters contains cosmological information \citep[see e.g.,][]{1999AN....320..189B, 2011MNRAS.413..386B,2012MNRAS.422...44P, 2014A&A...563A.141B,2023hxga.book..123C}; we can assign the relative effective bias using a single measurement for the cluster power spectrum from the the full cluster population.
    \item Some of the properties that we investigate in this work (such as the Mach number) have not  been explored with hydro-dynamical simulations (e.g., Illustris TNG\footnote{\url{http://www.tng-project.org}}) in the context of {{galaxy assembly bias}} (i.e., the secondary dependencies of galaxy clustering and occupancy at a fixed halo mass). Since the individual object-to-object bias has been successfully implemented in TNG \citep{2018MNRAS.476.3631P}, our work opens new routes to explore these effects. Also, the connection between galaxy assembly bias and the cosmic web has only been recently addressed with hydro-dynamical simulations (see e.g., \citealt{MonteroDorta2023}), so there are a number of aspects regarding the physical origins of the effect that have yet to be understood.
    \item Finally, the full applicability of the procedures described in \S\ref{sec:forth_mocks} is focused on the reconstruction of the halo bias and its dependencies with halo properties to generate halo mock catalogs. This application aims to extend the procedures shown in \citet[][]{2023A&A...673A.130B} which attempts to assign halo properties to mock tracers obtained from learning-based techniques, which is dependent on a precise signal for large-scale clustering.
\end{itemize}

%------------------------------------------------------------------------
\begin{acknowledgements}
  We would like to thank F.-S. Kitaura for discussions in the early stage of this project. We are grateful with Chia-Hsun Chuang, F.-S. Kitaura and Gustavo Yépez for granting access to the UNITSim. We thank the referee for his/her dedicated reading of the manuscript, comments and questions, which helped to improve the presentation of our results. ABA acknowledges the Spanish Ministry of Economy and Competitiveness (MINECO) under the Severo Ochoa program SEV-2015-0548 grants. The UNITSim has been run under at the MareNostrum Supercomputer hosted by the Barcelona Supercomputing Center, Spain, with computing time granted by PRICE under grant number 2016163937.
ADMD thanks the IAC facilities and Fondecyt for financial support through the Fondecyt Regular 2021 grant 1210612. GF is supported by a {\em Juan de la Cierva Incorporación} grant n.\,IJC2020-044343-I. No AI application has been used to generate neither text, data, code nor figures for this article. 

\end{acknowledgements}

\bibliographystyle{aa}
\bibliography{refs}  

%------------------------------------------------------------------------
%------------------------------------------------------------------------
%------------------------------------------------------------------------
%------------------------------------------------------------------------
%------------------------------------------------------------------------
%------------------------------------------------------------------------
%------------------------------------------------------------------------

%\pagebreak
%\newpage

\begin{appendix}
  
\section{Principal component analysis of halo properties}\label{sec:pca}

Halo mass is not the main halo property, as it can swap or share such stage with other quantities such as $V_{\rm max}$ \cite[see e.g.,][]{2011MNRAS.416.2388S, 2012ApJ...757..102W}, velocity dispersion or the peak velocity (as these are also probes of the depth of the halos gravitational potential) \citep[see e.g.,][]{2011MNRAS.415L..69J,2017MNRAS.466.3834L}.

\begin{figure}
\includegraphics[trim = .0cm 0cm 0cm 0cm ,clip=true, width=0.45\textwidth]{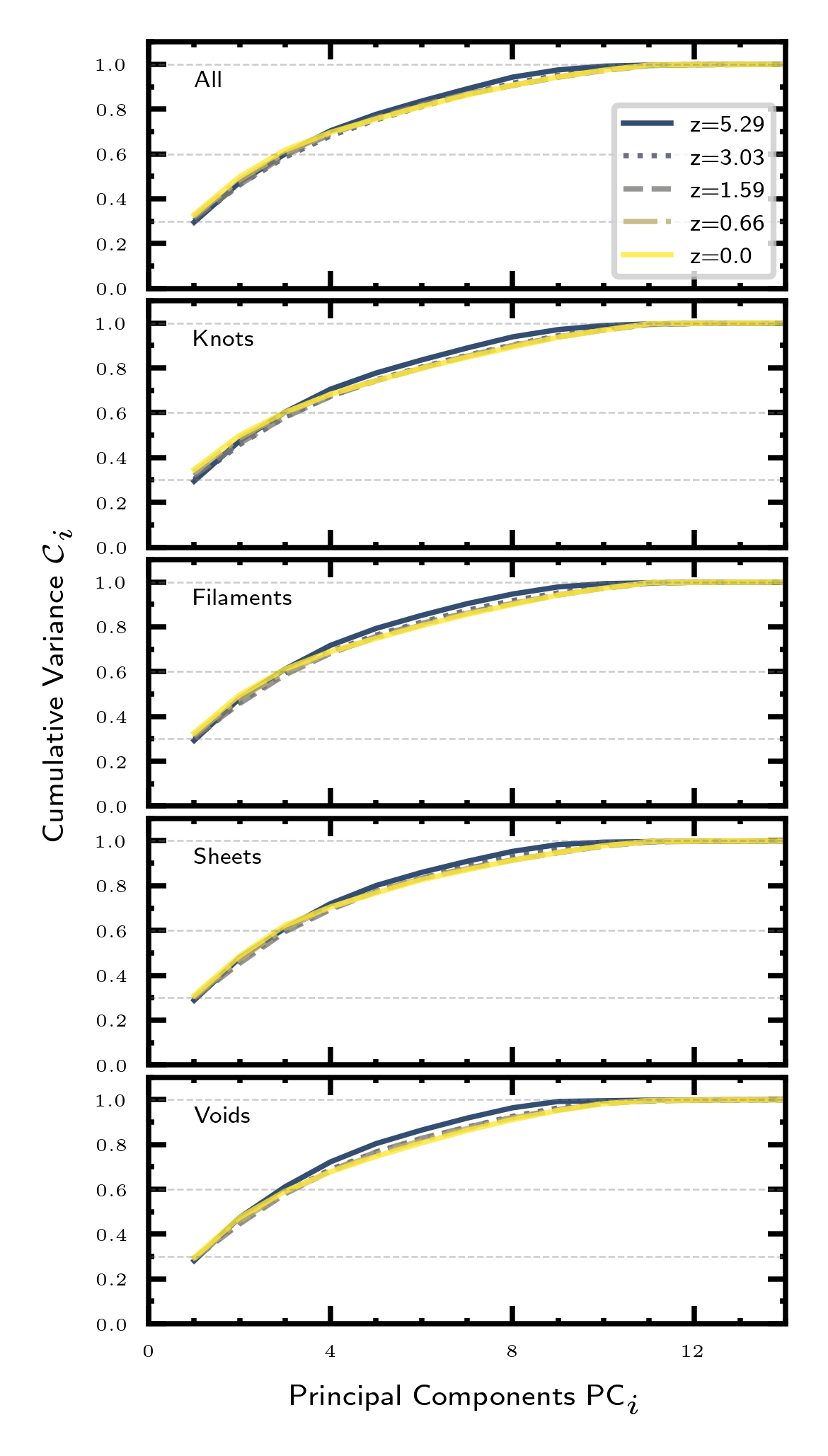}
\caption{\small{Cumulative variance $\mathcal{C}_{i}$ (Eq.(\ref{eq_cum_var}) for each principal component $PC_{i}$ at different redshifts and cosmic-web environments. The horizontal lines mark the $30$, $60$ and $100\%$ of the full variance. Almost $90\%$ of the information is retained by the $\sim 50\%$ of the highest principal components. The first component encodes $\sim 30\%$ of the full information.}}
\label{fig:cumvar}
\end{figure}

The need for a ``primary'' property of dark matter halos is key in the context of galaxy clustering, where typical approaches to the population of halos such as the 
halo occupation distribution (HOD) framework \citep[e.g.,][]{2002ApJ...576L.105C,2002PhR...372....1C,2002ApJ...575..587B,2004ApJ...609...35K}, or sub-halo abundance matching  \cite[see e.g.,][]{2004MNRAS.353..189V,2004ApJ...609...35K,2006ApJ...647..201C,2014MNRAS.443.3044Z,2016MNRAS.461.3421F, 2016MNRAS.460.2552H,2018ARA&A..56..435W}, which are historically based on halo masses, with extensions to other properties \cite[see e.g.,][]{2019ApJ...887...17Z,2020MNRAS.493.5506H}. This question is also present in the context of the generation of mock catalogs using calibrated methods \citep[][]{2023A&A...673A.130B}. 

\begin{figure}
\includegraphics[trim = .0cm 0.2cm 0cm 0cm ,clip=true, width=0.48\textwidth]{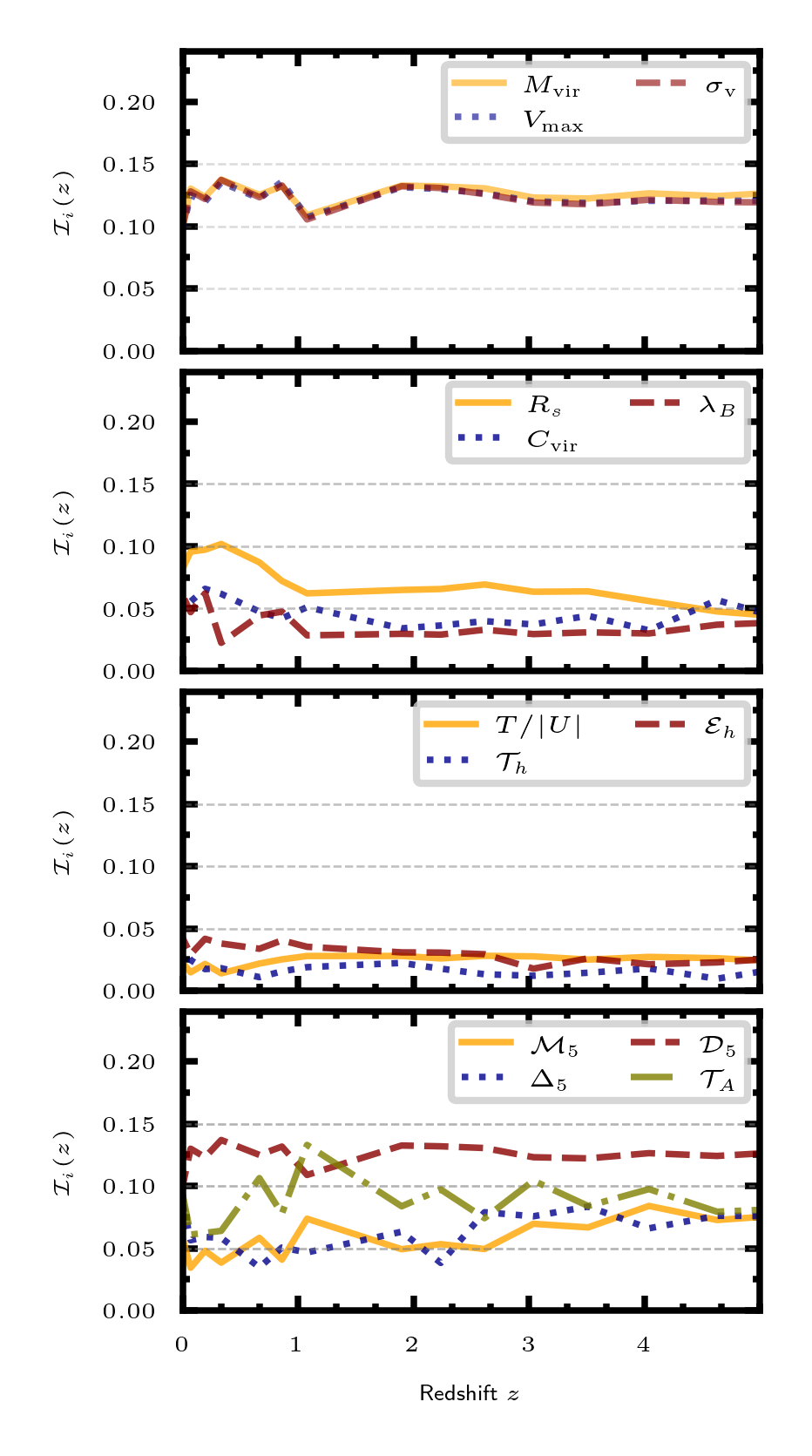}
\caption{Importance $\mathcal{I}_{i}(z)$ (Eq.(\ref{eq:importance}) for each halo property as a function of redshift. We have divided the set of halo properties in four panels in order to avoid clutter. The horizontal lines show the $5$ $10$ and $15\%$ importance levels. The level of importance is normalized to be unity when all the information from the shown properties is added.}
\label{fig:importance}
\end{figure}

To determine the main halo property in the UNITSim, we perform a principal component analysis (PCA) \cite[see e.g.,][]{article_pca, 2011MNRAS.416.2388S, 2012ApJ...757..102W,2023MNRAS.523.5789Z}, transforming each halo property $\eta_{\alpha}$ to its standardized form, $\eta_{\alpha}\equiv (\theta_{\alpha}-\langle\theta \rangle)/\sigma_{\theta}$, where $\langle \theta\rangle$ is the sample mean of the property and $\sigma_{\theta}=\langle (\theta_{\alpha}-\langle \theta\rangle)^{2}\rangle^{1/2}$ its corresponding sample variance.  We then compute the eigenvalues $\tilde{\lambda_{i}}$ (sorted in descending order) and their associated eigenvectors $\hat{\lambda}^{q}$ with elements $\hat{\lambda}^{(q)}_{i}$ ($i,q=1,2\cdots,N_{p}$) of the covariance matrix  whose elements are $\mathcal{C}_{ij}\equiv \langle \eta_{i}\eta_{j}\rangle$. The set of eigenvectors defines a new orthogonal basis in which $\hat{\lambda}^{(1)}$ represents the axis on which the dataset displays the largest variation. A new set of halo parameters (the principal components) can be then generated as as $\tilde{\eta}^{\alpha}_{q}=\sum _{i}(\hat{\lambda}^{(q)}_{i})\eta^{\alpha}_{i}$ ($p,q=1., N_{PCA}$) where $(\hat{\lambda}_{i})_{q}$ are the components of the eigenvector associated with the $q-$th eigenvalue. The cumulative variance retained in each principal component is
\be\label{eq_cum_var}
\mathcal{C}_{i}(z)=\frac{1}{\mathcal{V}(z)}\sum_{j<i}\tilde{\lambda}_{j}(z).  
\ee
and shown in Fig.~\ref{fig:cumvar} for each principal component, evaluated at number of redshifts and cosmic-web environments. Trends as a function of the cosmological redshift are consistent within the different cosmic-web environment, indicating that more principal components are requested as long as $z\to 0$ in order to retain the same amount of information on the halo properties. For example, the change in the number of PC's requested to have $90\%$ of full variability goes from $\sim 6$ at $z\sim 5$ to $\sim 8$ at $z=0$. Averaging over the values shown, the first PC contains $\sim 30\%$ of the total amount of information, the second holding $\sim 15\%$. We have verified that this conclusion are not modified when the cumulative variance is computed in bins of halo mass.

We define the ``importance'' of each halo property $\eta_{i}$ (as a function of redshift) as:
\be\label{eq:importance}
\mathcal{I}_{i}(z)\equiv \frac{1}{\mathcal{V}(z)}\sum_{p=1}^{N_{PCA}} \tilde{\lambda}_{p}(z) |\rho_{ip}(z)|,
\ee
normalized such that $\sum_{i}\mathcal{I}_{i=1}^{N_{p}}=1$ for all redshifts and cosmic web types. In Fig.~\ref{fig:importance} we show the ``importance'' of all halo properties as a function of cosmological redshift (for this calculation we have used all principal components). For each redshift, Eq.~(\ref{eq:importance}) is normalized adding the values from the properties shown in the plot. The main results can be summarized as follows:
\begin{itemize}
    \item The set of intrinsic properties $\{M_{\rm vir}, V_{\rm max},\sigma_{v},\nu\}$ display the highest level of importance within the set of halo properties,  very similar evolution across redshift ($10-15\%$ on the explored redshift range). We have verified that such behavior mildly changes in different cosmic-web environments. This set of properties amasses $\sim 40\%$ of the full information encoded in the set of halo properties explored.
       
    \item The scale radius $R_{s}$ displays an increasing trend of importance from high to low redshift, with $\mathcal{I}\sim 10\%$ at $z\sim 0$: The halo concentration with $\sim 5\%$ of average importance on average.
    \item The environmental properties $\{\mathcal{M}_{5},\mathcal{D}_{5},\Delta_{5}, \mathcal{T}_{A} \}$ display an average $\mathcal{I}\sim 8\%$. We have verified that its higher value is reached in voids, (specially $\mathcal{D}_{5}$). These properties amass $\sim 25\%$ of the total level of importance. 
    \item The set conformed by properties such as spin, the virial and halo geometry, $\{\lambda_{B}, T/|U|,\mathcal{T}_{h}\}$ show a $3\%$ importance over the redshift range explored.
\end{itemize}
In conclusion, most of the information is retained by probes of the potential well. In general, this claim is valid for all cosmic-web environments, reinforcing the idea to define the set $\{M_{\rm vir}, \nu,V_{\rm max}, \sigma_{v}\}$ as ``primary properties''.

%=====================================================================

\section{Secondary halo bias as a function of peak-height}\label{sec:sec_cwt}
In Fig.~\ref{fig:sec_bias_rat2} we present the measurements of relative bias (secondary bias in the fourth quartile divided by the same signal from the first quartile) using the for the halo ellipticity, relative Mach number, local dark matter density and tidal anisotropy as secondary properties.

\begin{figure}
\centering
%assembly_bias_delta.py
\includegraphics[trim = 0.2cm 0.8cm 0cm 0cm ,clip=true, width=.48\textwidth]{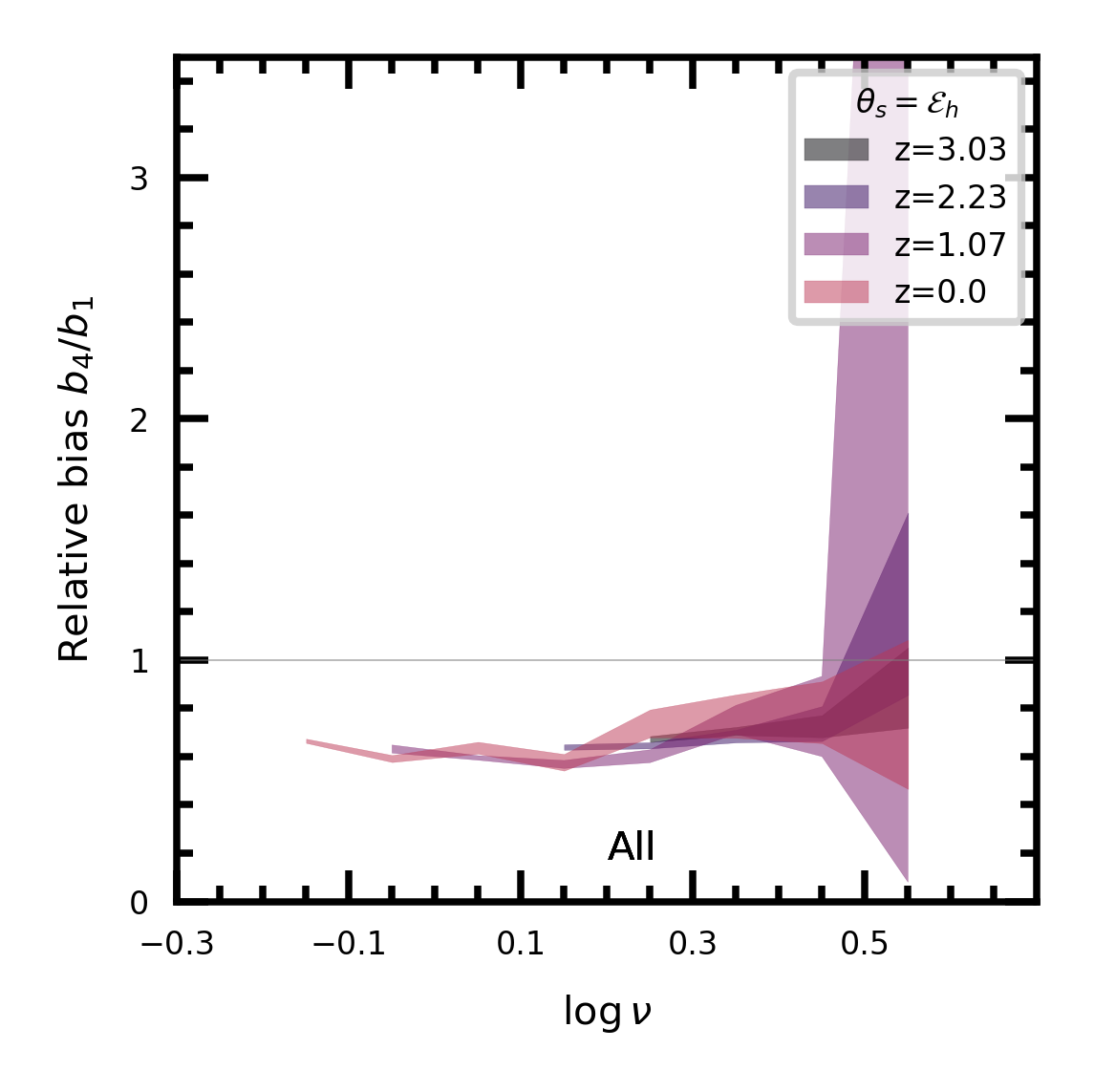}
\includegraphics[trim = 0.2cm 0.8cm 0cm 0cm ,clip=true, width=.48\textwidth]{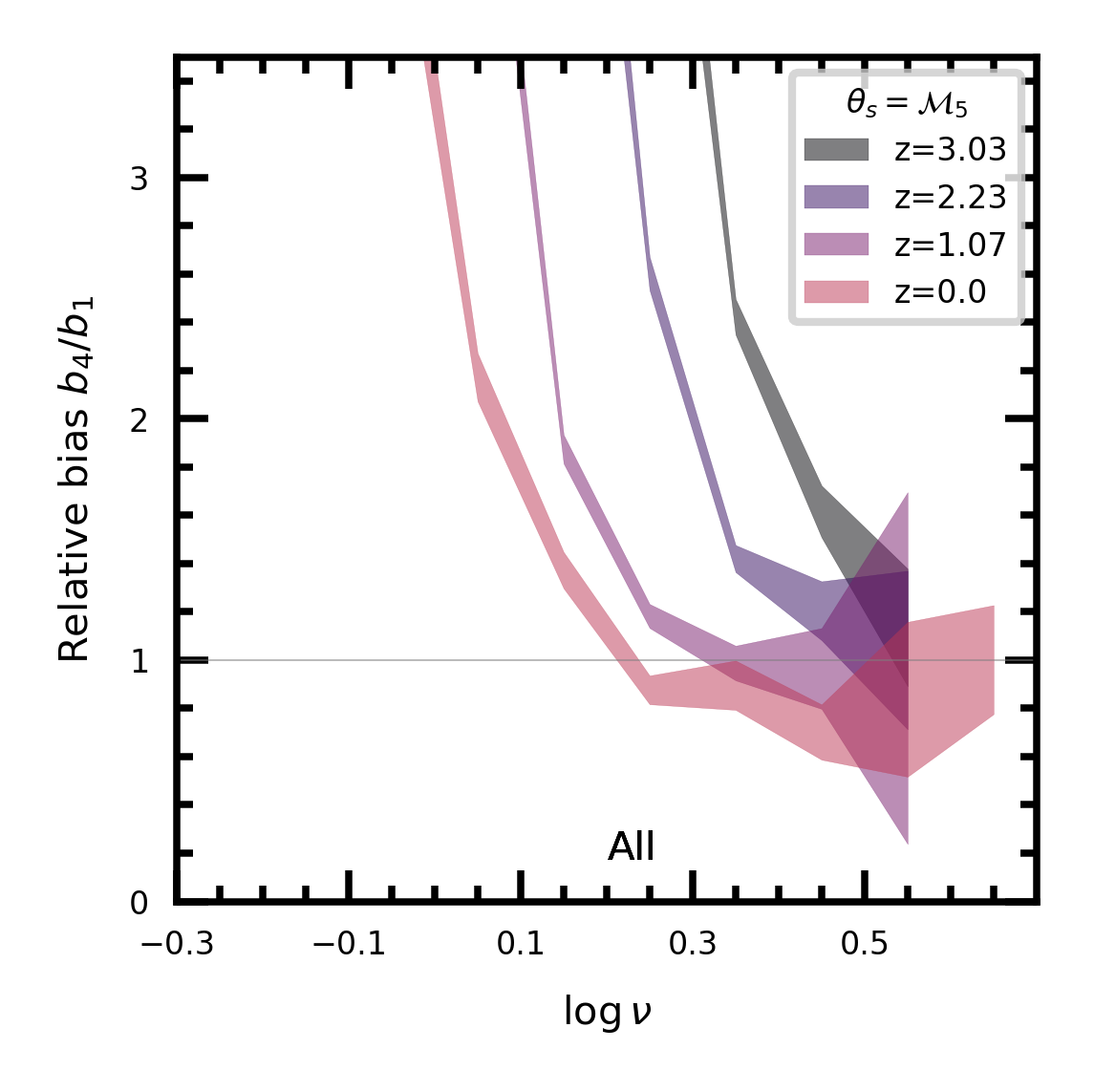}
\includegraphics[trim = 0.2cm 0.2cm 0cm 0cm ,clip=true, width=.48\textwidth]{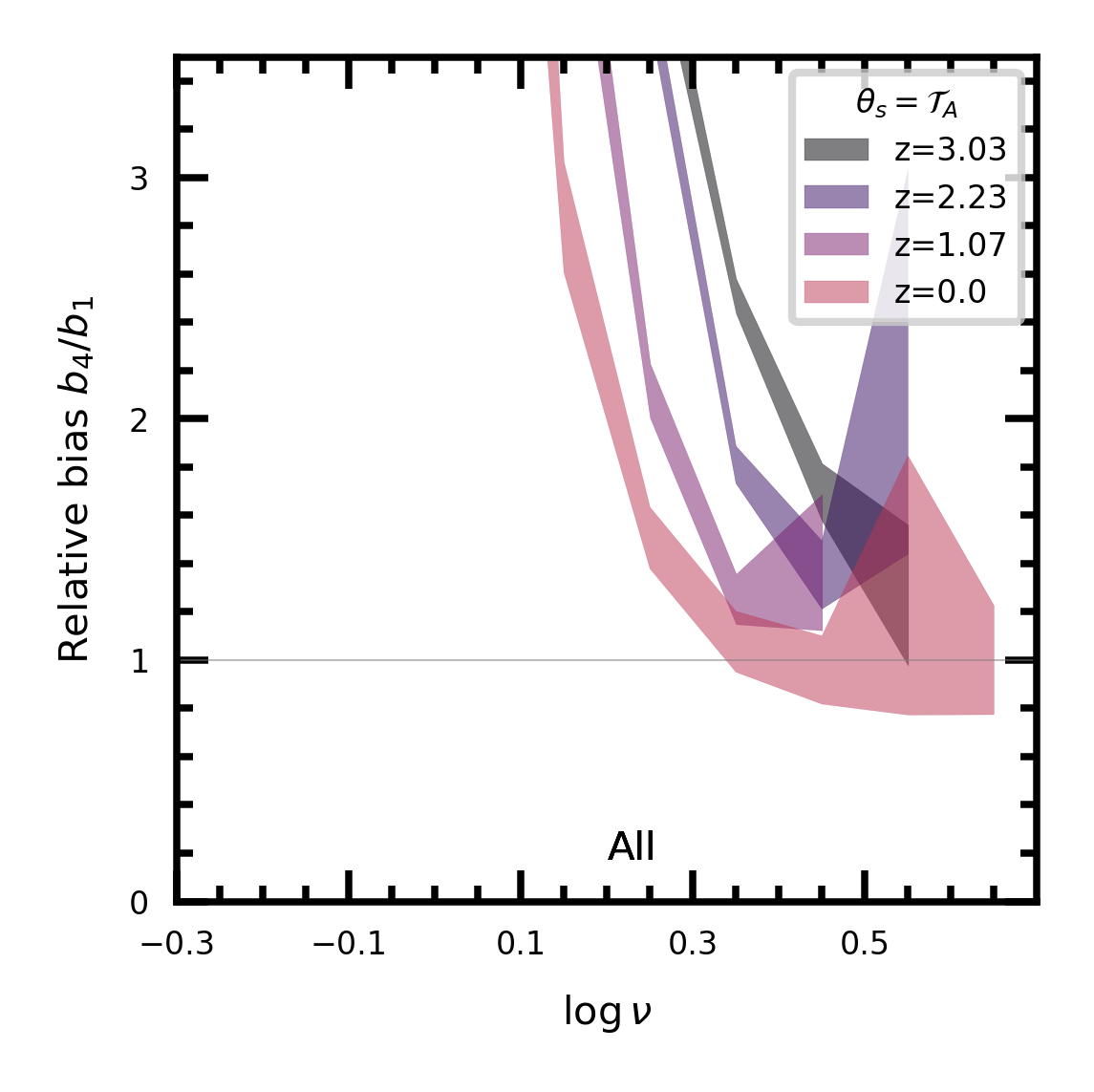}
\caption{\small{Ratio between halo bias as a function of the peak-height obtained in two different quartiles of halo ellipticity $\mathcal{E}_{h}$, relative Mach number $\mathcal{M}_{5}$, and tidal anisotropy $\mathcal{T}_{A}$, at different redshifts of the UNITSim.}}
\label{fig:sec_bias_rat2}
\end{figure}

%==========================================================
%==========================================================
\section{Rings of bias}\label{sec:scales_bias}
\begin{figure*}
\centering
\includegraphics[trim = 0.1cm 0.cm 0cm 0cm ,clip=true, width=1.03\textwidth]{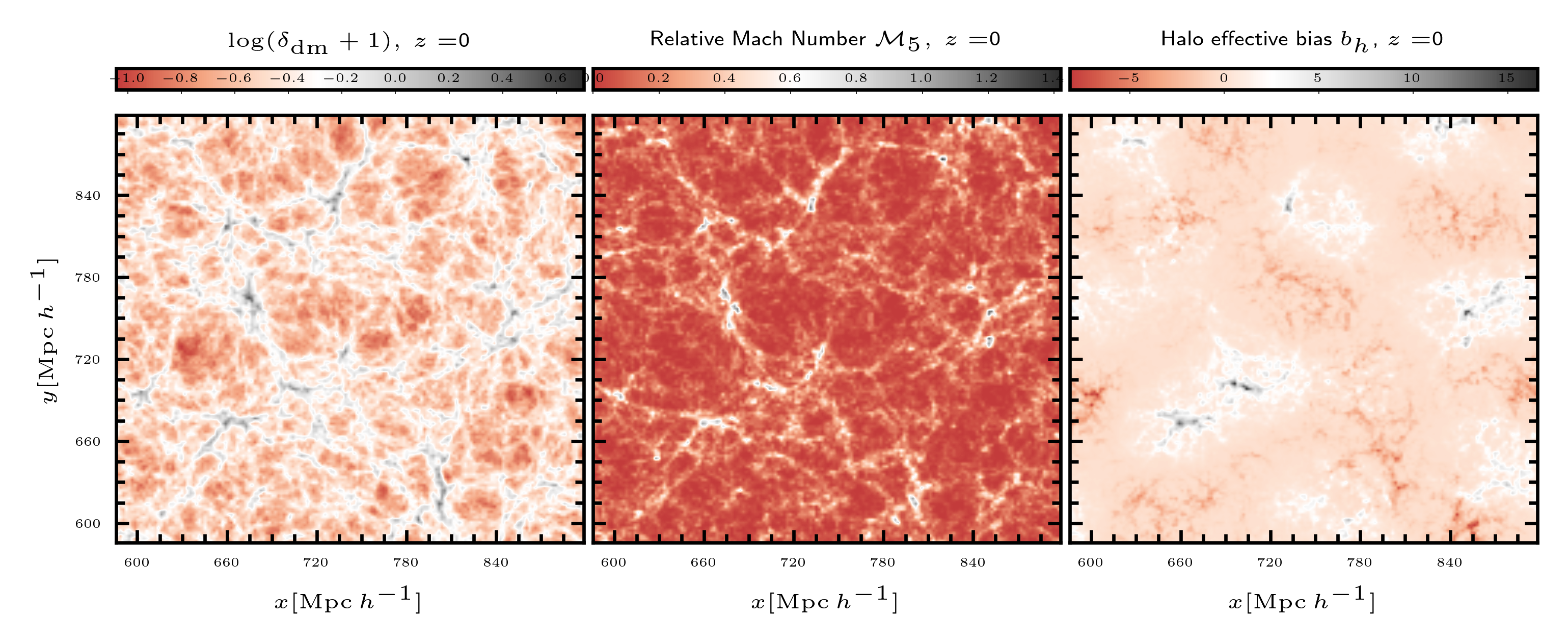}
\caption{\small{Slices of $50$ Mpc$h^{-1}$ thickness in an area of $\sim 300\,($Mpc$h^{-1})^{2}$ of the UNITSim volume, showing the dark matter density, the halo Mach number, and the effective bias (at $z=0$) interpolated on a $512^{3}$ mesh. It is evident how halo properties, in this case represented by the Mach number, follows the filamentary structure imposed by the dark matter density field, while the effective bias field filters out scales below $\sim 60$ Mpc $h^{-1}$.}}
\label{fig:field_bias}
\end{figure*}

In \S\ref{sec:individual_bias} we realized how the bias field of Fig.~\ref{fig:bias_field} depicts regularly placed large-scale structures. To see this closely, Fig.~\ref{fig:field_bias} shows a zoom-in into interpolations of the dark matter density field, the halo Mach number and the effective bias at $z=0$ in the UNITSim. This plot shows how the spatial distribution of a halo property (in this case, the Mach number) follows the filamentary structure depicted by the dark matter density field, while the effective halo bias acts as a filter, selecting regions of a typical size of $\sim 40-60$ Mpc $h^{-1}$ separated by a vast filamentary structure characterized by a bias $b_{h}\sim \langle b_{h}\rangle$ (notice though that inside these regions we can still distinguish the cosmic web). 
As pointed in \S\ref{sec:individual_bias}, this is in principle a consequence of the low-pass filter imposed to compute the halo bias in Eq.~(\ref{eq:bias_object}), which translates to a spatial resolution of $\sim 40$ Mpc$h^{-1}$, in rough agreement with the visual inspection of Fig.~\ref{fig:bias_field}.

One way to verify whether there is a particular scale set by thee effective halo bias is to use the ranked mark correlation function \citep[][]{2005astro.ph.11773S,2005MNRAS.364..796S, 2005astro.ph.11773S,2013MNRAS.429..458S} defined as
$M_{\eta}(r) \equiv (1+W_{\eta}(r))/(1+\xi(r))$,
where $\xi(r)$ denotes the usual two-point correlation function and $W_{\eta}(r)$ is the $\eta$-weighted correlation function. 
We use the ranks of effective bias as weights (or marks)\footnote{We use the simple estimator $WW/DD$ where $WW$ denotes the separation distribution of weighted pairs and $DD$ the unweighted counts.}. Similarly to Eq.(\ref{eq:bias_object}),  we can also assign to each tracer the relative bias $r_{h}$ (replacing in that expression the dark matter density field by the halo field) satisfying $\langle r_{h}\rangle=1$. The relative bias not only contains cosmological information, but also condenses information linked to the halo scaling relations, as shown by \citet[][]{2014A&A...563A.141B}.

To isolate the large-scale structures seen in  Figs.~\ref{fig:bias_field} and \ref{fig:field_bias}, we identify a set of $N_{\rm h}\sim 10^{5}$ halos with the highest and lowest values of effective bias in the sample. We dub these sets as $b_{i\,h}\gg \langle b_{h}\rangle$ or $b_{i\,h}\ll \langle b_{h}\rangle$ respectively. Due to the large scatter in the bias-mass relation (see Fig.~\ref{fig:bias_mass}), defining a subsample using a cut in effective (or relative) can induce  statistical incompleteness, especially on large separations \citep[see e.g.,][]{2011MNRAS.413..386B}. To account for this, we randomly redistribute the ranked marks within the tracers in the subsample and measure the marked correlation functions. We repeat this procedure $100$ times, generating a set from which we compute a mean and a variance, denoted $\langle M_{0}(r)\rangle$ and $\sigma_{r}(r_{j})$ respectively. We report the measurements of $\hat{M}(r)\equiv M(r)/\langle M_{0}(r)\rangle$, such that the value $\hat{M}(r)=1$ marks a scale in which the bias distribution is characterized by the mean bias of the sample. The uncertainty in this ratio is computed using $\sigma_{r}(r_{j})$ and the uncertainty in $M(r)$, which is obtained from a one-time deleted Jackknife technique \citep[see e.g.,][and references therein]{10.1214/aoms/1177706647,2009MNRAS.396...19N,2021MNRAS.505.5833F} using $64$ sub-volumes. These two are combined in the form of error propagation neglecting the covariance between these two components.\footnote{That is, the error in $r=x/y$ is given as $\sigma^{2}_{r}\approx r^{2}\lp \sigma^{2}_{x}/x^{2} + \sigma^{2}_{y}/y^{2}\rp$} 

\begin{figure*}[htb]
\centering
   \includegraphics[trim = 0cm 0.2cm 0.24cm 0cm ,clip=true, width=0.48\textwidth]{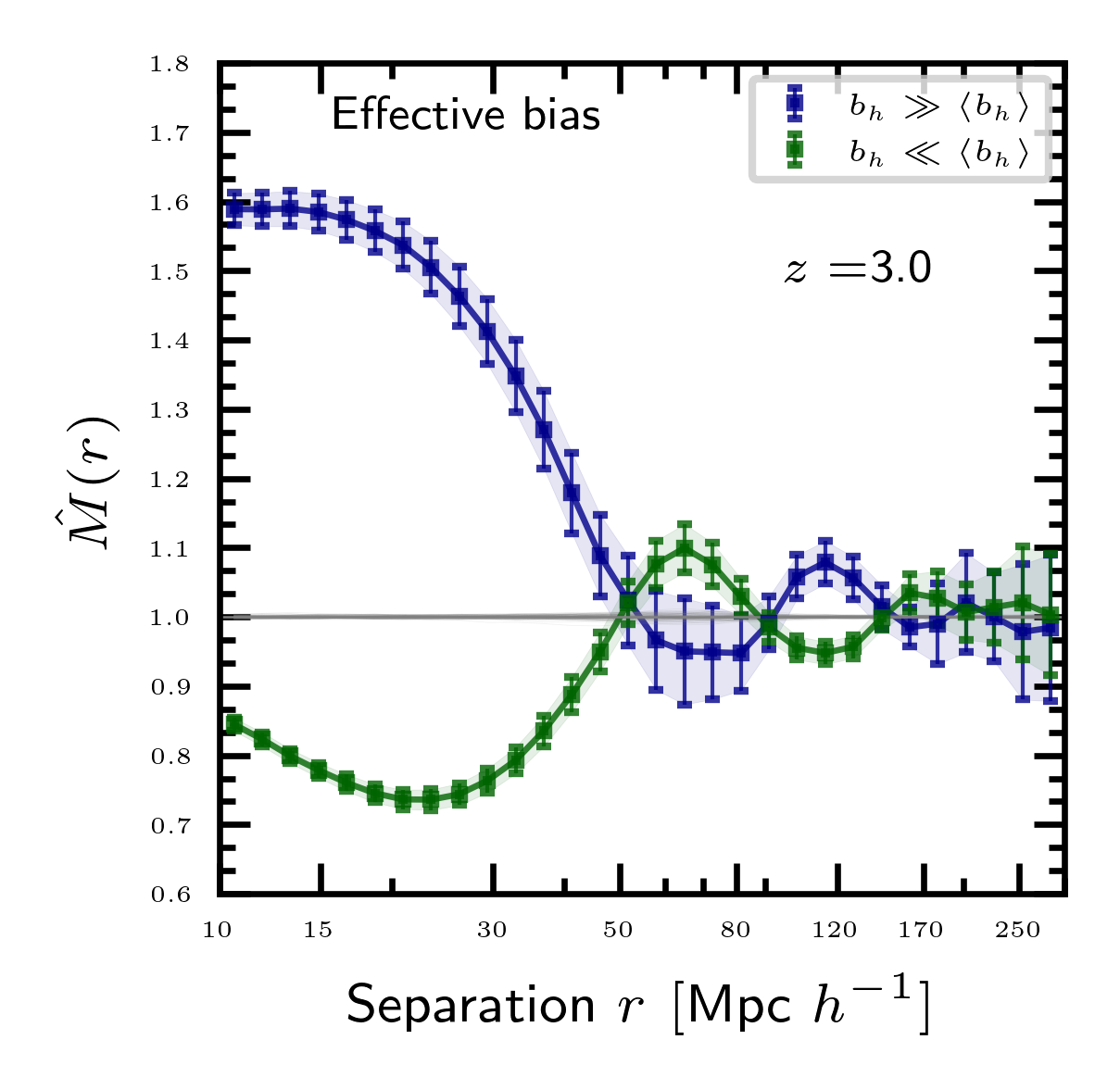}
    \includegraphics[trim = 0cm 0.2cm 0.24cm 0cm ,clip=true, width=0.48\textwidth]{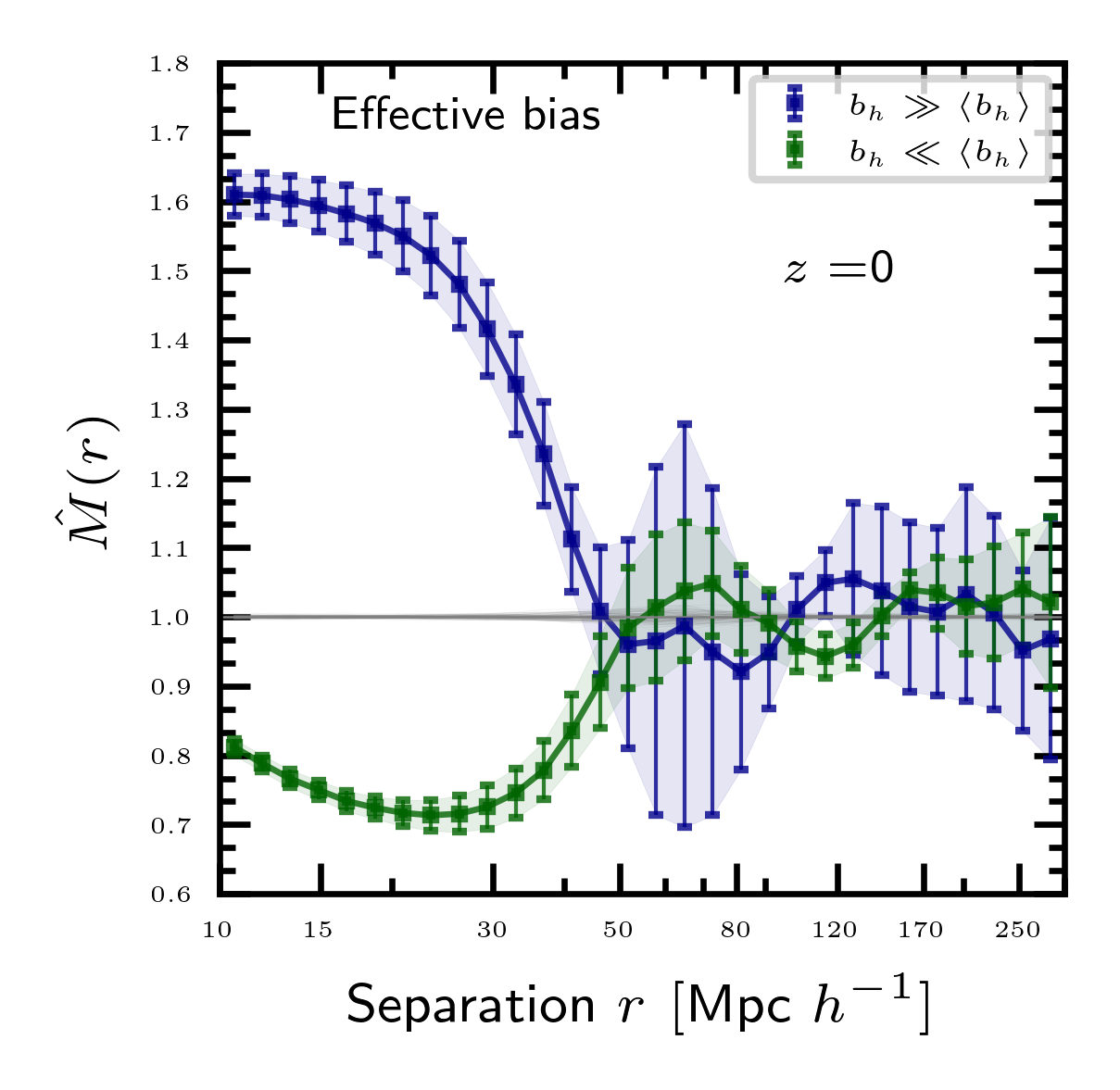}
\caption{\small{Ranked mark correlation function $\hat{M}(r)$ using the ranks of the effective bias as a mark for two different redshifts in the UNITSim. The dots with error bars show  both the result using tracers with effective bias much larger (or lower) than the mean bias of the sample. The gray lines denote the measurements of the ranked-mark correlation function after randomizing the marks among the tracers.}}
\label{fig:bias_scales}
\end{figure*}

In Fig.~\ref{fig:bias_scales} we show the measurements of the quantity $\hat{M}(r)$ probing separations in the range $10<r<250$ Mpc $h^{-1}$ at two different redshifts, using both the effective and relative halo bias. We present the measurements from the high-bias tracers $(b_{h}\gg \langle b_{h}\rangle)$ and the extremely low biased tracers  $(b_{h}\ll \langle b_{h}\rangle)$. All these follow a particular trend, namely, the bias of high (low) biased tracers depicts a positive (negative) deviation form the mean, reaching $\hat{M}(r)=1$ at $r\sim 50$ Mpc$h^{-1}$. An oscillatory pattern follows, with a period of nearly $40$ Mpc $h^{-1}$ and a decaying amplitude. This is more evident at high redshift, where actually low and high bias signals tend to be in phase, for both absolute and relative bias. In any case, we can interpret such oscillatory behavior as a reminiscent of Gibbs phenomena linked to the low pass filter used to define the effective (or relative) halo bias.

\end{appendix}
\end{document}